\documentclass[11pt,a4paper,francais]{book}
\usepackage{epsf,amsfonts,amsthm}
\usepackage{babel}
\usepackage{fontenc,indentfirst, delarray,amsfonts,amsmath,amssymb}
\usepackage{graphicx}
\usepackage{makeidx}
\DeclareGraphicsExtensions{ps,eps,ps.gz} \headheight 0.6cm
\makeatletter
\def\@citex[#1]#2{\if@filesw\immediate\write\@auxout
        {\string\citation{#2}}\fi
\def\@citea{}\@cite{\@for\@citeb:=#2\do
        {\@citea\def\@citea{,}\@ifundefined
        {b@\@citeb}{{\bf ?}\@warning
        {Citation `\@citeb' on page \thepage \space undefined}}
        {\csname b@\@citeb\endcsname}}}{#1}}
\newif\if@cghi
\def\cite{\@cghitrue\@ifnextchar [{\@tempswatrue
        \@citex}{\@tempswafalse\@citex[]}}
\def\citelow{\@cghifalse\@ifnextchar [{\@tempswatrue
        \@citex}{\@tempswafalse\@citex[]}}
\def\@cite#1#2{{\if@cghi\unskip$\null^{#1}$\else #1\fi\if@tempswa\typeout
        {warning: optional citation argument ignored: `#2'} \fi}}

\def\@biblabel#1{$\null^{#1}$}
\makeatother \makeindex
\begin{document}
\newtheorem{lem}{Lemme}[chapter]
\newtheorem{thm}[lem]{Th\'eor\`eme}
\newtheorem{prop}[lem]{Proposition}
\newtheorem{cor}[lem]{Corollaire}
\newtheorem{defi}[lem]{D\'efinition}
\newtheorem{cond}{Condition}

\newcommand{\norm}[1]{\left\lVert#1\right\rVert}
\newcommand{\abs}[1]{\lvert#1\rvert}
\newcommand{\scal}[1]{\langle#1,#1\rangle}
\newcommand{\scl}[2]{\langle#1,#2\rangle}
\newcommand{\suup}[1]{ \underset{#1}{\sup} }
\newcommand{\grad}[1]{\text{grad}\,#1}
\newcommand{\ket}[1]{\lvert#1\rangle}
\newcommand{\bra}[1]{\langle#1\lvert}
\newcommand{\tr}[1]{\text{Tr}(#1)}

\def\re{\text{Re}}
\def\im{\text{Im}}

\def\dm{\lp\begin{array}}
\def\fm{\end{array}\rp}
\def\dbb{\lb\begin{array}}
\def\fbb{\end{array}\rb}
\def\dbn{\left.\begin{array}}
\def\fbn{\end{array}\right.}

\def\ee{{\cal E}}
\def\lb{\left[}
\def\rb{\right]}
\def\lp{\left(}
\def\rp{\right)}
\def\alg{alg\`ebre }
\def\nc{non commutative}
\def\gnc{g\' eom\' etrie \nc}
\def\ms{mod\`ele standard}

\def\calg{$C^*$-\alg}
\def\dn{\partial_{\nu}}
\def\ds{\slash\!\!\!\partial}
\def\da{\left[ D,\pi(a) \right]}
\def\df{\left[ D,f \right]}

\def\m3{M_3 \lp \cc \rp}
\def\m2{M_2 \lp \cc \rp}
\def\mmp{M_p \lp \cc \rp}
\def\mn{M_n \lp \cc \rp}
\def\mnp{M_{np} \lp \cc \rp}
\def\mpn{M_{pn} \lp \cc \rp}
\def\mx{M_k}
\def\mz{M_{k'}}
\def\kk{{\mathbb{K}}}
\def\cc{{\mathbb{C}}}
\def\rr{{\mathbb{R}}}
\def\nn{{\mathbb{N}}}
\def\zz{{\mathbb{Z}}}
\def\ii{{\mathbb{I}}}

\def\aa{{\cal A}}
\def\bb{{\cal B}}
\def\ll{{\cal L}}
\def\bbb{{\aa_I}}
\def\dd{{\cal D}}
\def\hh{{\cal H}}
\def\jj{{\cal J}}
\def\oo{{\cal O}}
\def\ss{{\cal S}}
\def\mm{{M}}
\def\pp{{\cal P}}
\def\ccc{{\cal C}}
\def\hhh{{\mathbb H}}
\def\jj{{\cal J}}
\def\gg{{\cal G}}
\def\xx{{\cal X}}
\def\ginf{\Gamma^\infty}

\def\cinf{C^{\infty}\lp\mm\rp}
\def \difm{\text{Diff}({\mathcal{M}})}
\def\cm{C^{\infty}(\mm)}
\def\spin{\text{Spin}}
\def\del{\triangledown}
\def\L2{L_2(\mm)}
\def\LS{L_2(\mm , S)}

\def\cnp{\cc^{np}}
\def\cpn{\cc^{pn}}
\def\cn{\cc^{n}}
\def\cp{\cc^{p}}

\def\ot{\otimes}
\def\pof{\psi\otimes\xi}
\def\fop{\xi\otimes\psi}

\def\xox{{\xi}_x}
\def\yox{{\xi}_y}
\def\xoz{{\zeta}_x}
\def\yoz{{\zeta}_y}

\def\omx{\omega_x}
\def\omy{\omega_y}

\def\ox{\omega_{\xi}}
\def\oz{\omega_{\zeta}}
\def\xo0{\omega^0_x}
\def\yo0{\omega^0_y}
\def\tox{\tilde{\omega_{\xi}}}
\def\toz{\tilde{\omega_{\zeta}}}
\def\oe{\omega_{E}}
\def\oi{\omega_{I}}
\def\oei{\omega_{E}\ot\omega_{I}}
\def\oeip{\omega_{E}\ot\omega_{I}'}
\def\oepi{\omega_{E}'\ot\omega_I }
\def\oepip{ {\omega_E}' \ot {\omega_I}'}
\def\xoi{\omega_x\ot\omega_i}
\def\xoip{\omega_x\ot\omega_{i}'}
\def\yoi{\omega_y\ot\omega_i}
\def\yoip{\omega_y\ot\omega_{i}'}

\def\ou{{\omega_{1}}}
\def\od{{\omega_{2}}}
\def\odu{\omega_{21}}
\def\oud{{\omega_{12}}}
\def\opud{{{\omega'}_{12}}}
\def\opdu{{{\omega'}_{21}}}
\def\ou{{\omega_{1}}}
\def\ouu{{\omega_{11}}}
\def\odd{{\omega_{22}}}
\def\oc{{\omega_{c}}}
\def\oeu{\omega_E \ot \ou}
\def\oed{\omega_E \ot \od}
\def\oeo{\omega_E \ot \omega}

\def\oxe{\omega_{k_e}}
\def\oze{\omega_{{k'}_e}}
\def\oue{\omega_{1}\circ\alpha_e}
\def\ode{\omega_{2}\circ\alpha_e}
\def\o0{\omega_0}
\def\rx{\rho_\xi}
\def\rz{\rho_\zeta}
\def\sx{s_\xi}
\def\sz{s_\zeta}

\def\xox{x_{k}}
\def\yox{y_{k}}
\def\rk{\rho_k}
\def\rkp{\rho_{k'}}
\def\ux{{u_k}}
\def\uz{{u_{k'}}}
\def\xoz{x_{{k'}}}
\def\yoz{y_{{k'}}}
\def\xo0{x_\omega^0}
\def\yo0{y_\omega^0}
\def\tox{\tilde{\omega_{k}}}
\def\toz{\tilde{\omega_{{k'}}}}
\def\nx{{n_k}}
\def\nz{{n_{k'}}}

\def\ae{\alpha_e}
\def\aea{\alpha_e(\aa)}
\def\fu{\alpha_u}
\def\fue{\alpha_{u^*}}
\def\tu{\tau_1}
\def\td{\tau_2}
\def\futu{\fu(\tu)}
\def\futd{\fu(\td)}
\def\fuetu{\fue(\tu)}
\def\fuetd{\fue(\td)}
\def\tue{\tu\circ\ae}
\def\tde{\td\circ\ae}

\def\te{\tau_E}
\def\ti{\tau_I}
\def\tei{\tau_E\ot\tau_I}
\def\po{\pi_\omega}
\def\pto{\tilde{\pi}_\omega}

\def\auta{\text{Aut}(\aa)}
\def\ina{\text{In(\aa)}}
\def\outa{\text{Out}(\aa)}\def\auta{\text{Aut}(\aa)}
\def\inmn{\text{In}(\mn)}
\def\outmn{\text{Out}(\mn)}

\def\cl{\text{Cl}}
\def\ccl{\cc\text{l}}

\thispagestyle{empty}
\begin{center}

\huge{\bf THESE}\\

\vskip 2.5truecm

\large{pr\'esent\'ee par}\\

\vskip 0.5truecm

\Large{\bf Pierre Martinetti}\\

\vskip 1.5truecm

\large{pour obtenir le grade de docteur de
l'universit\'e de Provence, sp\'ecialit\'e}\\

\vskip 0.5truecm \large{physique des particules, physique
math\'ematique et mod\'elisation.}\\
\vskip 1.5truecm \Large{\bf DISTANCES EN G\'EOM\'ETRIE NON
COMMUTATIVE}\\

\vfill

\large{Soutenue le  1er octobre 2001, devant un jury compos\'e de}
\end{center}

\begin{center}
{\large
\begin{tabular}{l}
J. M. Gracia-Bondia\\
B. Iochum (directeur de th\`ese)\\
F. Lizzi (rapporteur)\\
J. Madore (rapporteur)\\
C. Rovelli (pr\'esident)\\
T. Sch\"ucker
\end{tabular}}
\end{center}
\newpage
\thispagestyle{empty} {\huge{\bf Remerciements}}
\newline

Je tiens \`a remercier toute l'\'equipe de g\'eom\'etrie non
commutative de Luminy, particuli\`erement mon directeur de
th\`ese Bruno Iochum, Thomas Sch\"ucker pour de nombreuses
discussions et conseils, Serge Lazzarini, Daniel Kastler et mon
compagnon de bureau Florian Girelli.
\newline

Je remercie \'egalement les membres du jury d'avoir accept\'e
d'\'evaluer ce travail, notamment les rapporteurs John Madore et
Fedele Lizzi, les personnes venant de loin, comme Jose Gracia
Bondia, et Carlo Rovelli pour avoir assum\'e la lourde t\^ache de
pr\'esident.
\newline

Ma gratitude va \'egalement \`a tous les membres du centre de
physique th\'eorique, en mentionnant Dolly pour sa patience dans
la restitution des livres emprunt\'es.
\newline

Enfin une mention sp\'eciale pour Thomas Krajewski et Denis
Perrot pour leur pr\'ecieuse aide sans laquelle ce travail aurait
abouti beaucoup plus difficilement...

\tableofcontents
{\renewcommand{\thechapter}{}\renewcommand{\chaptername}{}
\addtocounter{chapter}{-1}

\chapter{Introduction}\markboth{\sl INTRODUCTION}{\sl INTRODUCTION}}
La g\' eom\' etrie de notre espace pose probl\`eme en physique
car il n'en existe pas une description unique. Dans l'esprit de
la relativit\' e g\' en\' erale,  l'espace et le temps forment un
objet quadridimensionel dont la courbure est donn\' ee par la
distribution de masse. Quand un objet massif se d\'eplace, la
courbure change; la g\'eom\'etrie est un objet dynamique. Au
contraire la m\'ecanique quantique, et plus g\'en\'eralement la
th\'eorie quantique des champs, suppose la donn\'ee a priori d'un
espace dans lequel \'evoluent des champs. Pour reprendre une image
de [\citelow{rovelli}], la th\'eorie des champs prend l'espace
pour sc\`ene,  alors qu'en relativit\'e la sc\`ene elle-m\^eme
participe \`a l'action. La contradiction est d'autant plus
flagrante que chacune de ces th\'eories est valide et
v\'erifi\'ee avec pr\'ecision dans son domaine d'application: la
gravitation pour la relativit\'e; les interactions \'
electromagn\'etiques, faibles et fortes pour la th\' eorie
quantique des champs. Cette double approche de la g\'eom\'etrie
n'est pas forc\' ement scandaleuse. Rien n'interdit \`a deux
descriptions de cohabiter, tant que la cohabitation est
harmonieuse. Mais les ph\'enom\`enes  qui rel\`event \`a la fois
de la m\' ecanique quantique et de la gravitation, comme le tout
d\'ebut de l'univers dans la th\'eorie du big-bang, ou
l'effondrement gravitationel d'une \'etoile pass\'ee une certaine
\'echelle, brisent cette harmonie. L'hypoth\`ese r\'epandue au
jour d'aujourd'hui est, qu'\`a tout petite \'echelle, aucune des
descriptions g\'eom\'etriques classiques n'est valable. La
structure g\'eom\'etrique intime de l'espace-temps n'est pas
connue. Et la m\'ecanique quantique sugg\`ere que l'hypoth\`ese
du continu n'est pas justifi\'ee. On estime que cette structure
intime devrait \^etre visible \`a des \'echelles de l'ordre de
$10^{-33} cm$. C'est la longueur de Planck $l_p=\sqrt{\frac{{\cal
G}\hbar}{c^3}}$ obtenue par combinaison des constantes
fondamentales $\cal{G}$ (constante de Newton), $c$ (vitesse de la
lumi\`ere), $\hbar$ (constante de Planck). La g\'eom\'etrie non
commutative\cite{connes}, en \'etendant les concepts
g\'eom\'etriques usuels de mani\`ere compatible \`a la fois avec
la relativit\'e g\'en\'erale et avec la m\'ecanique quantique,
propose des outils math\'ematiques pour appr\'ehender la
g\'eom\'etrie \`a cette \'echelle.

Pour l'heure bien entendu, aucune th\'eorie ne d\'ecrit l'univers
\`a cet ordre de pr\'ecision. Parmi les candidats au titre de
th\'eorie de la gravitation quantique, aucun n'a jusqu'\`a
pr\'esent franchi avec succ\`es le cap de la v\' erification
exp\'erimentale. Une approche naturelle consiste \`a quantifier
le champ gravitationel comme les autres champs, mais la th\'eorie
obtenue est non renormalisable, c'est \`a dire sans int\' er\^et
physique.
 N\' eamoins cette optique, amener la relativit\' e \`a la th\'eorie des champs, reste
valable et a suscit\' e (et suscite)  des travaux consid\'erables
qui, dans les raffinements les plus r\'ecents,  aboutissent \`a
la th\' eorie des cordes et la supersym\'etrie. L'unification est
obtenue mais aux prix d'hypoth\`eses physiques fortes: l'espace
temps est \`a 11 dimensions et il existe deux fois plus de
particules que celles connues jusqu'\`a pr\' esent (\`a chaque
particule connue correspond un partenaire supersym\'etrique).
Pour l'instant, aucune de ces hypoth\`eses n'a \'et\'e
v\'erifi\'ee. Plut\^ot que de vouloir traiter la relativit\'e
comme une th\'eorie des champs, une autre approche consiste \`a
ne pas oublier l'\'el\' ement essentiel de la relativit\'e, \`a
savoir le caract\`ere dynamique de la g\'eom\'etrie. En clair, il
s'agit d'affranchir la th\'eorie quantique des champs d'un espace
donn\'ee a priori. On parle de th\'eorie des champs "background
independant", telle que la "loop quantum gravity"\cite{carlo}.
Malheureusement, pour l'instant, cette th\'eorie ne propose pas
de tests exp\'erimentaux.

Le probl\`eme r\'ecurrent est que la th\'eorie quantique des
champs n'est pas parfaitement comprise. Autant la relativit\'e
g\'en\'erale a une interpr\'etation g\'eom\'etrique simple,
autant ce que dit la m\'ecanique quantique de la g\'eom\'etrie
n\' ecessite des \' eclaircissements. Comment d\'efinir en effet
un point de l'espace en m\'ecanique quantique ? Ou plus
exactement comment donner une signification physique \`a la
notion de point ? Une mani\`ere simple consiste \`a appeler point
l'endroit occup\'e par une particule \`a un instant donn\' e. Mais
\`a supposer que l'on connaisse avec pr\'ecision un point, les
relations d'incertitude de Heisenberg indiquent que l'on ne peut
conna\^{\i}tre avec pr\'ecision la position de la particule \`a
un autre instant. Autrement dit, si une particule permet de
d\'efinir un point, elle ne permet pas d'en d\' efinir un autre.
Bien sur, on peut consid\'erer plusieurs particules au m\^eme
instant dont on connait les positions avec pr\'ecision, et on
d\'efinit ainsi plusieurs points. Mais pour savoir comment ces
points s'arrangent les uns par rapport aux autres, pour {\it
faire  la g\'eom\'etrie}, il faut pouvoir mesurer des distances.
Pour ce faire, il faut qu'un m\^eme objet, par exemple l'une des
particules, occupe \`a un instant donn\'e le point $a$, et \`a un
autre instant le point $b$. Connaissant sa vitesse, on mesure son
temps de vol et l'on en d\' eduit la distance. Mais plus on saura
avec pr\'ecision que la particule occupe le point $a$ \`a
l'instant $t$, moins on pourra \^etre sur qu'elle occupe le point
$b$ \`a l'instant suivant.  La solution sugg\'er\'ee par la
m\'ecanique quantique est de raisonner sur des valeurs moyennes.
Le point
 est d\'efini comme la valeur moyenne \`a un instant donn\'e de l'observable position appliqu\'ee sur l'\'etat repr\'esentant la
particule. En adoptant cette d\' efinition, on op\`ere un
changement de point de vue important: le point n'est plus
d\'efini en tant qu'objet abstrait de la g\'eom\'etrie (tel qu'on
l'apprend \`a l'\'ecole: "un point n'a pas d'\'epaisseur, une
ligne est un ensemble infini de points"), c'est un objet
alg\'ebrique, la valeur moyenne d'un op\'erateur sur un \'etat.

Or les math\'ematiciens savent traduire en langage alg\'ebrique
les propri\'et\' es g\'eom\'etriques d'un espace. Plus
pr\'ecis\'ement, les propri\'et\'es g\'eom\'etriques
(essentiellement la topologie, la mesure et la m\'etrique) d'un
espace ont une  traduction alg\'ebrique dans l'ensemble des
fonctions, \`a valeur complexe, d\'efinies sur cet espace. Par
exemple, la distance entre deux points $x,y$ est la longueur du
plus court chemin reliant $x$ \`a $y$. Mais c'est aussi le
supr\'emum, parmi toutes les fonctions dont la d\' eriv\'ee (en
valeur absolue) est toujours inf\' erieure \`a $1$,  du module de
la diff\'erence $f(x)-f(y)$. Ceci se v\'erifie sans difficult\'e
sur un exemple simple. Choisissons comme espace la droite
r\'eelle. La fonction $f$ d\' efinie sur $\rr$ par $f(x)=x$ a une
d\'eriv\' ee constante $f'(x)=1$, et on a bien
$\abs{f(x)-f(y)}=\abs{x-y}= \text{distance}(x,y).$ Si une
fonction $g$ est telle que $\abs{g(x)-g(y)}>\abs{x-y}$, alors par
le th\' eor\`eme de la valeur interm\' ediaire il existe n\'
ecessairement un r\'eel $c\in[x,y]$ tel que
$\abs{g'(c)}=\frac{\abs{g(x)-g(y)}}{\abs{x-y}}>1$. On voit ainsi
que les deux d\' efinitions de la distance, l'une comme plus
court chemin, l'autre comme supremum d'une diff\'erence
d'observables, co\"{\i}ncident.

Cet exemple \'el\'ementaire illustre comment faire de la g\'
eom\'etrie de mani\`ere alg\'ebrique. Ainsi qu'on le rappelle
longuement dans le premier chapitre, la g\'eom\'etrie au sens
usuel est commutative, c'est \`a dire que son expression alg\'
ebrique a pour cadre la th\'eorie des alg\`ebres commutatives.
L'id\'ee est que la g\'eom\'etrie ayant pour cadre les alg\`ebres
non commutatives (dont le principal artisan est le math\'
ematicien A. Connes\cite{connesfrance}) permet d'acc\'eder aux
espaces dans lesquels des ph\' enom\`enes physiques trouvent une
interpr\'etation g\' eom\' etrique qu'ils n'avaient pas jusque
l\`a. Par exemple le champ de Higgs (cf. chapitre 4)  apparait
comme le coefficient d'une m\'etrique dans une dimension
suppl\'ementaire, discr\`ete, qui rend compte des degr\'es de
libert\'e internes (spin ou isospin) d'une particule. Plus
g\'en\'eralement, l'espoir est que si cette interpr\' etation
g\'eom\'etrique est suffisamment riche, elle pourrait ouvrir la
voie \`a une unification, via la g\'eom\'etrie, de la
relativit\'e g\'en\'erale et de la th\'eorie quantique des champs.
\newline

Une alg\`ebre est un ensemble muni d'une loi d'addition et de
multiplication  par un scalaire, sur lequel est d\' efini en
outre une multiplication. Dans l'ensemble des fonctions \`a
valeur complexe sur un espace, ces lois sont d\'efinies point par
point. Pour la multiplication par exemple, si $f$ et $g$ sont
deux fonctions sur un espace $X$, alors
$$(f.g)(x)\doteq f(x).g(x)=g(x).f(x)=(g.f)(x).$$
Parce que le produit de deux nombres complexes est commutatif, le
produit de deux fonctions est commutatif, c'est \`a dire que
l'alg\`ebre des fonctions sur un espace est commutative.
Inversement,  \' etant donn\' ee une alg\`ebre commutative $\aa$,
on sait construire (construction de Gelfand-Naimark-Segal) un
espace $M$ tel que $\aa$ soit l'alg\`ebre des fonctions
(continues) sur $M$. Ainsi il est \' equivalent de se donner un
espace ou une alg\`ebre commutative: les propri\'et\' es g\'
eom\'etriques d'un espace ont une traduction dans l'alg\`ebre des
fonctions sur cet espace, et inversement les propri\'et\'es alg\'
ebriques d'une alg\`ebre commutative ont une traduction dans
l'espace associ\'e par la construction GNS:
$$
\text{espace } \Longleftrightarrow \text{alg\`ebre commutative.}
$$
La question naturelle est
$$
\text{?} \Longleftrightarrow \text{ alg\`ebre non commutative.}
$$

Naturellement, on ne saurait construire un espace tel qu'une
alg\`ebre non commutative soit son alg\`ebre de fonctions,
puisque l'alg\`ebre des fonctions sur un espace est n\'
ecessairement commutative. La g\'eom\'etrie non commutative est
une adaptation du dictionnaire qui permet de passer "d'alg\`ebre
commutative" \`a "espace" en remplacant, partout o\`u il y a
lieu, le mot commutatif par non commutatif. Evidemment les choses
ne sont pas si simples. Abandonner la commutativit\'e implique de
profonds changements dans les d\' efinitions du dictionnaire, et
requiert m\^eme la cr\' eation de notions nouvelles. Ce sont
d'ailleurs les plus int\' eressantes parce qu'elles traduisent
des effets qui n'ont pas d'\'equivalent dans le langage
commutatif.
\newline

La d\'efinition de cet  "espace non commutatif" est l'objet du
chapitre $1$. L'accent est mis sur l'aspect m\'etrique de ces
espaces, \`a travers la formule d\'efinissant la distance $d$
entre deux \'etats $\tau$, $\tau'$ (les d\'efinitions sont
discut\'ees longuement dans le premier chapitre) d'une alg\`ebre
$\aa$,
$$d(\tau, \tau') = \suup{a\in\aa} \left\{ \abs{\tau(a) - \tau'(a)}\, / \, \norm{[D, a]} \leq 1\right\}
$$
o\`u $D$ est {\it l'op\'erateur de Dirac} agissant sur un espace
de Hilbert support d'une repr\'esentation de l'alg\`ebre. Des
propri\'et\'es g\'en\'erales de cette formule sont mises en
\'evidence, ainsi que d'importantes simplifications quand $\aa$
est une alg\`ebre de von Neumann, en particulier l'invariance de
la distance par projection (proposition {\ref{projectionlem}).

Dans le deuxi\`eme chapitre, la formule de la distance est
\'etudi\'ee pour des alg\`ebres de dimension finie. On trouve que
le cas le plus simple, $\aa =\cc^n$ repr\'esent\'ee sur $\hh=
\cc^n$, n'est plus r\'esoluble explicitement d\`es $n=4$. Deux
cas particuliers de g\'eom\'etrie avec $\aa=\cc^n$ \-- $n$
quelconque \-- sont \'etudi\'es, ainsi que exemples avec des
alg\`ebres de matrices. Le cas $M_2(\cc)$ permet notamment de
munir la sph\`ere $S^2$ d'une m\'etrique.

Dans le troisi\`eme chapitre, on \'etudie la distance pour des
g\'eom\'etries obtenues par produit de l'espace-temps riemannien
avec une g\'eom\'etrie discr\`ete. Des conditions sont \'etablies
garantissant que l'espace discret soit orthogonal, au sens du
th\'eor\`eme de Pythagore, \`a l'espace continu. On obtient ainsi
une description compl\`ete de la m\'etrique pour un exemple de
base de la g\'eom\'etrie non commutative, le mod\`ele \`a deux
couches. On montre \'egalement en toute g\'en\'eralit\'e que la
m\'etrique d'une g\'eom\'etrie n'est pas perturb\'ee quand on
r\'ealise son produit avec une autre g\'eom\'etrie.

Le dernier chapitre \'etudie l'\'evolution de la m\'etrique
lorsque la g\'eom\'etrie est perturb\'ee par des champs de jauge.
En se limitant \`a la partie scalaire de ces champs, on calcule
les distances dans la g\'eom\'etrie du mod\`ele standard. Il
apparait alors que le champ de Higgs est le coefficient d'une
m\'etrique riemannienne dans un espace de dimension $4$
(continues) $+1$ (discr\`ete).

L'appendice contient des r\'esultats techniques interm\'ediaires,
isol\'es afin de ne pas alourdir le corps du texte.
\newline

On emploie la convention d'Einstein de sommation sur des indices
r\'ep\'et\'es, uniquement en position altern\'ee (haut-bas).
\vfill \pagebreak

\chapter{Espace non commutatif}

Avant de pr\'eciser ce qu'est un espace non commutatif, il n'est
pas inutile de rappeler en quoi un espace g\'eom\'etrique, au
sens usuel, est un espace commutatif. C'est une bonne mani\`ere
de pr\'esenter des r\' esultats fondamentaux comme le th¹eor\`eme
de Gelfand ou la construction GNS et d'introduire d\' efinitions
et propri\'et\'es dont nous ferons un usage intensif par la
suite. On trouvera les d\'emonstrations dans des trait\'es
d'alg\`ebres d'op\'erateurs tels que
[\citelow{takesaki,kadison,murphy}] ou dans [\citelow{jgb}] pour
un traitement plus orient\'e vers la g\'eom\'etrie non
commutative.

\section{Topologie de l'espace non commutatif}

\subsection{Espace commutatif}

Au sens le plus \'el\'ementaire, faire de la g\'eom\'etrie c'est
\^etre capable de d\'eterminer si deux  \'el\'ements sont voisins
l'un de l'autre. C'est en effet sous cette condition qu'un
ensemble prend le nom d'espace. Math\'ematiquement, il s'agit de
munir un ensemble $X$ d'une topologie, c'est \`a dire de
d\'efinir la notion de sous-ensemble ouvert (d'o\`u celle de
fonction continue). Quand  la topologie est suffisamment fine
pour distinguer les points, $X$ est dit s\' epar\'e (ou
Hausdorff).
 $X$ est compact signifie que de tout recouvrement infini d'ouverts $U_i$ \-- $\underset{i=1}{\overset{\infty}\bigcup} U_i = X$\-- on
peut extraire un recouvrement fini. On observe alors que
l'ensemble $C(X)$ \index{cx@$C(X)$} des fonctions \`a valeur
complexe continues sur $X$ est une alg\`ebre complexe commutative
qui, en tant qu'espace vectoriel, est compl\`ete pour la
m\'etrique induite par la norme
\begin{equation}
\label{normope} \norm{f}\doteq \suup{x\in X} \abs{f(x)}
\end{equation}
($C(X)$ est un espace de Banach).
 En tant qu'alg\`ebre $C(X)$ est munie d'une involution $^*$ naturelle h\'erit\'ee de la conjugaison complexe ainsi que d'une unit\'e (la fonction constante $1$). La
 norme v\'erifie
 $$\norm{fg}\leq\norm{f}\norm{g}$$
 ($C(X)$ une \alg de Banach) ainsi que
\begin{equation}
\label{c*} \norm{f}^2= \norm{f f^*}
\end{equation}
($C(X)$ est une $C^*$-alg\`ebre\index{cetoile@$C^*$}). A tout
espace topologique compact se trouve donc associ\'ee de mani\`ere
canonique une $C^*$-alg\`ebre complexe commutative avec unit\'e.
 Lorsque $X$ est seulement localement compact (tout point poss\`ede un voisinage compact), l'ensemble des fonctions continues est trop grand pour permettre de retrouver
 l'information topologique de l'espace. On lui pr\'ef\`ere la $C^*$-\alg  (sans unit\' e) $C_0(X)$\index{czerox@$C_0(X)$} des fonctions continues  s'annulant \`a l'infini.

R\'eciproquement, \`a toute \calg complexe $\aa$ commutative
correspond l'espace localement  compact (pour la topologie
*faible, cf. paragraphe I.\ref{voneuman}) $K(\aa)$
\index{ka@$K(\aa)$} des caract\`eres de $\aa$. Un caract\`ere est
un homomorphisme d'\alg (n\'ecessairement surjectif)  $$\mu: \aa
\rightarrow \cc.$$ Soulignons plusieurs propri\'et\'es
(\ref{caracun},\ref{caracspec},\ref{caracnorm},\ref{caracinvolution})
des caract\`eres, simples mais essentielles en ceci qu'elles
constituent le pivot de la g\'en\'eralisation au cas non
commutatif. Tout d'abord lorsque $\aa$ poss\`ede une unit\'e
$\ii$, $\mu(\ii)=\mu(\ii)^2$ d'o\`u
\begin{equation}
\label{caracun} \mu(\ii)= 1.
\end{equation}
Il s'en suit qu'un \'el\'ement inversible ne peut avoir pour
image z\'ero; donc $a - \mu(a)\ii$ n'est pas inversible.
Autrement dit, pour tout caract\`ere $\mu$ et tout $a$ de $\aa$,
\begin{equation}
\label{caracspec} \mu(a) \in \text{ sp} (a)
\end{equation}
ou $\text{ sp} (a)$, \index{spa@ $\text{sp} (a)$} le spectre de
$a$, est l'ensemble des valeurs $\lambda$ telles que
$a-\lambda\ii$ n'est pas inversible. Ensuite, sachant que pour
tout \'el\'ement $a$ d'une $C^*$-\alg complexe
\begin{equation}
\label{normspec} \suup{\lambda\, \in \text{ sp} (a)}
\abs{\lambda}\leq \norm{a},
\end{equation}
l'\'egalit\'e \'etant atteinte pour les \'el\'ements normaux
($a^*a = aa^*$), on observe que
\begin{equation}
\label{caracnorm} \norm{\mu}\doteq \suup{a\in\aa}
\frac{\abs{\mu(a)}}{\norm{a}}=1.
\end{equation}
Enfin, on montre qu'un caract\`ere \'evalu\'e sur un \'el\'ement
autoadjoint a valeur dans $\rr$. En d\'ecomposant tout $a$ en
\'el\'ements autoadjoints, $a = a_1 + i a_2$ avec $a_1 \doteq
\frac{1}{2}(a^* + a)$ et $a_2\doteq \frac{i}{2}(a^* - a)$, il
apparait qu'un caract\`ere pr\'eserve l'involution
\begin{equation}
\label{caracinvolution} \mu(a^*) = \mu(a_1 - i a_2) = \mu(a_1) -
i\mu(a_2) =  \bar{\mu}(a).
\end{equation}
Une forme lin\'eaire de ce type est dite {\it involutive}.

A l'aide de ces propri\'et\'es, on \'etablit (th\'eor\`eme de
Gelfand) que la transformation qui \`a tout $a\in\aa$ associe
l'application $\hat{a}\in C_0(K(\aa))$\index{achapeau@$\hat{a}$},
$$\hat{a}(\mu)\doteq \mu(a),$$
est un *isomorphisme isom\'etrique (i.e. pr\'eservant
l'involution et la norme) de $\aa$ dans  $C_0(K(\aa)).$ Lorsque
$\aa$ est munie d'une unit\'e, $K(\aa)$ est compact et $\aa$ est
*isomorphe \`a l'ensemble des fonctions continues sur $K(\aa)$.
Quand une \alg n'a pas d'unit\'e, on peut toujours lui en
adjoindre une en consid\'erant l'alg\`ebre augment\'ee. {\bf On
suppose donc dor\'enavant, sauf mention contraire, que les
alg\`ebres ont une unit\'e $\ii$.}\index{i@$\ii$} Avec cette
convention,  le th\'eor\`eme de Gelfand signifie que toute
$C^*$-alg\`ebre complexe commutative peut-\^etre vue comme
l'alg\`ebre des fonctions continues sur son espace des
caract\`eres.
\newline

Ainsi \`a toute $C^*$-\alg complexe commutative $\aa$ est
associ\'e un espace compact $K(\aa)$, tandis qu'\`a tout espace
compact $X$ est associ\'e une *\alg commutative $C(X)$.
 Le th\'eor\`eme de Gelfand assure que
$$\aa \longrightarrow K(\aa) \longrightarrow C(K(\aa))\sim \aa.$$
A l'inverse on montre que l'espace des caract\`eres de $C(X)$
n'est autre que $X$,
$$X \longrightarrow C(X)  \longrightarrow K(C(X))\sim X.$$
Dans un langage plus rigoureux\cite{jgb}, la cat\'egorie des
$C^*$-alg\`ebres commutatives complexes avec unit\'e est
\'equivalente \`a la cat\'egorie (oppos\'ee) des espaces
compacts. Sans entrer dans le d\'etail du langage des
cat\'egories, soulignons l'importante cons\'equence de cette
\'equivalence:

{\prop \label{homeoiso}
 Deux $C^*$-\alg complexes commutatives sont isomorphes si et seulement si leurs espaces de carac\-t\`eres sont hom\'eomorphes.}
\newline

De mani\`ere plus g\'en\'erale, toute l'information topologique
d'un espace compact est contenue dans $C(X)$. Soulignons que ceci
reste vraie pour l'alg\`ebre des fonctions lisses
$\cinf\index{cinf@$\cinf$}$ sur une vari\'et\'e compacte $\mm$:
bien que $\cinf$ ne soit pas une $C^*$-alg\`ebre mais seulement
une sous-alg\`ebre dense de $C(\mm)$, tout caract\`ere de $\cinf$
s'identifie \`a un point de $\mm$. Deux points de vue sont
possibles; classiquement on prend les points $x$ comme objet
premier et on interpr\`ete les r\'esultats de l'exp\'erience
comme des \'evaluations  d'observables sur ces points, ou bien on
consid\`ere les observables $f$ comme premi\`eres et les points
sont, par d\'efinition, les objets \'evaluant les observables.
Quand l'espace peut \^etre munie d'une topologie, c'est \`a dire
quand les observables (vues comme fonctions continues) commutent,
ces deux points de vue sont \'equivalents,
$$x(f) = f(x),$$
et les points sont les caract\`eres de l'\alg des observables.
Mais en m\'ecanique quantique la partie droite de l'\'equation,
l'\'evaluation d'une observable en un point, est mal d\'efinie.
En revanche l'ensemble des observables est bien d\'efini et c'est
une \alg non commutative. Pour donner sens \`a la partie gauche de
l'\'equation, il suffit de trouver l'objet \'equivalent au
caract\`ere pour une \alg non commutative.

 \subsection{Construction GNS}

Lorsque $\aa$ est une $C^*$-alg\`ebre complexe non commutative,
ses caract\`eres ne forment pas un ensemble localement compact.
Ils ne sont d'ailleurs pas int\'eressants en regard de la non
commutativit\'e puisqu'un caract\`ere, par nature, identifie $ab$
\`a $ba$ ($ab - ba$ a pour image zero). N\'eammoins, \`a la
lumi\`ere du th\'eor\`eme de Gelfand, les $C^*$-alg\`ebres non
commutatives sont le candidat id\'eal pour jouer le r\^ole d'\alg
des fonctions d'un "espace non commutatif". En tant qu'ensemble,
cet espace est compos\'e des formes lin\' eaires sur l'alg\`ebre
qui v\'erifient
 les propri\'et\'es des caract\`eres, except\'ees celles ayant trait \`a la commutativit\'e (\`a savoir la multiplicativit\'e:  $\mu(ab)=\mu(a)\mu(b)= \mu(b)\mu(a)=\mu(ba)).$

{\defi \label{etat} Un \' etat sur une $C^*$-alg\`ebre complexe
est une forme $\cc$-lin\' eaire positive de norme $1$.}
\newline

\noindent $\ss(\aa)$\index{sa@$\ss(\aa)$} d\'esigne l'ensemble
des \'etats de l'\alg $\aa$.

On rappelle qu'un \' el\' ement $a$ est {\it positif} s'il est
autoadjoint et $\text{sp(a)}\subset [0,+\infty[$ ou, de mani\`ere
\'equivalente, s'il existe un \' el\' ement $b$ tel que $a=b^*b$.
L'ensemble des \' el\'ements positifs est not\'e
$\aa_+$\index{aplus@$\aa_+$} et une forme lin\'eaire
$\tau$\index{tau@$\tau$} est positive si
$\tau(\aa^+)=\cc^+=\rr^+$. On montre [\citelow{kadison}, {\it Th.
4.3.2}] qu'une forme lin\'eaire $\tau$  sur une \alg de Banach
avec unit\'e est positive si, et seulement si,  elle est born\'ee
et $\norm{\tau}=\tau(\ii)$. Par cons\'equent un \'etat se
d\'efinit de mani\`ere \'equivalente comme une forme
$\cc$-lin\'eaire positive satisfaisant
\begin{equation}
\label{etatun} \tau(\ii) = 1,
\end{equation}
ou encore comme une forme $\cc$-lin\'eaire born\'ee telle que
\begin{equation}
\label{etatequiv} \norm{\tau} = \tau(\ii) = 1.
\end{equation}
La positivit\'e est une condition n\'ecessaire mais non
suffisante pour garantir l'involutivit\'e. Cependant quand
l'alg\`ebre a une unit\'e la positivit\'e implique
[\citelow{takesaki}, {\it Lem. 9.11}], et donc \'equivaut \`a,
\begin{equation}
\label{etatinvolution} \tau(a^*)= \bar{\tau}(a).
\end{equation}

La propri\'et\'e (\ref{caracspec}) n'est pas v\'erifi\'ee pour
les \'etats. A la place il apparait [\citelow{kadison}, {\it
Prop. 4.3.3}] que
 pour tout $\lambda \in \text{ sp}(a)$,  il existe un \'etat $\tau$ tel que $\tau(a) = \lambda$. Avec (\ref{normspec}), il vient pour tout
 \'el\'ement normal $a$
\begin{equation}
\label{normetat}
 \norm{a}= \suup{\tau\in\ss(\aa)}\abs{\tau(a)}.
\end{equation}
 Lorsque $\aa$
n'est pas commutative, la transformation de Gelfand (vue comme
une application de $\aa$ dans les fonctions sur $\ss(\aa)$) n'est
pas une isom\'etrie, sauf pour les \'el\'ements normaux dans la
mesure o\`u
$$
\norm{\hat{a}} \doteq \suup{\tau\in\ss(\aa)}\abs{\tau(a)} =
\norm{a}.
$$

L'espace des \'etats est convexe. Les points extr\'emaux, c'est
\`a dire les \'etats $\tau$ pour lesquels il n'existe pas
d'\'etats $\tau_1$, $\tau_2\neq \tau$ et de nombre $t\in [0,1]$
tels que  $\tau= t\tau_1 + (1-t)\tau_2$, sont appel\' es {\it
\'etats purs}.  Dans le cas commutatif, les caract\`eres
s'identifient aux \'etats purs. Par analogie ce sont les \'etats
purs de $\aa$, not\'e $\pp(\aa)$\index{pa@$\pp(\aa)$}, qui
tiennent lieu de "points" pour l'espace non commutatif. Il s'agit
d'une analogie, non d'une d\'efinition stricte. Pour certains
r\'esultats, on sera amen\'e en prendre en compte des \'etats non
purs.
\newline

 Les \'etats de $\aa$  constituent le socle de l'espace non commutatif en ceci aussi qu'ils garantissent, par la construction GNS (Gelfand-Naimark-Segal),
de pouvoir travailler concr\`etement avec $\aa$ vue comme sous
alg\`ebre de l'alg\`ebre des op\'erateurs born\' es sur l'espace
de Hilbert
$$\hh_\tau\doteq \overline{\aa / N_\tau},\index{htau@$\hh_\tau$}$$
o\`u la barre  sur $\aa / N_\tau$ signifie la compl\'etion au
sens de la norme d\'eduite du produit scalaire et
\begin{equation}
\label{ntau} N_\tau\index{ntau@$N_\tau$} \doteq \{ a\in\aa\,  /
\tau(a^*a)=0 \}
\end{equation}
est un id\'eal \`a gauche de $\aa$ appel\'e {\it noyau gauche} de
$\tau$. Les vecteurs de $\hh_\tau$ sont
 not\'es{\index{abar@$\underline{a}$} $\underline{a}\doteq a + N_\tau$, et le produit scalaire est
 $$\scl{\underline{a}}{\underline{b}} \doteq \tau({a^*b}).
 $$
L'*homomorphisme de $C^*$-\alg $\pi_\tau$
\index{pitau@$\pi_\tau$} associe \`a tout $a\in\aa$ l'op\'erateur
born\'e
$$\pi_\tau(a): \underline{b} \mapsto \underline {ab}.
$$
On note $\xi_\tau\doteq \underline{\ii}$
\index{xitau@$\xi_\tau$}de sorte que
\begin{equation}
\label{sclgns} \scl{\xi_\tau}{\pi_\tau(a)\xi_\tau}= \tau(a).
\end{equation}

Par d\'efinition, un *homorphisme d'une $C^*$-\alg $\aa$ dans
l'ensemble $\bb(\hh)$\index{bh@$\bb(\hh)$} des op\'erateurs
born\'es sur un espace de Hilbert $\hh$  est une {\it
repr\'esentation} de $\aa$ sur $\hh$. $\pi_\tau$ est appel\'ee la
repr\'esentation cyclique induite par $\tau$ car $\xi_\tau$ est
un vecteur
 cyclique ($\pi_\tau(\aa)\xi_\tau = \aa / N_\tau$). Une repr\'esentation $\pi$ sur $\hh$ est dite irr\'eductible si les seuls sous-espaces de $\hh$ invariants sous l'action
de $\pi(\aa)$ sont $0$ et $\hh$ lui-m\^eme. On montre que
$\pi_\tau$ est irr\'eductible si et seulement si $\tau$ est un
\'etat pur. Pour tout \'etat $\tau$, la repr\'esentation GNS ne
pr\'eserve a priori pas la norme. On a simplement
\begin{equation}
\label{normgns} \norm{\pi_\tau(a)}^2 \doteq
\suup{\underline{b}\in \hh_\tau}\frac{\norm{\pi_\tau(a)
\underline{b}}^2}{\norm{\underline{b}}^2}= \suup{b\in \aa}
\frac{\tau(b^*a^*ab)}{\tau(b^*b)}\leq\norm{a}^2,
\end{equation}
car la positivit\'e de $\tau$ implique $\tau(b^*a^*ab)\leq
\norm{a^*a}\tau(b^*b)$ [\citelow{jgb}, {\it p. 30}]. N\'eammoins
pour un $a$ donn\'e, il existe au moins une repr\'esentation GNS
isom\'etrique.
 En effet, d'apr\`es (\ref{normetat}) il existe un \'etat $\tau_a$ tel que $\abs{\tau_a(a^*a)}= \norm{a}^2$. Avec (\ref{sclgns}), on obtient $\norm{ \pi_{\tau_a}(a) \xi_{\tau_a}}= \norm{a}^2$. Comme
$\norm{\xi_{\tau_a}}=\tau_a(\ii)= 1$ par (\ref{etatun}),
l'in\'egalit\'e (\ref{normgns}) devient
$$\norm{\pi_{\tau_a}(a)}= \norm{a}.$$
En prenant la somme directe des $\pi_{\tau_a}$ lorsque $a$
parcourt $\aa$ on d\'efinit une repr\'esentation isom\'etrique
$\pi$ de $\aa$, d'o\`u le th\'eor\`eme de Gelfand-Naimark: {\thm
\label{gns} Toute $C^*$-\alg complexe a une repr\'esentation
isom\'etrique en tant que sous-alg\`ebre de l'\alg $\bb(\hh)$ des
op\'erateurs born\'es sur un espace de Hilbert.}
\newline

\noindent $\hh$ est appel\'e le {\it support} de la
repr\'esentation. Souvent on travaille avec l'\alg
repr\'esent\'ee plut\^ot qu'avec l'\alg d\'efinie abstraitement
et on omet le symbole $\pi$ quand il n'y a pas d'ambiguit\'e.

Terminons ce rapide survol par une remarque simple mais utile
d\`es la section suivante:
 \begin{equation}
 \label{kerpitau}
 \text{ ker } \pi_\tau \subset N_\tau \subset \text{ ker } \tau.
 \end{equation}
Pour montrer la seconde inclusion, il faut savoir que le noyau
gauche est d\'efini de mani\`ere \'equivalente \`a  (\ref{ntau})
par\cite{jgb}
\begin{equation}
\label{ntau2} \index{ntau@$N_\tau$}N_{\tau} = \left\{ a\in \aa\;
/ \; \tau(b^* a) = 0,\;\, \forall\, b\in\aa\right\}
\end{equation}
de sorte que $a\in N_\tau$ implique $\tau(\ii a) = 0$. La
premi\`ere inclusion est imm\'ediate car $\pi_\tau(a) = 0$
signifie que $\underline{a\ii} = \underline{a} = 0$. Une
repr\'esentation $\pi$ est dite {\it fid\`ele} lorsque  $\text{
ker } \pi = 0$. Un \'etat $\tau$ est fid\`ele quand $N_\tau = 0$.
La repr\'esentation GNS associ\'ee \`a un \'etat fid\`ele est
donc fid\`ele.

\subsection{Etat pur et projecteur}\label{voneuman}

A tout vecteur normalis\'e $\xi$ d'un espace de Hilbert $\hh$
est associ\'e de mani\`ere naturelle un projecteur orthogonal
$S\in\bb(\hh)$ ($S^* = S = S^2$) de rang $1$ d\'efini par
$$S\, \zeta= \scl{\xi}{\zeta}\, \xi$$  pour tout vecteur $\zeta$ de $\hh$. En particulier \`a tout \'etat $\tau$ correspond un projecteur (dor\'enavant on
sous entend orthogonal) $S_\tau\in\bb(\hh_\tau)$ de rang $1$ tel
que
$$S_\tau \pi_\tau(a) S_\tau \zeta = \scl{\xi_\tau}{\pi_\tau(a)\xi_\tau} \scl{\xi_\tau}{\zeta} \xi_\tau = \tau(a)S_\tau\zeta, $$
c'est \`a dire
\begin{equation}
\label{projecteur0} S_\tau \pi_\tau(a) S_\tau = \tau(a) S_\tau.
\end{equation}
Malheureusement comme $\pi_\tau$ n'est pas surjective, $S_\tau$
n'est pas forc\'ement l'image d'un \'el\'ement de $\aa$; or dans
les calculs de distance des sections suivantes, l'appartenance du
projecteur \`a l'\alg est un \'el\'ement simplificateur
extr\`emement utile. C'est pourquoi il est important de savoir
dans quels cas tout ou partie de l'espace des \'etats est en
correspondance avec des projecteurs de l'alg\`ebre. Cette
question trouve une r\'eponse dans la th\'eorie des alg\`ebres de
von Neumann. On commence par rappeler, sans preuve, les points
principaux de cette th\'eorie qui permettent d'\'ecrire
(\ref{projecteur0}) au niveau de l'alg\`ebre sans r\'ef\'erence
\`a une repr\'esentation. Ce sont des r\'esultats classiques
qu'on trouve en particulier, outre les ouvrages d\'ej\`a cit\'ees,
 dans [\citelow{sakai}]. Pour les notions de topologie plus g\'en\'erales on renvoie \`a [\citelow{reed}].
\newline

L'espace  vectoriel $\bb(X,\cc)$  des application lin\'eaires
born\'ees \`a valeur complexe  sur un espace de Banach $X$ est
lui m\^eme un espace de Banach pour la norme d'op\'erateur
(\ref{normope}). Cet espace not\'e $X^*$ est appel\'e espace dual
de $X$. Lorsque l'espace de Banach est une $C^*$-\alg $\aa$, ses
\'etats sont des formes lin\'eaires born\'ees donc
$$
\ss(\aa) \subset \aa^*\index{aetoilesup@$\aa^*$}.
$$
Lorsqu'un espace de Banach $X$ est le dual d'un autre espace de
Banach $Y$, alors $Y$ est appel\'e pr\'edual de $X$ et l'on note
$Y= X_*$. Quand $X$ est \'egalement un espace de Hilbert (i.e. sa
norme provient d'un produit interne), alors $X^*$ s'identifie \`a
$X$ par le th\'eor\`eme de Riesz de sorte qu'un espace de Hilbert
est son propre (pr\'e)dual. Mais de fa\c{c}on g\'en\'erale un
espace de Banach n'a pas forc\'ement de pr\'edual.

{\defi \label{w} Une $C^*$-\alg $\aa$ complexe est appel\'ee
$W^*$-\alg \index{wetoile@$W^*$}lorsque $\aa$ est le dual d'un
espace de Banach.}
\newline

\noindent Il apparait alors [\citelow{sakai}, {\it Cor. 1.13.3}]
qu'il n'y a qu'un seul pr\'edual
$\aa_*$\index{aetoileinf@$\aa_*$}, appel\'e {\it le} pr\'edual de
$\aa$.

$\aa$ est munie naturellement de la topologie uniforme d\'efinie
par la norme  (les ouverts sont les boules ouvertes).  De m\^eme
que la transformation de Gelfand est un *isomorphisme
isom\'etrique entre une $C^*$-alg\`ebre complexe commutative et
l'alg\`ebre des fonctions continues sur ses caract\`eres, en tant
qu'espace de Banach  $\aa$ est isom\'etriquement hom\'eomorphe
\`a un sous-espace de son double dual $\aa^{**}$ par la
correspondance
$$a\in\aa \; \longleftrightarrow \hat{a}\in\aa^{**}:\;  \hat{a}(f) = f(a) \text{ pour tout $f$ de $\aa^*$}. $$
Appliqu\'ee au pr\'edual, ceci implique que
 $$\aa_* \subset \aa^*.$$

On d\'efinit sur $\aa$ la topologie faible $\sigma(\aa, \aa^*)$
\index{sigmaaaetoile@$\sigma(\aa, \aa^*)$}comme la topologie la
plus faible pour laquelle tout \'el\'ement de $\aa^*$ est
continu. On rappelle qu'une topologie $T_1$ est plus faible
qu'une topologie $T_2$ lorsque tout ensemble ouvert au sens de
$T_1$ est \'egalement ouvert au sens de $T_2$. Comme $\aa^*$ est
par d\'efinition l'ensemble des formes lin\'eaires born\'ees sur
$\aa$, et que tout forme lin\'eaire born\'ee est continue pour la
topologie uniforme, la topologie faible est plus faible que la
topologie uniforme.
  En tant que dual de $\aa_*$, une $W^*$-\alg $\aa$ est \'egalement munie de la topologie *faible $\sigma(\aa, \aa_*)$\index{sigmaaaetoileb@$\sigma(\aa, \aa_*)$}
d\'efinie comme la plus faible topologie pour laquelle tout
\'el\'ement $x$ de $\aa_*$, vu comme forme lin\'eaire born\'ee
$\hat{x}$ sur $\aa$, est continu. A noter que c'est au sens de
cette topologie *faible que l'espace des caract\`eres de $\aa$,
vu comme sous ensemble du dual de $\aa$, est compact.

 Une propri\'et\'e importante des topologies $\sigma$ stipule que pour un espace de Banach $X$ et un ensemble $Y$ de formes
lin\'eaires sur $X$, alors l'ensemble des formes
$\sigma(X,Y)$-continues sur $X$ est pr\'ecis\'ement $Y$. En clair
$\aa_*$ est l'ensemble des formes lin\'eaires born\'ees
*faiblement continues sur $\aa$. De telles formes sont dites {\it
normales}  et on montre [\citelow{takesaki}, {\it Lem. III.3.6}]
qu'\`a toute forme normale positive non nulle  $\phi$ correspond
un unique projecteur non nul $s_\phi\in\aa$, appel\'e {\it
support} de $\phi$, tel que $\phi$ est fid\`ele sur $s_\phi\aa
s_\phi$ et $\phi = \phi (s_\phi .) =  \phi (. s_\phi) =
\phi(s_\phi . s_\phi)$. L'ensemble des \'etats normaux est
$$\ss(\aa)_* \doteq \ss(\aa) \cap \aa_*.\index{saetoile@$\ss(\aa)_*$}$$
Pour d\'eterminer si un \'etat est normal, il est utile de savoir
qu'une forme lin\'eaire born\'ee $\phi$  est normale si et
seulement si
$$\phi \lp \sum_{i\in I} e_i \rp = \sum_{i\in I} \phi(e_i)$$
pour toute famille $\{e_i\}_{i\in I}$ de projecteurs de $\aa$
orthogonaux deux \`a deux, $I\subset \nn$. Une telle famille est
lin\'eairement ind\'ependante de sorte que si $\aa$ est de
dimension finie, $I$ est de cardinalit\'e finie et la
propri\'et\'e est v\'erifi\'ee par lin\'earit\'e. En d'autres
termes tous les \'etats d'une $W^*$-\alg de dimension finie sont
normaux. D'apr\`es (\ref{kerpitau}) la repr\'esentation GNS
associ\'ee \`a un \'etat fid\`ele est fid\`ele si bien que le
support $s_\tau$\index{stau@$s_\tau$} d'un \'etat normal v\'erifie
\begin{eqnarray}
\label{support}
&\tau (s_\tau a s_\tau) = \tau(a s_\tau) = \tau( s_\tau a)  = \tau (a)\; \text{ pour tout } a\in\aa,&\\
\label{fidelite} & \pi_\tau \text{ est injective sur } s_\tau \aa
s_\tau.
\end{eqnarray}

Les projecteurs sont des \'el\'ements positifs et l'ensemble des
projecteurs d'une $W^*$-\alg est munie de la relation d'ordre
$q\leq p \Leftrightarrow p-q \in \aa_+$. Le support peut-\^etre
vu comme le compl\'ementaire orthogonal de la plus grande
projection $p$ annulant $\tau$, $s_\tau = \ii - p$, et on a
[\citelow{sakai}, {\it Def. 1.14.2}]
\begin{equation}
\label{supportnoyau} \index{ntau@$N_\tau$}N_\tau = \aa (\ii -
s_\tau).
\end{equation}
A noter que deux \'etats de support identique, bien qu'ayant le
m\^eme noyau gauche, ne sont pas n\'ecessairement \'egaux.

Gr\^ace \`a (\ref{supportnoyau}) il est clair que dans $\hh_\tau$
$$\underline{s_\tau}= \underline{\ii}= \xi_\tau.$$
On serait tent\'e d'identifier $\pi_\tau(s_\tau)$ au projecteur
$S_\tau$ de l'\'equation (\ref{projecteur0}). C'est vrai lorsque
l'\'etat est pur. Si $\tau$ n'est pas pur, $\pi_\tau$ n'est pas
irr\'eductible et tout projecteur de rang $1$ n'est pas
repr\'esentation d'un \'el\'ement de l'alg\`ebre. Pour \'etablir
la correspondance \'etat pur- projecteur de rang $1$, on a besoin
de quelques \'el\'ements de la th\'eorie des repr\'esentations
des $W^*$-alg\`ebres (th\'eorie des alg\`ebres de von Neumann).
\newline

Par d\'efinition une $W^*$-\alg est compl\`ete pour la norme,
donc ferm\'ee pour la topologie uniforme. En revanche elle n'est
pas n\'ecessairement ferm\'ee pour la topologie *faible.
Cependant toute $W^*$-\alg est isom\'etriquement *isomorphe \`a
une alg\`ebre *faiblement ferm\'ee. Avant de pr\'eciser ce
r\'esultat, soulignons que l'ensemble $\bb(\hh)$ des op\'erateurs
born\'es sur un espace de Hilbert est une $W^*$-alg\`ebre. De
m\^eme toute sous-alg\`ebre de $\bb(\hh)$ autoadjointe (i.e.
stable sous l'involution), munie d'une unit\'e  et ferm\'ee pour
la topologie *faible est une  $W^*$-alg\`ebre, appel\'ee {\it
alg\`ebre de von Neumann}.

Un  $W^*$-homomorphisme $\Phi$ entre deux $W^*$-alg\`ebres  est
par d\'efinition un  *homomorphisme continu pour les topologies
*faibles. Une $W^*$-repr\'esentation sur $\hh$  d'une $W^*$-\alg
$\aa$ est un  $W^*$-homomorphisme de $\aa$ dans $\bb(\hh)$. Comme
l'image de $\aa$ par un $W^*$-homomorphisme est *faiblement
ferm\'ee [\citelow{sakai}, {\it Prop. 1.16.2}], toute
$W^*$-repr\'esentation de $\aa$ sur $\hh$ est une alg\`ebre de
von Neumann. Si $\tau$ est un \'etat normal, la repr\'esentation
GNS $\pi_\tau$ est une $W^*$-repr\'esentation. La somme directe
des $\pi_\tau$ pour $\tau\in\ss(\aa)_*$ est une
$W^*$-repr\'esentation fid\`ele. Comme tout *isomorphisme de
$C^*$-alg\`ebres est isom\'etrique [\citelow{kadison}, {\it Thm.
4.1.8}], il apparait que toute $W^*$ alg\`ebre admet une
$W^*$-repr\'esentation isom\'etrique en tant qu'alg\`ebre de von
Neumann. Ce r\'esultat constitue une version "pour $W^*-$
alg\`ebres" du th\'eor\`eme de Gelfand-Naimark.

En notant $\index{paetoile@$\pp(\aa)_*$}\pp(\aa)_* \doteq
\pp(\aa) \cap \aa_*$ l'ensemble des \'etats purs normaux, on
montre alors que:

{\lem \label{projecteur1} Soient $\omega\in
\pp(\aa)_*$\index{omega@$\omega$} de support
$s_\omega$\index{somega@$s_\omega$}, et $S_\omega$ le projecteur
de rang $1$ sur $\xi_\omega$ \index{xiomega@$\xi_\omega$}. Alors
$\po(s_\omega)\index{piomega@$\pi_\omega$} = S_\omega$.}
\newline

\noindent{\it Preuve.} On rappelle que le commutant $B'$ d'une
sous-alg\`ebre $B$ de $\bb(\hh)$ est l'ensemble des \'el\'ements
de $\bb(\hh)$ commutant avec tous les \'el\'ements de $B$. Le
double commutant $B''$\index{bseconde@$B''$} est le commutant du
commutant. Puisque $\omega$ est pur $\pi_\omega$ est une
repr\'esentation irr\'eductible donc [\citelow{murphy}, {\it Thm.
4.1.12}]
$$\pi_\omega(\aa)''= \bb(\hh_\omega).$$
Puisque $\omega$ est normal, $\pi_\omega(\aa)$ est une alg\`ebre
de von Neumann et, th\'eor\`eme du bicommutant,
$$\po(\aa) = \pi_\omega(\aa)''.$$
Ainsi $\pi_\omega$ est surjective sur $\bb(\hh_\omega)$. Soit
$\{s_i\}$ l'ensemble des images inverses de $S_\omega$ par
$\pi_\omega$. Comme
$$\omega(s_i) = \scl{\xi_\omega}{s_\omega\xi_\omega}  = 1,$$
$\ii - s_i \in N_\omega$ donc, d'apr\`es la remarque
pr\'ec\'edent l'\'equation (\ref{supportnoyau}), $\ii- s_i \leq
\ii - s_\omega$ d'o\`u $s_\omega \leq s_i$ et
$\pi_\omega(s_\omega) \leq S_\omega$. Le sous-espace de $\hh$
invariant par $s_\omega$, inclus dans $\xi_\omega$ selon
[\citelow{kadison}, {\it Prop. 2.5.2}], ne peut-\^etre que
$\xi_\omega$ o\`u $0$. Comme $\pi(s_\omega)$ n'est pas nul
d'apr\`es (\ref{fidelite}), $\pi_\omega(s_\omega) = S_\omega$.
\hfill $\blacksquare$
\newline

\noindent Pour les calculs de distance \`a venir, on utilisera le
corollaire suivant qui prouve \'egalement que deux \'etats purs
normaux de m\^eme support sont identiques.

{\cor \label{projecteurc} Soit $s_\omega$ le support d'un \'etat
pur normal $\omega$ d'une $W^*$-\alg $\aa$, alors pour tout
$a\in\aa$
\begin{equation}
\label{projecteur} s_\omega a s_\omega = \omega(a) s_\omega.
\end{equation}}
\noindent{\it Preuve.} La preuve est imm\'ediate par application
du lemme \ref{projecteur1} sur (\ref{projecteur0}), en se
souvenant que $\po$ est injective sur l'alg\`ebre $s_\omega \aa
s_\omega$ dont $s_\omega a s_\omega - \omega(a) s_\omega$ est
\'el\'ement. \hfill $ \blacksquare$
\newline

L'\'equation (\ref{projecteur}) est bien l'\'equivalent de
(\ref{projecteur0}) au niveau de l'alg\`ebre, ind\'ependamment
d'un choix de repr\'esentation. Pour que ceci ait un sens il faut
bien entendu que $\pp(\aa)_*$ ne soit pas vide. Il apparait
qu'une $W^*$-alg\`ebre $\aa$ (sur un espace de Hilbert
s\'eparable) a des \'etats purs normaux d\`es lors que sa
d\'ecomposition int\'egrale relativement \`a son centre contient
un facteur de type I. Pour les exemples physiques des chapitres
suivants, il n'est pas n\'ecessaire d'aller si loin dans la
classification des alg\`ebres  de von Neumann. Le mod\`ele
standard met en jeu des alg\`ebres de dimension finie pour
lesquelles on sait que $\pp(\aa)_* = \pp(\aa)$.
\newline

En notation de Dirac, le corollaire \ref{projecteurc} ne dit rien
d'autre que
\begin{equation}
\label{diraceqtion}
 \ket{\xi}\bra{\xi} a \ket{\xi}\bra{\xi} = \bra{\xi} a \ket{\xi}\ket{\xi}\bra{\xi}.
 \end{equation}
Pour passer de (\ref{projecteur}) \`a (\ref{diraceqtion}), il
faut associer un vecteur $\xi$ au support $s_\omega$, c'est \`a
dire repr\'esenter le support comme projecteur sur un espace de
Hilbert. Le choix naturel est la repr\'esentation GNS
$\pi_\omega$ qui donne $\xi= \xi_\omega$.
 Lorsque plusieurs \'etats sont en jeu, chacune des repr\'esentations GNS est l\'egitime et il est important de conna\^{\i}tre l'image, par la repr\'esentation GNS
li\'ee \`a un \'etat, du support d'un autre \'etat.

{\lem \label{rangun} Soient $s_1$, $s_2$ les supports de $\ou$,
$\od \in \pp(\aa)_*$ et $\{\pi_1, \hh_1\}$, $\{\pi_2, \hh_2\}$ les
repr\'esentations GNS associ\'ees. Alors soit $\pi_1(s_2) =
\pi_2(s_1) = 0$, soit $\pi_1(s_2)$ et $\pi_2(s_1)$ sont des
projecteurs de rang  $1$.}
\newline

\noindent{\it Preuve.}  Puisque $\pi_2$ est un *homomorphisme,
$\pi_2(s_1)$ est un projecteur. S'il est non nul, il existe au
moins un vecteur norm\'e $\xi\in\hh_{2}$ tel que
\begin{equation}
\label{defxi} \pi_2(s_1)\xi = \xi.
\end{equation}
Comme $\pi_2$ est irr\'eductible, $\xi$ est cyclique
[\citelow{murphy}, {\it Th. 5.1.5}] et d\'efinit un \'etat
\begin{equation}
\label{etatxi} \omega_\xi (a) \doteq \scl{\xi}{\pi_2(a)\xi}.
\end{equation}
Cet \'etat est pur {\it [ibid. Thm. 5.1.7]} et $\pi_2$ est
unitairement \'equivalente \`a la repr\'esentation GNS $\pi_\xi$
associ\'ee \`a $\ox$. Par  (\ref{etatxi}) et (\ref{defxi})
$\ox(s_1) = \scl{\xi}{\xi} = 1$, si bien qu'en utilisant
(\ref{projecteur})
$$\ox(s_1 a s_1) =\ou(a) \ox( s_1) = \ou(a).$$
D'autre part comme $s_1^*=s_1$,
$$\ox(s_1 a s_1)= \scl{\pi_2(s_1)\xi}{\pi_2(a)\pi_2(s_1)\xi} =  \scl{\xi}{\pi_2(a)\xi} = \omega_\xi(a).$$
Autrement dit $\omega_\xi = \omega_1$, donc $\pi_1 = \pi_\xi$ qui
est unitairement \'equivalente \`a  $\pi_2$. Le rang d'un
op\'erateur est invariant par transformation unitaire et
$\pi_1(s_1)$ est de rang $1$, donc $\pi_2(s_1)$ est de rang $1$.
Ainsi $\pi_2(s_1)$ est soit nul soit de rang $1$.

Si $\pi_2(s_1)$ est non nul, alors $\pi_2$ est \'equivalente \`a
$\pi_1$ donc $\pi_1(s_2)$ est de rang $1$. En permutant les
indices on montre de la m\^eme mani\`ere que si $\pi_1(s_2)$ est
non nul, alors  $\pi_2(s_1)$ est de rang $1$. Autrement dit si
$\pi_2(s_1) = 0$ implique $\pi_1(s_2) = 0$. On montre
similairement l'\'equivalence dans l'autre sens, d'o\`u le
r\'esultat.
  \hfill $\blacksquare$
\newline

\noindent

Au m\^eme titre que  (\ref{diraceqtion}), la relation pour deux
vecteurs norm\'es distincts $\xi$ et $\zeta$
\begin{equation}
\label{dirac2}
 \ket{\xi}\bra{\zeta} a \ket{\xi}\bra{\zeta} = \bra{\zeta} a \ket{\xi}\ket{\xi}\bra{\zeta}\,
\end{equation}
admet une \'ecriture au niveau de l'alg\`ebre. Pour la
d\'eterminer, il convient associer un \'el\'ement de l'\alg \`a
tout couple d'\'etats purs normaux $\ou, \od$. Notons d'abord les
\'equivalences simples suivantes:

{\lem \label{equiv}  $ \ou(s_2)= 0  \Longleftrightarrow s_2 s_1
=0 \Longleftrightarrow s_1 s_2 = 0 \Longleftrightarrow \od(s_1) =
0$.}
\newline

\noindent{\it Preuve.} L'\'equivalence centrale est obtenu
gr\^ace \`a l'involution. Que $s_i s_j = 0$ entraine
$\omega_j(s_i) = 0$ vient de (\ref{projecteur}) (en remarquant
bien entendu que $s_j$ n'est pas nul sinon $\omega_j=0$). De
m\^eme $\omega_j(s_i)= 0$ signifie que $s_j s_i s_j = 0$, soit
encore $(s_i s_j)^* s_is_j =0$ d'o\`u, en prenant la norme,
$s_is_j = 0$.\hfill $\blacksquare$
\newline

\noindent  Ceci permet d'\'ecrire (\ref{dirac2}) dans
l'alg\`ebre, malheureusement de mani\`ere non univoque.

{\lem \label{projecteurcroise} Soient $\ou, \od\in\pp(\aa)_*$,
$s_1, s_2$ leurs supports et $\pi_1$ la repr\'esentation GNS
associ\' ee \`a $\ou$. Alors  il existe une forme lin\'eaire
$\omega_{12}$ sur $\aa$ et un \'el\'ement $s_{12}$ de l'alg\`ebre
tels que  pour tout $a\in\aa$
\begin{eqnarray*}
 s_1a s_2 &=& \omega_{12}(a) s_{12} + k_a \;\text{ si  $\pi_1(s_2) \neq 0$},\\
          &=& k_a\hspace{2.1truecm} \text{ si  $\pi_1(s_2) = 0$},
\end{eqnarray*}
o\`u $k_a$ est dans le noyau de $\pi_1$.}
\newline

\noindent{\it Preuve.} Si $\pi_1(s_2) = 0$, alors $s_1 a s_2
\in\text{ker }\pi_1.$ Si $\pi_1(s_2)$ est non nul, c'est un
projecteur de rang $1$ selon le lemme \ref{rangun} et on note
$\xi_2\in\hh_1$ son vecteur propre norm\'e (d\'efini \`a une
phase pr\`es).  On pose $\xi_1 = \underline{\ii}$. Alors $$\pi_1
(s_1 a s_2) = \omega_{12}(a) S_{12}$$ o\`u $\omega_{12}(a) =
\scl{\xi_1}{\pi_1(a)\xi_2}$ et $S_{12}\in \bb(\hh_1)$ est
l'op\'erateur tel que $$S_{12}\zeta = \scl{\xi_2}{\zeta}\xi_1$$
pour tout $\zeta\in\hh_1$. Comme $\pi_1$ est irr\'eductible et
$\pi_1(\aa)$ est une \alg de von Neumann, $\pi_1(\aa) =
\bb(\hh_1)$ si bien qu'il existe au moins un \'el\'ement
$s_{12}\in\pi_1^{-1}(S_{12}).$ Ainsi $s_1as_2 - \oud(a)s_{12} \in
\ker\,\pi_1$. \hfill $ \blacksquare$
\newline

\noindent $s_{12}$ est d\'efini ind\'ependamment de $a$ mais pas
$k_a$. Ce lemme est moins pr\'ecis que le corollaire
\ref{projecteurc} mais l'ambiguit\'e "modulo un \'el\'ement du
noyau"  est lev\'ee si $\pi_1$ est fid\`ele, ce qui est le cas
quand $\ou$ est fid\`ele. Requ\'erir d'un \'etat qu'il soit pur,
normal et fid\`ele r\'eduit dangereusement le champ
d'investigation. L\`a encore la difficult\'e n'apparait pas dans
les exemples physiques puisqu'ils font intervenir des alg\`ebres
de matrices pour lesquelles toute repr\'esentation irr\'eductible
est fid\`ele. A noter que lorsque $s_1s_2\neq 0$,
$$s_{12} = \frac{s_1s_2}{\scl{\xi_1}{\xi_2}}.$$


\subsection{ $C^*$-\alg r\'eelle}\label{algebrereelle}

Le plus souvent les $C^*$-alg\`ebres sont prises sur le corps des
complexes et tous les r\'esultats ci-dessus sont vrais sous cette
hypoth\`ese. Si $\cc$ peut-\^etre vue comme une \alg complexe ou
r\'eelle, l'alg\`ebre des quaternions $\hhh$ qui apparait dans la
description non commutative du mod\`ele standard est une \alg sur
$\rr$ mais pas sur $\cc$. Il est donc important de connaitre les
propri\'et\'es du cas complexe qui  restent vraies dans le cas
r\'eel. La plupart des d\'efinitions ont une traduction simple.
Ainsi une $C^*$-\alg r\'eelle $\aa$ est une  $*$-\alg norm\'ee
compl\`ete satisfaisant
$$\norm{ab}\leq\norm{a}\norm{b},\;\,  \norm{a^*a}=\norm{a}^2$$
ainsi que\cite{goodearl}
$$1 + a^* a \text{ est inversible pour tout $a$. }$$
Cette condition suppl\'ementaire, qui dans le cas complexe est
une cons\'equence de la d\'efinition, doit ici \^etre impos\'ee
\`a la main. Elle est importante puisqu'elle interdit par exemple
de voir $\cc$ muni de l'involution identit\'e ($a^*=a$ pour tout
$a$) comme une $C^*$-alg\`ebre  r\'eelle (puisqu'alors $1 + i^*i
=0$).

A priori un \'etat d'une \alg r\'eelle devrait \^etre une forme
positive $\rr$-lin\'eaire \`a valeur dans $\rr$ de norme $1$. Ce
n'est toutefois pas la d\'efinition consacr\'ee
[\citelow{goodearl}, {\it p. 44}].

{\defi \label{etatreel} Un \'etat r\'eel $\tau$ sur une
$C^*$-\alg r\'eelle est une forme positive, $\rr$-lin\'eaire, \`a
valeur dans $\rr$ telle que
$$ \tau(\ii) = 1\, \text{ et }\, \tau(a^*)= \tau(a).$$}
\noindent Cette d\'efinition est analogue \`a (\ref{etatun}), si
ce n'est qu'elle inclut l'involutivit\'e (\ref{etatinvolution})
alors que c'\'etait une cons\'equence  de la positivit\'e dans le
cas complexe. Par exemple l'application $x + iy \longmapsto x + y$
 est une forme $\rr$-lin\'eaire sur $\cc$, positive, envoyant $1$ sur $1$ mais elle n'est pas involutive. A noter [{\it ibid, p. 51}] qu'une forme
$\rr$-lin\'eaire $\tau$ telle que $\tau(\ii)=1$ est un
 \'etat si et seulement si elle est born\'ee et de norme $1$. Un \'etat r\'eel se d\'efinit donc aussi comme une forme
 $\rr$-lin\'eaire $\tau$ born\'ee telle que
\begin{equation}
\label{etatrequiv} \norm{\tau} = \tau(\ii) = 1.
\end{equation}
 La propri\'et\'e (\ref{normetat}) reste vraie dans le cas r\'eel, au moins pour les \'el\'ements positifs.
Concernant les supports des \'etats normaux, il n'est pas
n\'ecessaire pour ce qui nous concerne  de savoir
 si de pareils objets existent dans le cas r\'eel. On se contentera le cas \'ech\'eant d'exhiber des projecteurs \'el\'ements
  de l'alg\`ebre satisfaisant (\ref{projecteur}).

 En fait la d\'efinition (\ref{etatrequiv}) est un d\'ecalque de (\ref{etatequiv}). Plut\^ot que de voir un \'etat comme
 une forme positive de norme $1$ (ce qui n'est pas, rappelons le, la d\'efinition d'un \'etat r\'eel),  il est plus judicieux de
 prendre (\ref{etatequiv}) pour d\'efinir un \'etat sur une $C^*$-alg\`ebre, avec unit\'e, de corps de r\'ef\'erence quel\-conque: un \'etat sur une \alg
quaternionique est une forme
 $\hhh$-lin\'eaire $\tau$ de norme $\tau(\ii)=1$. Cette d\'efinition est encore
 imparfaite: $\hhh$ n'\'etant pas commutative, $\tau$ peut-\^etre lin\'eaire
 \`a gauche o\`u \`a droite. Nous reviendrons sur ce point dans le chapitre suivant.

La notion d'\'etat r\'eel reste valide pour une $C^*$-alg\`ebre
complexe $\aa$. Tout \'etat r\'eel $\tau_\rr$ de $\aa$ d\'efinit
une forme $\cc$-lin\'eaire positive
$$\tau(a) = \Xi (\tau_\rr) (a) \doteq\tau_\rr(a_1) + i\tau_\rr(a_2),$$
o\`u $a=a_1 + ia_2$ est la d\'ecomposition en \'el\'ements
autoadjoints, cf (\ref{caracinvolution}). Puisque $\tau(\ii)=1$,
$\tau$ est un \'etat au sens de la d\'efinition \ref{etat}. Si
$\Xi(\tau_\rr) = \Xi(\tau'_\rr)$, alors $\tau_\rr$ et $\tau'_\rr$
co\"{\i}ncident sur tout $a_1$ autoadjoint (et $ia_2$
antiautoadjoint) donc sur tout $\aa$ et $\Xi$ est une injection.
A l'inverse tout \'etat $\tau$ d\'efinit un \'etat r\'eel
$\re(\tau)$. Si $\re(\tau)= \re(\tau')$, $\tau$ et $\tau'$
co\"{\i}ncident sur tout $a_1$ autoadjoint et, par lin\'earit\'e,
sur tout $ia_2$. Donc $\tau=\tau'$ et $\Xi$ est une bijection. Si
$\tau$ est pur il en est de m\^eme pour $\re(\tau)$ -- $\re(\tau)
= t\re(\tau') + (1-t)\re(\tau'')$ implique $\tau(a_1) = t
\tau'(a_1) + (1-t)\tau''(a_1)$, idem pour $ia_2$ et $\tau$ n'est
pas pur-- et si $\tau$ n'est pas pur alors $\re(\tau)$ n'est pas
pur. Autrement l'ensemble des \'etats r\'eels purs d'une
alg\`ebre complexe est en bijection avec l'ensemble de ses
\'etats purs,
$$\pp_\rr(\aa) \simeq \pp(\aa).\index{par@$\pp_\rr(\aa)$}$$
En particulier $\pp_\rr(\cinf) = \{\re(\omega_x)\}$ o\`u
$\omega_x\in\pp(\cinf)\index{omegax@$\omega_x$}$ d\'esigne
l'evaluation au point $x$. Dans le mod\`ele standard (cf chapitre
4), $\cinf$ est vue comme alg\`ebre r\'eelle
$C^\infty_\rr(M)\index{cinfr@$C^\infty_\rr(M)$}$. L'involution,
l'identit\'e et les \'el\'ements positifs sont identiques au cas
complexe, en cons\'equence
\begin{equation}
\label{etatreelpur} \pp(C^\infty_\rr(M)) = \pp_\rr(C^\infty(M)).
\end{equation}
\section{G\'eom\'etrie spinorielle}

Outre la topologie, l'espace physique est muni d'une structure
diff\'erenti\-elle et d'un espace de degr\'es de libert\'e
internes (le spin en m\'ecanique quantique, l'isospin pour le
mod\`ele standard). De m\^eme que les propri\'et\'es topologiques
traduites en terme d'alg\`ebre commutative, l'objet
math\'ematique utilis\'e pour d\'ecrire l'espace de la th\'eorie
quantique des champs, la {\it vari\'et\'e \`a spin}, a une
d\'efinition alg\'ebrique. Privil\'egier cette derni\`ere rend
possible son adaptation aux espaces non commutatifs. On
pr\'esente ici la construction de la structure de spin par les
modules de Clifford (cf. [\citelow{jgb}] pour un expos\'e
d\'etaill\'e) plut\^ot que la construction en fibr\'e principal
souvent d\'evelopp\'ee\cite{choquet,jost,nakahara}.

\subsection{Module de Clifford}

On note $\mm$ une vari\'et\'e compacte, $T_x M$
\index{txm@$T_xM$} l'espace vectoriel r\'eel des vecteurs en $x$
(l'espace tangent) et $T_x ^* M$ \index{txmetoile@$T_x^*M$} son
dual, l'espace des $1$-formes en $x$ (espace cotangent). Dans la
carte $x^\mu$, on d\'esigne par $\{dx^\mu\}$ et
$\{\partial_\mu\}$ les bases des espaces cotangent et tangent.

 Un $\kk$-fibr\'e vectoriel  $E \overset{\pi}\longrightarrow M$ est un espace topologique localement hom\'eomorphe au produit $U_i\times F$, o\`u $U_i$ est un
ouvert de $\mm$ et $F$ un espace vectoriel sur un corps $\kk$,
$\pi$ d\'esignant la projection de $E$ sur $M$. Pour tout $x$ de
$\mm$, $E_x \doteq \pi^{-1}(x)$\index{ex@$E_x$}, la fibre au
dessus de $x$, est isomorphe \`a $F$. Une section locale
$\sigma_i$\index{sigmai@$\sigma_i$} de $E$ est une application de
$U_i$ dans $E$ telle que $\pi\circ\sigma_i$ soit l'identit\'e de
$U_i$. Si $r$ est la dimension de $F$, une section locale est la
donn\'ee de $r$ fonctions de $U_i$ dans $\rr$, appel\'ees
composantes de la section.  Une section locale est
diff\'erentiable (continue) quand ces composantes sont des
fonctions diff\'erentiables (continues) sur $\mm$.
 Une section diff\'erentiable (continue) est une collection de sections locales diff\'erentiables
(continues) $\{\sigma_i\}$ telle que l'union des $U_i$ soit un
recouvrement de $\mm$. On note
$\Gamma^{\infty}(E)$\index{gammainfinie@$\Gamma^{\infty}(E)$}
(resp. $\Gamma(E)$)\index{gammae@$\Gamma(E)$} l'ensemble des
sections diff\'erentiables (continues) de $E$. C'est le module
(par convention, \`a droite) sur l'alg\`ebre $\cinf$ (resp.
$C(\mm)$) des fonctions lisses (continues) sur $\mm$,
\begin{equation}
\label{modulesection} (\sigma_1 + \sigma_2f)(x) \doteq
\sigma_1(x) + \sigma_2(x)f(x)
\end{equation}
pour tout $\sigma_1, \sigma_2\in\Gamma^\infty(E)$. En prenant $F
= \rr^n$, o\`u $n$ est la dimension de $\mm$,  et $\pi^{-1}(x) =
T_x \mm$, on construit le fibr\'e vectoriel r\'eel
$TM$\index{tm@$TM$}, appel\'e {\it fibr\'e tangent}.  L'ensemble
des sections $\mathcal{X}(\mm) \doteq \Gamma^\infty(TM)$
\index{xm@$\mathcal{X}(\mm)$} est l'ensemble des champs de
vecteurs lisses sur $\mm$. De mani\`ere analogue, on construit le
{\it fibr\'e cotangent} $T^*M$\index{tetoilem@$T^*M$} dont les
sections $\Omega^1(\mm)\doteq \Gamma^\infty(T^*M)$
\index{omegaunm@$\Omega^1(\mm)$} sont les champs de $1$-forme.

Une {\it m\'etrique riemannienne} $g$\index{g} est une
application bilin\'eaire sym\'etrique ($g(X,Y) = g(Y,X)$),
d\'efinie positive ($g(X,X)> 0$ pour $X\neq 0$) de $\cal{X}(\mm)
\times \cal{X}(\mm)$ dans $\cinf$. Si $g$ est seulement non
d\'eg\'en\'er\'ee ($g(X,Y)= 0 \text{ pour tout } Y \, \Rightarrow
X = 0$), la m\'etrique est dite {\it pseudo-riemannienne}.
 Dans les deux cas, $g$ d\'efinit une bijection de $\cinf$-module entre ${\cal X}(M)$ et $\Omega^1(M)$,
la bijection {\it musicale} $\flat\sharp$
\begin{eqnarray*}
{\cal{X}}(\mm)\rightarrow  \Omega^1(\mm)&:& X \mapsto \index{xbemol@$X^\flat$}X^\flat\; \text{ tel que }\, X^\flat(Y)\doteq g(X,Y),\\
\Omega^1(\mm)\rightarrow \cal{X}(\mm)&:&
\varpi\index{omegabar@$\varpi$} \mapsto \varpi^\sharp \; \text{
tel que }\, g(\varpi^\sharp,Y) \doteq \varpi(Y),
\end{eqnarray*}
o\`u $\varpi(.), X^\flat(.)$ d\'esignent l'action par dualit\'e
de $\Omega^1(\mm)$ sur $\xx(\mm)$. Le {\it gradient} d'une
fonction $f\in\cinf$ est par d\'efinition
\begin{equation}
\label{gradientdiese} \index{gradient}\grad{f} \doteq df^\sharp.
\end{equation}
 La m\'etrique induit
 une forme bilin\'eaire sym\'etrique d\'efinie positive (ou seulement non d\'eg\'en\'er\'ee dans le cas pseudo-riemannien) sur $\Omega^1(M)$, pareillement not\'ee $g$,
 \begin{equation}
\label{scldiese} g(\varpi_1, \varpi_2)\doteq g(\varpi_1^\sharp,
\varpi_2^\sharp).
\end{equation}
En tout $x$, $T_xM$ et $T_x^*M$ sont munis de la norme de
\begin{equation}
\label{normetangent} \norm{\grad{f}}\doteq g(\grad{\bar{f}},
\grad{f}) = g (d\bar{f}, df) \doteq \norm{df}.
\end{equation}

L'{\it alg\`ebre ext\'erieure} $\Lambda V$\index{lambdav@$\Lambda
V$} sur un espace vectoriel r\'eel $V$ est l'alg\`ebre formelle
g\'en\'er\'ee par un \'el\'ement identit\'e $\ii$ et  les
produits $v_1\wedge... \wedge v_k$ avec $v_1, v_k \in V$, $k\leq
\text{dim }V$, $v_1\wedge v_2 = - v_2\wedge v_1$ et $\ii\wedge v
= v$.  Lorsque $V$ est munie d'une forme bilin\'eaire non
d\'eg\'en\'er\'ee $g$, sym\'etrique \`a valeur dans $\rr$, on
construit l'{\it alg\`ebre de Clifford}
$\text{Cl}(V,g)$\index{clvg@$\text{Cl}(V,g)$} en "quantifiant" la
relation d'anticommutation de l'alg\`ebre ext\'erieure \`a l'aide
de $g$. Concr\`etement, $\text{ Cl}(V,g)$ en tant qu'espace
vectoriel est identique \`a $\Lambda V$ mais le produit est
d\'efini de sorte que
\begin{equation}
\label{gmunuclifford} uv + vu = 2 g(u,v)\ii
\end{equation}
 pour tout $u,v \in V$.
Avec $V^\cc \doteq V + iV$ le complexifi\'e de $V$ et l'extension
de $g$,  $g(u, v+iw) = g(u,v) + i g(u,w)$, on construit de la
m\^eme mani\`ere l'alg\`ebre de Clifford complexe $\cc \text{l}
(V)$. On omet $g$ dans la notation car toutes les formes non
d\'eg\'en\'er\'ees sur $V + iV$ donnent des alg\`ebres de
Clifford isomorphes. On obtient ainsi [\citelow{jgb}, {\it Lem.
5.5}]
\begin{equation}
\label{cliffordmatrice} \ccl(\rr^{2m}) \simeq M_{2^m}(\cc)\,
\text{ et } \ccl(\rr^{2m+1}) \simeq M_{2^m}(\cc) \oplus
M_{2^m}(\cc).
\end{equation}
$\ccl(V)$ est munie d'une involution $*$, obtenue en \'etendant
\begin{equation}
\label{clinv} (\lambda v_1... v_k)^* = \bar{\lambda} v_k... v_1
\end{equation}
 avec $\lambda\in\cc$, $v_1,..., v_k \in V,$ par lin\'earit\'e \`a tout $\ccl(V)$ (restreint \`a $V$ l'involution co\"{\i}ncide avec l'identit\'e, ce qui est
coh\'erent puisque $V$ est un espace vectoriel r\'eel).

Un \'el\'ement de $\ccl(V)$ est pair lorsqu'il s'\'ecrit comme
combinaison lin\'eaire de produits d'un nombre pair de vecteurs
de $V$. On note $\ccl ^+ (V)$ \index{clvplus@$\ccl ^+ (V)$} la
sous-alg\`ebre g\'en\'er\'ee par les \'el\'ements pairs, et
$\ccl^-(V)$\index{clvmoins@$\ccl^-(V)$} le sous-espace vectoriel
des produits impairs de vecteurs. En tant qu'espace vectoriel,
$\ccl (V) = \ccl^+(V) \oplus \ccl^- (V)$. On note
$\chi$\index{chi@$\chi$} la $\zz_2$ graduation correspondante
\begin{equation}
\label{defchi} \chi(a) = \pm 1 \text{ pour } a\in \ccl^{\pm}(V).
\end{equation}

Lorsque $g$ est d\'efinie positive, on d\'efinit l'\'el\'ement
{\it chiralit\'e} de $\ccl(V)$
\begin{equation}
\label{defgamma} \gamma\index{gamma@$\gamma$}\doteq (-i)^m
e_1e_2... e_n
\end{equation}
o\`u  $\{e_i\}$ est une base de $V$ orthonorm\'ee pour $g$ et
$n=\text{dim } V = 2m$ o\`u $2m+1$. Modulo l'orientation,
$\gamma$ est ind\'ependant du choix de la base orthonorm\'ee. On
v\'erifie que $\gamma^2 = \gamma^*\gamma =\ii$. La chiralit\'e
anticommute ou commute avec $V$ selon que $n$ est pair ou impair.
Lorsque $n$ est pair, $\gamma v \gamma = -v$ pour tout $v$ de
$V$. Etendu \`a tout $\ccl(V)$, on montre que $\gamma . \gamma$
co\"{\i}ncide avec la $\zz_2$ graduation $\chi$. Lorsque $n$ est
impair, $\gamma . \gamma$ est l'identit\'e. La restriction, $g$
d\'efinie positive, est fondamentale car c'est elle qui par la
suite nous oblige \`a consid\'erer des vari\'et\'es riemanniennes.
%
\newline

La m\'etrique $g$ d'une vari\'et\'e $\mm$ est d\'efinie sur les
sections lisses du fibr\'e tangent $TM$. Les champs de vecteurs
lisses sont denses dans l'ensemble
 des champs de vecteurs continus,
et $g$ s'\'etend en une forme bilin\'eaire sur les sections
continues $\Gamma(TM)$. Par complexification,
 on obtient une forme bilin\'eaire, encore not\'ee $g$, sur les sections continues du fibr\'e vectoriel complexe de fibre $T_x M^\cc = T_x M + i T_x M$\index{txmc@$T_xM^\cc$}.
Sur chacune de ces fibres $g$ induit une forme bilin\'eaire
permettant de former en tout $x$ de $\mm$ l'alg\`ebre de
Clifford  $\ccl(T_x M)$.  Le fibr\'e vectoriel
 sur $\mm$ correspondant est not\'e $\ccl\, TM$. Le $C(\mm)$-module $\Gamma(\ccl\, TM)$
des sections continues de ce fibr\'e est une $C^*$-alg\`ebre,
produit et involution \'etant d\'efinis point par point
$$\sigma_1\sigma_2(x) \doteq \sigma_1(x)\sigma_2(x),\;\, \sigma^*(x)\doteq \sigma(x)^*\quad \forall x\in\mm\index{sigma@$\sigma$}$$
o\`u $*$ d\'esigne l'involution dans chaque $\ccl (T_xM)$, et la
norme est
$$\norm{\sigma}= \suup{x\in\mm} \{ \norm{\sigma(x)}\}$$
o\`u la norme de $\sigma(x)$ est celle de la $C^*$-alg\`ebre
$\ccl(T_xM)$. La construction est identique pour le fibr\'e
cotangent $T^*M$, ou pour n'importe quel fibr\'e vectoriel r\'eel
$E$ sur $\mm$, munie d'une forme bilin\'eaire non
d\'eg\'en\'er\'ee de $\Gamma^{\infty} (E) \times \Gamma^\infty
(E)$ dans $\cinf$. Pour disposer d'une chiralit\'e, on se limite
aux m\'etriques riemanniennes.

{\defi Le fibr\'e de Clifford sur une vari\'et\'e riemannienne
$M$ de m\'etrique $g$ est le fibr\'e $\ccl(M)\doteq
\ccl\,T^*M$\index{clm@$\ccl(M)$}.}
\newline

Evalu\'ee en un point $x$ de $\mm$, une section $\sigma$ d'un
fibr\'e de Clifford est un \'el\'ement $\sigma(x)$ de
$\ccl(T^*_xM)$. Si $F$ est un espace vectoriel complexe sur
lequel agissent chacune des alg\`ebres $\ccl(T^*_xM)$ via {\it
l'action de Clifford}
\begin{equation}
\label{actionc} c\index{c}:\, \ccl(T^*_xM) \rightarrow
\text{End}(F),
\end{equation}
alors une section $\sigma$ du module de Clifford agit (par
convention \`a gauche) sur une section $\sigma'$ d'un fibr\'e
vectoriel $E\overset{\pi}{\rightarrow} \mm$ de fibre $F$ (i.e.
$\pi^{-1}(x)\simeq F$ pour tout $x$) par
$$\lp c(\sigma)\sigma'\rp(x) \doteq c(\sigma(x))\sigma'(x).$$
Lorsque l'action de $c(\sigma)$ est continue, c'est \`a dire
lorsque $c(\sigma)\sigma' \in \Gamma(E)$ pour tout
$\sigma\in\Gamma(\ccl(M)$ et $\sigma'\in\Gamma(E)$, $\Gamma(E)$
est un $\Gamma(\ccl(M))$-module \`a gauche. $\Gamma(E)$ \'etant
d\'ej\`a un $C(\mm)$-module \`a droite, c'est un bimodule.

{\defi Un module de Clifford sur $\mm$ est la donn\'ee d'un
$C(\mm)$-module $\Gamma(E)$ des sections continues d'un fibr\'e
vectoriel complexe sur $M$ ainsi que d'un homorphisme
$C(\mm)$-lin\'eaire
$$c:\, \Gamma(\ccl(\mm) ) \rightarrow \text{End}(\Gamma(E)).$$
Autrement dit, un module de Clifford sur $\mm$ est un
$\Gamma(\ccl(M))$-$C(\mm)$-bimodule de sections d'un fibr\'e
vectoriel complexe sur $\mm$.}
\newline

\noindent Si $\text{dim }\mm = 2m$, d'apr\`es
(\ref{cliffordmatrice}) toutes les actions irr\'eductibles de
$\ccl(\mm)$ sont de dimension $2^m$. Si $\text{dim }\mm = 2m+1$,
il y a deux repr\'esentations irr\'eductibles in\'equivalentes de
dimension $2^m$. Quand le rang du fibr\'e $E$ (i.e. la dimension
de ses fibres en tant qu'espace vectoriel) n'est pas \'egale \`a
$2^m$ dans le cas pair, $2^{m+1}$ dans
 le cas impair, chaque fibre $E_x$ se d\'ecompose en somme directe de sous-espaces vectoriels invariant par l'action de l'alg\`ebre de Clifford.
Au contraire quand l'action de l'alg\`ebre de Clifford est
irr\'eductible sur chaque fibre, $\Gamma(E)$ est un {\it module
de Clifford irr\'eductible}.

\subsection{Structure de Spin}

Classiquement, le fibr\'e des spineurs sur une vari\'et\'e $\mm$
de dimension $n$ est construit \`a partir du fibr\'e tangent par
le rel\`evement du groupe $SO(n)$
 (groupe de structure du fibr\'e principal associ\'e au fibr\'e tangent)  \`a son recouvrement universel $\spin(n)$. L'approche alg\'ebrique construit
 directement un spineur comme support d'une action irr\'eductible du groupe $\spin$, vu
 comme sous groupe de l'alg\`ebre de Clifford.
 \newline

 Soit $V$ un espace vectoriel munie d'une forme bilin\'eaire non d\'eg\'en\'er\'ee $g$.
Un vecteur $u\in V$ est {\it unitaire} quand $g(u,u) = 1$. Vu
comme \'el\'ement de $\ccl(V)$,  $u^2 = \ii$ par
(\ref{gmunuclifford}) donc $u$ est inversible. On note $\phi(u)$
l'endomorphisme de $V$
$$\phi(u) v \doteq \chi(u) v u^{-1} = -u v u = (vu - 2 g(u,v))u = v- 2g(u,v)u$$
o\`u $\chi$ est la $\zz_2$ graduation d\'efinie en (\ref{defchi}).
Restreinte \`a $V$, qui est laiss\'e globalement invariant,
l'action de $\phi(u)$ est la r\'eflexion par rapport \`a
l'hyperplan orthogonal \`a $u$ (pour s'en convaincre on peut
regarder $\ccl(\rr^2)$ avec pour $g$ le produit scalaire usuel).
Par la multiplication
$$\phi_{u_1u_2}(v) \doteq u_2^{-1}u_1^{-1} v u_1 u_2 = \phi_{u_2}\circ\phi_{u_1}(v),$$
ces r\'eflexions g\'en\`erent le groupe orthogonal $O(V)$.
L'ensemble des produits pairs de r\'eflexions est le sous-groupe
des rotations $SO(V)$ (c'est la composante connexe de
l'identit\'e de $O(V)$). L'ensemble des produits pairs de
vecteurs unitaires $w$ de $V^\cc$ ($w = \lambda u$ avec $\lambda$
un nombre complexe de module $1$ et $u$ un unitaire de $V$) est
un sous groupe de $\ccl(V)$ not\'e $\text{Spin}^c(V)$.

Pour tout $w\in \spin^c(V)$\index{spinc@$\spin^c$}, l'application
$\phi(w):\, v \mapsto w v w$ ($\chi(w) = +1$) est une rotation
dans $V$. $\phi$ apparait comme un homomorphisme de $\spin^c(V)$
dans $SO(V)$. Un \'el\'ement du noyau de $\phi$ est un unitaire
central pair de $\ccl(V)$ et on montre [\citelow{jgb}, {\it p.
180}] qu'un tel \'el\'ement est n\'ecessairement un scalaire.
Autrement dit $\text{ker }\phi\simeq U(1)$. Pour
$w=w_1...w_{2k}\in \spin^c(V)$, on d\'efinit l'homorphisme $\nu$
\`a  valeur dans $U(1)$
$$\nu(w) = w_{2k} ... w_1 w_1 ... w_{2k} = \lambda_1 ... \lambda_{2k}
$$
o\`u $\lambda_i = w_i^2\in U(1)$. Le groupe {\it
\spin(V)}\index{Spin} est par d\'efinition le noyau de $\nu$. La
conjugaison complexe est d\'efinie sur tout $\ccl(V)$ en
\'etendant par lin\'earit\'e $\overline{\lambda v} \doteq
\bar{\lambda} v$ pour $\lambda\in\cc$, $v\in V$. $\text{Spin}(V)$
est l'ensemble des unitaires pairs $w$ de $\ccl(V)$  satisfaisant
$\overline{w^*} w = w^* w = \ii$, ou encore $\bar{w} = w$. En
d\'efinissant la conjugaison de charge $\kappa$
$$\index{kappa@$\kappa$}\kappa(a) \doteq \chi(\bar{a}) $$
pour tout $a\in\ccl(V)$, le groupe Spin apparait comme le sous
groupe de $\spin^c$ invariant par conjugaison de charge. Le noyau
de $\phi$ restreint \`a $\spin(V)$ est $\{-1, 1\}$. En prenant
$V=\rr^n$ et $g$ une m\'etrique (pseudo-)riemannienne, on
retrouve que le spin est le recouvrement universel \`a deux
feuillets du groupe des rotations.
\newline

Un spineur est une section d'un fibr\'e vectoriel $S$ sur une
vari\'et\'e $M$ dont chaque fibre est le support d'une
repr\'esentation irr\'eductible du groupe $\spin(M)\doteq
\spin(T^*_xM)$. $\Gamma(S)$ est donc un module de Clifford
irr\'eductible. Dans le cas o\`u $M$ est de dimension $n= 2m$
paire, un tel module est obtenu en demandant que $S$ impl\'emente
une {\it \'equivalence de Morita} entre $C(M)$ et
$\Gamma(\ccl(M))$.

{\defi\label{morita} Deux $C^*$-alg\`ebres $\aa$ et $\bb$ sont
Morita-\'equivalentes si et seulement si il existe un
$\aa$-module \`a droite plein $E$ tel que $\text{End}^{\,
0}_\aa(E) \simeq \bb$, o\`u $\text{End}^{\, 0}_\aa(E)$ est la
fermeture (pour la topologie de la norme d'op\'erateur) de
l'alg\`ebre des endomorphismes de $E$ de $\aa$-rang fini.}
\newline

\noindent Cette d\'efinition demande plusieurs pr\'ecisions. Un
$\aa$-module $E$ est plein lorsqu'il est muni d'un "produit
scalaire \`a valeur dans $\aa$", c'est \`a dire d'une forme de
$E\times E$ dans $\aa$ d\'efinie positive, $\aa$-lin\'eaire \`a
droite, antisym\'etrique (\,$(u \lvert v) = (v \lvert u)^*$), et
telle que $(E \lvert E) = \aa$. Un endomorphisme de $E$ est dit
de $\aa$-rang fini lorsqu'il est du type:
$$\lvert r) (s\lvert:\,t \longmapsto r(s\lvert t),$$
$r,s,t\in E$. Ces op\'erateurs forment une alg\`ebre qu'on munit
de la norme d'op\'erateur
$$\text{ sup}\{\norm{r(s\lvert t)}\, / \, \norm{t}= 1\}$$ o\`u la norme dans $E$ est
d\'efinie \`a partir de la norme de $\aa$ par $\norm{t}\doteq
\sqrt{\norm{(t\lvert t)}}.$

Si $\Gamma(S)$ impl\'emente l'\'equivalence de Morita entre
$C(M)$ et $\Gamma(\ccl(M))$, on montre qu'il existe un
isomorphisme de fibr\'e vectoriel $\text{End } S\simeq \ccl(M)$
o\`u $\text{End } S$ d\'esigne le fibr\'e vectoriel sur $M$ de
fibre $\text{ End}(S_x)$. $\ccl(M)$ est de rang $2^n$, donc
$\text{ End } S$ est de rang $2^n$, ce qui signifie que $S_x$ est
de dimension $\sqrt{2^n} = 2^m$. On peut donc choisir l'action de
Clifford de telle sorte que $S$ soit un module de Clifford
irr\'eductible. Rien n'assure en revanche qu'impl\'ementer
l'\'equivalence de Morita soit une condition n\'ecessaire pour
que $\Gamma(S)$ soit irr\'eductible. Mais il apparait que la
condition sur $M$ pour que $C(M)$ et $\Gamma(\ccl(M)$ soit Morita
\'equivalente (th\'eor\`eme de Plymen [\citelow{jgb}, {\it Th.
9.3}]) est tr\`es exactement la condition qui, dans l'approche
classique,  autorise le rel\`evement de $SO(n)$ au groupe
$\spin^c(n)$.

La possibilit\'e du rel\`evement \`a $\spin(n)$  correspond [{\it
ibid, Th. 9.6}] \`a l'existence d'une bijection antilin\'eaire
$J:\, S \rightarrow S$ telle que
\begin{eqnarray}
\nonumber
\index{J} J(\psi f ) &=& (J\psi) \bar{f} \text{ pour } f\in C(M),\\
\label{bijc}
J(a \psi ) &=& \chi(a) J \psi  \text{ pour } a\in \ginf(\ccl(M)),\\
\nonumber ( J\phi \lvert J \psi ) &=& ( \psi \lvert \phi) \text{
pour } \phi, \psi \in S
\end{eqnarray}
o\`u $\psi\in S$ et on identifie $a$ et $f$ \`a leurs actions sur
$S$. On montre [{\it ibid, Lem 9.7}] qu'un tel op\'erateur $J$
est n\'ecessairement de carr\'e $\pm 1$.

Le produit scalaire sur $\Gamma(S)$ \`a valeur dans $C(M)$ est
choisi de sorte que l'action de Clifford soit autoadjointe,
$$( \phi  \lvert c(a)\psi ) = (c(a^*)\phi \lvert \psi).$$
On note $c(a)^\dagger = c(a^*).$ Si $M$ est {\it orient\'ee}, il
existe un rep\`ere mobile de $1$-formes $\{e_i\}$ (i.e. une
section lisse du fibr\'e cotangent) tel qu'en tout $x$ de $M$ les
chiralit\'es $\gamma(x)$ d\'efinies par (\ref{defgamma}) sur
chaque fibre de $\ccl(M)$ s'\'ecrivent
$$\gamma(x) = (-i)^m e_1(x)... e_n(x).$$
$\gamma$ est une section de $\ccl(M)$ et $c(\gamma)$ est une
graduation (i.e. $c(\gamma)^\dagger c(\gamma) = c(\gamma)^2 =
\ii)$ de $\Gamma(S)$. On note
$\Gamma(S)^{\pm}$\index{gammaplusmoins@$\gamma^{\pm}$} les
sous-espaces propres de $c(\gamma)$ de sorte que
$$
\Gamma(S) = \Gamma(S)^+ \oplus \Gamma(S)^-.
$$
Si $M$ est de dimension paire, $c(\gamma)$ anticommute avec
$c(\varpi)$ pour toute $1$-forme $\varpi\in\ccl(M)$. Pour
$\psi^{\pm}\in\Gamma(S)^{\pm}$,
$$c(\gamma) c(\varpi) \psi^{\pm} = \mp c(\varpi)\psi^\pm,$$
autrement dit $c(\gamma)$ \'echange $\Gamma^+(S)$ et
$\Gamma^-(S)$.

 {\defi Une {\it structure de spin} sur une vari\'et\'e $M$ de dimension paire est la donn\'ee d'un bimodule
\index{S} $S$ garantissant l'\'equivalence de Morita
 $C(M)$-$\Gamma(\ccl(M))$, d'une bijection $J$ satisfaisant (\ref{bijc}) et d'une orientation de $M$.}
\newline

\noindent $M$ est alors dite {\it vari\'et\'e \`a spin}. Lorsque
$M$ est de dimension impaire, la construction est analogue en
rempla\c{c}ant $\ccl(M)$ par $\ccl^+(M)$ qui est le fibr\'e sur
$M$ de fibre
  $\ccl^+(T_x^*M)$.

\subsection{Op\'erateur de Dirac}

 Une {\it connexion} sur un fibr\'e vectoriel $E\overset{\pi}\rightarrow M$ est une application lin\'eaire
 \begin{equation}
\label{connexion} \index{nabla@$\triangledown$}\triangledown:\,
\Gamma^\infty(E) \longrightarrow \Gamma^\infty(E)\ot \Omega^1(M)
\end{equation}
 satisfaisant la r\`egle de Leibniz
 $$\triangledown(\sigma f) = (\triangledown \sigma)f + \sigma\ot df$$
 pour tout $\sigma\in\Gamma^\infty(E)$ et $f\in\cinf.$ $d$ d\'esigne la d\'eriv\'ee ext\'erieure de chaque $\Lambda T^*_x M$ \'etendue aux sections lisses.
  Les {\it coefficients de connexion} sont obtenus en \'ecrivant localement, dans une base $\{dx^\mu\}$ de $\Omega^1(\mm)$, l'action de la connexion sur une base $\{\sigma_i\}$ de $\Gamma^\infty(E)$,
\begin{equation}
\label{coeffconnex} \triangledown \sigma_i \doteq \Gamma^{j}_
{i\mu}\index{gammajimu@$\Gamma^{j}_ {i\mu}$} \sigma_j \ot dx^\mu.
\end{equation}
Ainsi
\begin{eqnarray}
\nonumber
\triangledown \sigma = \triangledown \sigma_i f^i &=& (\triangledown \sigma_i)f^i + \sigma_i\ot d(f^i),\\
\nonumber
 &=& f^i\Gamma^{j}_ {i\mu} \sigma_j \ot dx^\mu  + \sigma_i\ot d(f^i),\\
\label{connexlocale} &=& (d + \Gamma) \sigma
 \end{eqnarray}
 o\`u on note $d \sigma \doteq \sigma_i\ot d(f^i) \;\text{ et }\; \Gamma \sigma \doteq f^i\Gamma^j_{i\mu} \sigma_j \ot dx^\mu.$

Lorsque $E$ est le fibr\'e tangent $TM$ sur une vari\'et\'e
riemannienne ou pseudo-riemannienne, il existe une unique
connexion, la connexion de {\it Levi-Civita}, de torsion nulle
(cf. [\citelow{nakahara}] pour une d\'efinition de la torsion) et
compatible avec la m\'etrique de la mani\`ere suivante:
\begin{equation}
\label{lvg} g(\triangledown X,Y) + g(X, \triangledown Y) = d
(g(X,Y))
\end{equation}
pour tout $X,Y\in\cal{X}(\mm)$. $g$ agit sur
$(\xx(\mm)\ot\Omega^1(\mm)) \times \xx(\mm)$ par "contraction des
indices"
$$g( r^i_\nu \partial_i\ot dx^\nu, t^\lambda \partial_\lambda) \doteq  r^i_\nu t^\lambda g(\partial_i, \partial_\lambda)dx^\nu,$$
$r^{i}_\nu, t^\lambda$ \'etant des nombres r\'eels. Apr\`es
identification de $g(X, .)$ \`a $X^\flat = \varpi$, (\ref{lvg})
d\'efinit une connexion de Levi-Civita sur $E= T^*M$ \`a valeur
dans $\Gamma^\infty(T^*M)\ot\Omega^1(\mm)\simeq
\Omega^1(M)\ot\Omega^1(M)$,
\begin{equation}
\label{lv1f} \triangledown \varpi (Y) \doteq d (\varpi(Y)) -
\varpi(\triangledown Y),
\end{equation}
pour tout $Y$ de $\xx(\mm)$. Pour $\varpi = dx^i$,
$Y=\partial_\mu$, (\ref{lv1f}) avec (\ref{coeffconnex}) donne
$\triangledown dx^i (\partial_\nu) = -  \Gamma^i_{\nu\mu} dx^\mu$
d'o\`u
\begin{equation*}
\label{connex1forme} \triangledown dx^i  = -  \Gamma^i_{j\mu}
dx^j \ot dx^\mu.
\end{equation*}
Localement l'action de $\del$ sur un $\varpi = dx^i f_i\in T^*M$
s'\'ecrit
$$\triangledown \varpi = (d - \tilde{\Gamma}) \varpi$$
o\`u $d \varpi \doteq dx^i\ot d(f_i)$ et $\tilde{\Gamma}\varpi
\doteq   - f_i \Gamma^i_{j\mu} dx^j \ot dx^\mu$. Par application
r\'ecursive de la r\`egle de Leibniz, $\del$ s'\'etend \`a tout
$\Gamma^\infty(\ccl(\mm)$,
\begin{equation}
\label{connexioncl} \triangledown (uv) \doteq \triangledown(u) v
+ u \triangledown(v)
\end{equation}
pour tout $u,v\in \Gamma^\infty(\ccl(\mm))$ (le produit
$(\triangledown u)v$ consiste \`a multiplier les composantes dans
l'alg\`ebre de Clifford en laissant invariante la partie
$\Omega^1(\mm)$).

Pour une vari\'et\'e \`a spin $(M, S, C)$, il existe une unique
{\it connexion de spin} $\del^S$
\index{nablas@$\triangledown^S$}g\'en\'eralisant la connexion de
Levi-Civita tout en \'etant compatible avec la structure de spin
[\citelow{jgb}, {\it Th. 9.8}].

{\thm Soit $(M, S, J)$ une vari\'et\'e \`a spin de dimension $n$.
Il existe une unique connexion $\del^S:\, \ginf(S) \rightarrow
\ginf(S)\ot \Omega^1(M)$ hermitienne, i.e.
$$( \del^S \psi \lvert \phi) + (\psi \lvert \del^S \phi) = d(\psi \lvert \phi) ,$$
commutant avec $J$ et telle que
$$\del^S( c(a)\psi) = c(\del a)\psi + c(a) \del^S\psi \text{ pour } a\in\ccl(M), \psi\in \ginf(S) $$
o\`u $c$ d\'esigne l'action de $\Gamma^\infty(\ccl(M))$ sur
$\ginf(S)$ induite par (\ref{actionc}) et $\del$ la connexion
(\ref{connexioncl}).}
\newline

\noindent L'action de l'alg\`ebre de Clifford se r\'e\'ecrit
comme une application de $\ginf(S)\ot \ginf(\ccl(M)) $ dans
$\ginf(S)$ en posant
$$\hat{c}( \psi\ot a) \doteq c(a)\psi.$$
L'objet fondamental d'une g\'eom\'etrie spinorielle est l'{\it
op\'erateur de Dirac}, d\'efini comme suit.

{\defi \label{dirac} L'op\'erateur de Dirac d'une vari\'et\'e \`a
spin $(M, S, J)$ est l'endomorphisme de $\ginf(S)$
$$D\index{D}\doteq -i (\hat{c} \circ\del^S).$$}
\newline

 Cet objet co\"{\i}ncide bien avec l'op\'erateur de Dirac de
la th\'eorie quantique des champs. Pour s'en convaincre,
\'ecrivons localement l'action de la connexion de spin. Tout
espace de Hilbert de dimension finie admettant
 une base orthonorm\'ee, il existe en tout $x$ de $M$ une base orthonorm\'ee de $T_xM$, $\{\partial_\alpha = e_\alpha^\mu(x) \partial_\mu\}$,
ainsi qu'une base duale de $T^*_xM$, \'egalement orthonorm\'ee,
$\{dx^\alpha = e^\alpha_\mu dx^\mu\}$. Le {\it vielbein}
$\{e^\alpha_\mu\}$ d\'esigne la matrice inverse de
$\{e_\alpha^\mu\}$ et satisfait
\begin{equation}
\label{geed} g^{\mu\nu} e^\alpha_\mu e^\beta_\nu =
\delta^{\alpha\beta}
\end{equation}
 o\`u $g^{\mu\nu} = g(dx^\mu, dx^\nu)$. Soit
$\{\gamma^a\index{gamma@$\gamma^a$}, \gamma^b\}$ un champ de
matrices de Dirac, i.e. des matrices autoadjointes de $M_k(\cc)$
($k= 2^{[n/2]}$ est la dimension de la repr\'esentation
irr\'eductible de $\ccl(M)$ dans le cas pair, de $\ccl(M)^+$ dans
le cas impair) telles que
\begin{equation}
\label{diraclide} \gamma^a(x) \gamma^b(x) + \gamma^b(x)
\gamma^a(x) = 2\delta^{ab}\ii
\end{equation}
en tout $x$ de $M$. En d\'efinissant la repr\'esentation
\begin{equation}
\label{cliffaction} c(dx^\alpha) \doteq \gamma^a,
\end{equation}
l'action de $dx^\mu\in\ccl(M)$ sur $S$
\begin{equation}
\label{matgamma} c(dx^\mu) \psi \doteq \gamma^m \psi \doteq
e^\mu_\alpha \gamma^a \psi
\end{equation}
(on utilise un indice grec pour les coordonn\'ees de la
vari\'et\'e et un indice latin pour le fibr\'e, $a$ est
contract\'e avec $\alpha$) est bien une repr\'esentation
(irr\'eductible) de $\ccl(M)$ puisque
\begin{eqnarray*}
c(dx^\mu)c(dx^\nu) + c(dx^\nu)c(dx^\mu) &=& 2e^\mu_\alpha e^\nu_\beta  \delta^{ab} \ii,\\
                                &=& 2 g^{\lambda\rho} e^\mu_\alpha e^\alpha_\lambda e^\nu_\beta   e^\beta_\rho \ii,\\
                                &=& 2g(dx^\mu, dx^\nu)\ii = c( dx^\mu dx^\nu + dx^\nu dx^\mu).
\end{eqnarray*}
On montre alors que la connexion de spin s'\'ecrit
\begin{equation}
\label{connecspin} \del^S  = d - \frac{1}{4}{\Gamma}^i_{j\mu}
\gamma_i \gamma^j\ot dx^\mu
\end{equation}
o\`u $\gamma_i\doteq \gamma^i$ et $d$ agit sur un spineur $\psi=
s_if^i$ \--$f^i\in C(M)$, $s_i\in\ginf(S)$\-- selon $d \psi
\doteq s_i \ot d(f^i)$. Quand la vari\'et\'e est plate  (ce qui
est le cas en th\'eorie des champs quand on suppose que
l'interaction a un lieu dans une r\'egion o\`u la courbure est
localement n\'egligeable), les  coefficients de connexion sont
nuls et
\begin{equation}
\label{diracplat} iD\psi = \hat{c} (d\psi) = \hat{c} (s_i
\ot\partial_\alpha f^i dx^\alpha) = c(dx^\alpha) s_i
\partial_\alpha f^i  = \gamma^a s_i \partial_\alpha f^i = \ds
\psi.
\end{equation}

\subsection{Triplets spectraux}

Toute l'information g\'eom\'etrique d'une vari\'et\'e \`a spin,
en particulier la m\'etrique, est contenu dans l'op\'erateur de
Dirac. Cette remarque, dont nous rappelons dans cette section les
points cl\'es, est fondamentale puisqu'en donnant une
d\'efinition alg\'ebrique (i.e. en terme d'op\'erateur) des
objets de la g\'eom\'etrie spinorielle, elle permet de voir la
vari\'et\'e \`a spin commme un cas particulier, commutatif, d'une
th\'eorie beaucoup plus g\'en\'erale permettant de d\'efinir la
g\'eom\'etrie d'espaces non commutatifs. L'objet math\'ematique
d\'ecrivant ces g\'eom\'etries est le {\it triplet spectral
r\'eel}. Sa d\'efinition proc\`ede par \'etapes successives, en
commen\c{c}ant par isoler les propri\'et\'es essentielles (bien
s\^ur, elles n'apparaissent comme esssentielles qu'une fois la
construction achev\'ee)
 de l'op\'erateur de Dirac.

{\prop Si $D$ est l'op\'erateur de Dirac sur une vari\'et\'e \`a
spin $M$, alors
 \label{commutprop}
  $$
  [D, f] = -ic(df) \text{ pour tout } f\in\cinf.$$}

 \noindent{\it Preuve.} L'action de Clifford est $C(M)$-lin\'eaire, $c(af)\psi = c(a)\psi f$, donc
  \begin{equation*}
  i[D,f] \psi = \hat{c} (\del^S (\psi f)) - \hat{c}(\del^S\psi)f = \hat{c}\lp  \del^S (\psi f) - (\del^S\psi)f \rp = \hat{c} (\psi \ot df) = c(df)\psi,
  \end{equation*}
 pour tout $a\in\ginf(\ccl(M))$, $f\in\cinf$ et $\psi\in\ginf(S)$. \hfill $\blacksquare$
 \newline

 Gr\^ace au facteur $-i$ dans la d\'efinition \ref{dirac}, l'op\'erateur de Dirac est autoadjoint. $D$ \'etant non born\'e, cette affirmation
 n\'ecessite quelques pr\'ecautions. Notons tout d'abord qu'il existe un produit scalaire dans $\ginf(S)$,
 \begin{equation}
 \label{hilbspin}
 \scl{\psi}{\phi} \doteq \int_\mm (\psi \lvert \phi) \abs{\nu_g}
 \end{equation}
 o\`u
\begin{equation}
\label{nug} \index{nug@$\nu_g$} \nu_g = \sqrt{\text{det } g }\;
dx^1\wedge... \wedge  dx^n
\end{equation}
 est la forme volume de la vari\'et\'e $M$ et $g$ la matrice de composante $g(dx^\mu, dx^\nu)$.
 On renvoie aux ouvrages de g\'eom\'etries diff\'erentielles pour une \'etude
 de la th\'eorie de l'int\'egration sur une vari\'et\'e. Ici, il nous suffit de savoir que  (\ref{hilbspin}) co\"{\i}ncide localement avec l'int\'egrale de
 Lebesgue.  On note
 \begin{equation}
 \label{hld}
 \hh = L_2(M,S) \index{ldeuxms@$L_2(M,S)$}
 \end{equation}
 l'espace de Hilbert obtenue par compl\'etion de $\ginf(S)$ par rapport \`a la norme issue de ce produit scalaire. Dans le cas o\`u $M$ est plate, $L_2(M,S)$ est
l'espace des {\it spineurs de carr\'e sommable}
 de la m\'ecanique quantique. On conserve la m\^eme terminologie dans le cas g\'en\'eral. On montre alors que $D$ est formellement autoadjoint
 ($D=D^\dagger$) sur $\ginf(S)$, puis qu'il est essentiellement autoadjoint sur $\hh$, c'est \`a dire que
 $(D^\dagger)^\dagger$ est autoadjoint sur le sous-espace de $\hh$ compos\'e des spineurs $\psi_n$ pour lesquels \`a toute suite convergente $\psi_n\rightarrow \psi$
 correspond un spineur $\phi$ tel que $D\psi_n \rightarrow \phi$. Dans la suite, on identifie $D$ \`a $(D^\dagger)^\dagger$ en \'ecrivant simplement que $D$
 est autoadjoint.

 Quand $M$ est de dimension paire, la graduation $c(\gamma)$ anticommute avec $c(a)$ pour tout $a\in \Gamma^-(\ccl(M))$. Avec
(\ref{connecspin}), (\ref{diracplat}) et (\ref{matgamma}),
 $$ D \psi = -i c(\gamma) \lp \gamma^a \partial_\alpha  -  \frac{1}{4}\Gamma^{i}_{j\mu}
e^\mu_\alpha \gamma^a\gamma_i\gamma^j\rp \psi.$$
 Comme $\gamma^j$ et $\gamma^a\gamma_i\gamma^j$ appartiennent \`a $\Gamma^-(\ccl(M))$, $c(\gamma)$ anticommute avec l'op\'erateur de Dirac,
 \begin{equation}
 \label{chiralite}
 D\Gamma = - \Gamma D\index{Gamma@$\Gamma$}
 \end{equation}
 o\`u $\Gamma$ d\'esigne l'endomorphisme unitaire autoadjoint de $\hh$, extension de $c(\gamma)$, appel\'e {\it chiralit\'e}
 (on garde la m\^eme appellation pour $\gamma$ et $\Gamma$). A noter que $\Gamma$ est une $\zz_2$ graduation de $\hh$
 (pour \'eviter un conflit de notation, dans toute la suite $\Gamma$ d\'esigne la chiralit\'e et {\bf ne} d\'esigne {\bf plus} l'op\'erateur
 apparaissant dans la d\'efinition de la connexion de Levi-Civita).
\newline

 L'ensemble de ces propri\'et\'es est regroup\'e et g\'en\'eralis\'e dans les notions de {\it triplet spectral} et de $KR$-cycle. Pour tout
 $a\in\ginf(\ccl(M))$, $c(a)$ est un endomorphisme born\'e de $\hh$ car $c(a)$ est une collection d'op\'erateurs $c(a)_x$ agissant irr\'eductiblement sur
 les fibres $S_x$ de dimension $\text{dim }\ccl(T_xM^*)$ finie. Pour les m\^emes raisons, d'apr\`es la proposition  \ref{commutprop}, $[D,f]$ est born\'e,
 de m\^eme que $f$ qui agit simplement par multiplication sur $\hh$.

 {\defi \label{tripletspectral} Un triplet spectral $(\aa, \hh, D)$ pour une alg\`ebre $\aa$ est la donn\'ee d'un espace de Hilbert $\hh$, d'une repr\'esentation
 de $\aa$ dans l'alg\`ebre $\bb(\hh)$  des op\'erateurs born\'es sur $\hh$, et d'un op\'erateur autoadjoint $D$, de r\'esolvante compacte, tel
 que $[D,a]\in\bb(\hh)$ pour tout $a\in\aa$.}
\newline

\noindent Rappelons qu'un op\'erateur $D$ est \`a r\'esolvante
compacte\cite{reed} si et seulement si pour tout $\lambda\notin
\text{sp}(D)$, $(D- \lambda\ii)^{-1}$ est compact (un op\'erateur
$T$ sur $\hh$ est compact quand, pour $\epsilon>0$, $\norm{T}\leq
\epsilon$ sauf sur un sous-espace de $\hh$ de dimension fini).

 Lorsque $\aa$ est une alg\`ebre involutive, on d\'efinit la notion de {$KR$-cycle}.

 {\defi \label{krcycle}
 Soit $n\in\zz_8$; un $KR^n$-cycle pour une alg\`ebre involutive $\aa$ est un triplet spectral $(\aa, \hh, D)$ accompagn\'e de
 \begin{itemize}
 \item une bijection unitaire $J$ antilin\'eaire sur $\hh$ qui impl\'emente l'involution, i.e. $ J a J^{-1} = a^*$ pour tout $a$ de $\aa$;
 \item si $n$ est pair, une graduation $\Gamma$ de $\hh$ qui commute avec $\aa$ et anticommute avec $D$;
 \item la table de multiplication-commutation suivante
  \begin{center}
   \begin{tabular}{|c|cccccccc|}
   \hline
   n mod 8&0&1&2&3&4&5&6&7\\
   \hline
   $J^2 = \pm \ii$&+&+&\--&-&-&-&+&+\\
   $JD= \pm DJ$&+&-&+&+&+&-&+&+\\
   $J\Gamma = \pm \Gamma J$&+& &-& &+& &-&\\
   \hline
   \end{tabular}
 \end{center}
\end{itemize}
\noindent Pour $n$ impair, on pose $\Gamma =\ii$ (qui
naturellement commute avec $D$ et $J$) et on note de fa\c{c}on
g\'en\'erale $(\aa, \hh, D, \Gamma, J)$ un $KR$-cycle.}
\newline

Si $(M, S, J)$ est une vari\'et\'e \`a spin de dimension $n$, $D$
l'op\'erateur de Dirac et $\Gamma$ l'extension de $c(\gamma)$ aux
spineurs de carr\'e sommable, alors  $(\cinf, L_2(M,S), D, J,
\Gamma)$ est un $KR^n$-cycle. Que $(\cinf, \hh, D)$ soit un
triplet spectral est \'evident compte tenu de la discussion
pr\'ec\'edent  la d\'efinition \ref{tripletspectral} (on renvoie
\`a [\citelow{connes,jgb}] pour prouver que $D$ est \`a
r\'esolvante compacte); que $J$ impl\'emente l'involution (la
conjugaison complexe) de $\cinf$ d\'ecoule de (\ref{bijc});
l'action de Clifford est $C(M)$-lin\'eaire donc $\Gamma$ commute
avec la repr\'esentation de $\aa$;  l'anticommutation de $D$ et
$\Gamma$ est \'etablie en (\ref{chiralite}); reste la table de
commutation, montr\'ee en d\'etail dans [\citelow{jgb}, {\it Th.
9.19}]. A toute vari\'et\'e \`a spin est associ\'e un $KR$-cycle
mais l'inverse n'est pas vrai: la donn\'ee d'un $KR$-cycle ne
suffit pas
 \`a construire une vari\'et\'e \`a spin. Pour ce faire, il faut ajouter une s\'erie de conditions d\'etaill\'ees ci-dessous.
 \newline

  Nous donnons directement les conditions pour qu'un triplet spectral $(\aa, \hh, D)$ d\'efinisse une {\it g\'eom\'etrie non commutative}\cite{gravity}, en
  rappelant ensuite comment, adapt\'ees au cas commutatif, ces conditions tiennent lieu d'axiomes d'une vari\'et\'e \`a spin. Les trois premi\`eres
conditions sont plus analytiques qu'alg\'ebriques. Elles sont
importantes dans la d\'efinition axiomatique de la g\'eom\'etrie
commutative spinorielle mais dans les exemples
   \'etudi\'es dans cette th\`ese (g\'eom\'etrie de dimension z\'ero ou produit de g\'eom\'etries dont l'une est commutative)
   elles sont toujours remplies. Nous les donnons ici par exhaustivit\'e, en renvoyant \`a [\citelow{connes,gravity,jgb}] pour une d\'efinition pr\'ecise des objets
   qu'elles font intervenir.

{\cond[Dimension]  L'op\'erateur $D^{-1}$ est un infinit\'esimal
d'ordre $\frac{1}{n}$ o\`u $n\in\nn$ est la dimension (spectrale)
de la g\'eom\'etrie.}
\newline

\noindent $D$ \'etant \`a r\'esolvante compacte, $D^{-1}$
d\'efinit en restreignant $D$ \`a $\hh/\ker D$ est un op\'erateur
compact. Ainsi\cite{reed} la suite
 d\'ecroissante $\{\lambda_k\}$ des valeurs propres de $\abs{D^{-1}} \doteq  \sqrt{(D^{-1})^\dagger D^{-1}}$ tend vers z\'ero. $D^{-1}$ est un infinit\'esimal
 d'ordre $\frac{1}{n}$  signifie que cette suite d\'ecroit au moins aussi vite que $k^{-n}$,
 $$\lim_{k\rightarrow +\infty} \lambda_k = O(\frac{1}{k^n}).$$
Lorsque $\aa$ et $\hh$ sont de dimension finie, la dimension
spectrale est nulle.

{\cond[R\'egularit\'e] Pour tout $a\in\aa$, $a$ et $[D,a]$
appartiennent \`a l'intersection des domaines de toutes les
puissances $\delta^k$ de la d\'erivation $\delta(b)\doteq [
\abs{D}, b]$, o\`u  $b$ est \'el\'ement de l'alg\`ebre
g\'en\'er\'ee par $\aa$ et $[D, \aa]$.}
\newline

\noindent Cette condition est la version alg\'ebrique de la
diff\'erentiabilit\'e des coordonn\'ees.

{\cond[Finitude] $\aa$ est une pr\'e-$C^*$-alg\`ebre  et
l'ensemble $\index{hinfini@$\hh^{\infty}$}\hh^\infty\doteq
\underset{k\in \nn} \cap \text{ Dom } D^k$ des vecteurs lisses de
$\hh$ est un module projectif fini.}
\newline

\noindent Un $\aa$-module est libre quand il a une base
$\{e_i\}$, c'est \`a dire un ensemble de g\'en\'erateurs tels que
$a^i e_i = 0$ pour $a^i\in\aa$ implique $a^i = 0$ pour tout $i$.
Un module {\it projectif} est une somme directe de modules
libres. Un tel module n'est pas forc\'ement libre. Il est {\it
fini} lorsqu'il a une famille g\'en\'eratrice de cardinalit\'e
finie.  Une pr\'e-$C^*$-alg\`ebre est une sous alg\`ebre d'une
$C^*$-alg\`ebre, stable par le calcul fonctionnelle holomorphe
(cf [\citelow{jgb},  {\it Def. 3.26}]). En particulier $\cinf$
est une pr\'e-$C^*$-alg\`ebre.
\newline

Les quatre conditions restantes sont d'ordre alg\'ebriques et ce
sont elles qui seront discut\'ees dans les mod\`eles des
chapitres suivants. Au triplet spectral $(\aa, \hh, D)$ est
adjoint une {\it chiralit\'e} $\Gamma$ c'est \`a dire, lorsque la
dimension spectrale $n$ est paire, une $\zz_2$ graduation
$\Gamma=\Gamma^2$ de $\hh$, autoadjointe, qui anticommute avec
$D$ et commute avec  la repr\'esentation de $\aa$. On note
$\hh^{\pm}$ les sous-espaces propres de $\Gamma$. $D$ envoie un
sous-espace dense de $\hh^{\pm}$ dans $\hh^{\mp}$ si bien que,
dans la d\'ecomposition $\hh = \hh^+ + \hh^-$,
\begin{equation}
\label{dpm}D = \dm{cc} 0& D^-\\ D^+ & 0\fm\index{dplus@$D^+, D^-$}
\end{equation}
 o\`u
\begin{equation}
\label{dplus} D^+ \doteq \frac{\ii - \Gamma}{2}D\frac{\ii +
\Gamma}{2}
\end{equation}
 et $D^- = (D^+)^\dagger$. Quand $n$ est impair, $\Gamma=\ii$.

On demande \'egalement que $\hh$ soit le support d'une
repr\'esentation de l'alg\`ebre oppos\'ee $\aa^{\circ}$
(identique \`a $\aa$ en tant qu'espace vectoriel mais o\`u le
produit est invers\'e: $a^{\circ}b^{\circ} = (ba)^{\circ}$)
 impl\'ement\'ee par un op\'erateur unitaire antilin\'eaire  $J$,
$$b^{\circ} \mapsto Jb^*J^{-1},$$
tel que
\begin{equation}
\label{commutj} [a, Jb^* J^{-1}] =0.
\end{equation}
La repr\'esentation de $\aa^{\circ}$ commute avec la
repr\'esentation de $\aa$, et $\hh$ porte une repr\'esentation
$\pi$ de l'alg\`ebre involutive $\aa\ot\aa^{\circ}$
$$a\ot b^{\circ} \mapsto aJb^*J^{-1}$$
o\`u l'involution est donn\'e par $(a\ot b^{\circ})^* \doteq b^*
\ot (a^*)^\circ.$ De mani\`ere \'equivalente, on dit que $\aa$
est repr\'esent\'ee
 \`a gauche et $\aa^{\circ}$ \`a droite
$$a\psi b \doteq aJb^*J^{-1} \psi = Jb^*J^{-1}a \psi.$$
 Si $J^2= \pm \ii$, $J b^* J^{-1} = J^{-1} b^* J$ de sorte que
\begin{eqnarray*}
\pi ( (a\ot b^\circ)^*) &=& b^* J a J^{-1} = J a J^{-1} b^*,\\
                        &=& J (a J^{-1} b^* J) J^{-1} = J (a J b^* J^{-1})J^{-1},\\
                        &=& J \pi (a\ot b^\circ) J^{-1}
\end{eqnarray*}
et $J$ impl\'emente l'involution de $\aa\ot \aa^\circ$.

{\cond[R\'ealit\'e] $(\aa\ot\aa^{\circ}, \hh, D, \Gamma, J)$ est
un $KR^n$-cycle.  $J$ est appel\'ee {\it structure r\'eelle}.}

{\cond[Premier ordre] La repr\'esentation de $\aa^{\circ}$
commute avec $[D,\aa]$
$$[[D,a], Jb^*J^{-1}] = 0 \,\text{ pour tout } \, a,b\in\aa.$$}
Cette condition stipule que l'op\'erateur de Dirac est un
op\'erateur diff\'erentiel du premier ordre.

{\cond[Orientabilit\'e] Il existe un cycle de Hochschild $c\in
Z_n(\aa, \aa\ot \aa^{\circ})$ tel que $\pi(c) = \Gamma$.}
\newline

\noindent Cette condition est la g\'en\'eralisation de la non
d\'eg\'en\'erescence de la forme volume pour une vari\'et\'e
orient\'ee. Avant de d\'efinir l'homologie de Hochschild,
rappelons qu'un {\it complexe} est une suite de $\aa$-module
$E^i$, $i\in\nn$, et de morphismes $d_i$ de $E_i$ dans $E_{i+1}$,
\begin{equation}
\label{complexe} ... \rightarrow E_{i-1}
\overset{d_{i-1}}\rightarrow E_{i} \overset{d_i}\rightarrow
E_{i+1} \rightarrow ...
\end{equation}
tels que $d_i\circ d_{i-1} = 0$. L'image d'un morphisme est
incluse dans le noyau du morphisme suivant;
 quand cette inclusion est une \'egalit\'e, $\text{Im } d_{i-1} = \ker d_i$, le complexe est {\it exact}. Sinon on note $Z_i\index{zi@$Z_i$}\doteq
 \ker d_i$ le module des {\it i-cycles} et $B_i\index{bi@$B^i$}\doteq \text{ Im }d_{i-1}$ le module des {\it i-bords}. Le quotient
 $H_i\doteq Z_i/B_i$ est par d\'efinition le {\it $i^{\text{\`eme}}$ groupe d'homologie} du complexe ($H_i$ est en fait un $\aa$-module). L'ensemble des $H_i$
  forme l'homologie du complexe.
 En rempla\c{c}ant $E_i$ par $C_i(\aa, N)\doteq N\ot\aa\ot...\ot\aa$ o\`u $N$ est un bimodule sur $\aa$ et le produit tensoriel de $\aa$ par elle-m\^eme est
 r\'ep\'et\'e $i$ fois, on a
$$... \overset{b}{\rightarrow} C_i(\aa, N) \overset{b}{\rightarrow} C_{i-1}(\aa, N) \rightarrow ... \overset{b}\rightarrow C_0(\aa, N) \overset{b}\rightarrow
  \{ 0 \}$$
  o\`u l'application $b$ de $C_i(\aa, N)$ dans $C_{i-1}(\aa,N)$ d\'efinie par
$$b(n\ot a_1\ot...\ot a_i) \doteq na_1 \ot a_2 \ot ... \ot a_i + \sum_{p=1}^{i-1} (-1)în\ot a_1 \ot...\ot a_pa_{p+1}\ot ... \ot a_i + (-1)^i a_in \ot a_1\ot
  ... \ot a_{i-1}$$
  v\'erifie $b^2 = 0$, on d\'efinit {\it l'homologie de Hochschild de $\aa$ \`a valeur dans le bimodule
$N$}. On munit $\aa\ot \aa^{\circ}$ d'une structure de
$\aa$-bimodule
  $$x(a\ot b^{\circ})y \doteq xay\ot b^{\circ} \, \text{ pour tout }\, x,y,a,b\in\aa$$
  de mani\`ere \`a d\'efinir l'homologie de Hochschild de $\aa$ \`a valeur dans $N = \aa\ot \aa^{\circ}$. Un {\it cycle de Hochschild} $c\in
  Z_n(\aa, \aa\ot \aa^{\circ})$ est un \'el\'ement de $C_n(\aa, \aa\ot \aa^{\circ})$
  tel que $b(c) =0$. Un \'el\'ement $c= a\ot b^{\circ}\ot a_1\ot...\ot a_n$ de $C_n(\aa, \aa\ot\aa^{\circ})$ est repr\'esent\'e sur $\hh$ par
  \begin{equation}
\label{pic} \pi(c) \doteq aJb^*J^{-1}[D, a_1]...[D,a_n].
\end{equation}
 Cette repr\'esentation est coh\'erente avec la proposition \ref{commutprop} qui identifie $[D,f]$ \`a la $1$-forme $df$, $1$-bord dans la cohomologie de de Rham (cf. ci-dessous). Elle est \'etendue \`a tout $C_n(\aa,
\aa\ot\aa^{\circ})$ par addition.

{\cond[Dualit\'e de Poincar\'e] Le couplage additif sur $K_*(\aa)
$ d\'etermin\'e par l'indice de l'op\'erateur de Dirac est
non-d\'eg\'en\'er\'e.}
\newline

\noindent Cette condition est une version alg\'ebrique de la
dualit\'e de Poincar\'e. Rappelons qu'en rempla\c{c}ant dans
(\ref{complexe}) $E_i$ par l'espace vectoriel r\'eel
$\Omega^i(M)$ des $i$-formes sur une vari\'et\'e compacte $M$ de
dimension $n$, et $d^i$ par la diff\'erentielle ext\'erieure $d$,
on d\'efinit un complexe dont l'homologie  est appel\'ee {\it
cohomologie de de Rham}. La dualit\'e de Poincar\'e stipule que
pour tout entier positif $r\leq n$, les groupes de cohomologie de
de Rham $H^r(M)$ et $H^{n-r}(M)$ sont duaux; c'est \`a dire qu'il
existe une forme bilin\'eaire non-d\'eg\'en\'er\'ee de
$H^r(M)\times H^{n-r}(M)$ dans $\rr$
$$\scl{[\varpi^r]}{[\varpi^{n-r}]} \doteq \int_M \varpi^r \wedge \varpi^{n-r}$$
o\`u $[\varpi^r]$ d\'esigne la classe d'\'equivalence dans
$H^r(M)$ de $\varpi^r\in\Omega^r(M)$. Gr\^ace au {\it caract\`ere
de Chern}, cette forme bilin\'eaire se traduit par un couplage
additif (i.e. une forme bi-additive) des groupes de $K$-th\'eorie
de l'alg\`ebre $\cinf$. Pour une d\'efinition de ces objets, on
peut consulter [\citelow{wegge}]. Ici, contentons nous de
souligner que ce couplage, not\'e $\cap$\index{inter@$\cap$},
s'effectue gr\^ace \`a l'indice de l'op\'erateur de Dirac
(d\'efini ci-dessous) et ne fait pas r\'ef\'erence \a la
commutativit\'e de l'alg\`ebre, de sorte qu'en rempla\c{c}ant
$\cinf$ par une pr\'e-$C^*$-alg\`ebre quelconque (pour de tels
objets, la $K$-th\'eorie existe et est identique \`a la
$K$-th\'eorie de la $C^*$-alg\'ebre obtenue par compl\'etion), la
condition 7 appara\^{\i}t comme la d\'efinition abstraite de la
dualit\'e de Poincar\'e.

Sans entrer dans le cas g\'en\'eral, pr\'ecisons un exemple qui
sera utile pour l'\'etude des espaces non commutatifs finis
(chapitre 3). Quand la dimension spectrale $n$ est paire, la
dualit\'e de Poincar\'e pour $r$ pair se ram\`ene au couplage
additif de $K_0(\aa)\index{kzeroa@$K_0(\aa)$}\times K_0(\aa)$ \`a
valeur dans $\zz$ (car pour tout $r\leq n$ pair,  $K_r(\aa)\simeq
K_{n-r}(\aa)\simeq K_0(\aa)$) d\'efini de la mani\`ere suivante.
On note $P_l(\aa)$\index{pla@$P_l(\aa)$} l'ensemble des
projecteurs de $M_l(\aa)$\index{mla@$M_l(\aa)$} (alg\`ebre des
matrices $l\times l$ \`a coefficients dans $\aa$) et $GL_l(\aa)$
les \'el\'ements inversibles de $M_l(\aa)$. On a les plongements
\'evidents
$$m\in M_l(\aa) \mapsto \dm{cc} m & 0 \\ 0 & 0 \fm \in M_{l+1}(\aa) \; \text{ et }\; v\in GL_l(\aa) \mapsto \dm{cc} v & 0 \\ 0 & 1\fm \in GL_{l+1}(\aa)$$
et on d\'efinit
$$M_\infty(\aa)\doteq \underset{l=1}{\overset{\infty}{\bigcup}} M_l(\aa)\, ,\;P_\infty(\aa)\doteq \underset{l=1}{\overset{\infty}{\bigcup}} P_l(\aa)\, , \;
GL_\infty(\aa)\doteq \underset{l=1}{\overset{\infty}{\bigcup}}
GL_l(\aa).$$ Deux projecteurs $p,q\in P_l(\aa)$ sont
\'equivalents, $p\sim q$ si et seulement si ils sont conjugu\'es
via un $v\in GL_{\infty}(\aa)$, c'est \`a dire s'il existe
$k\in\nn$ et $v\in GL_{k+l}(\aa)$ tels que
\begin{equation}
\label{equivpoinc} v\dm{cc} p & 0 \\ 0 & 0_k\fm v^{-1} = \dm{cc}
q & 0 \\ 0 & 0_k \fm.
\end{equation}
Le quotient $K_0^+(\aa)\doteq P_{\infty}(\aa) / \sim$ est un
semi-groupe (ie. les \'el\'ements ne sont pas inversibles) pour
l'addition
$$[p] + [q] \doteq \left[ \dm{cc} p& 0 \\ 0& q\fm\right] = \left[ \dm{cc} q& 0 \\ 0& p\fm\right].$$
$K_0(\aa)$ est par d\'efinition le groupe de
Grothendieck\cite{lang} de $K_0^+(\aa)$ dont les \'el\'ements
sont les classes d'\'equivalence de $K_0^+(\aa)\times K_0^+(\aa)
/\sim$, o\`u $(p,q)\sim(p',q')$ si et seulement si $p+q' = q +
p'$. $K_0(\aa)$ est un groupe pour l'addition $(p,q) + (p', q')
\doteq (p+p', q+q')$ avec l'\'el\'ement nul $(0,0)$ et l'inverse
$-(q,p)\doteq (p,q)$. C'est le m\^eme proc\'ed\'e qui permet de
construire $\zz$ \`a partir de $\nn$: $(p,q)$ s'identifiant \`a
$p-q$, on utilise la notation $\pm p$, $\pm q$ pour d\'esigner
les \'el\'ements de $K_0(\aa)$. Si $p\in M_k(\aa)$ et $q\in
M_l(\aa)$, alors $P\doteq p\ot (J\ot \ii_l)q(J^{-1}\ot\ii_l)$ est
un projecteur agissant sur $\hh\ot \cc^{kl}$ ($\aa$ est
suppos\'ee complexe) et
\begin{equation}
\label{intersection} \cap ([p], [q] ) \doteq \text{ indice} \lp
P(D\ot\ii_{kl})P\rp \doteq \dim\lp\ker P(D^+\ot\ii_{kl})P\rp -
\dim\lp\ker P(D^-\ot\ii_{kl})P\rp
\end{equation}
o\`u $D^+$, $D^-$ sont d\'efinis dans (\ref{dpm}).
\newline

{\defi \label{gnc} Un triplet spectral satisfaisant les sept
conditions ci-dessus est un {\it triplet spectral r\'eel}, ou
encore {\it une g\'eom\'etrie non commutative (spinorielle)},
not\'e $(\aa, \hh, D, \Gamma, J)$.}

\subsection{G\'eom\'etrie commutative}

On appelle {\it g\'eom\'etrie commutative} une g\'eom\'etrie non
commutative au sens de la d\'efinition \ref{gnc} o\`u l'alg\`ebre
$\aa$ est commutative. Parmi les g\'eom\'etries commutatives, les
{\it g\'eom\'etries de Dirac} sont les triplets spectraux r\'eels
\begin{equation}
\label{td} T_D = (\cinf, L_2(M,S), D, J, \Gamma)
\end{equation}
dans lesquels $(M, S, J)$ est une vari\'et\'e riemannienne
compacte orient\'ee \`a spin, $D$ est l'op\'erateur de Dirac
d\'efini par la connexion de spin et $\Gamma$ la chiralit\'e
(\ref{chiralite}) si $\text{dim }M = n$ est paire, $\Gamma=\ii$
si $n$ est impair. Les g\'eom\'etries de Dirac sont bien des
g\'eom\'etries non  commutatives, c'est \`a dire que $T_D$
v\'erifient les $7$ conditions de la section pr\'ec\'edente. On
renvoie \`a [\citelow{jgb}] pour la preuve de cette affirmation.
Contentons nous de rappeler que la dimension spectrale de la
g\'eom\'etrie de Dirac est \'egale \`a la dimension de la
vari\'et\'e. Il est \'egalement int\'eressant de s'attarder sur
la condition d'orientabilit\'e dont l'appellation trouve son
origine dans les g\'eom\'etries de Dirac.

Notons tout d'abord les simplifications dues \`a la
commutativit\'e de l'alg\`ebre.  $\cinf$ est identique \`a son
alg\`ebre oppos\'ee. $L_2(M,S)$ est donc le support de deux
repr\'esentations distinctes, la multiplication \`a gauche par
une fonction $f$, et l'action \`a droite correspondant \`a la
multiplication par la fonction complexe conjugu\'ee $\bar{f}$.
Ainsi $JfJ^{-1}\psi = \bar{f}\psi$, ce qui est coh\'erent avec
(\ref{bijc}) puisque $J f \psi = JfJ^{-1}J\psi  = \bar{f}J{\psi}$
(\`a noter le changement de convention lors du passage du
$\cinf$-module droit $S$ \`a l'espace de Hilbert $L_2(M,S)$
$\cinf$-lin\'eaire \`a gauche). Dans (\ref{pic}), identifier
$J\bar{f}J^{-1}$ \`a $f$ permet de voir le cycle de Hochschild
$c$ comme un \'el\'ement de $Z_n(\aa)$, ensemble des $n-cycles$
dans l'homologie de Hochschild du complexe
$$... \overset{b}{\rightarrow} C_i(\aa) \overset{b}{\rightarrow} C_{i-1}(\aa) \rightarrow ... \overset{b}\rightarrow C_1(\aa) \overset{b}\rightarrow
  \{ 0 \}$$
o\`u $C_i(\aa)\doteq \aa \ot ... \ot \aa$ ($\aa$ apparait $i$
fois), repr\'esent\'e par
\begin{equation}
\label{piccom} \pi( f_0 \ot f_1\ot f_i) \doteq
f_0[D,f_1]\,...\,[D,f_i].
\end{equation}

Soit $\{U_j, x_j\}$ un atlas de $M$. $x_j$ est une fonction de
$U_j \rightarrow \rr^n$ et chacune de ses composantes $x_j^\mu$
est \'el\'ement de $C^{\infty}(U_j)$. La forme volume (\ref{nug})
est l'unique $n$-forme qui, \'evalu\'ee sur toute base
orthonorm\'ee et orient\'ee de $TM$, vaille $1$. Localement,
 $$\nu_g = \sqrt{\text{det }{g}}\; dx_j^1 \wedge ... \wedge dx_j^n$$
o\`u $g$ est la matrice de composante $g(\partial_\mu,
\partial_\nu)$ et (en omettant l'indice $j$) $\{\partial_\mu\}$
est la base locale de $TM$. La forme volume est ind\'ependante du
choix des coordonn\'ees sur l'ouvert $U_j$. Lorsque $\{
\theta_j^\alpha = e_{j\mu}^\alpha dx_j^\mu\}$
 est une base locale orthonorm\'ee de $1$-formes, $g$ est la matrice identit\'e et
$$\nu_g = \theta_j^1 \wedge ... \wedge \theta_j^n = e_j dx_j^1 \wedge ... \wedge dx_j^n$$
o\`u $e_j\doteq \text{det }\left\{  e_{j\mu}^\alpha \right\}$. On
pose $e^0_j\doteq i^{n-m}f_j e_j\in C^{\infty}(U_j)$, o\`u
$m\doteq [n/2]$ et $f$ est une {\it partition de l'unit\'e},
c'est \`a dire un ensemble $\{f_j\in C^{\infty}(U_j)\}$ de
fonctions telles que
$$0\leq f_j(x) \leq 1,\quad f_j(x) = 0 \, \text{ pour } x\notin U_j,\quad \sum_j f_j(x) = 1 \text{ pour tout } x\in M.
$$
On en d\'eduit l'\'ecriture non-locale de l'\'el\'ement de volume,
$$i^{n-m}\nu_g = \sum_j e^0_j dx_j^1 \wedge... \wedge dx_j^n.$$
Le $n$-cycle de Hochschild $c\in Z_n(\aa)$ correspondant est, par
d\'efinition,
$$c\doteq \frac{1}{n!} \sum_{\sigma\in S_n} (-1)^\sigma \sum_j e^0_j\ot x_j^{\sigma(1)}\ot ... \ot x_j^{\sigma(n)},$$
o\`u $S_n$ est le groupe des permutations de $\{1,..., n\}$ et
$\sigma$ d\'esigne \`a la fois un \'el\'ement de $S_n$ et sa
parit\'e (l'exposant $\sigma$ \'egale $\pm 1$ selon que la
permutation $\sigma$ est paire ou impaire).  Par (\ref{piccom}),
en utilisant la proposition \ref{commutprop}, on v\'erifie que
\begin{eqnarray*}
\pi(c) &=& \frac{1}{n!} \sum_{\sigma\in S_n} (-1)^\sigma \sum_j e_j^0 [D, x_j^{\sigma(1)}]\, ... \, [D, x_j^{\sigma(n)}]\\
       &=& \frac{(-i)^n}{n!} \sum_{\sigma\in S_n} (-1)^\sigma \sum_j e_j^0  c(d x_j^{\sigma(1)})\, ... \, c(dx_j^{\sigma(n)})\\
       &=& \frac{(-i)^m}{n!} \sum_j f_j \sum_{\sigma\in S_n} (-1)^\sigma c(\theta_j^{\sigma(1)})\, ... \, c(\theta_j^{\sigma(n)})\\
       &=& (-i)^m \sum_j f_j  c(\theta_j^{1})\, ... \, c(\theta_j^{n}) \\
       &=& c(\gamma) \sum_j f_j = \Gamma.
\end{eqnarray*}
Ainsi la chiralit\'e  est bien l'image du cycle de Hochschild
correspondant \`a l'\'el\'ement de volume.
\newline

Si $\aa$, $D$, $J$ et $\Gamma$ commutent avec un projecteur $p$
de $\hh$, alors la g\'eom\'etrie non commutative $(\aa, \hh, D,
J, \Gamma)$ peut s'\'ecrire comme somme directe de deux
g\'eom\'etries non commutatives d\'efinies sur $p\aa$ et
$(\ii-p)\aa$ (pour des g\'eom\'etries de Dirac, ceci correspond
\`a une vari\'et\'e non connexe). Pour \'eviter ces cas, on dit
qu'une g\'eom\'etrie non commutative  est {\it irr\'eductible}
lorsqu'il n'y a pas de projecteur non nul commutant avec $\aa$,
$D$, $J$ et $\Gamma$. Ainsi toute vari\'et\'e \`a spin connexe de
dimension $n$ d\'efinit par (\ref{td}) une g\'eom\'etrie de Dirac
irr\'eductible, c'est \`a dire une g\'eom\'etrie non commutative
irr\'eductible, de dimension spectrale $n$, d\'efinie sur
l'alg\`ebre $\cinf$. A l'inverse, selon le th\'eor\`eme suivant
\'enonc\'e dans [\citelow{gravity}] et dont on trouve une
d\'emonstration d\'etaill\'ee dans [\citelow{jgb}], toute
g\'eom\'etrie non commutative sur $\cinf$, irr\'eductible et de
dimension spectrale $n$ est une g\'eom\'etrie de Dirac pour une
vari\'et\'e \`a spin.

{\thm \label{thconnes}Soit $T=(\aa, \hh, D, J, \Gamma)$ une
g\'eom\'etrie non commutative irr\'eductible sur $\aa = \cinf$,
de dimension spectrale $n =\text{dim } M$ o\`u $M$ est une
vari\'et\'e compacte  orient\'ee connexe sans bord. Alors
\begin{itemize}
\item Il existe une unique m\'etrique riemannienne $g=g(D)$ sur $M$ telle que la distance g\'eod\'esique sur $M$ soit donn\'ee par
\begin{equation}
\label{distancecommut} d(x,y) =\suup{f\in C(M)} \{ f(x) - f(y)\,
/ \, \norm{[D, f]}\leq 1 \}.
\end{equation}
\item $M$ est une vari\'et\'e \`a spin et les op\'erateurs $D'$ pour lesquelles $g(D') = g(D)$ forment une union d'espaces affines identifi\'es par les
structures de spin sur $M$.
\item La fonctionnelle $S(D)\doteq {\int\!\!\!\!\!\!{\--}} \abs{D}^{-n+2}$ d\'efinit une forme quadratique sur chacun de ces espaces affines, atteignant
son minimum pour $D = D_s$, l'op\'erateur de Dirac correspondant
\`a la structure de spin; ce minimum est proportionnel \`a
l'action d'Einstein-Hilbert, c'est \`a dire l'int\'egrale de la
courbure scalaire $s$
$$S(D_s) = -\frac{n-2}{24}\int_M s\sqrt{\text{det }g}\,d^nx.$$
\end{itemize}}

\noindent Les deux et troisi\`eme points n\'ecessitent quelques
explications. La structure de spin de $M$ est donn\'ee par le
bimodule $S = \hh^{\infty}$ d\'efini par l'op\'erateur $D$ (cf.
condition de finitude) et l'op\'erateur $J$. En r\`egle
g\'en\'erale, l'op\'erateur de Dirac $D_s$ correspondant \`a cette
structure de spin n'est pas l'op\'erateur $D$. La seule chose
qu'on puisse affirmer est que
$$D = D_s + \rho$$
pour $\rho\in\text{End}(\ginf(S))$ v\'erifiant
\begin{equation}
\label{condrho} \rho^\dagger = \rho,\; \Gamma\rho = (-1)^n \rho
\Gamma,\; J\rho J^{-1} = \pm \rho,
\end{equation}
le signe \'etant n\'egatif lorsque et seulement lorsque $n=1$ ou
$5$ mod $8$. Tout $\rho'$ satisfaisant (\ref{condrho}) d\'efinit
un op\'erateur $D'\doteq D_s + \rho'$ tel que $g(D') = g(D)$.
Ainsi pour la structure de spin donn\'ee par $T$, l'ensemble des
op\'erateurs d\'eterminant la m\^eme m\'etrique que $D$ est un
espace affine. Si maintenant on consid\`ere une vari\'et\'e
riemannienne $M$ o\`u la  m\'etrique $g$ est fix\'ee, il existe
plusieurs structures de spin sur $M$ (le nombre de structure de
spin est fini et est d\'etermin\'e par la cohomologie  de Cech de
$M$). Fixer une structure de spin d\'etermine de mani\`ere unique
l'op\'erateur $D_s$, et  les op\'erateur $D'$ du type $D_s +
\rho$ d\'efinissent un ensemble de g\'eom\'etries \'equivalentes.
Ainsi les structures de spin d'une vari\'et\'e riemannienne $M$
permettent de classifier, au regard de la topologie,  les
g\'eom\'etries sur $M$.

La fonctionnelle du troisi\`eme point est d\'efinie par
$${\int\!\!\!\!\!\!{\--}} \abs{D}^{-n+2} \doteq \frac{1}{2^{[n/2]}\Omega_n} \text{Wres } \abs{D}^{-n+2}
$$
o\`u $\Omega_n$ est l'int\'egrale de la forme volume sur la
sph\`ere $S^n$ et $\text{ Wres}$ est le {\it r\'esidu de
Wodzicki} (cf [\citelow{jgb}, {\it Th. 7.5}] pour une
d\'efinition).
\newline

Le triplet spectral r\'eel est un outil permettant de classifier
les g\'eom\'etries spinorielles sur une vari\'et\'e compacte
(sans bord). L'avantage de cette formulation alg\'ebrique est que
la d\'efinition \ref{gnc} est valable pour des
pr\'e-$C^*$-alg\`ebres quelconques, pas forc\'ement
commutatives.  Dans la premi\`ere partie de ce chapitre, on a vu
que les $C^*$-alg\`ebre non commutatives \'etaient des candidats
s\'erieux pour jouer le r\^ole de fonctions sur un espace non
commutatif. De m\^eme que le th\'eor\`eme de Gelfand, par
analogie avec le cas commutatif, justifie le choix des \'etats
purs d'une $C^*$-alg\`ebre comme points d'un espace non
commutatif, de m\^eme le th\'eor\`eme \ref{thconnes} sugg\`ere
que le triplet spectral r\'eel est un bon outil pour faire la
g\'eom\'etrie de ces espaces non commutatifs.  En particulier, et
c'est l'objet de cette th\`ese, la formule (\ref{distancecommut})
dans sa
 formulation g\'en\'erale d\'efinit une distance sur l'espace des \'etats d'une alg\`ebre.

\section{La distance}

\subsection{La formule de la distance}

Classiquement, la distance entre deux points $x$, $y$ est la
longueur du plus court chemin reliant $x$ \`a $y$. Physiquement
cette mani\`ere de voir n'est pas acceptable
 car la m\'ecanique quantique invalide l'id\'ee d'un chemin entre deux points. D'autre part un point n'est pas accessible \`a l'exp\'erience autrement que par l'interm\'ediaire
 d'une observable. Pour concilier g\'eom\'etrie et m\'ecanique quantique, il faudrait donc d\'efinir une distance $d(x,y)$ qui ne fasse r\'ef\'erence qu'aux
 valeurs prises par les observables sur $x$ et $y$. Qu'apparaissent des valeurs d'observables sur d'autres points $p$ est tol\'er\'e,
 \`a condition que les-dits points soient caract\'eris\'es autrement que par une appartenance \`a un chemin entre $x$ et $y$. Par ailleurs une distance est par d\'efinition
 une fonction de deux variables \`a valeur r\'eelle, positive, sym\'etrique, r\'eflexive ($d(x,x)=0$) et qui  v\'erifie l'in\'egalit\'e triangulaire.
 La mani\`ere la plus simple d'impl\'ementer ces propri\'et\'es au niveau des observables est de consid\'erer une quantit\'e du type
$\abs{f(x) - f(y)}$ o\`u $f$ est une fonction complexe  sur
l'espace. Dans le cas le plus simple de la droite r\'eelle,
$d(x,y)=\abs{x-y}$. Pour que $\abs{f(x) - f(y)} = \abs{x-y}$ il
faut au moins que
\begin{equation}\label{principe}
\abs{f'(p)} =1 \text{ pour un point $p$ du segment $[x,y]$,}
\end{equation}
$f'$ d\'esignant la d\'eriv\'ee de $f$. Caract\'eriser $p$ par
son appartenance au segment $[x,y]$ viole les principes d'une
"bonne" distance au sens quantique. Heureusement la condition
(\ref{principe}) peut s'exprimer ind\'epen\-damment de $x$ et
$y$. En posant
\begin{equation}
\label{distreelle} d(x,y) = \suup{f\in C^1(\rr)} \left\{
\abs{f(x) - f(y)}\; / \, \abs{f'(p)}\leq 1\, ,\forall p\in \rr\,
\right\}
\end{equation}
o\`u $C^1(\rr)$ est l'ensemble des fonctions d\'erivables sur
$\rr$, on v\'erifie ais\'ement que $d(x,y)= \abs{x-y}$, le
supr\'emum \'etant atteint par la fonction de d\'eriv\'ee
constante (\'egale \`a $1$): $x\mapsto x$.

Dans cette formule les points demeurent les objets premiers (non
seulement parce qu'il s'agit d'une distance entre points, mais
aussi parce que sont privil\'egi\'ees  des valeurs d'observables
en des points pr\'ecis). Cependant le th\'eor\`eme de Gelfand
assure qu'un points $x$ de l'espace
 n'est rien d'autre qu'un \'etat pur $\omega_x$ de l'alg\`ebre commutative des observables continues sur cet espace. En repr\'esentant $\aa =
C^1(\rr)$ sur l'espace des fonctions r\'eelles de carr\'e
sommable $\hh = L_2(\rr)$ par simple multiplication point par
point, (\ref{distreelle}) s'\'ecrit
\begin{equation}
\label{distcorse} d(x,y) = \suup{f\in\aa} \left\{
\abs{\omega_x(f) - \omega_y(f)}\, / \, \norm{\left[\frac{d}{dx},
f\right]} \leq 1 \right\}
\end{equation}
o\`u  $\frac{d}{dx}$ est l'op\'erateur de d\'erivation sur
$L_2(\rr)$. Pour s'en convaincre, il suffit de remarquer que pour
tout $\psi\in\hh$,
$$[\frac{d}{dx}, f]\psi = \frac{d}{dx} f\psi - f\frac{d}{dx}\psi = -\lp \frac{d}{dx}f\rp \psi = -f'\psi$$
d'o\`u
$$\norm{\left[\frac{d}{dx}, f\right]}= \suup{\psi\in\hh} \frac{\norm{f'\psi}}{\norm{\psi}} = \suup{\psi\in\hh}
\lp \frac{\int_\rr \abs{f'(x)}^2\abs{\psi(x)}^2 dx}{\int_\rr
\abs{\psi(x)}^2dx}\rp^{\frac{1}{2}} = \suup{x\in
\rr}\abs{f'(x)}.$$ En rempla\c{c}ant $\rr$ par une vari\'et\'e
riemannienne \`a spin $M$, $C^1(\rr)$ par $\aa = C(M)$,
$L_2(\rr)$ par $\hh= L_2(M,S)$ et $\frac{d}{dx}$ par un
op\'erateur $D$ tel que $(\aa, \hh, D)$ soit un triplet spectral
au sens de la d\'efinition \ref{tripletspectral},
(\ref{distcorse}) est identique \`a (\ref{distancecommut}). Cette
d\'efinition de la distance g\'eod\'esique, en apparence plus
complexe que la d\'efinition usuelle, est en fait plus pr\'ecise
puisqu'elle se g\'en\'eralise imm\'ediatement \`a tout triplet
spectral.\cite{metrique,connes}

{\defi Soit $(\aa, \hh, D)$ un triplet spectral. La distance $d$
entre deux \'etats $\tau_1$ et $\tau_2$  est
\begin{equation}
\label{distance} d(\tau_1,\tau_2)\doteq \sup_{a\in\aa} \left\{ \,
\abs{\tau_1(a) -\tau_2(a)}\, /\, \norm{[D,a]}\leq 1 \right\}.
\end{equation}}
On v\'erifie imm\'ediatement que cette distance est positive,
sym\'etrique
  et reflexive, et presque imm\'ediate\-ment
qu'elle satisfait l'in\'egalit\'e triangulaire puisque
\begin{eqnarray*}
d(\tau_1, \tau_2) &=& \sup_{a\in\aa} \left\{ \, \abs{\tau_1(a) -\tau_2(a)}\, /\, \norm{[D,a]}\leq 1 \right\},\\
                    &\leq&  \sup_{a\in\aa} \left\{ \,\abs{\tau_1(a) - \tau_3(a)} + \abs{\tau_3(a) -\tau_2(a)}\, /\, \norm{[D,a]}\leq 1\right\},\\
                    &\leq&  \sup_{a\in\aa} \left\{ \,\abs{\tau_1(a) - \tau_3(a)}\, /\, \norm{[D,a]}\leq 1 \right\} + \sup_{a\in\aa} \left\{ \,
\abs{\tau_1(a) - \tau_3(a)}\, /\, \norm{[D,a]}\leq 1\right\},\\
    &\leq& d(\tau_1, \tau_3) + d(\tau_3, \tau_2).
\end{eqnarray*}

A noter que cette d\'efinition n'impose pas au triplet spectral
d'\^etre r\'eel, la seule condition indispensable est que le
commutateur $[D,a]$ reste born\'e pour tout $a$. Dans le chapitre
suivant, on \'etudiera des exemples de distance associ\'ee \`a
des triplets r\'eels et \`a d'autres non r\'eels. De m\^eme $\aa$
n'est pas n\'ecessairement une (pr\'e)-$C^*$-alg\`ebre. Cependant
les propri\'et\'es des $C^*$-alg\`ebres, et \`a plus forte raison
celles des $W^*$-alg\`ebres, permettent de mener bon nombre de
calculs \`a terme. De plus c'est ce type d'alg\`ebre qui
s'interpr\`ete comme fonction sur l'espace non commutatif, on
s'int\'eresse donc dans la suite essentiellement aux triplets
spectraux sur des $C^*$-alg\`ebres.

Dans le cas commutatif, soulignons que (\ref{distancecommut}),
qui fait intervenir l'alg\`ebre des fonctions continues, n'est
pas la traduction exacte de (\ref{distance}) appliqu\'ee au
triplet (\ref{td}) construit sur l'alg\`ebre des fonctions
lisses. La formulation de (\ref{distancecommut}) est emprunt\'ee
\`a [\citelow{jgb}] qui reprend [\citelow{connes}] o\`u cette
formule est donn\'ee avec pour alg\`ebre $\aa$ l'alg\`ebre des
fonctions born\'ees mesurables sur $M$ (dense dans $C(M)$). Dans
[\citelow{gravity}], la formule de la distance est donn\'ee
directement pour $\aa=\cinf$. Ce point est discut\'e dans la
section I du chapitre 3.

\subsection{Positivit\'e et condition sur la norme}

L'id\'ee que la distance pour la droite r\'eelle est
"r\'ealis\'ee" par une fonction positive de d\'eriv\'ee partout
\'egale \`a $1$ prend dans le cas g\'en\'eral la forme  d'un
lemme extr\`emement utile pour les calculs explicites. On montre
que la distance non commutative est r\'ealis\'ee par un
\'el\'ement $a$ positif de l'alg\`ebre tel que la norme du
commutateur $[D,a]$ \'egale $1$. Quelques pr\'eliminaires simples
sont n\'ecessaires. Dans tout ce qui suit $(\aa, \hh, D)$ est un
triplet spectral dans lequel $\aa$ est une $C^*$-alg\`ebre,
$\tu$, $\td$ deux \'etats de $\aa$ et $d(\tu, \td)$ la distance
d\'efinie par (\ref{distance}). On note $\dd$ l'ensemble des
\'el\'ements de
 l'\alg satisfaisant la condition sur la norme:
 $$\dd \doteq \left\{ a \in \aa\, / \, \norm{D, a} \leq 1\right\}.
 $$

 {\lem 1)
 \label{tiaf} $d(\tau_1, \tau_2) = 0$ si et seulement si $\tau_1 = \tau_2.$

 \hspace{1.8truecm} 2) S'il existe $a$ tel que $\, [D,a] = 0\; $ et $\tu(a)\neq \td(a)$ alors  $d(\tu, \td)  = +\infty$.}
\newline

 \noindent{\it Preuve.}
 1) Si $d(\tau_1, \tau_2) = 0$, alors $\tau_1$ et $\tau_2$ coincident sur  $\dd$. Pour $a\notin \dd$,
$\norm{[D,a]}\neq 0$ et $\frac{a}{\norm{[D,a]}} \in\dd$ donc
$$\tu(\frac{a}{\norm{[D,a]}}) =  \td(\frac{a}{\norm{[D,a]}})$$
 d'o\`u, par lin\'earit\'e, $\tu(a) = \td(a)$. A l'inverse il est \'evident que si $\tu = \td$ alors $d(\tu, \td) = 0$.

2) 
Posons $\lambda=\tu(a) - \td(a)\neq 0$. Soit $r$ est un r\'eel
positif non nul. $[D,ra] = 0$ donc $ra\in\dd$. Comme
$\abs{\tu(ra) - \td(ra)}=r\abs{\lambda}$,
\begin{equation}
\label{plusgd} d(\tu, \td)\geq r\abs{\lambda}.
\end{equation}
Le r\'esultat est prouv\'e en faisant tendre $r$ vers l'infini.


{\lem \label{positif} Si $d({\tau_1} , \tau_2)$ est finie alors
$d({\tau_1} , \tau_2) = \underset {a \in \mathcal{A}_{+}} {\sup}
\, \{  \, \abs{\tau_1(a) - \tau_2(a)}\;\, / \;\, \norm{[D,a]} = 1
\}.$}
\newline

{\it Preuve.} Si $D$ commute avec l'\alg toutes les distances
sont infinies donc il existe au moins un \'el\'ement $b$ tel que
$[D, b]\neq 0$. A noter que si $[D, b + b^*]= 0$, alors $[D, i(b
- b^*)]\neq 0$ de sorte qu'il existe au moins un \'el\'ement
autoadjoint $b_h$ qui ne commute pas avec $D$. On note  $\ccc$
l'ensemble des $a$ de $\aa_+$ tel que $\norm{[D,a]}=1$ . Puisque
$\ccc\subset\dd$,
\begin{equation}
\label{inegall1} \suup{a\in\ccc}\, \abs{\tau_1(a) -
\tau_2(a)}\leq \suup{a\in\dd}\, \abs{\tau_1(a) - \tau_2(a)}=
d(\tu, \td).
\end{equation}

Si $\tu = \td$, la preuve est imm\'ediate. ÒPour $\tu \neq \td$,
on note $K_{12}$ l'ensemble des $a\in\aa$ tels que $\tau_{12}(a)
\doteq \tu(a) - \td(a) \neq 0$ et $\bb \doteq \dd\setminus
K_{12}.$  Clairement
\begin{equation}
\label{inegall2} d(\tu, \td) = \suup{a\in \bb} \left\{
\abs{\tau_{12}(a)}\right\}.
\end{equation}
Pour tout $a\in\bb$ posons
$$a_h=\frac{1}{2}({a e^{-i\theta} + a^*e^{i\theta}})$$
o\`u $\theta\doteq \text{arg}\left( \tau_{12}(a) \right)$. Noter
que
$$\tau_{12}(a_h) = \frac{1}{2} \lp \abs{\tau_{12}(a)} + \abs{\tau_{12}(a)}\rp = \abs{\tau_{12}(a)} \neq 0$$
de sorte que
$$0 < \norm{[D, a_h]} \leq \norm{[D,a]} = 1$$
(la partie droite est l'in\'egalit\'e triangulaire,
l'in\'egalit\'e de gauche est stricte sinon $d(\tu, \td)$ serait
infinie en vertu du lemme \ref{tiaf}). Comme
$$a_+\doteq \frac{a_h }{\norm{[D,a_h]}} + \norm{ \frac{a_h }{\norm{[D,a_h]}}}\ii  \in \ccc$$
satisfait
$$\abs{\tau_{12}(a_+)} = \frac{\abs{\tau_{12}(a_h)}}{\norm{[D,a_h]}} = \frac{\abs{\tau_{12}(a)}}{\norm{[D,a_h]}} \geq  \abs{\tau_{12}(a)},$$
il apparait qu'\`a tout \'el\'ement $a$ de $\bb$ est associ\'e un
\'el\'ement $a_+$ de $\ccc$ tel que $\abs{\tau_{12}(a)}\leq
\abs{\tau_{12}(a_+)}$. Ainsi
$$\suup{a\in\bb}\abs{\tau_{12}(a)}\leq \suup{a\in\ccc}\abs{\tau_{12}(a)}.$$
Avec (\ref{inegall1}) et (\ref{inegall2}), il vient $d(\tau_1,
\td) = \suup{a\in\ccc}\abs{\tau_{12}(a)}$, ce qui est
pr\'ecis\'ement le r\'esultat. \hfill $\blacksquare$
\newline
\newline

Pour clore cette pr\'esentation g\'en\'erale de la formule de la
distance, citons un corollaire du lemme \ref{tiaf} pour un
triplet spectral $(\aa, \hh, D)$
 o\`u $\aa$ est une $W^*$-alg\`ebre. Il permet d'isoler certains \'etats purs normaux pathologiques.

{\cor \label{finitude2} Soient $\omega$ un \'etat pur normal
d'une $W^*$-\alg $\aa$ et $s_\omega$ son support. Si $[D,
s_\omega]=0$, alors $\omega$ est \`a une distance infinie de tous
les autres \'etats purs normaux.}
\newline

\noindent{\it Preuve.} On note $\ou = \omega$, $s_1= s_\omega$.
D'apr\`es le lemme \ref{tiaf}, il suffit de montrer que $\ou(s_1)
\neq \od(s_1)$ pour tout \'etat pur $\od$ distinct de $\ou$.
Selon (\ref{support}) (avec $a= \ii$) c'est simplement prouver que
$\od(s_1)\neq1$. Supposons donc que $\od(s_1) = 1$ et montrons
qu'alors $\ou = \od$. La preuve est analogue \`a  celle du lemme
\ref{rangun}.
 Soit $\{\pi_2, \hh_2\}$ la repr\'esentation GNS de $\od$ et $\xi_2 = \underline{\ii}$. Puisque $\od(s_1) = \scl{\xi_2}{\pi_2(s_1)\xi_2} = 1$, $\pi_2(s_1)$ est
non nul et donc un projecteur de rang $1$. Soit $\xi$ un vecteur
propre normalis\'e de $\pi_2(s_1)$. On sait que l'\'etat pur
$\ox$ induit par $\xi$ n'est autre que $\ou$.

Par ailleurs comme $\pi_2(s_1)\eta = \scl{\xi}{\eta}\xi$ pour
tout vecteur $\eta$ de $\hh_2$,
$$\scl{\xi_2}{\pi_2(s_1)\xi_2} = \abs{\scl{\xi_2}{\xi}}^2 = 1 = \norm{\xi}^2\norm{\xi_2}^2.$$
Par l'in\'egalit\'e de Cauchy-Schwarz\cite{kadison} il existe un
nombre complexe $\lambda$ de module $1$ tel que $\xi_2 = \lambda
\xi$. Ainsi
$$\od(a) = \scl{\lambda\xi}{\pi_2(a)\lambda\xi} = \scl{\xi}{\pi_2(a)\xi} = \ox(a)= \ou(a)$$
pour tout $a\in \aa$. D'o\`u le r\'esultat.
 \hfill$\blacksquare$
\newline

\section{ Isom\'etrie}

\subsection{Sym\'etrie de l'espace non commutatif}\label{symetrie}

Au sens usuel, une sym\'etrie d'un espace est une transformation
sous laquelle l'espace est globa\-lement invariant. Au sens
topologique, il s'agit d'un hom\'eomorphisme. L'ensemble $\text{
Hom} (X)$ des hom\'eomorphismes d'un espace topologique $X$ est
un groupe pour la loi de composition des applications. Comme  la
proposition \ref{homeoiso} introduit une \'equivalence entre
l'identification topologique de deux espaces (compacts) $X$ et $Y$
\-- $X$ hom\'eomorphe \`a $Y$ \-- et l'identitification de leur
alg\`ebre de fonctions continues  \--$C(X)$ isomorphe \`a
$C(Y)$\-- il est naturel de chercher une correspondance entre les
sym\'etries de $X$ et des "sym\'etries" de $C(X)$. Une alg\`ebre
n'\'etant a priori pas un espace topologique, une sym\'etrie
d'\alg n'est pas d\'efinie en terme d'hom\'eomorphisme mais
plut\^ot en terme de bijection pr\'eservant la structure
alg\'ebrique (et l'involution s'il y a lieu). Plus
pr\'ecis\'ement on note $\text{Aut}(\mathcal{A})$ l'ensemble des
automorphismes d'une \alg involutive $\aa$ sur le corps $\kk$,
c'est \`a dire l'ensemble des applications $\alpha$ de
$\mathcal{A}$ dans $\mathcal{A}$,  $\kk$-lin\'eaires,
inversibles, telles que  $\alpha(ab)=\alpha(a)\alpha(b)$ et
$\alpha(a^*)=\alpha(a)^*$. $\text{Aut} (\aa)$ est un groupe pour
la composition des applications. Par application imm\'ediate de
la proposition \ref{homeoiso}, on obtient la correspondance
d\'esir\'ee entre sym\'etries topologiques et sym\'etries
alg\'ebriques.

{\cor Le groupe des automorphismes d'une $C^*$-\alg commutative
est isomorphe au groupe des hom\'eomorphismes de son espace de
caract\`eres.}
\newline

\noindent Lorsque $\mathcal{A}=\cinf$ est l'alg\`ebre des
fonctions lisses sur une vari\' et\' e compacte $\mathcal{M}$, on
a\cite{jgb}
$$ \difm \simeq \text{ Aut}(\cinf).$$
$\text{ Aut}(\mathcal{A})$ pour une \alg $\aa$ non commutative
s'interpr\`ete ainsi comme le groupe des "diff\'eomor\-phismes"
de l'espace non commutatif.

L'action d'un automorphisme sur l'espace non commutatif se traduit
par un changement de repr\'esentation dans le triplet spectral
$$ \pi \longrightarrow \pi\circ\alpha.$$
Dire que $\alpha$ est une sym\'etrie, c'est dire que $(\aa, \hh,
D, \pi)$ et $(\aa, \hh, D, \pi\circ\alpha)$ d\'ecrivent des
"espaces non commutatifs identiques". Les guillements sont de
rigueur car la notion "d'espaces non commutatifs identiques" n'a
pas \'et\'e d\'efinie.  En terme alg\'ebrique, on pr\'ef\`ere
parler d'\'equivalence (unitaire).

{\defi \label{equivalencetriplet} Deux triplets spectraux r\'eels
$(\aa,\hh,D, J , \Gamma)$ et $(\aa',\hh',D', J', \Gamma')$ sont
dits (unitai\-rement) \'equivalents s'il existe un op\'erateur
unitaire $U$ de $\hh$ dans $\hh'$ et un isomorphisme d'alg\`ebre
$\phi$ de $\aa$ sur $\aa'$ tels que
\begin{eqnarray*}
&UDU^*= D',\;  U\pi U^*= \pi'\circ \phi,&\\
&UJU^*=J',\;  U\Gamma U^*=\Gamma'.&
\end{eqnarray*}}

L'\'equivalence de triplets spectraux non r\'eels est d\'efinie
pareillement en omettant les conditions sur $J$ et $\Gamma$. En
posant $\phi=\alpha^{-1}$ et $U=\ii$, on v\'erifie que tout
automorphisme $\alpha$ est bien une sym\'etrie au sens non
commutatif, mais pas n\'ecessairement une isom\'etrie. En effet,
les distances calcul\'ees dans la g\'eom\'etrie $(\aa, \hh, D,
\pi, J, \Gamma)$ ou dans la g\'eom\'etrie \'equivalente  $(\aa,
\hh, D, \pi\circ\alpha, J, \Gamma)$ ont peu de chance d'\^etre
\'egales dans la mesure o\`u
\begin{equation}
\label{conorm1} \norm{[D, \pi(a)]}=1
\end{equation}
 n'est pas \'equivalent \`a
\begin{equation}
\label{conorm2} \norm{[D, \pi\circ\alpha(a)]}=1,
\end{equation}
 sauf si $\alpha$ est l'identit\'e de $\text{Aut}(\aa)$.

\subsection{Invariance de la distance}\label{invdeladist}

Pour trouver des sym\'etries de l'espace non commutatif qui
pr\'eservent les distances, plusieurs approchent sont possibles.
La plus naturelle, mais non la plus facile, consiste \`a s'en
tenir strictement aux d\'efinitions et \`a d\'eterminer les
automorphismes pour lesquels (\ref{conorm2}) est \'equivalent \`a
$(\ref{conorm1})$. Cette question est l'objet du paragraphe
suivant. L'autre approche s'appuie sur la remarque suivante: bien
que l'action d'un automorphisme n'ait \'et\'e envisag\'ee qu'au
niveau de la repr\'esentation, il existe une autre action de
$\alpha$, tout aussi naturelle, \`a l'int\'erieur de l'espace des
\'etats
$$\index{alphatau@$\alpha(\tau)$}\alpha(\tau) \doteq \tau\circ\alpha.$$
Comme $\alpha$ pr\'eserve l'involution, il pr\'eserve la
positivit\'e et $\alpha(\ii)=\ii$, donc $\alpha(\tau)$ est bien
un \'etat. En notant
$$d_\alpha \text{ la distance associ\'ee au triplet }
(\aa, \hh, D , \pi\circ\alpha),$$ il est imm\'ediat que tout
automorphisme, vu comme agissant \`a la fois sur l'espace des
\'etats et sur le triplet spectral, pr\'eserve les distances.
{\prop \label{distint} Pour tout automorphisme $\alpha$ de $\aa$
et tout \'etat $\tau_1$, $\tau_2$ dans $\ss(\aa)$,
$$d_\alpha(\alpha(\tu), \alpha(\td))= d(\tu, \td).$$}
\noindent{\it Preuve.} En posant $b\doteq \alpha(a)$,
$$  d_\alpha(\alpha(\tu), \alpha(\td)) = \suup{b\in\alpha(\aa)=\aa} \left\{  \abs{\tu(b) - \td(b)}\,/\,\norm{[D,b]}\leq 1 \right\} = d(\tu, \td).
\hspace{2cm} \blacksquare$$

D\'eterminer tous les automorphismes isom\'etriques au sens
$d_\alpha(\tu, \td) = d(\tu, \td)$ n'est pas ais\'e, nous allons
simplement en exhiber une certaine classe que nous emploierons
ensuite pour des calculs explicites de distance. La non
commutativit\'e de $\aa$ met en \'evidence un sous groupe normal
de $\text{Aut}(\aa)$ masqu\'e dans le cas commutatif, \`a savoir
l'ensemble des $\alpha$ pour lesquels il existe un unitaire $u$
de $\aa$ tel que
$$\alpha(a) = u a u^*.$$
Un tel automorphisme, not\'e $\alpha_u\index{alphau@$\alpha_u$}$,
est dit {\it int\'erieur}. L'ensemble
$\text{In}(\aa)\index{ina@$\text{In}(\aa)$}$ des automorphismes
int\'erieurs est un groupe pour la composition des applications,
\begin{equation}
\label{autint} \alpha_u\alpha_v=\alpha_{uv}\,,\;
\alpha_u^{-1}=\alpha_{u^*}.
\end{equation}
Quand $\aa$ est commutative $\text{In}(\aa)$ ne pr\'esente aucun
int\'eret puisqu'il ne contient que l'identit\'e. Quand ce n'est
pas l'\alg qui commute avec un unitaire $u$, mais l'op\'erateur
de Dirac, on dit que l'automorphisme int\'erieur associ\'e
$\alpha_u$ commute avec $D$. Alors $\alpha_u$ est une isom\'etrie
au sens pr\'ecis\'e au d\'ebut du paragraphe aussi bien qu'au
sens de son action sur l'espace des \'etats.

{\prop \label{isometrieunitaire} Un automorphisme int\'erieur
$\alpha_u$ commutant avec $D$ est une isom\'etrie aux sens
suivants:

$$ d_{\alpha_u}(\tu, \td) = d(\tu, \td) = d(\alpha_u(\tu), \alpha_u(\td)).$$}

\noindent {\it Preuve.} En utilisant  $\norm{[D,\alpha_{u}(a)]}=
\norm{[D,u a u^*]}= \norm{u^*[D,a]u}=\norm{[D,a]},$
\begin{eqnarray*}
d_{\alpha_u}(\tu, \td)            &=&\suup{a\in \aa}           \left\{  \abs{ \tu(a) - \td(a)}\, / \, \norm{ [D,\alpha_{u}(a)] }\leq 1\right\},\\
                                  &=&\suup{a\in \aa}           \left\{  \abs{ \tu(a) - \td(a)}\, / \, \norm{       [D,a]       }\leq 1\right\} = d(\tu, \td),\\
                                  & &\\
d( \alpha_u(\tu), \alpha_u(\td) ) &=&\suup{b\in \alpha_u(\aa)} \left\{  \abs{ \tu(b) - \td(b)}\, / \, \norm{[D,\alpha_{u^*}(b)]}\leq 1\right\},\\
                                  &=&\suup{b\in  \aa}          \left\{  \abs{ \tu(b) - \td(b)}\, / \, \norm{[D,b]              }\leq 1\right\} = d(\tu, \td),
\end{eqnarray*}
o\`u $b\doteq \alpha_{u}(a)$. \hfill $\blacksquare$
\newline

\noindent A noter que par la proposition \ref{distint},
$\alpha_u$ est aussi une isom\'etrie pour l'espace des \'etats
dans la g\'eom\'etrie $(\aa, \hh, D, \pi\circ\alpha_u)$ puisque
$d_{\alpha_u}(\alpha_u(\tu), \alpha_u(\td)) = d_{\alpha_u}(\tu,
\td)$.
\newline

Outre les automorphismes int\'erieurs, d'autres applications
remarquables de l'\alg dans elle-m\^eme sont les projections
\begin{equation}
\label{isopartielle} \ae(a)\doteq eae,\index{alphae@$\alpha_e$}
\end{equation}
o\`u $e=e^*=e^2\in \aa$. La projection du triplet spectral $(\aa,
\hh, D, \pi)$ est, par d\'efinition, le triplet
\begin{equation}
\label{tripletrestreint} (\aa_e\doteq \aea,\quad \hh_e\doteq
e\hh,\quad D_e\doteq eDe\big|_{\hh_e},\quad \pi_e\doteq
\pi\big|_{\hh_e})
\end{equation}
et on note $d_e$ la distance associ\'ee. En g\'en\'eral
$\alpha_e$ n'est pas injective si bien que la forme lin\'eaire
$\tau\circ\alpha_e$ n'est pas n\'ecessairement un \'etat, par
exemple si $e$  est dans le noyau de $\tau$ (pour une
$W^*$-alg\`ebre, il suffit de consid\'erer un \'etat dont le
support est orthogonal au projecteur). En revanche tout \'etat
(pur) $\tau_e$ de $\ae(\aa)$ est un \'etat (pur) de $\aa$ (pour
s'en convaincre il suffit d'\'ecrire $\tau_e=
\tau_e\circ\alpha_e$). En clair
$$\ss(\aa_e)=\ss(\aa_e)\circ\alpha_e\subset\ss(\aa)\, \text{ et }\,  \pp(\aa_e)=\pp(\aa_e)\circ\alpha_e\subset\pp(\aa).$$
Lorsque $e$ commute avec l'op\'erateur de Dirac, la projection
$\alpha_e$ est une isom\'etrie pour $\ss(\aa_e)$ au sens suivant.
{\prop \label{projectionlem} Soient $e$ un projecteur de $\aa$ et
$\tu, \td$ deux \'etats de $\aa_e$. Si $[D,\pi(e)]=0$ alors
$$d_e(\tu,\td)=d(\tue,\tde).$$}
\noindent {\it Preuve.} Pour tout  $a_e\in\aa_e$, $\, \norm{[D_e,
\pi_e(a_e)]}=\norm{[\pi(e) D \pi(e), \pi(a_e)]}= \norm{[ D,
\pi(a_e)]}$ d'o\`u
\begin{eqnarray*}
d_e(\tu,\td) &=&\suup{a_e\in \aa_e} \left\{\abs{(\tu-\td)(a_e)}\,
/\,
\norm{[D, \pi(a_e)]}\leq 1 \right\},\\
&\leq&\suup{a\in\aa} \left\{ \abs{(\tue-\tde)(a)}\, / \,\norm{[D,
\pi(a)]}\leq 1 \right\}= d(\tue,\tde).
\end{eqnarray*}
Cette borne sup\'erieure est atteinte par $a_e \doteq
\alpha_e(b)$ o\`u $b\in\aa$ r\'ealise le supr\'emum pour la
distance  $d$,
$$d(\tue, \tde)= \tue(b) - \tde (b)\,  \text{ et }\, \norm{[D,\pi(b)]} = 1,$$
car $\norm{[D_e, \pi_e(a_e)]}= \norm{[D, \pi(a_e)]}=
\norm{\pi(e)[D, \pi(b)]\pi(e)]}\leq \norm{[D,\pi(b)]}.$ \hfill
$\blacksquare$
\newline

\noindent Soulignons que cette proposition n'implique pas que la
distance entre deux \'etats de $\aa$ dans la g\'eom\'etrie $T =
(\aa, \hh, D)$ \'egale la distance dans la g\'eom\'etrie $T_e =
(\aa_e, \hh_e, D_e)$, ni m\^eme que la distance entre $\tu$ et
$\td$ dans $T$ soit \'egale \`a la distance entre
$\tu\circ\alpha_e$ et $\td\circ\alpha_e$  dans $T_e$ (ce qui
n'aurait aucun sens puisque, rappelons le, $\ss(\aa)\circ\alpha_e
\subset\!\!\!\!\!\!/ \, \,   \ss(\aa_e$) ). Si on d\'efinit
l'action de  la projection $\alpha_e$ dans l'espace des \'etats
par
$$\alpha_e( \tau) = \tau \text{ si } \tau\in \ss(\aa_e)\,, \;\, \alpha_e(\tau) = 0 \text{ sinon,}$$
alors une projection qui commute avec l'op\'erateur de Dirac est
une isom\'etrie,
$$d_e(\alpha_e(\tu), \alpha_e(\td)) = d (\tu, \td),$$
mais seulement sur $\alpha_e(\ss(\aa))$.

\vfill \pagebreak

\chapter{Espace fini}

Les exemples les plus simples d'espaces non commutatifs sont
associ\'es \`a des alg\`ebres de dimension finie.  On peut
r\'esoudre de mani\`ere syst\'ematique les contraintes impos\'ees
par les axiomes de la g\'eom\'etrie non commutative et \'etablir
une classification compl\`ete des triplets  spectraux
finis\cite{krajew,paschke}. L'objet de ce chapitre est de
d\'eterminer explicitement la m\'etrique de ces espaces. Il
apparait rapidement que les distances ne sont pas calculables
exactement, sauf dans certains cas dont nous pr\'esentons ici un
\'eventail. Les r\'esultats de ce chapitre ont fait l'objet de
l'article\cite{finite}. Pr\'ecisons encore une fois que la
formule (\ref{distance}) d\'efinit une distance sur l'ensemble
des \'etats d'une \alg ind\'ependamment des axiomes de la
g\'eom\'etrie non commutative, aussi dans un premier temps nous
consid\'erons des triplets spectraux $(\aa, \hh, D)$ qui ne sont
pas r\'eels. En dimension fini, $D$ et $[D,a]$ pour tout $a\in
\aa$ sont born\'es. La seule contrainte qu'on impose \`a
l'op\'erateur de Dirac est d'\^etre autoadjoint. Dans la mesure
du possible, on s'est efforc\'e de tenir \`a ce souci de
g\'en\'eralit\'e  mais les calculs devenant rapidement
impraticables, on s'est limit\'e pour les espaces finis
commutatifs aux op\'erateurs \`a entr\'ees r\'eelles. De m\^eme,
pour l'espace \`a deux points on est revenu aux axiomes afin de
s\'electionner un op\'erateur simple permettant de mener les
calculs \`a terme. Quoi qu'il en soit, l'importance des axiomes
est discut\'ee dans la derni\`ere partie du chapitre.

S'appuyant sur l'\'equivalence dans le cas commutatif entre
caract\`eres et \'etats purs, on a choisi de ne consid\'erer que
les distances entre \'etats purs.

\section{Topologie des espaces finis}

\subsection{Espace des \'etats purs}

Toute $C^*$-\alg $\aa$ de dimension finie est isomorphe \`a une
somme directe finie d'alg\`ebres de matrices\cite{goodearl} \`a
entr\'ees complexes si $\aa$ est une \alg complexe, \`a entr\'ees
r\'eelles, complexes ou quaterni\-oniques si $\aa$ est une \alg
r\'eelle.
Ainsi
\begin{equation}
\label{algfini} \aa = \underset{k= 1}{\overset{N} \bigoplus}\,
\aa_k
\end{equation}
o\`u $k, N \in\nn$, $\aa_k =\kk$ ou $M_n(\kk)$ et $\kk$ est un
symbole g\'en\'erique pour d\'esigner les corps $\rr, \cc$ ou
$\hhh$.

{\lem  \label{etatsommedirecte} Soient $\aa_1$, $\aa_2$ deux
$C^*$-alg\`ebres, alors $\pp(\aa_1 \oplus \aa_2) = \pp(\aa_1)
\cup \pp(\aa_2)$.}
\newline

\noindent {\it Preuve.} Tout \'etat pur $\ou$ de $\aa_1$
s'\'etend en un \'etat $\tau$ sur $\aa_1\oplus\aa_2$
\begin{equation*}
\label{deftau} \tau( a_1 \oplus a_2) \doteq \omega_1(a_1)
\end{equation*}
 (la positivit\'e est imm\'ediate puisque $(\aa_1 \oplus \aa_2)_+ = {\aa_1}_+ \oplus {\aa_2}_+$). Si $\tau$ n'est pas pur, il existe deux \'etats $\tau'$ et
$\tau''$ et un r\'eel $t\in ]0,1[$ tels que $\tau = t \tau' + (1-
t) \tau''.$ Ainsi
$$\omega_1(a_1) = \tau( a_1 \oplus \ii_2) =  t \tau'_1(a_1) + (1- t) \tau''_1(a_1)$$
o\`u $\tau'_1 (a_1) \doteq \tau'(a_1 \oplus \ii_2)$ et $\tau''_1
(a_1) \doteq \tau''(a_1 \oplus \ii_2)$ sont des \'etats de
$\aa_1$. Comme $\omega_1$ est pur, $\tau'_1 = \tau''_1$ d'o\`u
$$
 \tau'(0 \oplus \ii_2) = \tau'' (0 \oplus \ii_2) =  \frac{t-1}{t} \tau''(0 \oplus a_2) = 0$$
c'est \`a dire  $\tau' = \tau''$. $\tau$ est donc pur et
$\pp(\aa_1)\cup \pp(\aa _2) \subset \pp(\aa_1 \oplus \aa_2).$

Soit $\omega\in\pp(\aa_1 \oplus \aa_2)$. Si $\lambda_1 \doteq
\omega(\ii_1 \oplus 0) = 0$ alors $\tau_2 (a_2)\doteq \omega(a_1
\oplus a_2)$ est un  \'etat de $\aa_2.$ S'il n'est pas pur, alors
c'est une combinaison lin\'eaire convexe de deux \'etats
${\tau'}_2$ et ${\tau''}_2$ que l'on \'etend \`a $\aa_1
\oplus\aa_2$ par
$$\tau'(a_1 \oplus a_2) \doteq {\tau'}_2(a_2)\,,\; \tau''(a_1 \oplus a_2) \doteq {\tau''}_2(a_2).$$
 $\tau'=\tau''$ car  $\omega =t \tau' + (1- t) \tau''$ est pur. Donc ${\tau'}_2 = {\tau''}_2$ et
$\tau_2$ est pur. Un raisonnement analogue montre que  si
$\lambda_2 = \omega(0 \oplus \ii_2) = 0$, $\omega$ restreint \`a
$\aa_1$ est un \'etat pur de $\aa_1$.

Reste la possibilit\'e que $\lambda_1$ et $\lambda_2$ soient tous
deux non nuls. En ce cas $\omega = \lambda_1\, \tau' +
\lambda_2\, \tau''$ o\`u
$$\tau'(a_1\oplus a_2)  \doteq \frac{1}{\lambda_1}\omega(a_1\oplus 0)\,\text{ et }\, \tau''(a_1\oplus a_2)  \doteq \frac{1}{\lambda_2}\omega(0\oplus a_2)$$
sont tous deux des \'etats de $\aa_1 \oplus \aa_2$. Comme
$\lambda_1 + \lambda_2 = 1$ et que $\omega$ est pur, ${\tau'} =
{\tau''}$ d'o\`u
$$\omega(0\oplus a_2) = {\lambda_2} \tau''(0\oplus a_2) =  {\lambda_2} \tau'(0 \oplus a_2) = \frac{{\lambda_2}}{\lambda_1} \omega(0\oplus 0) = 0.$$  De m\^eme
$\omega(a_1 \oplus 0)= 0 $, donc $\omega$ est nul.\hfill
$\blacksquare$

\noindent Ce lemme appliqu\'e r\'ecursivement sur $\aa$ donne
\begin{equation}
\label{etatpurespacefini} \pp(\aa) = \underset{k= 1}{\overset{N}
\bigcup} \pp(\aa_k),
\end{equation}
si bien que pour connaitre $\pp(\aa)$ il suffit de connaitre les
\'etats purs de $\kk$ et $M_n(\kk)$.

\subsection{Etat de $\rr$, $\cc$ et $\hhh$}

Dans la section I.\ref{algebrereelle}, un \'etat sur une
alg\`ebre de corps de r\'ef\'erence $\kk$ est d\'efini comme une
forme $\kk$-lin\'eaire $\tau$
 \`a valeur dans $\kk$ et de norme $\norm{\tau}= \tau(\ii) = 1$. D\`es lors $\kk$ vu comme alg\`ebre sur $\kk$ n'a qu'un \'etat: l'identit\'e. Dit autrement les
alg\`ebres r\'eelle $\rr$, complexe $\cc$ et quaternionique
$\hhh$ n'ont chacune qu'un \'etat (pur). Ceci reste vrai pour
$\cc$ et $\hhh$ vues comme \alg r\'eelle.  Pour $\cc$ tout
d'abord, un \'etat ${\omega_c} \index{omegac@$\omega_c$}$ est, au
sens de la d\'efinition \ref{etatreel}, une application
$\rr$-lin\'eaire positive \`a valeur dans $\rr$ satisfaisant
${\omega_c}(1)=1$ et ${\omega_c}(x- iy)= {\omega_c}(x+iy)$, c'est
\`a dire $\omega_c(i) = -\omega_c(i) = 0$. En clair ${\omega_c}$
coincide avec la partie r\'eelle.

{\lem \label{etatquaternion} L'unique \'etat
$\omega_h\index{omegah@$\omega_h$}$ de l'\alg r\'eelle $\hhh$ est
$$\omega_h(q)= \frac{1}{2}\text{Tr} (q)$$
o\`u $ q = \dm{cc} x &-\bar{y}\\ y & \bar{x}\fm$\,  avec
$x,y\in\cc$ d\'esigne un quaternion quelconque.}
\newline

\noindent {\it Preuve.}  La repr\'esentation de $\hhh$ sur
l'espace vectoriel r\'eel de dimension quatre, dont la base
canonique $\{1,i,j,k\}$ v\'erifie $i^2=j^2=k^2=-1$, $ij=-ji=k$,
$jk=-kj=i$ et $ki=-ik=j$, est
$$q=\alpha + \beta i + \gamma j + \delta k$$
avec $\alpha,\beta,\gamma, \delta\in\rr.$ Par d\'efinition,
$$\bar{q}\doteq\alpha-\beta i - \gamma j - \delta k$$
de sorte que
$$q\bar{q} = \alpha^2 + \beta^2 + \gamma^2 + \delta^2 \in\rr^+.$$ En cons\'equence toute forme $\rr$-lin\'eaire est
positive.  Par lin\'earit\'e, un \'etat $\tau$ est compl\`etement
d\'etermin\'e par les valeurs qu'il prend sur $i, j, k$. Comme
$\tau(q)= \tau(\bar{q})$,
$$\alpha + \beta \tau(i) + \gamma \tau(j) + \delta \tau (k) =\alpha - \beta \tau(i) - \gamma \tau(j) - \delta \tau (k)$$
pour tout $\alpha, \beta, \gamma , \delta \in \rr$. Autrement dit
$$\tau (i)= \tau(j)= \tau(k) =0$$
et $\tau (q) = \alpha \doteq \omega_h (q).$  Dans la
repr\'esentation matricielle de $\hhh$ sur $\cc^2$, $x
\doteq\alpha + i \beta$ d'o\`u $\text{Tr}(q) \doteq 2 \text {Re}
(x) = 2 \alpha = 2\omega_h(q).$ \hfill $\blacksquare$
\newline

\noindent Il peut sembler peu satisfaisant de ne disposer avec
$\hhh$, sous \alg des matrices $ 2 \times 2 $ dont l'ensemble des
\'etats purs est de cardinalit\'e infini (voir paragraphe
suivant), que d'un seul \'etat pur. Afin d'\'elargir le champ
d'investigation, on peut modifier la d\'efinition d'un \'etat
r\'eel et ne plus exiger que l'involution soit conserv\'ee.
$\hhh$ a alors plusieurs \'etats $\tau$ mais toujours un seul
\'etat pur. En effet, si on note $\tau_{\tau(i)}$ la forme
lin\'eaire telle que $\tau_{\tau(i)}(i)=\tau(i)$,
$\tau_{\tau(i)}(1)=\tau_{\tau(i)}(j)=\tau_{\tau(i)}(k)= 0$ et
qu'on adopte des d\'efinitions similaires pour $\tau_{\tau(j)}$
et $\tau_{\tau(k)}$, il vient
\begin{eqnarray}
\nonumber
\tau &=& \omega_h + \tau_{\tau(i)} +  \tau_{\tau(j)} + \tau_{\tau(k)}\\
\label{quatetat}
       &=& t (\omega_h + \tau_{r\tau(i)} + \tau_{r\tau(j)} +  \tau_{r\tau(k)} )+ (1-t)
(\omega_h + \tau_{r'\tau(i)} +  \tau_{r'\tau(j)} +
\tau_{r'\tau(k)}),
\end{eqnarray}
 o\`u $t,r\in\rr/\{1\}$ et $r'\doteq
\frac{1-rt}{1-t}$. Chacun des deux facteurs de la moiti\'e droite
de (\ref{quatetat}) envoyant $1$ en $1$ est un \'etat, si bien
que $\tau$ n'est pas pur sauf si $\tau(i)=\tau(j)=\tau(k)=0$.

\subsection{Etats des alg\`ebres de matrices}

L'espace des \'etats de $\cc$ et $\hhh$ ne d\'epend pas du corps
de r\'ef\'erence. En revanche $\ss(M_n(\kk))$ en d\'epend. On note
$$\index{mnkprimek@$M_n^{\kk'}(\kk)$}M_n^{\kk'}(\kk)$$
pour d\'esigner $M_n(\kk)$ vue comme \alg sur le corps $\kk'$.
Consid\'erons d'abord les cas $\kk = \kk'$.
\newline

Une forme $\kk$-lin\'eaire $\tau$ sur $M_n^\kk(\kk)$ se
d\'ecompose de mani\`ere unique sur la base $\tau_{ij}$,
$$\tau_{ij} (m_{kl}) \doteq \delta_{il}\delta_{jk}$$
o\`u $\left\{ m_{ij}\right\}$, $i,j=1...n$ est la base canonique
de $M_n^\kk(\kk)$. Pour tout $a = a^{kl} m_{kl}$, $a^{kl}\in \kk$,
\begin{equation*}
\tau(a)= \lambda^{ij}\tau_{ij}(a) = \lambda^{ij}\tau_{ij}( a^{kl}
m_{kl}) =\lambda^{ij} \delta_{il}\delta_{jk} a^{kl} = \text{ Tr}
(\Lambda a)
\end{equation*}
o\`u $\Lambda\in M_n^\kk(\kk)$ d\'esigne la matrice de composante
$\lambda^{ij}$. A noter qu'\`a cause de la non commutativit\'e,
la notion de forme $\hhh$-lin\'eaire est ambigu\"e
(conform\'ement \`a l'usage\cite{adler} on consid\`ere la
$\hhh$-lin\'earit\'e \`a droite). Dans les trois cas r\'eel,
complexe et quaternionique un \'etat envoie $\ii_n$ en $\ii$ donc
$\text{ Tr}(\Lambda) = \ii$.

Pour $\kk= \rr$ ou $\cc$
\begin{equation}
\label{noncommuth} \tr{\Lambda a^*} = \tr{(a \Lambda^*)^*} =
\overline{\text{Tr}}(a \Lambda^*).
\end{equation}
Un \'etat est involutif, $\tr{\Lambda a^*} =
\overline{\text{Tr}}(\Lambda a)$, d'o\`u $\tr{(\Lambda^* -
\Lambda)a} = 0$ pour tout $a$. La trace est un produit scalaire
pour $M^\kk_n(\kk)$ donc $\Lambda^* = \Lambda$. On note $\left\{
\xi_i \right\}$ une base orthonorm\'ee de vecteurs propres de
$\Lambda$, $\lambda^i\in \rr$ les valeurs propres correspondantes
et $s_i$ les projecteurs associ\'es. Les $s_i$ sont des
\'el\'ements positifs de l'alg\`ebre donc
$$\tr{\Lambda s_i} = \lambda^i \geq 0.$$
De plus
\begin{equation}
\label{etatmdeux} \tr{\Lambda a} = \lambda^1 \tr{ s_1 a} + (1 -
\lambda^1) \frac{\lambda^j\tr{ s_j a }}{1 - \lambda^1}
\end{equation}
o\`u $j$ va de $2$ \`a $n$. Comme $\tr{s_1} = 1$ et $\tr{s_1
a^*a}= \tr{ (a s_1)^*( a s_1)}\geq 0$, le premier facteur du
terme de droite de (\ref{etatmdeux}) est un \'etat. De m\^eme,
parce que les  $\lambda^i$ sont positifs et leur somme vaut $1$,
le deuxi\`eme facteur est aussi un \'etat. En cons\'equence
$\tau$ n'est pur que si $\Lambda$ est de rang $1$. On pose
$\Lambda= s_\omega$ et on note $\xi$ un vecteur propre norm\'e de
$s_\omega$. Alors
\begin{equation}
\label{projecteurfini} s_\omega a s_\omega = \scl{\xi}{a\xi}
s_\omega = \tr{s_\omega a} s_\omega = \omega(a) s_\omega.
\end{equation}
A tout \'etat pur $\omega$ de $M_n^\kk(\kk)$ est associ\'e un
projecteur de rang $1$ $s_\omega$, et tout projecteur de rang $1$
d\'efinit un \'etat pur. A noter que pour $\kk= \cc$ ce
r\'esultat est conforme au chapitre $1$ (tout \'etat d'une
alg\`ebre complexe de dimension finie est normal). Un vecteur
propre norm\'e n'est d\'efini qu'\`a un facteur de module $1$
pr\`es ( $\pm 1$ dans le cas r\'eel, une phase dans le cas
complexe) ce qui prouve que l'espace des \'etats purs n'est autre
que l'espace des droites de $\kk^n$. {\lem \label{etatprojectif}
Pour $\kk = \rr$ ou $\cc$, $\; \pp(M_n^\kk(\kk)) = \kk P^{n-1}.$}
\newline

\noindent
Dans le cas r\'eel $\rr P^n$ n'est autre que la sph\`ere $S^n$
o\`u les points antipodaux sont identifi\'es. En basse dimension
on trouve\cite{jost} que $\pp(M_2(\rr)) = \rr P^1$ est
diff\'eomorphe au cercle $S^1$. Dans le cas complexe $\cc P^n$
peut-\^etre vu comme un quotient de la sph\`ere  $S^{2n+1}$ dans
la mesure o\`u tout \'el\'ement de $\cc P^n$ rencontre $S^{2n+1}$
en un cercle $S^1$. En basse dimension il apparait que
$P(M_2^\cc(\cc))=\cc P^1$ est diff\'eomorphe  \`a $S^2$.

Pour le cas quaternionique les arguments ci-dessus ne sont pas
valides pour plusieurs raisons. D'une part la notion d'\'etat
quaternionique n'a pas \'et\'e clairement identifi\'ee. Si on
s'en tient \`a la remarque de la fin de la section
I.\ref{algebrereelle},  un \'etat sur $M_n^\hhh(\hhh)$ est une
forme $\hhh$-lin\'eaire born\'ee $\tau$ de norme $\tau(\ii) = 1$.
Rien ne garantit qu'une telle forme soit involutive. D'autre part
m\^eme si on impose l'involutivit\'e, la non commutativit\'e de
$\hhh$ emp\^eche d'\'ecrire l'\'equation (\ref{noncommuth}).
Cependant on peut voir qu'un vecteur quaternionique d\'efinit
bien un \'etat. Dans le d\'etail, consid\'erons l'espace
vectoriel (\`a droite) quaternionique $\hhh^n$ muni du produit
scalaire \`a valeur dans $\hhh$
$$\scl{\xi}{\zeta}\doteq \sum_{i=1}^{n} \bar{\xi}_i \zeta_i = \overline{\scl{\zeta}{\xi}}$$
ou $\xi_i, \zeta_i \in \hhh$ sont les composantes des vecteurs
$\xi$ et $\zeta$. Ce produit scalaire d\'efinit une norme \`a
valeur r\'eelle
$$\norm{\xi}^2= \scl{\xi}{\xi}$$
et v\'erifie l'in\'egalit\'e de Cauchy-Schwarz\cite{adler}, de
sorte que pour $a\in M_n^\hhh(\hhh)$
$$\abs{\scl{\xi}{a \xi}}^2 = \overline{\scl{\xi}{a \xi}}\scl{\xi}{a \xi} = \scl{ a \xi}{\xi} \scl{\xi}{a \xi} \leq \scl{a \xi}{a\xi} \scl{\xi}{\xi} =
\norm{a\xi}^2\norm{\xi}^2.
$$
Ainsi pour un vecteur normalis\'e $\xi$ l'application
$$\tau: \; a \mapsto \scl{\xi}{a\xi}$$
est de norme
$$\norm{\tau} = \suup{ \norm{a}= 1} \norm{\tau(a)} = \suup{ \norm{a}= 1} \abs{\scl{\xi}{a\xi}}\leq  \suup{ \norm{a}= 1} \norm{a\xi} \leq \suup{ \norm{a}= 1}
\norm{a} \leq 1,$$ cette borne sup\'erieure \'etant atteinte pour
$a= \ii_n$. Par cons\'equent $\tau$ est un \'etat quaternionique
au sens de la section I.\ref{algebrereelle}. En d\'esignant par
$S^n_\hhh$ la sph\`ere quaternionique de dimension $n$, on a
$$\ss(M_n^\hhh(\hhh) \subset S_\hhh^n.$$
Contrairement aux cas r\'eel et complexe, deux vecteurs norm\'es
$\xi$ et $\xi q$ avec $\abs{q} = 1$ ne d\'efinissent pas le m\^eme
\'etat car $\scl{ \xi q}{ a \xi q} = \bar{q} \scl{\xi}{a \xi}q
\neq \scl{\xi}{a \xi}$. En particulier l'alg\`ebre
$M_2^\hhh(\hhh)$ ne permet pas de munir $\hhh P^1 \sim S^4$ d'une
m\'etrique.
\newline
\newline

Prendre $\kk\neq \kk'$ recouvre deux exemples, $M_n^\rr(\cc)$ et
$M_n^\rr(\hhh)$. Comme ce dernier n'apparait dans aucun mod\`ele
physique, nous ne l'\'etudierons pas. L'alg\`ebre r\'eelle
$M_n^\rr(\cc)$ est de dimension $2n^2$. Tout \'el\'ement $a=
a^{kl} m_{kl}$ \--$\left\{m_{kl}\right\}$ est la base canonique
de $M_2(\cc)$\-- s'\'ecrit sous forme d'une matrice r\'eelle pour
peu qu'on voit chaque nombre complexe $a^{kl}$ sous la forme
$$a^{kl} = \dm{cc} \re(a^{kl}) & \im(a^{kl}) \\  - \im(a^{kl}) & \re(a^{kl}) \fm.$$
On note
$$r_{kl}\doteq \ii_2 \ot m_{kl} \,\text{ et }\, i_{kl}\doteq \dm{cc} 0 & 1 \\ -1 & 0 \fm \ot m_{kl}.$$
 L'ensemble $\left\{ r_{kl}, i_{kl}\right\}$ est une base de l'\alg $M_n^\rr(\cc)$ et tout \'el\'ement  $a$ s'\'ecrit
$$a = \re(a^{kl})r_{kl} + \im(a^{kl})i_{kl}.$$
Une forme $\rr$-lin\'eaire $\tau$ \`a valeur dans $\rr$ se
d\'ecompose sur la base $\left\{ \tau^R_{ij},\tau^I_{ij}\right\}$
d\'efinie par
\begin{eqnarray*} \tau^R_{ij} (r_{kl}) = \delta_{il}\delta_{jk}, & & \tau^R_{ij} (i_{kl}) = 0,\\
\tau^I_{ij} (r_{kl}) = 0, & & \tau^I_{ij} (i_{kl}) =
\delta_{il}\delta_{jk}
\end{eqnarray*}
de sorte que
\begin{eqnarray*}
\tau (a)  &=& \lp \gamma^{ij} \tau^R_{ij} + \lambda^{ij} \tau^I_{ij} \rp  \lp \re(a^{kl}) r_{kl} + \im(a^{kl})i_{kl}\rp,\\
          &=& \gamma^{ij}\re(a^{kl})\delta{il}\delta_{jk} + \lambda^{ij} \im(a^{kl})\delta_{il}\delta_{jk},\\
          &=& \tr{\Gamma a_R} + \tr{\Lambda a_I}
\end{eqnarray*}
o\`u $\Gamma$ et $\Lambda$  sont les matrices r\'eelles $n\times
n$ de composantes $\gamma^{ij}$, $\lambda^{ij}$ respectivement et
$a_R$, $a_I$ les matrices de composantes $\re(a_{kl})$,
$\im(a_{kl})$. Noter que $r_{kl}^* = r_{lk}$ tandis que $i_{kl}^*
= - i_{lk}$ d'o\`u
$$a^* = \re( a^{lk}) r_{kl} - \im(a^{lk}) i_{kl}.$$
 Si $\tau$ est un \'etat r\'eel alors
$\tau(a^*) = \tau(a)$, c'est \`a dire
$$\tr{\Gamma a_R^*} - \tr{\Lambda a_I^*} = \tr{\Gamma a_R} + \tr{\Lambda a_I}$$
pour toute matrice $a^I$ et $a^R$. Comme les parties imaginaires
et r\'eelles des $a^{kl}$ sont ind\'ependantes, on peut prendre
$a_I = 0$ et $a_R$ quelconque. Il vient $\tr{(\Gamma^*-\Gamma)
a_R}=0$ d'o\`u
$$\Gamma^* = \Gamma.$$
 De mani\`ere analogue
$$\Lambda^* = -\Lambda.$$
 On note $\gamma^i\in\rr$ les valeurs propres de
$\Gamma$, $s_i$ les projecteurs associ\'es, $\lambda^i\in\rr$ les
valeurs propres de $i\Lambda$ et $t_i$ les projecteurs
correspondants, c'est \`a dire
\begin{equation}
\label{trace1} \tr{\Gamma s_i} = \gamma^i \; \text{ et } \;
\tr{i\Lambda t_i}= \lambda^i.
\end{equation}
Quand $a= \ii_n$, $a_I=0$ et $a_R$ est l'identit\'e des matrices
$n\times n$. Requ\'erir $\tau(\ii) = 1$ revient \`a imposer
\begin{equation}
\label{trace2} \tr{\Gamma} =1.
\end{equation}
Pour \'ecrire la condition de positivit\'e, notons d'abord la
table de multiplication
\begin{eqnarray*}
r_{kl}r_{k'l'}= \delta_{lk'}r_{kl'}, & & r_{kl}i_{k'l'} = \delta_{lk'}i_{kl'},\\
i_{kl}r_{k'l'}= \delta_{lk'}i_{kl'}, & & i_{kl}i_{k'l'} =
-\delta_{lk'}r_{kl'}
\end{eqnarray*}
permettant d'obtenir
\begin{eqnarray*}
a^* a &=& \lp\re(a^{lk})r_{kl} - \im(a^{lk})i_{kl}\rp \lp \re(a^{k'l'})r_{k'l'} + \im(a^{k'l'})i_{k'l'}\rp,\\
      &=& \re(a^{k'k})\re(a^{k'l'})r_{kl'} +  \re(a^{k'k})\im(a^{k'l'})i_{kl'} - \im(a^{k'k})\re(a^{k'l'})i_{kl'}   +
\im(a^{k'k})\im(a^{k'l'})r_{kl'},\\
      &=& (a_R^* a_R  + a_I^*  a_I)_{kl'}    r_{kl'} +  (a_R^* a_I - a_I^* a_R)_{kl'} i_{kl'},\\
\end{eqnarray*}
d'o\`u
$$\tau(a^*a)= \tr{\Gamma(a_R^* a_R  + a_I^*  a_I) + \Lambda(a_R^* a_I - a_I^* a_R)}.$$
Cette expression doit \^etre r\'eelle positive pour tout $a$. En
particulier lorsque $a_I = 0$ et $a_R = s_i$ on a avec
(\ref{trace1} $\gamma^i\geq 0$, ce qui prouve que $\Gamma$ est
une matrice positive.  Lorsque $a_R$ est l'identit\'e et $a_I =
t_i$, on a avec (\ref{trace2}) $1 - i\lambda^i \in \rr^{*+}$, ce
qui n'est pas possible sauf si $\lambda^i=0$.

En cons\'equence un \'etat $\tau$ de $M_n^\rr(\cc)$ d\'efinit, et
est d\'efini par, une matrice $n\times n$ r\'eelle sym\'etrique
$\Gamma$ telle que
$$\tau(a) = \tr{\Gamma a_R} = \gamma^1 \tr{s_1 a_R} + (1 - \gamma^1) \frac{\gamma^j \tr{s_j a_R}}  {1 - \gamma^1} $$
o\`u $j$ varie de $2$ \`a $n$. De m\^eme que dans
(\ref{etatmdeux}), chacun des deux facteurs du terme de droite
est un \'etat si bien que $\tau$ est pur si et seulement si
$\Gamma$ est de rang un. Deux vecteurs r\'eels norm\'es et
\'egaux \`a un signe pr\`es d\'efinissent le m\^eme \'etat pur
d'o\`u
$$\pp(M_n^\rr(\cc)) = \rr P^{n-1}.$$

Pour clore cette \'etude des \'etats des alg\`ebres de dimension
finie, pr\'ecisons que les projecteurs mis en \'evidence pour les
alg\`ebres r\'eelles et quaternioniques ne sont a priori pas des
supports au sens donn\'e dans le chapitre $1$ puisque dans toutes
les r\'ef\'erences cit\'ees la notion de $W^*$-\alg est d\'efinie
pour des alg\`ebres complexes. Par exemple le projecteur
$\ii_\hhh$ associ\'e \`a l'\'etat r\'eel de $\hhh$ n'est pas de
rang $1$ et il existe deux vecteurs $\xi$ (ceux de la base
canonique de $\cc^2$) permettant d'\'ecrire $\omega(q) =
\scl{\xi}{q \xi}.$

\section{Espace \`a 1 point}

\subsection{D\'efinition et propri\'et\'es g\'en\'erales}

 Les espaces finis apparaissent comme des espaces de $N$ points ($N$ d\'esigne dans (\ref{algfini}) le
nombre de composantes de $\aa$) muni chacun d'une fibre
identique \`a $\pp(\aa_k)$. Lorsque $N=1$ on parle d'espace \`a
un point. Les cas $\aa=\kk$ sont sans int\'eret. L'exemple le plus
simple est $M_n^\kk(\kk)$
 repr\'esent\'e de mani\`ere irr\'eductible sur $\kk^n$. L'op\'erateur $D$ est alors \'el\'ement de l'alg\`ebre. Pour utiliser les
propri\'et\'es des $W^*$-alg\`ebres, on prend $\kk=\cc$ de sorte
que tous les \'etats sont normaux. Dans toute la section $\aa =
M_n^\cc(\cc)$, $\hh = \cc^n$ et $D$ est un \'el\'ement
autoadjoint de $\aa$. Les \'etats et leur support sont indic\'es
par les vecteurs de $\cc P^{n-1}$:
\begin{equation}
\label{notationcp} \omega_\xi (a) =\tr{s_\xi a} = \scl{\xi}{a\xi}.
\end{equation}

Le fait que $D$ soit dans $\aa$ permet d'isoler facilement des
points pathologiques. Si $\psi$ est un vecteur propre de $D$,
$\omega_\psi$ est appel\'e {\it \'etat propre}. Son support
$s_\psi$ commute avec $D$. Par le corollaire \ref{finitude2} on
obtient imm\'ediatement que de tels \'etats sont isol\'es.

{\cor \label{pointsisoles} Tous les \'etats propres de $D$ sont
\`a une distance infinie des autres \'etats purs.}
\newline

\noindent A noter que si $p$ est un projecteur propre de $D$ de
rang sup\'erieur \`a $1$, tous les \'etats purs de support $s\leq
p$ sont \'egalement des points isol\'es.
\newline

Pour exploiter l'identit\'e entre $\pp(\aa)$ et $\cc P ^{n-1}$,
il convient d'associer \`a tout \'etat pur un $n$-uplet norm\'e
de nombres complexes. Pour $n=2$ ceci permet d'\'ecrire la
distance non commutative comme une m\'etrique sur la sph\`ere
$S^2$. Le choix de la base de $\hh$ dans laquelle on \'ecrit la
repr\'esentation de l'\alg fixe la correspondance entre $\cc
P^{n-1}$ et l'ensemble des $n$-uplets norm\'es de complexes: si
\`a l'op\'erateur $a$ est associ\'e la matrice $\hat{a}$, au
vecteur $\xi$ est associ\'e (modulo un facteur de phase) le
n-uplet $\hat{\xi}$ de mani\`ere \`a ce que
$\scl{\hat{\xi}}{\hat{a}\hat{\xi}} = \ox(a).$ On dit que le choix
de la base de la repr\'esentation induit une {\it orientation} de
l'espace des \'etats purs et, de m\^eme qu'on identifie
l'op\'erateur \`a sa matrice, on identifie le vecteur au n-uplet
de ses composantes. Ce point n'est pas aussi transparent qu'il
parait puisque, pour le calcul des distances, on peut changer de
base dans $\hh$ (vu comme espace de repr\'esentation de $\aa$)
sans changer de base dans $\hh$ (vu comme espace o\`u vivent les
repr\'esentants des classes d'\'equivalence de $\cc P^{n-1}$)
sous r\'eserve que ce changement de base commute avec
l'op\'erateur de Dirac. En effet supposons qu'\`a l'op\'erateur
$a$ soit associ\'e dans une nouvelle base la matrice $uau^*$
tandis qu'\`a $\xi$ est toujours associ\'e le n-uplet
$\hat{\xi}$. Alors, pour peu que le changement de base $u$
commute avec $D$ et bien que
$$\scl{\xi}{uau^*\xi}= \scl{u^*\xi}{au^*\xi}= \alpha_u(\ox)(a)\neq\ox(a),
$$
le calcul de la distance entre deux \'etats purs quelconques ne
sera pas modifi\'e en vertu  de la proposition
\ref{isometrieunitaire}.
 Les orientations les plus naturelles sont celles induites par les bases orthonorm\'ees de vecteurs propres de $D$.
Elles se d\'eduisent toutes les unes des autres par un unitaire
qui commute avec $D$, de sorte qu'elles sont toutes
\'equivalentes pour les distances et on peut sans ambiguit\'e
\'evoquer l'orientation induite par la diagonalisation de $D$, ou
simplement l'orientation induite par $D$.

Dans l'orientation induite par $D$, on note $\xi_i$ les
composantes du vecteur $\xi$ et $s_{ij} = \xi_i\bar{\xi_j}$
celles du support $s_\xi$. Les support des \'etats propres, les
{\it supports propres}, sont simplement les \'el\'ements $e_{ii}$
de la base canonique de $\aa$. Par le lemme \ref{tiaf} si deux
\'etats purs $\ou$, $\od$ ne coincident pas sur l'ensemble des
supports propres alors ils sont \`a une distance infinie. En
composantes, $\ox(e_{ii}) = \tr{s_\xi e_{ii}}= \abs{\xi_i}^2,$ ce
qui permet de caract\'eriser facilement des couples d'\'etats \`a
distance infinie.

{\cor \label{pointsinfinis} Si $\abs{\xi_i}\neq\abs{\zeta_i}$
pour au moins une valeur de $i$ alors $d(\ox, \oz) = +\infty$.}

\subsection{ L'exemple de $M_2(\cc)$}

Pour $n=2$, l'espace des \'etats purs $\mathbb{C}P^{1}$ est
isomorphe \`a la sph\`ere $S^2$. Un isomorphisme explicite est
donn\'e par la projection de Hopf qui \`a tout vecteur complexe
$\xi$ de dimension deux norm\'e \`a une phase pr\`es associe le
point $p_\xi$ de $S^2$ \-- vue comme une surface dans $\rr^3$ \--
de coordonn\'ees cart\'esiennes
\begin{equation}
\label{hopf} x_{\xi}\doteq 2\re(\xi_1\bar{\xi}_2), \quad
y_{\xi}\doteq 2\im(\xi_1 \bar{\xi}_2)\,  \text{ et } \,
z_{\xi}\doteq  |\xi_1|^2 -|\xi_2|^2.
\end{equation}
On dira que deux \'etats $\ox$, $\oz$ sont de m\^eme altitude
quand  $z_{\xi} = z_{\zeta}$. L'altitude est une
caract\'eristique intrins\`eque d'un \'etat dans l'orientation
induite par $D$. En effet si $\xi_i$ sont les composantes de
$\xi$ dans une base propre $\left\{\psi_j\right\}$ et $u$ un
unitaire, puisque
$$\ox(a) = \scl{\xi}{a\xi} = \scl{u\xi}{uau^*\, u\xi},$$
les composantes de $\xi$ dans l'orientation induite par la base
$\left\{u\psi_j\right\}$ sont ${\xi'}_i \doteq u_{ij}\xi_j$. On
suppose que les deux valeurs propres $D_1$, $D_2$ de
l'op\'erateur de Dirac sont distinctes (sinon $D$ est
proportionnel \`a l'identit\'e et les distances sont toutes
infinies) et que $u$ commute avec la forme diagonale de $D$,
\begin{equation}
\label{unitaireoriente}
 u =\dm{cc} e^{i\theta_1}& 0 \\ 0 & e^{i\theta_2} \fm
\end{equation}
o\`u $\theta_1, \theta_2\in [0, 2\pi]$. Alors ${\xi'}_1 =
e^{i\theta_1} \xi_1$, ${\xi'}_2 = e^{i\theta_2} \xi_2$ et
l'altitude de $\ox$ dans l'orientation induite par $\left\{
u\psi_j\right\}$ est $z_{\xi'}= \abs{\xi'_1}^2 - \abs{\xi'_2}^2=
z_\xi$. Ainsi on peut l\'egitimement esp\'erer que l'altitude
soit une  donn\'ee pertinente pour caract\'eriser les
propri\'et\'es m\'etriques des \'etats purs. C'est effectivement
le cas.

{\prop \label{propmercrediun} Deux points de la sph\`ere $S^2$
d'altitude diff\'erente sont \`a une distance infinie. En
particuliers les p\^oles sont des points isol\'es.}
\newline

\noindent{\it Preuve.} Les deux \'etats propres de $D$
correspondent aux vecteurs
$$\dm{c}1\\0 \fm \,\text{ et }\,\dm{c} 0 \\ 1 \fm$$
qui par (\ref{hopf}) sont envoy\'es sur les p\^oles de la
sph\`ere (points de coordonn\'ees $(0,0,1)$ et $(0,0,-1)$). Par
le corollaire \ref{pointsisoles}, ces points sont \`a distance
infinie de tous les autres. Si maintenant deux points $(x_\xi,
y_\xi, z_\xi)$ et $(x_\zeta, y_\zeta, z_\zeta)$ ne sont pas de
m\^eme altitude, par d\'efinition de $z$ on a n\'ecessairement
$\abs{\xi_1}\neq\abs{\zeta_1}$ ou/et
$\abs{\xi_2}\neq\abs{\zeta_2}$. Le r\'esultat est alors
imm\'ediat par le corollaire \ref{pointsinfinis}. \hfill
$\blacksquare$
\newline

Le vecteur repr\'esentant $\alpha_u(\ox)$ est $\tilde{\xi}\doteq
u^*\xi$ de composantes $\tilde{\xi}_i = e^{-i\theta_i}\xi_i$. Le
point de $S^2$ associ\'e $p_{\tilde{\xi}}$ a pour coordonn\'ees
\begin{equation}
\label{rotation} \dm{c} x_{\tilde{\xi}}\\  y_{\tilde{\xi}} \\
z_{\tilde{\xi}}\fm = \dm{ccc} \cos \theta & - \sin\theta & 0\\
\sin\theta & \cos \theta & 0
\\ 0&0& 1\fm
\dm{c} x_\xi \\ y_\xi \\ z_\xi\fm
\end{equation}
 o\`u $\theta \doteq \theta_1 - \theta_2$.  Autrement dit pour deux points quelconques $p_\xi$, $p_\zeta$ de $S^2$ et leur image $p_{\tilde{\xi}}$,
 $p_{\tilde{\zeta}}$ par la rotation (\ref{rotation}), la proposition \ref{isometrieunitaire} assure que
$$d(p_\xi, p_\zeta) = d(p_{\tilde{\xi}}, p_{\tilde{\zeta}}).$$
La g\'eom\'etrie $(M_2(\cc), \cc^2, D)$ munie la sph\`ere $S^2$
d'une m\'etrique invariante par rotation d'axe $z$, cet axe
\'etant d\'etermin\'e de mani\`ere non ambig\"ue par la
diagonalisation de l'op\'erateur de Dirac.  Les sym\'etries mises
en place dans le chapitre pr\'ec\'edent  ne suffisent pas \`a
d\'eterminer compl\`etement cette m\'etrique, il faut
entreprendre des calculs explicites qui ici ne sont pas
difficiles.
\newline

\noindent{\bf Proposition \ref{propmercrediun}'.}\; {\it La
distance entre deux \'etats purs $\omega_\xi, \omega_\zeta$ est
finie si et seulement si ils sont de m\^eme altitude. Alors la
distance non commutative est la distance euclidienne sur le
cercle \`a un facteur multiplicatif pr\`es
$$d(p_{\xi}, p_{\zeta})={\frac{2}{|D_1-D_2|}}\sqrt{(x_{\xi} - x_{\zeta})^2+(y_{\xi}-y_{\zeta})^2}.$$
Le r\'esultat est exprim\'e de mani\`ere \'equivalente, au niveau
alg\'ebrique, par
$$d(\omega_{\xi},\omega_{\zeta})= {2\sqrt{ \frac{1-\tr{\sx\sz}}{|D_1-D_2|^2}}}. =
{2\sqrt{ \frac{1-|\langle\xi,\zeta\rangle|^{2}}{|D_1-D_2|^2}}}.$$}

\noindent{\it Preuve.}  Dans la base qui diagonalise $D$ on note
$a_{ij}= \bar{a}_{ji}$ les composantes de $a$ ($a$ est pris
autoadjoint en vertu du lemme \ref{positif}). Si $\ox$ et $\oz$
ne sont pas de m\^eme altitude alors la distance est infinie. On
suppose donc que $z_\xi = z_\zeta$. Comme en outre $\abs{\xi_1}^2
+ \abs{\xi_2}^2 = 1 = \abs{\zeta_1}^2 + \abs{\zeta_2}^2$, cette
condition impose que $\abs{\xi_1} = \abs{\zeta_1}$ et
$\abs{\xi_2} = \abs{\zeta_2}$. On prend $\ox$ distinct de $\oz$,
donc $\abs{\xi_1}\neq 0$. Le vecteur $\xi$ n'\'etant d\'efini
qu'\`a une phase pr\`es, on suppose que $\xi_1\in \rr^{*+}$.
Alors
\begin{equation}
\label{sxmoinssz}
\sx- \sz = \dm{cc}0  & \xi_1(\xi_2 - \zeta_2) \\
\xi_1(\bar{\xi_2} - \bar{\zeta_2}) &0\fm
\end{equation}
et
$$
\abs{\ox(a) - \oz(a)} =  \abs{\tr{(\sx - \sz)(a)}} = \abs{
2\xi_1\re \lp a_{12}\lp \bar{\xi_2} - \bar{\zeta_2}\rp\rp}   \leq
2\xi_1\abs{a_{12}}\abs{\xi_2 -\zeta_2}.
$$
Un rapide calcul indique $\norm{[D,a]}= \abs{a_{12}} \abs{D_1 -
D_2}$ d'o\`u
$$d(\ox, \oz) \leq \frac{2}{\abs{D_1-D_2}}\xi_1\abs{\xi_2 -\zeta_2},$$
cette borne sup\'erieure \'etant atteinte par n'importe quel
\'el\'ement $a$ tel que
$$\abs{a_{12}} = \frac{2}{\abs{D_1-D_2}} \,\text{ et }\, \arg(a_{12}) = - \arg(\xi_2 -\zeta_2).$$
Il suffit alors de remarquer que $\xi_1\abs{\xi_2 - \zeta_2}=
\abs{x_\xi + i y_\xi - x_\zeta - i y_\zeta} = \sqrt{ (x_\xi -
x_\zeta)^2  + ( y_\xi - y_\zeta)^2}$ pour obtenir le r\'esultat.

En calculant
$$\tr{(\sx - \sz)^2} = \tr{\sx + \sz - \sx\sz - \sz\sx} = 2 - 2\abs{\scl{\xi}{\zeta}}^2$$
explicitement \`a l'aide de (\ref{sxmoinssz}), on trouve que
$\xi_1\abs{\xi_2 - \zeta_2} = \sqrt{1-\abs{\scl{\xi}{\zeta}}^2}$,
d'o\`u la seconde expression de la proposition.\hfill
$\blacksquare$

\section{Espace \`a deux points}\label{deuxpoints}

Pour $N=2$ l'espace le plus simple correspond \`a l'alg\`ebre
$\mathcal{A}=M_{n}(\mathbb{C})\oplus\mathbb{C}$ represent\'ee par
une matrice diagonale par bloc sur
$\mathcal{H}=\mathbb{C}^{n}\oplus\mathbb{C}$
\begin{equation}
\label{rep2points} a=\dm{cc}x & 0\\ 0 &y\fm,
\end{equation}
avec $x \in M_n(\mathbb{C})$ et $y\in \mathbb{C}$. A noter que
cette d\'efinition de la repr\'esentation $\pi$ suppose qu'une
base $\left\{ \psi_j\right\}$ de $\hh$ est d\'ej\`a fix\'ee,
modulo un unitaire du type
\begin{equation}
\label{udeuxpoints} U= \dm{cc} u & 0\\ 0 &e^{i\theta}\fm.
\end{equation}
En effet rien ne garantit que dans d'autres bases $a$ soit
toujours repr\'esent\'e sous une forme diagonale par bloc. Pour
une d\'efinition intrins\`eque de $\pi$, il faut se donner une
$\zz_2$-graduation $\chi$ de $\cc^{n+1}$, c'est \`a dire un
op\'erateur qui laisse globalement invariant deux sous espaces
$\hh_n$ et $\hh_1$ de $\hh$ de dimension respective $n$ et $1$.
$\pi$ est alors d\'efinie comme la somme directe de la
repr\'esentation fondamentale de $M_n(\cc)$ sur $\hh_n$ et de
$\cc$ sur $\hh_1$. Contrairement \`a l'espace \`a un point o\`u
la repr\'esentation est surjective, l'orientation induite par la
diagonalisation de $D$ n'est pas pertinente car dans une base qui
diagonalise l'op\'erateur de Dirac, r\'ep\'etons le, $a$ n'est
plus forc\'ement de la forme (\ref{rep2points}). En clair, il
faut consid\'erer que $D$ est une matrice autoadjointe
quelconque, ou abandonner l'aspect diagonal par bloc de la
repr\'esentation. Dans les deux cas la d\'etermination du
supr\'emum dans la formule de la distance reste ardue.  La prise
en compte des axiomes de la g\'eom\'etrie non commutative, en
restreignant le choix de l'op\'erateur de Dirac, permet de mener
les calculs \`a terme.

\subsection{Triplet spectral et orientation}
Trois des axiomes,  relatifs \`a l'analyse fonctionnelle, sont
syst\'ematiquement v\'erifi\'es par les triplets spectraux
finis\cite{finite}. La dualit\'e de Poincar\'e est discut\'ee de
mani\`ere g\'en\'erale pour les triplets finis dans la derni\`ere
partie de ce chapitre. Restent la r\'ealit\'e, la condition
d'ordre un et l'orientabilit\'e. Puisque la dimension spectrale
est nulle, il faut montrer qu'il existe (r\'ealit\'e) un
op\'erateur antilin\'eaire $J= J^*= J^{-1}$ de $\hh$ dans
lui-m\^eme tel que $J^2= \ii$, $[a, JbJ^{-1}]=[J,
\Gamma]=[J,D]=0.$ Tout \'el\'ement de $C_0(\aa, \aa\ot
\aa^{\circ})$ est un cycle. Les g\'en\'erateurs de $Z_0(\aa,
\aa\ot \aa^{\circ})$ sont
 les \'el\'ements de $\aa\ot \aa^{\circ}$. Il doit donc exister $a^i$, $b_i$ dans $\aa$ tels que  (orientabilit\'e) la graduation s'\'ecrive $\Gamma = a^i J b_i
J^{-1}$. Enfin l'op\'erateur de Dirac satisfait (condition du
premi\`ere ordre)  $[[D,a], J b J^{-1}] = 0$. Rappelons que par
d\'efinition la graduation commute avec $\aa$ et anticommute avec
$D$.

Si $\hh = \cc^{n+1}$, $J$ apparait comme une matrice unitaire
compos\'ee avec la conjugaison complexe, $J = U\circ c$. La
relation de commutation $[a,JbJ^{-1}] = 0$ s'\'ecrit $[a,
U\bar{b} U^*] = 0$ ce qui ne peut \^etre vrai pour tout $a$ et
$b$ puisque l'alg\`ebre n'est pas commutative. En revanche si
$\aa$ est repr\'esent\'ee sur $\hh= M_{n+1}(\cc)$ par simple
multiplication matricielle, alors un $J$ possible est
l'op\'erateur  d'involution puisque $J^2=\ii$,
$$[a, JbJ^{-1}]\psi = aJbJ^{-1}\psi - JbJ^{-1}a\psi = a J b \psi^* - Jb \psi^* a^* = a \psi b^* - a \psi b^* = 0$$
et on v\'erifie, pour n'importe quel op\'erateur de Dirac,
$$[[D,a], J b J^{-1}] \psi = [D,a] \psi b^* - [D,a]\psi b^* =0.$$
Pour que $[D,J]=0$, on peut prendre
$$D\psi = \Delta \psi + \psi \Delta$$
o\`u $\Delta=\Delta^*\in M_{n+1}(\cc)$. En effet $[D,J]\psi =
D\psi^* - JD\psi = \Delta\psi^* + \psi^*\Delta -J(\Delta\psi +
\psi\Delta) = 0.$

Concernant la graduation, le choix le plus simple est de prendre
$b_i = a^i = 0$ sauf $b_1=a^2 =\ii$ et $a^1 = b_2= K$ o\`u
$$K \doteq  \dm{cc} \ii_n & 0 \\ 0 & -1 \fm$$
n'est autre que la graduation de $\cc^{n+1}$. Ainsi $\Gamma \psi
= K\psi + \psi K$ et  $[\Gamma, J]=0$. Comme $K$ commute avec
tout $a\in\aa$, on v\'erifie que
$$[\Gamma,a]\psi = \Gamma a \psi - a\Gamma\psi = Ka\psi + a\psi K - aK\psi - a\psi K  =0.$$
Enfin,
$$(D\Gamma + \Gamma D)\psi = D(K\psi + \psi K) + \Gamma(\Delta\psi + \psi\Delta) = (\Delta K + K\Delta)\psi + \psi(\Delta K + K\Delta)$$
est nul pour tout $\psi$ si et seulement si $(\Delta K +
K\Delta)=0$. $\Delta$ s'\'ecrit donc, selon la graduation de
$\cc^{n+1}$,
$$
\Delta=\dm{cc} 0_n& m\\ m^{*}&0 \fm
$$
o\`u  $m$ un vecteur non nul de $\cc^n$.

A priori, repr\'esenter l'alg\`ebre sur $M_{n+1}(\cc)$ ne
facilite pas le calcul de la norme du commutateur $[D, a]$.
Cependant la
 norme d'op\'erateur sur $M_{n+1}(\cc)$ est \'egale \`a la norme  d'op\'erateur sur $\cc^{n+1}$\cite{murphy} si bien que, pour le calcul des distances, tout ce
passe comme si  on travaillait avec le triplet spectral $(\aa,
\hh=\cc^{n+1}, \Delta)$ au lieu de $(\aa, M_{n+1}(\cc), D).$
\newline

L'ensemble des \'etats purs de $\aa$ est l'union de $\pp(\mn)$
avec l'\'etat $\omega_c$ de $\cc$. La correspondance entre un
\'etat $\ox$ de $\pp(\mn)$ et les composantes du vecteur
$\xi\in\cc P ^{n-1}$ \--l'orientation de $\cc P^{n-1}$ \-- est
fix\'ee par l'op\'erateur de Dirac de la mani\`ere suivante. On
suppose  que
$$\norm{D} = \norm{m} = 1$$
 car diviser $D$ par une constante revient \`a multiplier les distances par
cette m\^eme constante. En notant $e_1$ le premier vecteur de la
base canonique de $\cc^n$, il existe un op\'erateur unitaire
$v\in\mn$ tel que
$$v m = e_1.$$ On appelle orientation induite par $D$ le choix de la base $\left\{ V\psi_j \right\}$ o\`u $V$ est la matrice du type
(\ref{udeuxpoints}) correspondant \`a $v$. Dans cette base $a$
est toujours diagonal par bloc et $D$ s'\'ecrit
\begin{equation}
\label{dfi} D = \dm{cc} 0 & e_1 \\ e_1^* & 0 \fm.
\end{equation}
Cette base n'est pas unique.  Dans toute base se d\'eduisant de
$\left\{ V\psi_j\right\}$ par un unitaire $U$ du type
(\ref{udeuxpoints}) commutant avec $D$, la repr\'esentation est
diagonale par bloc et l'op\'erateur de Dirac s'\'ecrit $UDU^* =
D$. Comme dans l'espace \`a un  point, les orientations induites
par ces choix de base sont toutes \'equivalentes pour le calcul
des distances, on peut donc sans ambig\"uit\'e parler de
l'orientation induite par $D$ et la repr\'esentation, ou plus
simplement de l'orientation induite. A la diff\'erence du cas \`a
un point, le choix de l'orientation fait intervenir non seulement
l'op\'erateur de Dirac mais aussi la repr\'esentation car la
pr\'eservation de (\ref{rep2points}) ne va pas de pair avec la
pr\'eservation de $(\ref{dfi})$, c'est \`a dire, tout unitaire
commutant avec  (\ref{dfi}) n'est pas n\'ecessairement du type
(\ref{udeuxpoints}).

\subsection{Distances}

Dans l'orientation induite, les vecteurs propres orthonorm\'es
\`a une phase pr\`es de $D$ sont
$$\psi_1\doteq \frac{1}{\sqrt{2}}\dm{c} e_1\\ 1 \fm,\; \psi_{n+1}\doteq \frac{1}{\sqrt{2}}\dm{c} e_1\\ -1 \fm,\; \psi_j=\dm{c} e_j \\ 0 \fm$$
o\`u  $e_j$, $j=2...n$ sont les vecteurs de la base canonique de
$\cc^n$. Ils correspondent aux valeurs propres $\lambda_1 = 1 ,\,
\lambda_{n+1} = -1\, \text{ et } \lambda_j =0$ et les projecteurs
propres s'\'ecrivent
$$p_{1} = \frac{1}{2} \dm{cc} e_{11} &e_1 \\ e_1^*& 1\fm,\;  p_{n+1} = \frac{1}{2} \dm{cc} e_{11} &-e_1 \\ -e_1^*& 1\fm,\;  p_{j} = \dm{cc} e_{jj} &0
\\ 0& 0\fm$$
o\`u $\{ e_{ij} \}$ est la base canonique de $\mn$. La
diff\'erence essentielle avec l'espace \`a un point est que seul
$p_{j}$ appartient \`a l'alg\`ebre. Les \'etats propres
$\tau_{\psi_1}$ et $\tau_{\psi_{n+1}}$ sont bien des \'etats de
$\aa$ mais ne sont pas purs. On pourrait y voir une contradiction
avec les conclusions de la section \ref{voneuman} o\`u il est
indiqu\'e que tout \'etat d'une $W^*$-\alg de dimension finie est
normal, donc de support inclus dans l'alg\`ebre. En fait $p_{1}$
est le support de $\tau_{\psi_1}$ vu comme \'etat (pur) de
$M_{n+1}(\cc)$, de m\^eme pour $p_{n+1}$. Les supports de
$\tau_{\psi_1}$ et $\tau_{\psi_{n+1}}$ vus comme \'etats de $\aa$
coincident et valent
$$p'_1 = p'_{n+1} \doteq \dm{cc} e_{11} & 0 \\ 0& 1\fm$$
qui appartient bien \`a $\aa$ et n'est pas de rang $1$. Le
projecteur propre $p_j$ quant \`a lui est bien support d'un
\'etat propre pur qui, par le corollaire \ref{finitude2}, est \`a
une distance infinie des autres \'etats purs. Bien que n'\'etant
pas un support d'\'etats purs, $p'_1$ v\'erifie l'\'equation
(\ref{projecteur}) \`a une constante multiplicative pr\`es: $p'_1
a p'_1 = 2\tau_{\psi_1}(a)$ (on utilise la notation
\ref{notationcp} pour les \'etats).  On peut consid\'erer
$\tau_{\psi_1}$ "presque comme" un \'etat pur et penser qu'il est
aussi \`a distance infinie des autres \'etats purs. C'est un cas
particulier d\^u au fait que $\psi_1$, tout comme $\psi_{n+1}$,
projette sur deux directions $\psi$, $\psi'$ en somme directe par
rapport \`a l'alg\`ebre (i.e. il n'existe pas d'\'el\'ement $a$
de l'\alg pour lesquels $\scl{\psi}{a\psi'}$ serait non nul). Pour
des questions plus g\'en\'erales sur les liens entre m\'etriques
sur les \'etats purs et m\'etriques sur les \'etats, on renvoie
\`a [\citelow{rieffel2}].

Avoir s\'electionner gr\^ace aux axiomes un op\'erateur de Dirac
simple permet d'exprimer facilement sa norme, puis de calculer
les distances, en dimension $n$ quelconque. On note $\xi_i$ les
composantes d'un vecteur $\xi$ de $\cc P^{n-1}$ dans
l'orientation induite.

{\prop Si $\ox$, $\oz$ sont tels que $\xi_j= e^{i\theta}\zeta_j$
pour tout $j\in[2,n]$,
$$d(\ox, \oz) = \frac{2}{\norm{m}} \sqrt{1 - \abs{\scl{\xi}{\zeta}}^2}.$$
Par ailleurs $w_c$ est \`a distance infinie de tous les \'etats
purs except\'e $\omega_{e_1}$ et
$$d(\oc, \omega_{e_i}) = \frac{1}{\norm{m}}. $$}

\noindent{\it Preuve.} Soit $a = x\oplus y \in \aa$ autoadjoint
en vertu du lemme \ref{positif}, $x_{ij}$ les composantes de $x$
et $s^{ij}$ celles de $\sx - \sz = \bar{\xi_i}\xi_j -
\bar{\zeta_i}\zeta_j$.
$$[D,a] =  \dm{cc} 0  & (y\ii_n - x) e_1\\ e_1^* (x - y\ii_n)  &   0 \fm$$
est nul si et seulement si, pour tout $i\in[2,n]$,
\begin{equation}
\label{commutdeuxpoints} x_{11} = y\, \text{ et } x_{i1} =0.
\end{equation}

 Deux \'etats purs $\ox$ et $\oz$ coincident sur l'ensemble  des \'el\'ements commutant avec $D$ lorsque
\begin{equation}
\label{eeqmercredie} \ox(a) - \oz(a) = \ox(x) - \oz(x) = \tr{(\sx
- \sz)x} = s^{ij} x_{ji} = 0
\end{equation}
pour tout $a$ remplissant les conditions
(\ref{commutdeuxpoints}). En particulier si
\begin{equation}
\label{elementparticulier} y= x_{11} = x_{1i}= x_{i1} = 0
\;\text{ et }\; x_{ij} = \delta_{ik}\delta_{jl}
\end{equation}
o\`u  $k,l$ sont deux \'el\'ements fix\'es dans $[2,n]$, alors
(\ref{eeqmercredie}) se r\'eduit \`a
\begin{equation}
\label{normedeuxpoints} s^{kl} = 0.
\end{equation}
On obtient une \'equation de ce type pour tout couple d'indice
$k,l\in[2,n]$. En d\'esignant par $\tilde{\xi}$, $\tilde{\zeta}$
les vecteurs de dimension $n-1$ de composantes $\xi_k$,
$\zeta_k$, (\ref{normedeuxpoints}) indique que
$\tilde{\norm{\xi}} = \tilde{\norm{\zeta}}.$ Il existe donc un
facteur de phase $\theta$ tel que $\xi = e^{i\theta}\zeta$, ou
encore
\begin{equation}
\label{coinciddp} \xi_k = e^{i\theta}\zeta_k.
\end{equation}
En cons\'equence si $\ox$ et $\oz$ ne satisfont pas
(\ref{coinciddp}) pour tout $k\in[2,n]$, ils ne coincident pas
sur $\dd'$ et par le lemme \ref{tiaf}, $d(\ox,\oz)= +\infty$.

Si $\abs{\xi_j}= \abs{\zeta_j}$ pour tout $j\in[2,n]$, alors
$\abs{\xi_1}= \abs{\zeta_1}$ puisque $\norm{\xi}=
\norm{\zeta}=1$. Donc $s^{11}=0$. De plus $s^{ji}=
\overline{s^{ij}}$ et $s^{ij} = 0$ pour $i,j\in[2,n]$, d'o\`u
$$\ox(a) - \oz(a) =  s^{ij}  x_{ji} = 2\re\lp s^{1j} x_{j1}\rp$$
o\`u la somme porte de $j= 2$ \`a $n$. On note $S$ et $X$ les
deux vecteurs de $\cc^{n-1}$ de composantes respectives
$\overline{s}^{j1}$, $x_{j1}$. Par l'in\'egalit\'e de
Cauchy-Schwarz,
\begin{equation}
\label{cauchsch} \abs{\ox(a) - \oz(a)}\leq  2\abs{\scl{S}{X}}
\leq 2 \norm{S}\norm{X}.
\end{equation}
D'autre part
\begin{equation}
\label{NGCnorme}\norm{[D,a]} =  \abs{x_{11} - y}^2 + \norm{X}^2.
\end{equation}
Ins\'er\'ee dans (\ref{cauchsch}),
 $$
d(\ox, \oz) \leq  2 \norm{S}.
$$
Cette borne sup\'erieure est atteinte par tout $a$ du type $y=
x_{11}$ et $X=\frac{S}{\norm{S}}$. Pour conclure, il suffit de
remarquer, comme \`a la fin de l'espace \`a un point, que
\begin{eqnarray*}
\tr{(\sx - \sz)^2} &=& \sum_{i,j = 1}^n s^{ij}s^{ji} = \sum{j=2}^{n} 2\abs{s^{j1}}^2 = 2 \norm{S}^2,\\
                   &=& 2 - 2\tr{\sx\sz} = 2 - 2\abs{\scl{\xi}{\zeta}}^2
\end{eqnarray*}
et de multiplier par $\norm{D}= \norm{m}.$

En consid\'erant $\oc$ plut\^ot que $\oz$, (\ref{eeqmercredie})
devient
$$
\ox(a) - \oc(a) = \ox(x) - \oc(y) = \tr{\sx x} - y.$$ Pour des
\'el\'ements du type (\ref{elementparticulier}) on obtient
$\tilde{\norm{\xi}} = 0$. Autrement dit $\xi$ est colin\'eaire
\`a $e_1$. Si tel n'est pas le cas alors $\oc$ et $\ox$ ne
coincide pas sur les \'el\'ements commutant avec $D$ donc la
distance est infinie.

Pour finir $\omega_{e_1}(a) - \oc(a)  = x^{11} - y$. En vertu de
(\ref{NGCnorme}), $d(\omega_{e_1}, \oc) \leq 1$,
\hfill$\blacksquare$ \newline
\newline

Appliquons ces r\'esultats \`a $M_2(\cc)\oplus \cc$. L'espace des
\'etats purs est l'union disjointe de la sph\`ere $S^2$ et du
point $\oc$. Le point isol\'e correspond au vecteur
$$\dm{c} 0 \\ 1\fm$$
qui, par la fibration de Hopf, est envoy\'e sur le p\^ole sud
$(0,0, -1)$ de la sph\`ere. Le point correspondant \`a
$\omega_{e_1}$ est le p\^ole nord, et c'est le seul point qui se
trouve \`a distance finie de $\oc$. Les conditions sur la
finitude des autres distances sont identiques \`a celles du cas
\`a $1$ point et on retrouve que la distance sur des plans de
m\^eme altitude est, \`a une constante pr\`es, la distance
euclidienne du cercle (selon [\citelow{mario}], on peut rendre
finie la distance entre plans d'altitude constante en
agrandissant l'espace de repr\'esentation).

 A noter que l'ajout du point $\oc$ donne un sens \`a
l'orientation induite. Dans l'espace \`a un point rien ne permet
de distinguer les deux points isol\'es, tandis que dans l'espace
\`a deux points le p\^ole sud est par d\'efinition l'unique point
isol\'e.

D'autres espaces \`a deux points du type $M_p(\mathbb{C})\oplus
M_q(\mathbb{C})$ ne sont pas \'etudi\'es ici, pas plus que les
sommes de plus de deux alg\`ebres comprenant au moins une \alg
non commutative car les calculs deviennent rapidement
impraticables. En revanche les sommes d'alg\`ebres commutatives
constituent une classe d'exemples int\'eressants. Leur \'etude
est l'objet du reste de ce chapitre.

\section{Espaces finis commutatifs}\label{espacefinicommut}
Un espace fini commutatif de $n$ points est d\'ecrit par un
triplet spectral ($\mathcal{A},\mathcal{H},D)$ o\`u
$\mathcal{A}={\overset{n}{\underset{1}{\bigoplus}}} \,
\mathbb{C}$ est repr\'esent\'e sur $\mathcal{H}= \mathbb{C}^n$
par les matrices diagonales
$$
\mathcal{A} \ni a = \left(
\begin{array}{cccc}
a_1    & 0 & \ldots  & 0 \\
\vdots & a_2 &       & \vdots \\
\vdots  &    & \ddots    &  \vdots   \\
0       &\ldots&\ldots  & a_n \\
\end{array}
\right),
$$
avec $a_i \in \mathbb{C}$. Cependant  on sait (lemme
\ref{positif}) que pour les calculs de distance on peut supposer
que les $a_i$ sont des r\'eels positifs. Pour simplifier les
calculs on se limite aux op\'erateurs de Dirac \`a entr\'ee
r\'eelle. Comme $D$ n'intervient qu'au travers du commutateur
$[D,a]$, on peut sans perte de g\'en\'eralit\'e supposer qu'il
est de la forme:
$$
D= \left(
\begin{array}{ccccc}
0    &D_{12}&\ldots&\ldots&D_{1n} \\
D_{12}&      &D_{23}&      &0 \\
\vdots&D_{23}&0    &\ddots&\vdots   \\
\vdots&     &\ddots&\ddots& D_{n-1,n}   \\
D_{1n}&\ldots&\ldots &D_{n-1,n}& 0 \\
\end{array}
\right) \text{ avec }D_{ij} \in \mathbb{R}.
$$
L'espace des \'etats purs est compos\'e de $n$ fois l'\'etat pur
$\oc$ de $\cc$. Si $\oc^i$ d\'esigne la $i^{\text{\`eme}}$
occurence de $\oc$, on a
$$\oc^i(a) = a_i.$$
Pour all\'eger les notations on \'ecrit cette \'equation
$i(a)\doteq a_i$, de sorte que la formule de la distance devient
\begin{equation}
\label{distance2} d(i, j) = \underset {a \in \mathcal{A}^+}
{\sup}  \, \{  \, |a_i - a_j|  \;\,  / \;\, \|[D,a]\| = 1 \}.
\end{equation}

$D$ s'interpr\`ete comme la matrice d'incidence d'un
r\'eseau\cite{dimakis1}: deux points $i$ and $j$ sont reli\'es
d'un trait si et seulement si l'\'element de matrice
correspondant $D_{ij}$ n'est pas nul. Par exemple un espace de
quatre points avec $D_{13} = D_{24} = 0$ est repr\'esent\'e par
le graphe cyclique
\begin{figure}[h]
\begin{center}
\mbox{\rotatebox{270}{\scalebox{0.45}{\includegraphics{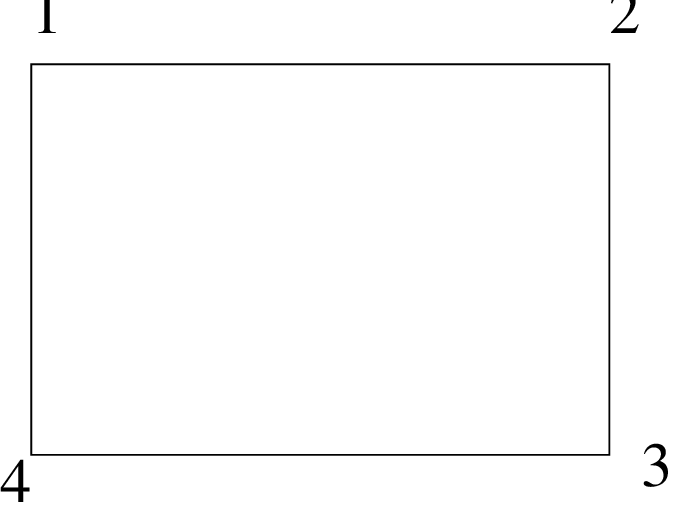}}}}
\end{center}
\end{figure}

Un chemin $\gamma_{ij}$ est une suite de $p$ points distincts
$(i,i_2,...,i_{p-1},j)$ tels que
$$D_{i_{k}i_{k+1}}\neq 0\; \text{ pour tout } k\in\{1,p-1\}.$$
Dans l'espace commutatif de deux points l'unique distance est
$$d(1,2)=\frac{1}{|D_{12}|}.$$
Ainsi il est naturel de d\'efinir la longueur d'un chemin
$\gamma_{ij}$ par
\begin{equation*}
L(\gamma_{ij})\doteq \underset
{k=1}{\overset{p-1}\Sigma}{\frac{1}{|D_{i_ki_{k+1}}|}}.
\end{equation*}
Deux points $i$,$j$ sont dits connect\'es lorsqu'il existe au
moins un chemin $\gamma_{ij}$. La distance g\'eod\'esique
$L_{ij}$ est par d\'efinition la longueur du plus court chemin
reliant $\gamma_{ij}$.

{\prop\label{chemin} 1) Soit $D'$ l'op\'erateur obtenu en
annulant une ou plusieurs lignes, ainsi que les colonnes
correspondantes, de l'op\'erateur $D$ et $d'$ la distance
associ\'ee. Alors $d' \geq d$.

\hspace{2.3truecm} 2) La distance entre deux points $i$ et $j$ ne
d\'epend que des \'el\'ements de matrice correspondant \`a des
points situ\'es sur un chemin $\gamma_{ij}$.

\hspace{2.3truecm} 3) La distance entre deux points est finie si
et seulement si ils sont connect\'es.}
\newline

\noindent{\it Preuve.} 1) Soit $e\in\mathcal{A}$ tel que
$e_{i}=0$ si les $i^{\text{\`eme}}$ lignes et colonnes sont
annul\'ees, $e_{i}=1$ autrement. $e$ est un projecteur qui
commute avec $\mathcal{A}$ ainsi qu'avec $D'\doteq eDe$. D\`es
lors $||[D',a]||\leq ||[D,a]||$ et
$$\sup\{
|a_i-a_j|\,/\, \|[D,a]\|\leq 1 \} \geq \sup\{ |a_i-a_j|\,/\,
\|[D',a]\|\leq 1 \}.$$

2) Soit $\Gamma_{ij}$ le graphe associ\'e \`a l'ensemble des
points appartenant au moins \`a un chemin $\gamma_{ij}$, et
$I_{ij}$ l'ensemble des points qui n'appartiennent \`a aucun
chemin $\gamma_{ij}$. Tout point de $I_{ij}$ est connect\'e \`a
$\Gamma_{ij}$ par au plus un chemin. Pour un \'el\'ement $l$ de
$I_{ij}$, il existe au plus un point $m_l\in\Gamma_{ij}$ tel que
$l$ et $m_l$ soient connect\'es par un chemin $\gamma_{lm_l}$
dont tous les points  (except\'e $m_l$) sont dans $I_{ij}$. Soit
$D'$ l'op\'erateur obtenue en annulant toutes les lignes et
colonnes correspondant aux points de $I_{ij}$, et $a'$
l'\'el\'ement qui r\'ealise le supr\'emum pour la distance $d'$.
Soit $a\in\aa$ d\'efini par $a_p=a'_p$, sauf pour les points de
$I_{ij}$ o\`u l'on pose $a_l=a_{m_l}$ ou $a_l=0$ si $m_l$
n'existe pas. Alors
$$[D,a] = \left\{ d_{ij}(a_i - a_j) \right\}= [D',a']$$
et $d(i,j) \geq d'(i,j)$. D'apr\`es ${\it 1)}$, $d(i,j)\leq
d'(i,j)$ d'o\`u le r\'esultat.
\newline

3) Supposons que $i$ et $j$ soient connect\'es. Il existe au
moins un chemin $\gamma_{ij}=(i,i_2,...,i_{p-1},j)$ dont la
longueur est la distance g\'eod\'esique $L_{ij}$. Soit  $D'$
l'op\'erateur obtenu en annulant les lignes et colonnes ne
correspondant \`a aucun point de $\gamma_{ij}$. Alors $d(i,j)\leq
d'(i,j)$. L'in\'egalit\'e triangulaire indique que
$$
d'(i,j)\leq \underset{k=1}{\overset{p-1}{\Sigma}}d'(i_k,i_{k+1})
=\underset{k=1}{\overset{p-1}{\Sigma}}{\frac{1}{|D_{i_ki_{k+1}}|}}\doteq
L_{ij}.
$$
La distance $d(i,j)$ est plus petite que la distance
g\'eod\'esique, donc elle est finie. Quand $i$ et $j$ ne sont pas
connect\'es, on d\'efinit $a$ par $a_i=t>0, a_k=a_i$ si $k$ et
$i$ sont connect\'es, $a_k=0$ autrement. Alors $[D,a]=0$ et
$|a_i-a_j|=t$. Puisque $t$ est arbitraire, $d(i,j)$ est infinie.
\hfill $\blacksquare$
\newline

 Afin de simplifier les notations, on pose
\begin{equation}
\label{x} a_{ij}\doteq a_j - a_i \quad \text{ et } \quad x\doteq
a_{21} \, , \, x_i \doteq a_{i+1,1},\quad 2\leq i \leq n-1.
\end{equation}
Dans les espaces de $n=3$ et $n=4$ points, ces notations se
r\'eduisent \`a
\begin{equation}
\label{x4} y\doteq a_{31} \, \text{ and } \,  z\doteq a_{41}.
\end{equation}

\subsection{Espace r\'egulier}

Un espace commutatif de $n$ points est dit r\'egulier lorsque
tous les coefficients de l'op\'erateur $D$ sont \'egaux
$$D_{ij} =k(1 - \delta_{ij}) \;  ,\; k \in \mathbb{R}.$$

{\prop\label{espacereg} 1) La distance entre deux points $i,j$
d'un espace r\'egulier de constante $k$ est
$$ d(i,j) ={\frac{1}{|k|}}{\sqrt{\frac{2}{n}}}.$$

\hspace{2.4truecm} 2) Si le lien entre deux points $i_1,i_2$ \--
et uniquement ce lien\--  est coup\'e, $D_{i_1i_2}=0$, alors
$$ d(i_1,i_2) ={\frac{1}{|k|}}{\sqrt{\frac{2}{n-2}}}.$$}

\noindent {\it Preuve.} Un espace r\'egulier est sym\'etrique par
rapport \`a toutes les permutations d'indices. Toutes les
distances sont \'egales et pour fixer les notations on calcule
$d(1,2)$. De m\^eme lorsque un lien est coup\'e, on peut sans
perte de g\'en\'eralit\'e poser $i_1=1,i_2=2$. ${D'}$ d\'esigne
l'op\'erateur obtenu de $D$ en posant $D_{12}=0$. Pour $D$ comme
pour $D'$,  (\ref{distance2}) et (\ref{x}) donnent
\begin{equation}
\label{nptsdistance} d(1,2) = \underset {a \in \mathcal{A}_{+}}
{\sup} \{\; |x|\, /\, \|[\,D \text{ ou } {D'}, a]\| =1 \}.
\end{equation}
Pour calculer cette distance, il faut d'abord calculer la norme
du commutateur pour ensuite d\'eterminer le supr\'emum.
\newline

{\lem\label{lemmespacereg1}

$ \|[D,a]\|^2 =|k|^2
                         \overset{n}{\underset{i=1}{\Sigma}}
                         \,
                          \overset{n}{\underset{j=i+1}{\Sigma}}
                         a_{ij}^{2}
             =|k|^2  \left[
                        x^2
                        + \overset{n-1}{\underset{i=2}{\Sigma}} \left( x_i^2 + (x-x_i)^2  + \overset{n-1}{\underset{j=i+1}{\Sigma}}(x_i - x_j)^2\right)
                      \right]
                       .$

\begin{eqnarray*}
\hspace{3truecm} \|[{D'},a]\|^2 &=&\frac{|k|^2}{2}
                   \left[
                   \overset{n}{\underset{i=1}{\Sigma}}\, \overset{n}{\underset{\underset{-(1,2)}{j=i+1}}{\Sigma}}a_{ij}^2
                   +\sqrt{
                          ( \overset{n}{\underset{i=1}{\Sigma}} \,  \overset{n}{ \underset { \underset{-(1,2)}{j=i+1} } {\Sigma} }  a_{ij}^2) ^2
                          -4a_{12}^2 \overset{n}{\underset{i=3}{\Sigma}}\; \overset{n}{ \underset{j=i+1}{\Sigma}} a_{ij}^2
                          }
                   \right]\\
                     &=&\frac{|k|^2}{2}\left[
                                 \overset{n-1}{\underset{i=2}{\Sigma}} \left( x_i^2 + (x-x_i)^2  + \overset{n-1}{\underset{j=i+1}{\Sigma}}(x_i -
x_j)^2\right)\right] \\
               & & + \left[ \sqrt{
                                          (\overset{n-1}{\underset{i=2}{\Sigma}} \left( x_i^2 + (x-x_i)^2  + \overset{n-1}{\underset{j=i+1}{\Sigma}}(x_i -
x_j)^2\right))^2 - 4x^2 \overset{n-1}{\underset{i=2}{\Sigma}}\;
\overset{n-1}{\underset{j=i+1}{\Sigma}}(x_i - x_j)^2
                                 }
                                  \right].
\end{eqnarray*}}
\newline

\noindent{\it Preuve.}
 $C \doteq i[D,a]$ est une matrice carr\'ee de dimension $n$
$$
C = k\left(\begin{array}{cccc}
0      &ia_{12}&      &    \\
ia_{21}&\ddots &ia_{ij}&   \\
       &ia_{ji}& \ddots &    \\
        &      &       & 0 \\
\end{array}\right)
$$
et de rang $\leq 2$ puisque son noyau est g\'en\'er\'e par les
$(n-2)$ vecteurs ind\'ependants
\begin{equation*}
 \Lambda_{k} =( {\frac{a_{k2} }{a_{21} }};  {\frac{ a_{1k} }{ a_{21}
}}; 0;...;1;...; 0)
\end{equation*}
o\`u $1$ est \`a la $k^{\text{i\`eme}}$ position, $3\leq k\leq
n$. De plus $C$ est une matrice autoadjointe de trace nulle; elle
a donc deux valeurs propres non nulles $\pm\lambda$. Autrement
dit $\lambda=\sqrt{\frac{Tr(C^2)}{ 2}}$ et un calcul direct donne
$\lambda= k{\sqrt  { \overset {n}     {\underset {i=1}
{\Sigma}}\, \overset {n} {\underset {j=i+1} {\Sigma} }
(a_{ij})^{2}}}.$ Ainsi
 \begin{eqnarray*}
\|[D,a]\| &=&\|i[D,a]\|=|\lambda|=|k|{\sqrt  { \overset {n}
{\underset {i=1}    {\Sigma}}
 \, \overset {n} {\underset {j=i+1} {\Sigma} }
(a_{ij})^{2}}},\\
&=& |k|\sqrt{ x^2 +  { \overset {n-1}     {\underset {i=2}
{\Sigma}}\,  x_i^2 +  { \overset {n-1}     {\underset {i=2}
{\Sigma}} \, \overset {n-1} {\underset {j=i+1} {\Sigma} } (x_i -
x_j})^{2}}}.
\end{eqnarray*}

 Soit ${C'} \doteq i[{D'},a]$ la matrice carr\'ee de dimension $n$
$$
{C'} = k\left(\begin{array}{cccc}
0     & 0    &ia_{13} &  \\
0     & 0     &        & ia_{ij} \\
ia_{31}&       &\ddots  & \\
      &ia_{ji}&        &0   \\

\end{array}\right)
$$
et de rang $\leq 4$ puisque $ker({C'})$ est g\'en\'er\'e par les
$(n-4)$ vecteurs ind\'ependants
\begin{equation*}
{\Lambda'}_{p}
=(0;0;{\frac{a_{p4}}{a_{43}}};{\frac{a_{3p}}{a_{43}}};
0;...;1;...; 0)
\end{equation*}
o\`u  $1$ est \`a la $p^{\text{i\`eme}}$ position, $5\leq p\leq
n.$ Puisque $C'$ est autoadjointe et que $\bar{C'}=-C'$, la
matrice $C'$ a quatre valeurs propres r\'eelles
$\pm{\lambda'_1}$, $\pm{\lambda'_2}$ et son polyn\^ome
caract\'eristique est
\begin{equation}
\label{carc}
\mbox{\Large$\chi$}({C'})=X^{n-4}(X^2-{\lambda'_1}^2)(X^2-{\lambda'_2}^2)
=X^n -({\lambda'_1}^2 + {\lambda'_2}^2)X^{n-2}
+{\lambda'_1}^2{\lambda'_2}^2X^{n-4}.
\end{equation}
Par calcul direct il vient
$${\lambda'_1}^2 + {\lambda'_2}^2=\frac{1}{2}Tr({C'}^2)=
k^2{\overset{n}{\underset{i=1}{\Sigma}}\,\overset{n}{\underset{\underset{-(1,2)}{j=i+1}}
{\Sigma}}(a_{ij})^{2}}.$$ Le coefficient en $X^{n-4}$ est la
somme des mineurs de ${C'}$ de degr\'e $4$. Un mineur
$M(1,k,l,p)$ form\'e de la premi\`ere (ou la seconde) colonne, de
trois autres colonnes $k,l,p \notin \{1,2\}$ et des lignes
associ\'ees est \'egalement un mineur de $C$. Comme $C$ est de
rang $\leq 2$, ses mineurs de degr\'e sup\'erieur \`a $2$ sont
nuls et $M(1,k,l,p)=M(2,k,l,p)=0$. Il en est de  m\^eme pour les
mineurs $M(q,k,l,p)$ avec $q\notin \{1,2\}$. Au final, les seuls
mineurs non nuls sont
$$M(1,2,l,p)= k^4 Det\left(
\begin{array}{cccc}
0     & 0      & ia_{1l} & ia_{1p} \\
0     & 0      & ia_{2l} & ia_{2p} \\
ia_{l1}& ia_{l2} & 0      & ia_{lp}\\
ia_{p1}& ia_{p2} & a_{pl} & 0
\end{array}
\right) =k^4 Det\left(
\begin{array}{cc}
a_{1l} & a_{1p} \\
a_{2l} & a_{2p}
\end{array}
\right)^2=k^4 a_{21}^2 a_{pl}^2.$$ L'addition de ces mineurs
donnent
$${\lambda'_1}^2{\lambda'_2}^2 =a_{12}^2 \underset{l=3}{\overset{n}\Sigma}\;
\underset{p=l+1}{\overset{n}\Sigma} a_{pl}^2.$$ On peut
r\'esoudre (\ref{carc}) et obtenir
$$\|[{D'},a]\|^2 =\frac{|k|^2}{2}
                   \left(
                   \overset{n}{\underset{i=1}{\Sigma}}\;  \overset{n}{\underset{\underset{-(1,2)}{j=i+1}}{\Sigma}}a_{ij}^2
                   +\sqrt{
                          ( \overset{n}{\underset{i=1}{\Sigma}} \,  \overset{n}{ \underset { \underset{-(1,2)}{j=i+1} } {\Sigma} }  a_{ij}^2) ^2
                          -4a_{12}^2 \overset{n}{\underset{i=3}{\Sigma}}\;  \overset{n}{ \underset{j=i+1}{\Sigma}} a_{ij}^2
                          }
                   \right). \qquad \square$$

{\lem \label{lemmespacereg2} Dans l'espace r\'egulier, le
supr\'emum des $x$ dans (\ref{nptsdistance}) est atteint lorsque
tous les $x_i$ sont \'egaux.}
\newline

\noindent{\it Preuve.} On pose
$$f(x,x_2,...,x_{n-1})\doteq x^2+{ \overset{n-1}{\underset{i=2}{\Sigma}}x_i^2 + (x-x_i)^2 +{\overset {n-1}{\underset {i=2}{\Sigma}}
\overset {n-1}{\underset{j=i+1}{\Sigma}}(x_i-x_j})^{2}}.$$
Supposons que $(x,x_2,...,x_{n-1}) \in \mathbb{R}^{n-1}$
r\'ealise le supr\'emum, c'est \`a dire
$$
f(x,x_2,...,x_{n-1})=\frac{1}{|k|^2} \; \text{ et } \; d(1,2) =
|x|.
$$
Alors

- $x$ est positif: $f$ est globalement paire, on peut donc
choisir $x$ positif.

- $x_i\leq {\frac{x}{2}},\; \forall i \in \{2,...,n-1\}:$
supposons que $p$ des $x_i$ soient plus grands que $\frac{x}{2}$
et d\'esignons les de mani\`ere g\'en\'erique par $x_p$. Soit
maintenant le (n-1)-uplet o\`u les $x_p$ sont remplac\'es par
$\frac{x}{2}$. $f$ d\'ecroit car
$$x_p^2 + (x-x_p)^2 \geq \frac{x^2}{4} + (x-\frac{x}{2})^2$$
 et
$$(x_i - x_p)^2 \geq (x_i - \frac{x}{2})^2.$$
Fixer les valeurs des autres $x_i$ permet de voir $f$ comme
fonction de la seule variable $x$,
$$
f(x)= x^2 + p(\frac{x}{2})^2 + {\underset {i}{\Sigma}}
x_i^2             + p(x-\frac{x}{2})^2 +  {\underset
{i}{\Sigma}}(x-x_i)^2+ {\underset {i}{\Sigma}}p(\frac{x}{2}-
x_i)^2 + {\underset {i}{\Sigma}} {\underset {j}{\Sigma}}(x_j -
x_i)^2,$$ dont la d\'eriv\'ee est
$$f'(x)= 2x + 2px +
2{\underset {i}{\Sigma}}(x-x_i) + {\underset
{i}{\Sigma}}p(\frac{x}{2}-x_i).
$$
Puisque  $x_i \leq \frac{x}{2} \leq x$ ,
$$f'(x) > 0$$
d\`es que $x > 0$. $f$ est continue et ${\underset {x\rightarrow
\infty}{\lim}}f(x) = +\infty$,
 donc il existe $x_0 > x$ avec $f(x_0)=\frac{1}{|k|^2}$. Ceci signifie que le n-uplet initial en $x_p$ n'atteint pas le supr\'emum en contradiction avec
l'hypoth\`ese. Par cons\'equent $p=0$.

- $x_i \geq 0$ pour tout  $i \in \{2,...,n-1\}$: en remplacant
$x_i \leq 0$ par $\frac{x}{2}$, la preuve est identique au
paragraphe pr\'ec\'edent.

- Tous les $x_i$ sont \'egaux:
 soient $\lambda$ et $\Lambda$ les deux plus petites valeurs des $x_i$, choisies telles que $\lambda \leq \Lambda$. $\lambda=\Lambda$ signifie
que tous les $x_i$ sont \'egaux. Si $\lambda\neq\Lambda$, alors
$$0 \leq \lambda < \Lambda \leq x_i \leq \frac{x}{2} \; ,\; \forall i \in \{2,...,n-1\}.$$
Supposons que $m$ des $x_i$ soient \'egaux \`a $\lambda$. La
somme sur tous les $x_i \neq \lambda$ donne:
\begin{eqnarray*}
f(x,x_2,...,x_{n-1}) = x^2 + m\lambda^2 + {\underset {i} \Sigma}
x_i^2 + {\underset {i} \Sigma} (x-x_i)^2+ m(x-\lambda)^2 +
{\underset {i}\Sigma} m(\lambda -x_i)^2 + {\underset {i,j}\Sigma}
(x_i -x_j)^2.
\end{eqnarray*}
En fixant les valeurs des $x_i \neq \lambda$ et en consid\'erant
$\lambda$ non plus comme une constante mais comme la valeur de la
variable $x_{min}$,  $f$ apparait comme une fonction $f_m$ de
deux variables $x_{min}$ et $x$. Il vient
$${\frac{\partial f_m}{\partial x_{min}}}(x_{min},x) = 2mx_{min} + 2m(x_{min} - x) + {\underset {i} \Sigma} 2m(x_{min} -x_i).$$
Comme
$${\frac{\partial f_m}{\partial x_{min}}}(x_{min},x)< 0$$
pour $x_{min} \in [\lambda,\Lambda[$, on a $f_m(\Lambda,x) <
f_m(\lambda,x)=\frac{1}{|k|^2}.$ De plus
$${\frac{\partial f_m}{\partial x}}(\Lambda,x) = 2x + 2m(x-\Lambda) + {\underset {i} \Sigma} 2(x-x_i) > 0,$$
donc il existe $x_0 > x$ tel que
$f_m(x_0,\Lambda)=\frac{1}{|k|^2}$, ce qui contredit notre
hypoth\`ese.  Par cons\'equent $\lambda = \Lambda$.
 \hfill$\square$
\\

\noindent{\bf Preuve de la proposition \ref{espacereg}}

1) Gr\^ace au lemme \ref{lemmespacereg2}, $x_i = x_2$ pour $2\leq
i \leq n-1$. La condition sur la norme de l'\'equation
(\ref{nptsdistance}) s'\'ecrit
 $$ 2(n-2)x_2^2 + [2(2-n)x]x_2 + [(n-1)x^2 - \frac{1}{|k|^2}] = 0.$$
Ce polyn\^ome en $x_2$ n'a pas de solution r\'eelle sauf si
$$|x|\leq \frac{1}{|k|}\sqrt{\frac{2}{n}}.$$ Cette borne sup\'erieure est atteinte lorsque
$$x_2=\frac{x}{2} =\frac{1}{2|k|}\sqrt{\frac{2}{n}}.$$

2) Posons
\begin{eqnarray*}
&h_1(x,x_i) \doteq \overset{n-1}{\underset{i=2}{\Sigma}}
x_i^2+(x-x_i)^2,\quad
h_0(x_i)   \doteq \overset{n-1}{\underset{i=2}{\Sigma}}\, \overset{n-1}{\underset{j=i+1}{\Sigma}}(x_i - x_j)^2,&\\
&g(x,x_i)\doteq h_1(x,x_i)-2x^2.&
\end{eqnarray*}
Par le lemme \ref{lemmespacereg1}
\begin{equation}
\label{h} \|[{D'},a]\|^2={\frac{|k|^2}{2}}\left(  h_1 +h_0
+\sqrt{h_1^2 + h_0^2 + 2g.h_0}  \right) .
\end{equation}
Soit $x_0=\underset{x,x_i \in \mathbb{R}}{\sup} \{ x /
h_1(x,x_i)= \frac{1}{|k|^2} \}$. Comme $g$ et $h_0$ sont tous
deux positifs, (\ref{h}) implique $d(1,2)\leq x_0$. En
r\'ep\'etant toute la proc\'edure du lemme \ref{lemmespacereg2}
on trouve que cette borne sup\'erieure est atteinte lorsque tous
les $x_i$ sont \'egaux et
$x_0=\frac{1}{|k|}\sqrt{\frac{2}{n-2}}$.\hfill $\blacksquare$
\newline

\noindent Ces deux exemples ne doivent pas laisser croire que les
distances dans les espaces finis sont toujours  explicitement
calculables. C'est le cas pour $n=3$ mais pas pour $n=4$.

\subsection{Espace \`a trois points}\label{troispoints}
On consid\`ere un espace de trois points avec comme op\'erateur
de Dirac
$$
D = \left(
\begin{array}{ccc}
0 & D_{12} & D_{13}\\
D_{12} & 0 & D_{23}\\
D_{13} & D_{23} & 0
\end{array}
\right)$$ ou  $D_{ij}\in \rr$. Par permutation des coefficients
on obtient toutes les distances \`a partir de $d(1,2)$.

{\prop \label{distancetroispoints} Dans un espace \`a trois
points avec l'op\'erateur de Dirac ci-dessus,
$$d(1,2)=\sqrt{\frac{D_{13}^2+D_{23}^2}{D_{12}^{2}D_{13}^{2}+D_{12}^{2}D_{23}^{2}+D_{23}^{2}D_{13}^{2}}}.$$}
\newline

\noindent {\it Preuve.} L'\'equation (\ref{distance2}) et les
notations (\ref{x4}) donnent
$$
d(1,2) = \underset {a \in \mathcal{A}_{+}}{\sup} \,\{x\;/\;
\|[D,a]\|=\left\| \left(
\begin{array}{ccc}
0 & -D_{12}x & -D_{13}y\\
D_{12}x & 0 & D_{23}(x-y)\\
D_{13}y & D_{23}(y-x) & 0
\end{array}
\right) \right\|=1\}.
$$
Un calcul direct donne
$$\|[D,a]\|=\sqrt{ D_{23}(x-y)^2+D_{13}y^{2}+D_{12}x^{2} }.$$
$d(1,2)$ est la plus grande valeur de $x$ pour laquelle il existe
un point $(x,y)$ appartenant \`a l'ellipse
\begin{equation}
\label{ellipse}
(D_{23}^{2}+D_{12}^{2})x^{2}+(D_{13}^2+D_{23}^{2})y^2-2D_{23}^2xy=1.
\end{equation}
$d(1,2)$ est la valeur positive de $x$ pour laquelle
l'\'equation en $y$ (\ref{ellipse}) a un discriminant nul, c'est
\`a dire
$$
x=\sqrt{\frac{D_{13}^2+D_{23}^2}{D_{12}^{2}D_{13}^{2}+D_{12}^{2}D_{23}^{2}+D_{23}^{2}D_{13}^{2}}}.$$
\hfill $\blacksquare$

Les trois distances v\'erifient une in\'egalit\'e triangulaire
"au carr\'e"
\begin{equation}
\label{trgcarre} d(1,2)^2 + d(2,3)^2 \geq  d(1,3)^2,
\end{equation}
ainsi que deux autres in\'egalit\'es obtenues par permutations
des indices. Disposant d'une formule exacte pour chaque distance
d'un espace de trois points, on peut raisonnablement se demander
s'il est possible d'inverser la m\'etrique pour remonter \`a
l'op\'erateur de Dirac. Plus exactement, \'etant donn\'es trois
nombres positifs $(a,b,c)$ v\'erifiant (\ref{trgcarre}), existe
t-il une g\'eom\'etrie dans laquelle $(a,b,c)$ sont les distances
d'un espace commutatif de trois points ?

{\prop Soient $a,b,c$ trois nombres r\'eels strictement positifs
qui v\'erifient $a^2+b^2\geq c^2,\, b^2+c^2 \geq a^2, \,
a^2+c^2\geq b^2.$ Il existe un op\'erateur $D$ tel que $d(1,2)=
a$, $d(1,3)= b$, $d(2,3)=c$. Les coefficients de $D$ sont
$$
D_{12}=\sqrt{{\frac{2(b^2 + c^2 -
a^2)}{(a+b+c)(-a+b+c)(a-b+c)(a+b-c)}}},$$ $D_{13}$ and $D_{23}$
sont d\'eduits par permutation de $a,b,c$.}
\newline

\noindent{\it Preuve.} En posant ${\frac{1}{D_{12}^2}}=R_{12}$,
${\frac{1}{D_{23}^2}}=R_{23}$, ${\frac{1}{D_{13}^2}}=R_{13}$,  la
proposition \ref{distancetroispoints} donne
\begin{equation*}
{\frac{1}{d(1,2)^2}}={\frac{1}{R_{12}}}+{\frac{1}{R_{23}+R_{13}}}.
\end{equation*}
 $d(1,2)^2 $ apparait comme la r\'esistance \'electrique entre les points 1 et 2 du circuit en triangle.
\begin{figure}[h]
\begin{center}
\mbox{\rotatebox{0}{\scalebox{0.7}{\includegraphics{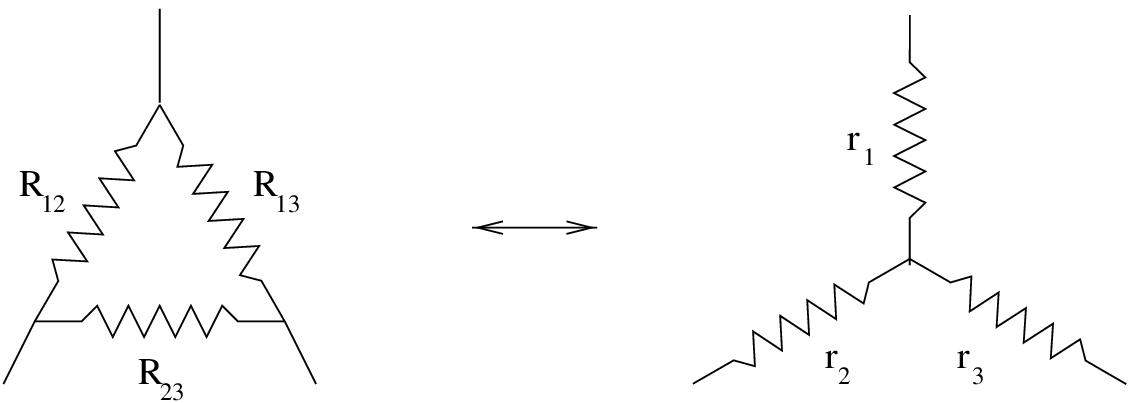}}}}
\end{center}
\end{figure}
Trouver le coefficient $D_{ij}$ signifie d\'eterminer trois
r\'esistances $R_{kp}$ induisant  une imp\'edance $d(i,j)^2$
entre les points $i,j$. Un r\'esultat classique d'\'electricit\'e
\cite{ed}  pr\'ecise que le circuit en triangle est \'equivalent
au circuit en \'etoile $(r_1,r_2,r_3)$ o\`u
\begin{equation}
\label{resistance} R_{12}={\frac{1}{ r_3}}({r_1r_2 + r_1r_3 +
r_2r_3}),
\end{equation}
$R_{13}$ et $R_{23}$ ayant des formules analogues obtenues par
permutations des indices. Dans le circuit en \'etoile,
\begin{eqnarray*}
&d(1,2)^2= r_1 + r_2,&\\
&d(1,3)^2= r_1 + r_3,&\\
&d(2,3)^2= r_2 + r_3
\end{eqnarray*}
d'o\`u
\begin{eqnarray*}
&2r_1=d(1,2)^2 + d(1,3)^2 - d(2,3)^2,&\\
&2r_2=d(1,2)^2 + d(2,3)^2 - d(1,3)^2,&\\
&2r_3=d(1,3)^2 + d(2,3)^2 - d(1,2)^2.
\end{eqnarray*}
Inser\'e dans (\ref{resistance}) ce syst\`eme d'\'equations
conduit \`a
 $$D_{12}=\sqrt{{   \frac {2(d(1,3)^2+d(2,3)^2-d(1,2)^2)}{2 (d(1,2)^2d(1,3)^2
+d(1,2)^2d(2,3)^2+d(1,3)^2d(2,3)^2)-d(1,2)^4-d(1,3)^4-d(2,3)^4}}}.$$
\hfill $\blacksquare$

\subsection{Espace \`a quatre points}

Calculer les distances dans un espace de $n$ points est une
t\^ache sans fin.  Le calcul de la norme du commutateur n'est a
priori plus g\'en\'eriquement possible d\`es que $n\geq 10$
($[D,a]$ \'etant une matrice antisymm\'etrique r\'eelle, son
polyn\^ome caract\'eristique est alors d'ordre 5 et il n'y pas de
formule explicite pour les racines des polyn\^omes de degr\'e
sup\'erieur \`a 4). Bien que pour $n\leq 9$ la norme soit
toujours calculable,  il apparait que la d\'etermination du
supr\'emum n'est d\'ej\`a plus toujours possible d\`es $n=4$.

On utilise les notations (\ref{x4}) ainsi que
\begin{equation*}
\label{d4} d_1 \doteq \frac{1}{D_{12}},\; d_2 \doteq {\frac{1}{
D_{13}}},\; d_3 \doteq { \frac{1}{ D_{14}}},\; d_4 \doteq
{\frac{1}{ D_{23}}},\; d_5 \doteq {\frac{1}{D_{24}}},\; d_6\doteq
{\frac{1}{D_{34}}}
\end{equation*}
o\`u $D_{ij}$ sont les composantes d'une matrice $4 \times 4$
antisymm\'etrique r\'eelle. Ainsi
$$[D,a] = \left(
\begin{array}{cccc}
0 &  -{\frac{x}{d_1}}  &-{\frac{y}{d_2}} & -{\frac{z}{d_3}}\\
\frac{x}{d_1}& 0 & {\frac{x-y}{d_4}}&{\frac{x-z}{d_5}} \\
\frac{y}{d_2} &\frac{y-x}{d_4}& 0 &{\frac{y-z}{d_6}} \\
 \frac{z}{d_3}      &{\frac{z-x}{d_5}}&\frac{z-y}{d_6}& 0
\end{array} \right).$$

{\prop \label{fuck} 1)  $d(1,2)$ est la racine d'un polyn\^ome de
degr\'e $\delta\leq 12$.

\hspace{2.4truecm} 2)  G\'en\'eralement $d(1,2)$ n'est pas
calculable par radicaux.

\hspace{2.4truecm} 3) Il existe des cas o\`u $d(1,2)$ est
calculable. Par exemple quand
${\frac{1}{d_2}}={\frac{1}{d_5}}=\infty$, on a le r\'esultat
suivant:
$$ \hspace{0.6truecm}d(1,2)\, = \, d_1 \;\text{ quand }\quad d_1^2\leq d_6^2,\hspace{98mm}$$
$$\hspace{1truecm}=\frac{{d_1}\,{\sqrt{{{\left( {{{d_3}}^2} + {d_1}\,{d_6} \right) }^2}}}}{ {\sqrt{{{{d_1}}^2} + {{{d_3}}^2}}}\,{\sqrt{{{{d_3}}^2} +
                  {{{d_6}}^2}}}} \quad\text{ quand } \quad d_1d_6 = d_3d_4,\hspace{42mm}$$
$$\hspace{1truecm} = \sqrt{\frac{ d_1^2(d_3^2 + d_6^2)(d_4^2 + d_6 ^2)}{(d_3d_4 - d_1d_6)^2}}\quad \text{ quand } \quad C \leq 0,\hspace{54mm}$$
$$\hspace{1truecm} = \max\lp\sqrt{\frac{ d_1^2(d_3^2 + d_4^2)}{(d_3+d_4)^2 + (d_1-d_6)^2}},\; \sqrt{ \frac{ d_1^2(d_3^2 - d_4^2)}{(d_3-d_4)^2
+(d_1+d_6)^2}}\rp \quad\text{autrement},$$ avec $\; C \doteq
({{({d_3} + {d_4}) }^2}{d_6} + ({d_1} - {d_6})({d_3}\,{d_4} -
{{{d_6}}^2})) ({{( {d_3} - {d_4} ) }^2}{d_6} +({d_1}+{d_6}) \,(
{d_3}{d_4} + {{{d_6}}^2})).$
\newline
$$\hspace{0.9truecm}d(1,3)\,  = {\sqrt{{{{d_3}}^2} + {{{d_6}}^2}}} \quad\text{ quand }\quad(d_3^2 + d_6^2) \leq (d_1d_6 - d_3d_4)^2,\hspace{54mm}$$
$$\hspace{1truecm} = {\sqrt{{{{d_1}}^2} + {{{d_4}}^2}}} \quad\text{ quand }\quad(d_1^2 + d_4^2) \leq (d_1d_6 - d_3d_4)^2,\hspace{40mm}$$
$$\hspace{1truecm} = \max\lp{\frac{{\sqrt{{{\left( {d_1}\,{d_3} +{d_4}\,{d_6} \right)}^2}}}}{{\sqrt{{{\left( {d_3} + {d_4}\right) }^2} +
            {{\left( {d_1} - {d_6} \right)}^2}}}}},{\frac{{\sqrt{{{\left( {d_1}\,{d_3} +{d_4}\,{d_6} \right) }^2}}}}
{{\sqrt{{{\left( {d_3} -{d_4}\right) }^2} + {{( {d_1} + {d_6}
)^2}}}}}} \rp\quad\text{ autrement}.$$ Par permutation on
d\'eduit $d(2,3)$, $d(3,4)$, $d(1,4)$ (resp. $d(2,4)$) de
$d(1,2)$ (resp.$d(1,3)$). }
\newline

\vspace{1mm}

\noindent La suite de cette section est enti\`erement consacr\'ee
\`a la preuve de la proposition \ref{fuck}. Pour $a$ dans
$\aa_+$, on note $\overrightarrow{r_a}$  le vecteur colonne de
composantes $(x,y,z)$. Les fonctions de $\rr^3$ dans $\rr$
\begin{eqnarray*}
\alpha (\overrightarrow{r}) &\doteq&
\frac{x^2}{{d_1}^2}+\frac{y^2}{{d_2}^2} + \frac{z^2}{{d_3}^2} +
\frac{(x-y)^2}{{d_4}^2} +\frac{(x-z)^2}{{d_5}^2} +
\frac{(y-z)^2}{{d_6}^2},\\
\beta({\overrightarrow r}) &\doteq&{\frac{ x(y-z)} {d_1 d_6}} + {
\frac{ z(x-y)}{d_3 d_4}}
+{ \frac{ y(z-x)}  {d_2 d_5}},\\
n(\overrightarrow{r}) &\doteq& \alpha(\overrightarrow{r}) + \sqrt{\alpha(\overrightarrow{r})^2-4\beta(\overrightarrow{r})^2},\\
f(\overrightarrow{r}) &\doteq& \alpha(\overrightarrow{r}) -
\beta(\overrightarrow{r})^2 -1
\end{eqnarray*}
permettent de d\'efinir les surfaces $\mathcal{N}$ et
$\mathcal{F}$ dans $\rr^3$
\begin{eqnarray*}
\mathcal{N} &\doteq& \{\overrightarrow{r}\in
\mathbb{R}^3 \; / \; n(\overrightarrow{r}) = 2\},\\
\mathcal{F}&\doteq& \{ \overrightarrow{r} \in \mathbb{R}^3 \; /
\; f(\overrightarrow{r})=0 \},\quad \text{ avec }
\mathcal{N}\subset \mathcal{F}.
\end{eqnarray*}
\newline

{\lem \label{fuck1} 1) Pour $a \in \mathcal{A}_{+}$,
$\|[D,a]\|^2= \frac{1}{2}n(\overrightarrow{r_a}).$\newline

\hspace{1.5truecm} 2) Si $\overrightarrow{r} \in \mathcal{N}$ est
tel que $\alpha(\overrightarrow{r})=2$, alors $(\grad{f})
(\overrightarrow{r})= 0$.}
\newline

\noindent {\it Preuve.}
 1) Les quatre valeurs propres de $i[D,a]$ sont $\lambda_i = \pm \frac{1}{  \sqrt{2}} \sqrt{\alpha \pm \sqrt{ \alpha ^2 - 4 \beta
^2}}(\overrightarrow{r_a}),$ d'o\`u
$$\|[D,a]\|^2=\frac{1}{2}(\alpha + \sqrt{ \alpha ^2 - 4 \beta ^2})(\overrightarrow{r_a}).$$

2) Montrons que ${\frac{\partial f} {\partial y}}
(\overrightarrow{r}) = 0,$ la preuve \'etant identique pour les
autres composantes de $\grad{f}$. Comme $\overrightarrow{{r}} \in
\mathcal{N}\subset \mathcal{F}$ et $\, \alpha(\overrightarrow{r})
=2$, alors $\beta(\overrightarrow{r}) =\pm 1$. Si $\;
\beta(\overrightarrow{r}) =  1$
\begin{eqnarray*}
&\alpha(\overrightarrow{{r}}) = 2 \beta(\overrightarrow{{r}}),&\\
&{\frac{\partial f}  {\partial y}}(\overrightarrow{r}) =
{\frac{\partial \alpha}  {\partial y}}(\overrightarrow{r})
-2\beta(\overrightarrow{{r}}){\frac{\partial \beta}  {\partial
y}}(\overrightarrow{r}) = {\frac{\partial \alpha}  {\partial
y}}(\overrightarrow{r}) - 2{\frac{\partial \beta}  {\partial
y}}(\overrightarrow{r}).&
\end{eqnarray*}
Un calcul explicite de $\alpha(\overrightarrow{r}) - 2
\beta(\overrightarrow{r})=0$ montre que
$$\frac{x}{d_1} = {\frac{y-z}{d_6},\; \frac{y}{d_2}} = {\frac{z-x}{d_5}},\; \frac{z}{d_3} = {\frac{x-y}{d_4}}$$
d'o\`u ${\frac{\partial \alpha}{\partial y}}(\overrightarrow{r})
- 2{\frac{\partial \beta}{\partial y}}(\overrightarrow{r})=0$. La
preuve est identique lorsque
$\beta(\overrightarrow{r})=-1$.\hfill $\square$
\newline

\noindent Comme corollaire imm\'ediat, l'\'equation
(\ref{distance2}) s'\'ecrit
\begin{equation}
\label{rajout} d(1,2) = \sup \;\{ \scl{e_1}{ {\overrightarrow
{r_a}}}\, / \, {\overrightarrow{r_a}}\in\mathcal{ N}\}
\end{equation}
ou $e_1$ est le premier vecteur de la base canonique de $\rr^3$
et $\scl{.}{.}$ d\'esigne le produit scalaire euclidien usuel.
Cette formule n'est pas utilisable dans la mesure o\`u
$\mathcal{N}$ n'est pas d\'efinie par une forme quadratique. Il
est plus facile de travailler avec $\mathcal{F}$.
\newline

\noindent{\lem \label{fuck2} $ d(1,2) \in  \left\{
\scl{e_1}{{\overrightarrow{r}}} \; / \; \overrightarrow{r}\in
\mathcal{F} \,\text{ et }\, \frac{\partial f}{\partial
y}(\overrightarrow{r})=\frac{\partial f}{\partial
z}(\overrightarrow{r})=0\right\}$.}
\newline

\noindent {\it Preuve.} Le supr\'emum dans (\ref{rajout}) est
atteint en un point $\overrightarrow{r}$ tel que
$(\grad{n})(\overrightarrow{r})$, s'il est d\'efini, est
parall\`ele \`a l'axe des $x$. Si $\alpha(\overrightarrow{r})=2$,
alors $(\grad{n})(\overrightarrow{r})$ n'est pas d\'efini mais
$${\frac{\partial f}
{\partial y}}(\overrightarrow{r})={\frac{\partial f}  {\partial
z}}(\overrightarrow{r})=0$$ par le lemme \ref{fuck1}.  Si
$\alpha(\overrightarrow{r})\neq 2$, alors
$(\grad{f})(\overrightarrow{r})$ est colin\'eaire \`a
$(\grad{n})(\overrightarrow{r})$, de sorte que
$$ {\frac{\partial f} {\partial y}}(\overrightarrow{r})={\frac{\partial f} {\partial z}}(\overrightarrow{r})=0.$$
Pour conclure, il suffit de remarquer que pour tout
$\overrightarrow{r} \in \mathbb{R}^3$, il existe $a
\in\mathcal{A_+}$ tel que $\overrightarrow{r} =
\overrightarrow{r_a}$, par exemple $a=(\xi, \xi-x,\xi-y,\xi - z)$
o\`u $\xi \doteq \sup\{|x|,|y|,|z|\}.$\hfill $\square$
\newline

\noindent Ainsi la distance est une racine commune \`a un
polyn\^ome de plusieurs variables et \`a ses polyn\^omes
d\'eriv\'es. Avant d'entreprendre des calculs explicites,
rappelons quelques r\'esultats concernant les syst\`emes
d'\'equations polyn\^omiales.

\subsubsection*{Remarques sur les syst\`emes d'\'equations polyn\^omiales}

Soient $P$ et $Q$ deux polyn\^omes de la forme
\begin{eqnarray*}
P(x) &=& a_n x^n + a_{n-1} x^{n-1} + ... + a_0\\
Q(x) &=& b_m x^m + b_{m-1} x^{m-1} + ... + b_0
\end{eqnarray*}
avec $a_n,b_m \neq 0$. Sans conna\^{\i}tre explicitement les
racines $p_i$, $q_j$ de $P$, $Q$, on peut calculer \cite{lang}
par une s\'erie de manipulations alg\'ebriques des coefficients
$a_i$, $b_j$ le r\'esultant de $P$ et $Q$
\begin{equation}
\label{resultante} \index{Res(P)}Res(P,Q)\doteq a_n^m
b_m^n\,{\underset {i,j}{\Pi}} (p_i - q_j), \quad1\leq i \leq n,
\; 1 \leq j \leq m.
\end{equation}
$Res(P,Q)$ est un polyn\^ome en $a_i$ et $b_j$. $P$ et $Q$ ont
une racine commune si et seulement si leur r\'esultant est nul. Un
r\'esultant bien connu est le discriminant
\begin{equation*}
\index{Dis(P)}Dis(P)\doteq Res(P,P')
\end{equation*}
dont les racines sont les racines doubles de $P$.

Lorsque $P$ et $Q$ sont des polyn\^omes de plusieurs variables
$x,y,z$, on d\'esigne par $Res[P,Q,y]$ le r\'esultant de $P$ et
$Q$ vu en tant que polyn\^ome en $y$. De m\^eme  $Dis[P,z]$
d\'esigne le discriminant de $P$ vu commme polyn\^ome en $z$.

{\lem\label{fuck3}
 Soit $P(x,y,z)$ un polyn\^ome de degr\'e 2 en $z$ dont les coefficients sont des fonctions r\'eelles de $x$ et $y$. Si
$$P(x_0,y_0,z_0)={\frac{\partial P}{\partial y}}({x_0,y_0,z_0}) = {\frac{\partial P}{\partial z}}({x_0,y_0,z_0}) =0$$
pour $(x_0,y_0,z_0) \in\mathbb{R}^3$, alors $x_0$ est une racine
du polyn\^ome $Dis[Dis(P,z),y]$ et $y_0$ est une racine de
$Dis(P,z)(x_0,y)$.}
\newline

\noindent {\it Preuve.} En posant $P(x,y,z) = a(x,y)z^2 + b(x,y)z
+ c(x,y)$, on obtient
\begin{eqnarray*}
V                                &\doteq  & Dis(P,z) = a(4ac - b^2),\\
{\frac{\partial V}{\partial y}}&=       & {\frac{\partial
a}{\partial y}}(8ac - b^2)- 2ab{\frac{\partial b}{\partial y}} +
4a^2{\frac{\partial c}{\partial y}},\\ Res({\frac{\partial
P}{\partial y}},{\frac{\partial P}{\partial z}},z)&=&
b^2{\frac{\partial a}{\partial y}} -2ab{\frac{\partial
b}{\partial y}} +  4a^2{\frac{\partial c}{\partial
y}}.\end{eqnarray*} $P(x_0,y_0,z_0)= {\frac{\partial P}{\partial
z}}({x_0,y_0,z_0})$ signifie que $V(x_0,y_0)=0$, c'est \`a dire
$(8ac-b^2)(x_0,y_0)=b^2(x_0,y_0)$ d'o\`u
\begin{equation*}
Res({\frac{\partial P}{\partial y}},{\frac{\partial P}{\partial
z}},z)({x_0,y_0})={\frac{\partial V}  {\partial y}}({x_0,y_0}).
\end{equation*}
Ainsi ${\frac{\partial P}{\partial
y}}({x_0,y_0,z_0})={\frac{\partial P}{\partial
z}}({x_0,y_0,z_0})$ implique ${\frac{\partial V} {\partial
y}}({x_0,y_0})=0 = V(x_0,y_0),$ si bien que $y_0$ est une racine
double de $V(x_0,y)$ et
$$Dis(V,y)(x_0)=Dis[Dis[P,z],y](x_0)=0.$$\hfill $\square$

\subsubsection*{Preuve de la proposition \ref{fuck}}

 Gr\^ace aux lemmes \ref{fuck1} et \ref{fuck2}
$$d(1,2) \in  \{  x \,  / \, Dis[V(x,y),y]=0\} \text{ avec } V(x,y)=Dis[f(x,y,z),z].$$
Au lieu de $V(x,y)$, on utilise une forme effective pour
\'eliminer le terme correctif $a_n^m b_m^n$ apparaissant dans
(\ref{resultante}) de mani\`ere \`a ce que les racines de
$V_{eff}$ correspondent exactement \`a l'existence d'une racine
commune \`a $f$ et $\frac{\partial f}{\partial y}$:
$$V_{eff}(x,y) \doteq \text{Num\'erateur}(\frac{Dis(f,z)}{n^nf_n^{2n-1}} )$$
o\`u $f_n$ est le coefficient de plus haut degr\'e de $f$ vu
comme polyn\^ome en $z$ et $n=\text{deg}(f)$. Bien entendu le
num\'erateur est pris apr\`es simplification (pas toujours
possible) de la fraction.
\newline

1)  Par calcul direct, $V_{eff}(x,y)=V_i\,y^i, \; 0\leq i\leq 4$.
L'expression exacte des $V_i$ est donn\'ee en annexe. Ce sont des
polyn\^omes en $x$ de la forme
\begin{eqnarray*}
&V_4(x)=v_{4_0},\;V_3(x)=v_{3_1}x,\;V_2(x)=v_{2_2}x^2+v_{2_0},&\\
&V_1(x)=v_{1_3}x^3+v_{1_1}x,\;V_0(x)=v_{0_4}x^4+v_{0_2}x^2+v_{0_0}.&
\end{eqnarray*}
Le discriminant $J$ d'un polyn\^ome $C=C_i y^i$ de degr\'e quatre
est
\begin{eqnarray*}
J(C)\doteq Res[C,C'] &=& {C_4}({{{C_3}}^2}({{{C_1}}^2}{{{C_2}}^2}
- 4{{{C_1}}^3}{C_3}+ 18{C_0}{C_1}{C_2}{C_3} - {C_0}( 4{{{C_2}}^3}
+
27{C_0}{{{C_3}}^2} ))\\
  &+& 2( -2{{{C_2}}^3}({{{C_1}}^2} - 4{C_0}{C_2} ) + {C_1}{C_2}( 9{{{C_1}}^2} - 40{C_0}\,{C_2} ){C_3} - 3{C_0}( {{{C_1}}^2} - 24{C_0}{C_2} ){{{C_3}}^2}){C_4}\\
&-&( 27{{{C_1}}^4} - 144{C_0}{{{C_1}}^2}{C_2} +
128{{{C_0}}^2}{{{C_2}}^2} +
       192{{{C_0}}^2}{C_1}{C_3}){{{C_4}}^2} + 256{{{C_0}}^3}{{{C_4}}^3}).
\end{eqnarray*}
En rempla\c{c}ant $C_i$ par $V_i(x)$ on voit que  $J$ est un
polyn\^ome en $x$ de degr\'e $\delta\leq 12$.
\newline

2) Il suffit d'exhiber un contre-exemple. Prenons
$$d_1 = d_2 = d_3 = d_4 = d_6 = 1 \text{ et } \frac{1}{ d_5}=0$$
qui donne
$$f(x,y,z)= x^2 + y^2 + z^2 + (x-y)^2 + (y-z)^2 -(x(y-z) + z(x-y))^2.$$
C'est un polyn\^ome de degr\'e 2 en $x,y$ et $z$. Les
r\'esultants sont facilement calculables et l'on obtient
\begin{eqnarray*}
V_{eff}(x,y) &=& 2 - 6\,{y^2} + 3\,{y^4} +
 4\,{x^2}\,\left( -1 + {y^2} \right)  -
  4\,x\,y\,\left( -1 + {y^2} \right),\\
Dis(V_{eff},y) &=& -768\,\left( -54 - 54\,{x^2} + 135\,{x^4} +
    296\,{x^6} - 368\,{x^8} + 128\,{x^{10}}
    \right).
\end{eqnarray*}
Posons maintenant
 $$p(x)=-128 x^5 + 368x^4 - 296 x^3 - 135x^2 +54x + 54.$$
$p$ a une racine r\'eelle $x_1$, deux racines complexes
distinctes $x_2$, $x_4$ et leurs conjugu\'ees $x_3$ \& $x_5$.

La th\'eorie de Galois permet de montrer que $p$ ne peut \^etre
r\'esolu par radicaux. On indique ici les \'etapes essentielles
du raisonnement, renvoyant au livre [\citelow{galois}] pour une
pr\'esentation d\'etaill\'ee de la th\'eorie. On remarque que
$p(-11)$, $p(-5)$ et $p(1)$ sont premiers. Si $p(x)= f(x) g(x)$
o\`u $f,g$ sont dans $\mathbb{Q}[x]$ avec $g$ de degr\'e au plus
$2$, alors $g(-11) \in \{ p(-11), 1\}$ (on peut supposer ce
coefficient positif), $g(-5) \in \{\pm p(-5), \pm 1\}$ et $g(1)=
\{\pm p(1), \pm 1\}$. Par interpolation, on d\'etermine pour
chacun des $32$ triplets $\lp \alpha_1\doteq g(-11),
\alpha_2\doteq g(-5), \alpha_3 \doteq g(1)\rp$  possibles le
polyn\^ome $g$ correspondant,
$$g(x) \doteq \sum_{i=1}^3 \frac{\alpha_i}{\underset{j\neq i}{\Pi}(\alpha_i - \alpha_j)} \underset{j\neq i}{\Pi}(x-\alpha_j).$$
 On v\'erifie ensuite qu'aucun de ces polyn\^omes ne divisent $p$.  Donc $p$ est irr\'eductible sur $\mathbb{Q}$.
Soit  $E/\mathbb{Q}$ une "splitting field" extension de $p$.
Puisque $p$ a cinq racines distinctes, son groupe de Galois
$$\mathbb{G}=Gal(E/\mathbb{Q})$$
est isomorphe \`a un sous groupe du groupe de symm\'etrie $S_5$
(groupe des permutations des racines
$X\doteq\{x_1,x_2,x_3,x_4,x_5\})$. Comme $p$ n'a pas de racines
doubles, $p$ est s\'eparable et
$$|\mathbb{G}|=[E/\mathbb{Q}]$$
o\`u $|\mathbb{G}|$ est l'ordre de $\mathbb{G}$ et
$[E/\mathbb{Q}]$ son indice, c'est \`a dire le nombre de cosets
$\mathbb{Q}$ dans $\mathbb{G}$. Si $\alpha$ est une racine de $p$
alors
$$[\mathbb{Q}(\alpha),\mathbb{Q}]=5$$
donc
$|\mathbb{G}|=[E/\mathbb{Q}]=[E/\mathbb{Q}(\alpha)][\mathbb{Q}(\alpha),\mathbb{Q}]$
est divisible par 5. Donc $\mathbb{G}$ contient un \'el\'ement
d'ordre 5, \`a savoir le 5-cycle $\tau=(12345)$. Un autre
\'el\'ement, not\'e $\sigma$, de $\mathbb{G}$ est donn\'e par la
restriction \`a $X$ de la conjugaison complexe
$$\sigma=(23)(45).$$
$\sigma$ est d'ordre deux, donc $|\mathbb{G}|$ est divisible par
2. Comme $$\tau\sigma = (124)\in\mathbb{G}$$ est d'ordre trois,
$|\mathbb{G}|$ est un multiple de $5\times 2\times 3=30$ et
divise $|S_5|=120$. Dans la mesure o\`u $S_5$ n'a pas de sous
groupe d'ordre 30, $|\mathbb{G}|\in\{60,120\}.$  Puisque les
sous-groupes r\'esolubles de $S_5$ sont d'ordre au plus $24$
[\citelow{galois}, {\it Thm. A 38}], le th\'eor\`eme de Galois
indique que $p$ n'est pas r\'esoluble par radicaux, tout comme
$Dis(V_{eff},y)=768 p(x^2).$
\newline

3) Lorsque $d_2 = d_5 = 0$ et $d_1d_6\neq d_3d_4$
\begin{eqnarray*}
Dis(V_{eff})=-16d_1^{16}d_3^{14}d_4^{12}d_6^{14}(d_4^2+d_6^2)(x^2-d_1^2)(x^2(d_3d_4-d_1d_6)^2-d_1^2(d_3^2+d_6^2)(d_4^2+d_6^2))\\
             (x^2((d_3-d_4)^2+(d_1+d_6)^2)-d_1^2(d_3-d_4)^2)^2(x^2((d_3+d_4)^2+(d_1-d_6)^2)-d_1^2(d_3+d_4)^2)^2.
\end{eqnarray*}
Ce  polyn\^ome a quatre racines simples  $\pm x_0 , \pm x_1$ et
quatre racines doubles $\pm x_2 , \pm x_3$
$$
x_0=|d_1|,\; x_1=|d_1|{
\frac{\sqrt{(d_3^2+d_6^2)(d_4^2+d_6^2)}}{|d_3d_4 - d_1d_6|}},$$
$$
x_2=|d_1|\sqrt{ {\frac{(d_3^2 + d_4^2)}{(d_3+d_4)^2 +
(d_1-d_6)^2}}},\; x_3=|d_1|\sqrt{ {\frac{(d_3^2 -
d_4^2)}{(d_3-d_4)^2+(d_1+d_6)^2}}}.$$ On sait d'apr\`es les
lemmes \ref{fuck1} et \ref{fuck2} que $d(1,2)$ est l'un de ces
$x_i$ et que la valeur $y_i$ associ\'ee est une racine double de
$V_{eff}(x_0,y)$. La valeur $z_i$ associ\'ee est d\'etermin\'ee
en r\'esolvant $f(x_i,y_i,z)=0$. On d\'etermine ensuite sous
quelles conditions chacun des $x_i$  v\'erifie $n(x_i,y_i,z_i)=2$
et on prend le plus grand d'entre eux. Dans le d\'etail, cette
m\'ethode appliqu\'ee \`a $x_0$ donne $y_0 = d_1, z_0 = 0$ et
$$
n(x_0,y_0,z_0)=1 + \frac{d_1^2}{d_6^2} + \sqrt{ \frac{(d_1^2 -
d_6^2)^2}{d_6^4}} = \left\{
\begin{array}{ccc}
2 &\text{si}& d_1^2 \leq d_6^2,\\
2\frac{d_1^2}{ d_6^2} > 2  & \text{si}& d_1^2 > d_6^2.
\end{array} \right.
$$
Ainsi $x_0$ ne peut \^etre solution que si $d_1^2 \leq d_6^2$. En
utilisant les valeurs de $y_1$ et $z_1$ donn\'ees en annexe, on
trouve que $x_1$ ne peut pas \^etre solution sauf si $C \leq 0$.
Au contraire $x_2$ et $x_3$ sont toujours susceptibles d'\^etre
solution. Maintenant appliquons la proposition \ref{chemin} en
annulant tous les liens sauf $d_1$. On trouve  $d(1,2) \leq x_0$
de sorte que si $d_1^2 \leq d_6^2$ alors $d(1,2)=x_0.$ Comme $x_1
\geq x_2$ et $x_1 \geq x_3$,  $d(1,2) = x_1$ lorsque $C \leq 0$,
et $d(1,2)= \max(x_1,x_2)$ autrement. Lorsque  $d_1d_6=d_3d_4$,
$x_1$ n'est pas d\'efini mais la preuve est identique.

Le calcul de $d(1,3)$ est semblable. On cherche maintenant le
maximum de $y$. $Dis(V,x)$ est un polyn\^ome en $y$ de degr\'e
douze avec les racines simples $\pm y_0, \pm y_1$ et les racines
doubles $\pm y_2, \pm y_3$
\begin{eqnarray*}
y_0 = {\sqrt{{{{d_3}}^2} + {{{d_6}}^2}}} &,& y_1 = {\sqrt{{{{d_1}}^2} + {{{d_4}}^2}}},\\
y_2 = {d_1}{\frac{{| {d_1}\,{d_3} +{d_4}\,{d_6}|}}{{\sqrt{{{(
{d_3} + {d_4})}^2} +{{(d_1-{d_6}) }^2}}}}} &,& y_3 =
{\frac{{|d_1d_3+d_4d_6|}}{{\sqrt{{{( {d_3} - {d_4}) }^2} +  {{(
{d_1} + {d_6}) }^2}}}}}.
\end{eqnarray*}
Les valeurs des $x_i$ et $z_i$ en appendice permettent de
v\'erifier que $y_0$  (resp. $y_1$) ne peut-\^etre solution que si
${{{({{{d_3}}^2}+{{{d_6}}^2})}^2}\leq{{(
{d_3}{d_4}-{d_1}{d_6})}^2}}$ (resp. ${{({{{d_1}}^2}+{{{d_4}}^2})
}^2}\leq{{({d_3}{d_4}- {d_1}{d_6})}^2})$. Comme pr\'ec\'edemment,
$y_2$ et $y_3$ sont susceptibles d'\^etre toujours solution. On
obtient le r\'esultat en remar\-quant que $y_2, y_3 \leq y_0$,
$y_2$, $y_3 \leq y_1$ et que si $y_0$ et $y_1$ sont susceptibles
d'\^etre solution, alors $y_0=y_1$.\hfill$\blacksquare$
\newline

Contrairement \`a l'espace \`a trois points, les distances dans
l'espace \`a quatre points ne peuvent pas \^etre lues directement
dans l'op\'erateur de Dirac \`a l'aide d'un algorithme fini. Le
calcul des distances rel\`eve donc d'une approche plus
pragmatique et doit \^etre envisag\'e cas par cas. Il n'est pas
non plus possible de remonter des distances vers l'op\'erateur de
Dirac: caract\'eriser les m\'etriques qui proviennent d'un
op\'erateur de Dirac est une question sans r\'eponse claire dans
la mesure o\`u il n'y a pas ici de formules \`a inverser comme
cela a pu \^etre le cas dans l'espace \`a trois points. Il y a
cependant une solution pour peu que l'on rel\^ache une contrainte
sur le triplet spectral, \`a savoir le choix de l'espace de
repr\'esentation. On montre en effet, et c'est l'objet de la
derni\`ere section de ce chapitre, que pour toute m\'etrique dans
un espace de $n$ points il est possible de construire un triplet
spectral, satisfaisant aux axiomes de la g\'eom\'etrie non
commutative,  tel que la distance associ\'ee soit pr\'ecis\'ement
la m\'etrique d\'esir\'ee.

\section{Distance et axiomes de la g\'eom\'etrie non commutative}

Dans les sections pr\'ec\'edentes nous n'avons que peu tenu
compte des axiomes de la g\'eom\'etrie non commutative
pr\'esent\'es dans le premier chapitre. Ces axiomes sont
introduits afin que la g\'eom\'etrie (\ref{td}) co\"{\i}ncide
avec la g\'eom\'etrie spinorielle riemanni\-enne. Dans l'optique
o\`u les espaces finis seraient des \'equivalents discrets de
vari\'et\'e riemannienne, il peut \^etre important que les
triplets consid\'er\'es soient des triplets spectraux r\'eels
(cf. d\'efinition \ref{gnc}). Mais dans le cas finis, ces
axiomes\cite{krajew} obligent \`a travailler avec des matrices
dont la taille croit rapidement avec le nombre de points. Les
calculs explicites sont vite impraticables sauf dans quelques cas
simples. Telle est la raison pour laquelle nous n'en avons pas
tenu compte dans la section consacr\'ee aux espaces commutatifs
finis. Cet oubli volontaire n'est pas lourd de cons\'equence car
on peut montrer que pour les espaces commutatifs finis les
axiomes n'imposent pas de contraintes sur les m\'etriques
susceptibles d'\^etre d\'ecrites par un triplet spectral.
Autrement dit, \'etant donn\'es $n^2 - n$ nombres r\'eels
$d_{ij}$ ($i\neq j$) strictement positifs tels que
\begin{equation}
\label{hypo} d_{ij}=d_{ji}\, \text{ et }\, d_{ij}\leq
d_{ik}+d_{kj},
\end{equation}
 il existe un triplet spectral r\'eel $(\aa,\hh,D, J, \Gamma)$ avec
$\aa=\mathbb{C}^{n}$ tel que la distance associ\'ee sur
l'ensemble des \'etats purs de  $\aa$  soit donn\'ee par les
nombres $d_{ij}$. Pour le montrer nous aurons besoin du lemme
suivant.

On note $a_i$ les composantes d'un \'el\'ements $a$ de $\aa$ et
$\pi$ la repr\'esentation de $\aa$ sur $\hh$. Les $d_{ij}$ sont
des r\'eels satisfaisant (\ref{hypo}).

{\lem Il existe un triplet spectral r\'eel $(\aa,\hh,D, J,
\Gamma)$ tel que
\begin{equation}
\label{comm} \|[D,\pi(a)]\|=\mathop{\sup}\limits_{1\leq i,j\leq
n, i\neq j}\frac{|a_{i}-a_{j}|}{d_{ij}}.
\end{equation}}

\noindent{\it Preuve.} Raisonnons par r\'ecurrence sur  $n$. Pour
$n=2$ prenons $\aa_{2}=\mathbb{C}^{2}$, $\hh_2=\mathbb{C}^{3}$,
$$\pi_{2}(a_1, a_2)=\dm{ccc} a_1&0&0\\ 0& a_2 &0\\ 0&0&a_2\fm \text{ et }\, \Gamma_{2}=\dm{ccc} 1&0&0\\ 0&-1&0\\ 0&0&1 \fm.$$
L'op\'erateur $D_{2}$ et la structure r\'eelle $J_{2}$ sont
$$
D_{2}=\left(
\begin{array}{ccc}
0       &{\frac{1}{d_{12}}}&0\\
\frac{1}{{d_{12}}}&0       &\frac{1}{d_{12}}\\
 0      &\frac{1}{d_{12}}&0
\end{array}
\right),\quad J_{2}=\left(
\begin{array}{ccc}
0&0&1\\
0&1&0\\
1&0&0
\end{array} \right)C
$$
o\`u $C$ d\'esigne la conjugaison complexe.

Dans le cas fini les axiomes se r\'eduisent \`a la {\it
r\'ealit\'e}, le {\it premier ordre}, l'{\it orientabilit\'e} et
la {\it dualit\'e de Poincar\'e}\cite{krajew}. Dans le cas
pr\'esent les deux premiers axiomes sont des relations de
commutations ($[\pi_2(a),
J_2\pi_2(b^*)J_2^{-1}]=[J_2,\Gamma_2]=[J_2, D_2]=0$,
$[[D_2,\pi_2(a)], J\pi_2(b)J^{-1}]=0$) et de multiplication
($J_2^2=\ii$) faciles \`a v\'erifier. A noter que la
repr\'esentation de l'alg\`ebre oppos\'ee $\aa_2^\circ = \cc^2$
est
$$\pi_2^\circ (a_1, a_2) = J \pi_2(a_1, a_2) J^{-1} = \dm{ccc} a_2 & 0 &0 \\ 0&a_2 & 0 \\ 0&0&a_1\fm.$$
$\Gamma_2$ est bien une chiralit\'e puisque $\Gamma^2 = \ii$,
$[\Gamma_2, \pi_2(a)] = 0$ et $\Gamma_2$ anticommute avec $D_2$.
La condition d'orientabilit\'e est vrai pour peu que l'on \'ecrive
$$\Gamma_{2}=\pi_{2}(1,-1)J_{2}\pi_{2}(-1,1)J_{2}^{-1}. $$
Concernant la dualit\'e de Poincar\'e, on sait par la
p\'eriodicit\'e de Bott\cite{jgb} que tous les groupes $K_r$ pour
$r$ impair sont isomorphes \`a $K_1$, qui est nul pour
l'alg\`ebre $\cc^2$, tandis que pour $r$ pair
$$K_r(\cc^2)\simeq K_0(\cc^2)\simeq \zz^2.$$ Le seul couplage est donc de $K_0(\cc^2) \times K_0(\cc^2)$ dans
$\zz$.  Pour $l=1$, les projecteurs de $P_l(\cc^2)$ sont
$p_1\doteq (1,0)$ et $p_2\doteq (0,1)$. Tout autre projecteur
pour $l$ quelconque est \'equivalent, au sens de
(\ref{equivpoinc}), \`a $p_1$ ou $p_2$, qui sont donc les
g\'en\'erateurs de $K_0(\cc^2)$. Le couplage (\ref{intersection})
est donn\'ee par la {\it matrice de la forme d'intersection} de
coefficients
$$\cap_{ij}  = \text{dim } (\text{ker } P_{ij} {D_2}^+ P_{ij}) - \text{dim }( \text{ker } P_{ij} {D_2}^- P_{ij}) $$
o\`u
$$P_{ij} \doteq  \pi_2(p_i) J \pi_2(p_j) J^{-1}$$ et
${D_2}^+ = ({D_2}^-)^{\dagger}$
est donn\'e en (\ref{dplus}). En dimension finie, un op\'erateur
$\oo$ d'un espace de dimension $m$ dans un espace de dimension
$n$ et son adjoint ont des images de m\^eme dimension, de sorte
que
$$\text{indice } \oo \doteq \dim \ker \oo - \dim \ker \oo^\dagger  = m - n.$$
Ainsi
\begin{eqnarray}
\nonumber \cap_{ij} &=& \text{indice } (P_{ij} {D_2}^+ P_{ij}) =
\text{dim } \frac{\ii + \Gamma_2}{2}P_{ij} -
\text{dim } P_{ij}\frac{\ii - \Gamma_2}{2},\\
\label{matintersection}
            &=& \tr{\Gamma_2 P_{ij}}.
\end{eqnarray}
o\`u l'on utilise que la dimension d'un projecteur est donn\'e
par sa trace. On trouve alors
$$
\cap_2= \left( \begin{array}{cc}
0&1\\
1&-1
\end{array} \right)
$$
qui est de d\'eterminant non nul, donc la dualit\'e de Poincar\'e
est satisfaite. Enfin un rapide calcul garantit que
$$
||\lb D_{2},\pi_{2}(a_1, a_2)\rb||=\frac{|a_{1}-a_{2}|}{d_{12}}.
$$

Supposons maintenant que $(\aa_{n},\hh_{n},D_{n}, \pi_{n},
\Gamma_{n}, J_{n})$ aient \'et\'e construits pour $n>2$.
Construire le triplet spectral \`a l'ordre $n+1$ consiste tr\`es
exactement \`a r\'ep\'eter la construction de l'ordre $n=2$. Soit
$\aa=\mathbb{C}^{n+1}$ et
$$
\hh_{n}=\hh_{n-1}\oplus \left(
\mathop{\oplus}\limits_{i=1}^{n-1}\hh_{n}^{i}\right)$$ avec
$\hh_{n}^{i}=\mathbb{C}^{3}$ pour tout $i$ et $n$. D\'esignons de
mani\`ere g\'en\'erique par $\oo$ les op\'erateurs $D,\pi,
\Gamma$ ou $J$. $\oo$ est diagonal par bloc et d\'efini de
mani\`ere r\'ecursive par
$$\oo_{n}=\oo_{n-1}\oplus\left(\mathop{\oplus}\limits_{i=1}^{n-1}\oo_{n}^{i}\right). $$
Comme pour $n=2$, on d\'efinit
$$
\pi_{n}^{i}(a_{i},a_{n})=\mathrm{diag}(a_{i},a_{n},a_{n})\,, \;
\Gamma_{n}^{i}=\mathrm{diag}(1,-1,1).
$$
L'op\'erateur de Dirac $D_{n}$ et la conjugaison de charge
$J_{n}$ sont donn\'ees par
$$
D_{n}^{i}= \dm{ccc}
0               & \frac{1}{d_{in}}  &0                \\
\frac{1}{d_{in}}& 0                 &\frac{1}{d_{in}}\\
 0              & \frac{1}{d_{in}}  &0
\fm, \quad J_{n}^{i}=\left(
\begin{array}{ccc}
0&0&1\\
0&1&0\\
1&0&0
\end{array} \right)C. $$
On v\'erifie alors ais\'ement tous les axiomes,  sauf la
dualit\'e de Poincar\'e qui n\'ecessite quelques pr\'ecautions.
Chaque it\'eration ajoute un g\'en\'erateur au groupe $K_0(\aa_n)
\simeq \zz^n$. La matrice de la forme d'intersection n'est autre
que la matrice de multiplicit\'e\cite{krajew} du triplet spectral
et on trouve
$$
\cap_{n}=\left(
\begin{array}{cccc}
0&1&\dots&1\\
1&-1&\ddots&\vdots\\
\vdots&\ddots&\ddots&1\\
1&\dots&1&-(n-1)
\end{array} \right).
$$
Pour tout entier $N$ on peut toujours consid\'erer le triplet
spectral trivial
$(\mathbb{C}^{n},\mathbb{C}^{n}\oplus\mathbb{C}^{N},0)$ avec
d'\'evidentes representation et conjugaison de charge et une
chiralit\'e \'egale \`a  -1. La somme directe de ce triplet
spectral avec $(\mathcal{A}_{n},\mathcal{H}_{n},D_{n})$, conduit
\`a une matrice de multiplicit\'e $\mu_n+NI_{n}$ qui n'est pas
d\'eg\'en\'er\'ee pour $N$ suffisamment grand.

Pour finir, le calcul de la norme du commutateur $||\lb
D_{n},\pi_{n}(a)\rb||$ d\'ecoule de la structure diagonale par
bloc des op\'erateurs et de l'hypoth\`ese de r\'ecurrence. \hfill
$\blacksquare$
\newline

On peut ainsi montrer le r\'esultat annonc\'e.

{\prop Etant donn\'es $n^2- n$ nombres r\'eels $d_{ij}$ positifs
satisfaisant (\ref{hypo}), il existe un triplet spectral
$(\aa,\hh,D)$ avec $\aa=\mathbb{C}^{n}$ satisfaisant aux axiomes
tel que la distance non commutative soit donn\'ee par les nombres
$d_{ij}$.}
\newline

\noindent{\it Preuve.} Gr\^ace au lemme ci-dessus on construit un
triplet spectral $(\aa,\hh,D)$ ob\'eissant aux conditions
(\ref{comm}). Lorsque $a$ v\'erifie la condition sur la norme,
alors $|a_{i}-a_{j}|\leq d_{ij}$, d'o\`u
$$d(i,j)\leq d_{ij}.$$

Fixons maintenant deux points tels que $d_{ij}<\infty$ et prenons
$a_{k}=d_{ik}$, fini pour tout $k$ gr\^ace \`a l'in\'egalit\'e
triangulaire); en particulier, $|a_{i}-a_{j}|=d_{ij}$.  Par
(\ref{hypo}), $|d_{ik}-d_{il}|\leq d_{kl}$ pour tout $k$ et $l$.
Ainsi $||\lb D,\pi(a)\rb||\leq 1$ par (\ref{comm}). D'o\`u
$$d_{ij}\leq d(i,j)$$
et le r\'esultat pour peu que $d_{ij}$ soit finie. Si tel n'est
pas le cas alors $d_{ik}$ et $d_{jk}$ sont \'egalement infinis
pour tout $k$. L'in\'egalit\'e (\ref{comm}) n'impose aucune
contrainte sur $a_{i}$ et $a_{j}$ puisque les \'el\'ements de
matrice de $D$  correspondant s'annulent. On peut donc envoyer
$|a_{i}-a_{j}|$ \`a l'infini et on a bien
$d(i,j)=d_{ij}$.\hfill$\blacksquare$
\newline

En conclusion, une fois donn\'ee $\aa =\cc^n$, les axiomes de la
g\'eom\'etrie non commutative n'apportent aucune contrainte sur
les m\'etriques susceptibles d'\^etre obtenues par un op\'erateur
de Dirac. De telles contraintes apparaissent quand on impose une
autre condition, comme fixer $\hh= \cc^n$ ainsi que nous l'avons
fait dans la discussion des cas \`a trois et quatre points. Il
est important de souligner que la fonction qui associe une
m\'etrique \`a un op\'erateur de Dirac est surjective.  Dans les
mod\`eles de gravit\'e quantique bas\'e sur les valeurs propres
de l'op\'erateur de Dirac\cite{landi}, on a \'egalement besoin de
savoir combien d'op\'erateurs de Dirac correspondent \`a une
m\'etrique donn\'ee, ainsi que les relations entre leurs spectres.

La contrainte (\ref{hypo}) est une condition n\'ecessaire et
suffisante pour que les nombres $d_{ij}$ soient des distances.
Rien n'emp\^eche de poser une condition plus forte. Par exemple
si l'on d\'esire voir l'espace non commutatif comme une
disrc\'etisation de l'espace euclidien,  les $d_{ij}$ doivent,
selon la dimension de cet espace, v\'erifier des conditions
suppl\'ementaires\cite{pyr1,pyr2} (pour que six nombres positifs
soient les distances euclidiennes entre sommets d'une pyramide,
il ne suffit pas qu'ils v\'erifient l'in\'egalit\'e triangulaire
trois par trois).

\vfill \pagebreak

\chapter{Produit du continu par le discret}

L'un des int\'erets de la g\'eom\'etrie non commutative en
physique est de proposer des mod\`eles simples d'espace-temps
o\`u cohabitent le discret et le continu. Un m\^eme formalisme
permet de rendre compte des sym\'etries de l'espace-temps continu
de la relativit\'e g\'en\'erale (malheureusement uniquement avec
signature euclidienne), \`a savoir l'invariance par
diff\'eomor\-phisme, ainsi que des sym\'etries de jauge des
interactions \'electrofaibles et fortes. Ces derni\`eres sont
interpr\'et\'ees g\'eom\'etri\-quement comme sym\'etries d'un
espace interne discret. Nous reviendrons longuement dans le
dernier chapitre sur cette interp\'etation qui donne une
justification g\'eom\'etrique au champ de Higgs. Ici nous
\'etudions l'aspect m\'etriques de ce type de g\'eom\'etries,
d\'ecrites par des produits tensoriels de triplets spectraux.

Pour \'etablir les notations on donne une preuve simple, extraite
de [\citelow{krajew2}], de l'\'egalit\'e (\ref{distancecommut})
entre distance non commutative et distance g\'eod\'esique pour
une vari\'et\'e riemannienne compacte \`a spin. Ensuite sont
\'etablies des propri\'et\'es g\'en\'erales de la distance
associ\'ee au produit de triplets spectraux. En particulier il
est connu\cite{connes,cham1} que dans le mod\`ele \`a deux
couches,
$$\aa= \cinf\ot\cc^2 = \cinf \oplus \cinf,$$
la distance sur chaque copie de la vari\'et\'e est la distance
g\'eod\'esique tandis que la distance entre les couches ne
d\'epend que du triplet spectral interne. Ce r\'esultat est
g\'en\'eralis\'e \`a tout produit tensoriel de g\'eom\'etries.
Pour finir, la distance est explicitement calcul\'ee pour des
mod\`eles d'espace temps euclidien. A noter que la distance dans
le continu a \'et\'e \'etudi\'ee dans l'optique voisine des
alg\`ebres de Lie dans [\citelow{gyros}]. Les r\'esultats de ce
chapitre ont donn\'e lieu \`a l'article [\citelow{kk}].

\section{Distance pour une vari\'et\'e \`a spin}

Le triplet spectral associ\'e \`a une vari\'et\'e riemannienne
compacte \`a spin avec une m\'etrique $g$ est donn\'e par
(\ref{td}).  D'ap\`es la proposition \ref{commutprop}, la partie
connexion de spin de l'op\'erateur de Dirac commute avec $\cinf$
si bien que
 pour tout $f\in\cinf$,
\begin{eqnarray*}
[D,f] &=& - i c(df) = -i \partial_\mu f c( dx^\mu) =  -i \partial_\mu f c( e^\mu_\alpha dx^\alpha),\\
      &=&  -i \partial_\mu f  e^\mu_\alpha\gamma^a = -i \gamma^m\partial_\mu f,\\
      &=& [-i\gamma^m\partial_\mu, f]
\end{eqnarray*}
o\`u $c$ est l'action de Clifford (\ref{cliffaction}),
$e^\alpha_\mu$ les vielbein, $\gamma^a$ les matrices de Dirac
euclidiennes (\ref{diraclide}) et $\gamma^m \doteq e^\mu_\alpha
\gamma^a$ les matrices de Dirac riemanniennes (\ref{matgamma})
qui v\'erifient, gr\^ace \`a (\ref{geed}),
$$
\gamma^m \gamma^n + \gamma^n \gamma^m = 2 g^{\mu\nu}\ii$$
(conform\'ement au chapitre I, on utilise un indice grec pour la
vari\'et\'e et un indice latin pour les degr\'es de libert\'e de
spin; $a,m,n$ se contractent avec $\alpha, \mu, \nu$).   Pour le
calcul des distances, on peut consid\'erer que le triplet
spectral d'une vari\'et\'e \`a spin est
\begin{equation}
\label{triplettvariete} \aa=\cm,\qquad \hh=\LS,\qquad
D=-i\gamma^{m}\partial_{\mu}=-i\ds.
\end{equation}
La dimension spectrale est la dimension de la vari\'et\'e qu'on
prend \'egale \`a  $4$. Le triplet spectral est pair donc la
chiralit\'e (\ref{chiralite}) s'\'ecrit
$$\Gamma = c(\gamma) = (-i)^2 c( \overset{4}{\underset{\alpha=1}{\Pi}}dx^\alpha )  =
-\overset{4}{\underset{\alpha=1}{\Pi}}\gamma^a = -\gamma^5
$$
o\`u $\gamma$ est donn\'ee en (\ref{defgamma}) et $\gamma^5$
d\'esigne traditionnellement le produit des matrices gamma
euclidiennes. Le produit scalaire  de $\hh$ est donn\'e par
(\ref{hilbspin}) ou le couplage $(.\lvert.)$ de $S$ \`a valeur
dans $C(M)$ est le produit scalaire euclidien des spineurs vus
comme vecteurs colonnes dont les entr\'ees sont des fonctions
d'onde,
$$(\psi\lvert\phi) = \psi^\dagger \phi$$
o\`u  $\psi^\dagger$ d\'esigne le vecteur ligne complexe
conjugu\'e de $\psi$.

Comme \'enonc\'e dans le th\'eor\`eme \ref{thconnes}, la distance
non commutative (\ref{distance})
\begin{equation}
\label{distance1,5} d(x,y)=\sup_{f\in C(M)} \left\{ \, \abs{f(x)
-f(y)}\, /\, \norm{[D,f]}\leq 1 \right\}\,,
\end{equation}
coincide avec la distance g\'eod\'esique $L(x,y)$ entre les
points $x,y$ de $\mm$. C'est un r\'esultat classique
\cite{connes} mais la preuve introduit id\'ees et notations dont
nous ferons un usage intensif par la suite, aussi nous donnons en
d\'etail la d\'emonstration.

Par la d\'efinition (\ref{clinv}) de l'involution dans $\ccl(M)$,
$(dx^\alpha)^* = dx^\alpha$ si bien qu'en choisissant les
matrices de Dirac autoadjointes, on choisit en fait une action de
Clifford autoadjointe,
$$(\gamma^a)^\dagger = \gamma^a = c(dx^\alpha) =c({dx^\alpha}^*).$$
Ainsi la norme d'op\'erateur de $[D,f]$, pour une fonction $f$
r\'eelle selon le lemme \ref{positif}, s'\'ecrit
\begin{eqnarray*}
\norm{[D,f]}^2 &=& \displaystyle \norm{c(df)^2} = \sup_{\psi \in
\hh}
\frac{ \int_{\mm} ( c(df)\psi \lvert c(df)\psi)\abs{\nu_g}}{\int_{\mm}(\psi\lvert \psi)\abs{\nu_g}},\\
&=& \displaystyle \sup_{\psi \in \hh} \frac{\int_M \psi^\dagger c(df^*) c(df)\psi \abs{\nu_g}}{\int_{\mm}\psi^\dagger \psi\abs{\nu_g}},\\
&=& \displaystyle \sup_{x\in\mm} \left\{
g^{\mu\nu}(x)\partial_{\mu}f(x)\partial_{\nu}f(x)\right\} =
\suup{x\in M} g(df, df) = \norm{\grad{f}}^2
 \end{eqnarray*}
o\`u on utilise $c(df^*)c(df) = \partial_\mu {f} \partial_\nu f
\gamma^m\gamma^n =  g^{\mu\nu} \partial_\mu f \partial_\nu f \ii$
ainsi que l'\'equation (\ref{normetangent}).
D'o\`u
$$
\norm{[D,f]}=\suup{x\in\mm}\norm{(\grad{f})(x)}.
$$

Soit maintenant  $c\!:\!t\in [0,1]\!\!\rightarrow\!\!\mm$ une
g\'eod\'esique minimale entre  $x$ et $y$. On d\'esigne par un
point la d\'eriv\'ee totale par rapport au param\`etre $t$.  Pour
tout $f\in\cinf$
$$f(x)-f(y) = \int_{0}^{1} \dot{f}(c(t))\; dt=\int_{0}^{1}
\partial_{\mu}f(p)\; \dot{c^{\mu}}(t) dt$$
avec $p\doteq c(t)$. Les fonctions $\dot{c^{\mu}}$ sont les
composantes d'un champ de vecteur $X\in\xx(M)$. On note
$\dot{c}_\nu$ les composantes de $\dot{c}\doteq X^\flat$, si bien
que
$$
{\partial_{\mu}f(p)\, \dot{c^{\mu}}(t)}={g^{\mu\nu}(p)\,
\partial_{\mu}f(p) \, \dot{c_{\nu}}(t)}= g( df(p), \dot{c}(p)).
$$
Par l'in\'egalit\'e de Cauchy-Schwarz,
$$\abs{{\partial_{\mu}f(p)\,\dot{c}^{\mu}(t)}} \leq \norm { df(p)} \norm{\dot{c}(t)}.$$
Si $f$ atteint le supr\'emum,
$\norm{df(p)}=\norm{(\grad{f})(p)}\leq1$ en tout $p$ et
\begin{equation*}
\label{firstineq} d(x,y) =\abs{f(x)-f(y)}\leq \int_0^1
\norm{\dot{c}(t)}\; dt = L(x,y)\,.
\end{equation*}

Cette borne sup\'erieure est atteinte par la fonction
\begin{equation}
\label{defl} \index{L@$L$}L: q \mapsto L(q,y).
\end{equation}
En effet, $L(x)-L(y) = L(x,y)$ et
\begin{equation}
\label{normedel} \sup_{q\in\mm}\norm{{\grad{L}}(q)}\leq 1\,.
\end{equation}
Pour montrer cette derni\`ere in\'egalit\'e,  choisissons
$q,q'\in\mm$, de coordonn\'ees $q^{\mu},{q'}^\mu$ dans une carte
donn\'ee, o\`u $q'$ est l'image de $q$ par la transformation
infinit\'esimale  $\sigma(\epsilon)$, $\epsilon<\!\!< 1$,
$\sigma$ d\'esignant le flot g\'en\'er\'e par le champ de
vecteurs $g^{\mu\nu}(\partial_{\nu} L)\partial_{\mu}$ avec la
condition initiale $\sigma(0)=q$. Alors, avec $dq^{\mu}{\doteq
q'}^{\mu}-q^{\mu}$,
$$
q^{\mu}+dq^{\mu}= {q'}^{\mu} = \sigma^{\mu}(\epsilon) =
\sigma^{\mu}(0) + \epsilon\, \frac{d \sigma^{\mu}}{dt}(0) + {\cal
O}(\epsilon^2) = q^{\mu} +\epsilon\, g^{\mu\nu}(q)\partial_{\nu}
L(q) +{\cal O}(\epsilon^2)\,,
$$
c'est \`a dire
\begin{equation}
\label{flot} dq^{\mu}=\epsilon\,g^{\mu\nu}(q)\partial_{\nu} L(q)
+{\cal O}(\epsilon^2)\,.
\end{equation}
Comme $L(q',y)$ est le plus court chemin de $q'$ \`a $y$,
$L(q',y)\leq L(q',q) + L(q,y)$. Ainsi
\begin{equation}
\label{inegtrg} L(q+dq) \leq L(q',q) + L(q)\,.
\end{equation}
Par (\ref{flot})
$$
L(q',q)\doteq \sqrt{g_{\lambda\rho}(q) dq^{\lambda}dq^{\rho}}=
\sqrt{\epsilon^2 g_{\lambda\rho}(q)  g^{\lambda\mu}(q)
\partial_{\mu}L(q)\, g^{\rho\nu}(q) \partial_{\nu} L (q)}=
\epsilon \sqrt{ g^{\mu\nu}\partial_{\mu}L(q)\, \partial_{\nu} L
(q)}\,.
$$
Inser\'e dans le membre de droite de (\ref{inegtrg}) dont la
partie gauche est d\'evelopp\'ee par rapport \`a $\epsilon$,
cette \'equation donne
$$
L(q) + \partial_{\mu}  L(q) \, dq^{\mu}= L(q) + \,
\epsilon\,g^{\mu\nu}(q)\partial_{\mu}L(q)\partial_{\nu} L(q)
+{\cal O}(\epsilon^2) \leq
\epsilon\sqrt{g^{\mu\nu}\partial_{\mu}L(q)\,  \partial_{\nu} L
(q)} + L(q) +{\cal O}(\epsilon^2),
$$
qui est vraie quel que soit $q$, d'o\`u (\ref{normedel}) et
finalement $$d(x,y)=L(x,y).$$

A noter que $L$ n'est pas lisse en $y$ mais seulement
continue\cite{jgb}. Pour \'ecrire (\ref{distance1,5}) en
rempla\c{c}ant $C(M)$ par $\cinf$, il faudrait exhiber une suite
de fonctions lisses $f_n$ qui converge vers $L$ et v\'erifie
$\norm{[D, f_n]}\leq 1$ pour tout $n$. La preuve est identique
avec l'alg\`ebre r\'eelle $C^\infty_\rr(M)$ puisqu'alors, les
\'etats purs \'etant donn\'es par (\ref{etatreelpur}),
$\re(\omega_x(L)) - \re(\omega_y(L)) =  L(x) - L(y).$

\section{Produit de g\'eom\'etries}

\subsection{Produit tensoriel de triplets spectraux}

Le produit tensoriel d'un triplet spectral r\'eel pair
$T_I=(\aa_I, \hh_I, D_I,\pi_I)$ muni d'une chiralit\'e
$\Gamma_I$, par le triplet spectral r\'eel $T_E=(\aa_E, \hh_E,
D_E,\pi_E)$ est le triplet spectral $T_I\ot
T_E\doteq(\aa',\hh',D')$ d\'efini par
\begin{equation*}
\label{pdt0} \aa'\doteq \aa_I\otimes\aa_E,\quad
\hh'\doteq\hh_I\otimes\hh_E,\quad  D'\doteq D_I\otimes\ii_E +
\Gamma_I\otimes D_E.
\end{equation*}
La repr\'esentation est $\pi'\doteq \pi_I\ot \pi_E$ (dans ce
chapitre nous n'utiliserons ni la chiralit\'e ni la structure
r\'eelle du triplet produit mais toutes deux sont d\'efinies, cf
[\citelow{vanhecke}]). Dans la mesure o\`u les triplets spectraux
ne forment pas un espace vectoriel, la notation $T_I\ot T_E$ est
essentiellement une convention. Ce produit est commutatif car
lorsque $T_E$ est pair et muni d'une chiralit\'e $\Gamma_E$,
alors le triplet spectral $T_E\otimes T_I\doteq(\aa,\hh,D)$ est
\'egalement d\'efini (il suffit de permuter les facteurs)
\begin{equation}
\label{pdt1} \aa\doteq \aa_E\otimes\aa_I,\quad
\hh\doteq\hh_E\otimes\hh_I,\quad  D\doteq D_E\otimes\ii_I +
\Gamma_E\otimes D_I\, ,
\end{equation}
$\pi=\pi_E\ot\pi_I$ et il est \'equivalent \`a $T_I\otimes T_E$
via l'op\'erateur unitaire
\begin{equation*}
U\doteq \frac{\ii_I+ \Gamma_I}{2}\otimes\ii_E + \frac{\ii_I-
\Gamma_I}{2}\otimes\Gamma_E\,.
\end{equation*}

En physique ce produit tensoriel est utilis\'e pour d\'ecrire un
espace continu dont chaque point est muni d'une fibre discr\`ete.
Dans le mod\`ele standard l'espace interne $T_I$ est choisie de
mani\`ere \`a ce que le groupe des unitaires de $\aa_I$, modulo
le rel\`evement aux spineurs\cite{spingroup,farewell}, soit le
groupe de jauge des interactions. $\aa_I$ est une \alg de
matrices, $\hh_I$ est l'espace des fermions et l'op\'erateur de
Dirac interne a pour coefficients les masses des fermions,
\'eventuellement pond\'er\'ees par la matrice unitaire de
Cabibbo-Kobayashi-Maskawa. Nous reviendrons longuement sur le
mod\`ele standard dans le chapitre consacr\'e aux fluctuations de
la m\'etrique. Ici  nous \'etudions de mani\`ere plus
g\'en\'erale la distance pour des produits tensoriels de triplets
spectraux sans pr\'esupposer, dans un premier temps au moins, que
l'un est fini et l'autre relatif \`a une vari\'et\'e
diff\'erentiable.

N\'eammoins, afin de ne pas multiplier les notations, tous les
objets relatifs au premier terme du produit tensoriel sont
appel\'es {\it externes} alors que ceux relatifs au second sont
dits {\it internes}.

\subsection{Distance dans l'espace interne et dans l'espace externe}

Pour fixer les notations on suppose que $T_E$ est pair et l'on
travaille avec $T_E\ot T_I$. Pour \'etudier cette g\'eom\'etrie,
le premier point est de calculer l'espace des \'etats. Soient
$\te$ et $\ti$ des \'etats de $\aa_E$ et $\aa_I$ respectivement.
La paire $(\te, \ti)$ agissant comme $\te\ot\ti$ est un \'etat de
$\aa$. En effet
$$(\te\ot\ti) (\ii_E \ot \ii_I) = \te(\ii_E)\ti(\ii_I) = 1.$$
Pour montrer la positivit\'e, on note $a^*a = (f^i\ot
m_i)^*(f^j\ot m_j) = {f^i}^*f^j \ot m_i^*m_j$ un \'el\'ement
positif de $\aa$, $F^{ij}\doteq\te({f^i}^*f^j)$ et
$M_{ij}\doteq\ti({m_i}^*m_j)$.  Alors
$$(\te\ot \ti)(a^*a) = F^{ij}\ot M_{ij}.$$
La somme sur $i,j$ est finie. La matrice $F$ de composantes
$F^{ij}$ est autoadjointe et se d\'ecompose sur ses projecteurs
propres $t_k$. La valeur propre $\lambda$ associ\'e au vecteur
propre $\psi$ est positive puisque $\lambda = \scl{\psi}{F\psi}=
\bar{\psi_i}F^{ij}\psi_j =\te\lp (\psi_if^i)^*(\psi_if^i)\rp\geq
0.$ La composante $t^k_{ij}$ du $k^{\text{i\`eme}}$ projecteur
propre est ${\bar{\psi}^k}_i\psi^k_j$ de sorte que
$$(\te\ot\ti)(a^*a) = \lambda_k \sum_{i,j} t^k_{ij} M_{ij} = \lambda_k\, \oi\lp(\sum_{ij} \psi^k_jm_j)^*(\sum_{ij}\psi^k_jm_j)\rp \geq 0.$$
Le produit de deux \'etats purs $\oei$ est un \'etat mais pas
n\'ecessairement pur. Qui plus est, il peut exister des \'etats
(purs) de $\aa$ qui ne s'\'ecrivent pas comme produit tensoriel.
Quoi qu'il en soit, nous ne consid\'erons ici que des \'etats du
type $\tei$. Cette restriction n'est pas g\^enante dans la mesure
o\`u, d\`es qu'une des alg\`ebres est commutative, tout \'etat
pur et de cette forme  [\citelow{kadison}, {\it p. 857}].

{\prop \label{etatproduit} Soit $\aa_E$ et $\aa_I$ deux
$C^*$-alg\`ebres dont l'une au moins est ab\'elienne. Alors
$$\pp({\aa_E}\ot{\aa_I})\simeq\pp({\aa_E})\times \pp({\aa_I}).$$}
 \noindent En ce cas tout \'etat pur $\omega$ de $\aa$ s'\'ecrit $$\omega =\oe\ot\oi.$$
$\oe$ et $\oi$ sont appel\'es respectivement partie externe et
interne de $\omega$. Dans le mod\`ele \`a deux couches $\aa_E=
\cinf$ est ab\'elienne si bien que tout \'etat pur de $\aa$ est du
type  $\xoi$ o\`u $\omega_i$, $i=1,2$, d\'esignent les \'etats
purs de $\cc^2$ et $\omega_x\index{omega_x@$\omega_x$}$ ceux de
$\cinf$. Il est\cite{connes} connu que $d(\xoi,\, \yoi)$ est la
distance g\'eod\'esique $L(x,y)$ tandis que $d(\xoi,\,
\omega_x\ot\omega_j)$ est une constante. Ce r\'esultat s'\'etend
\`a tout produit de deux triplets spectraux. Une fois fix\'e la
partie externe $\te$, $d(\tei, \tei')$ ne d\'epend que de  $T_I$;
de m\^eme $d(\tei,\te'\ot\ti)$ ne d\'epend que de $T_E$.

{\thm\label{sansfluct} Soient $d_E$, $d_I$ et $d$ les distances
associ\'ees \`a $T_E$, $T_I$, $T_E\ot T_I$ respectivement. Quels
que soient $\te$, $\te'$ dans $\ss(\aa_E)$ et $\ti$, $\ti'$ dans
$\ss(\aa_I)$,
\begin{eqnarray*}
&d\lp\tei, \tei'\rp= d_I\lp\ti,\ti'\rp,&\\
&d\lp\tei, \te'\ot\ti\rp= d_E\lp\te,\te'\rp.&
\end{eqnarray*}}

\noindent {\it Preuve.} Notons $f_j$ les \'el\'ements de $\aa_E$
et $m_i$ ceux de  $\aa_I$. Tout \'el\'ement de  $\aa$ s'\'ecrit
$$a=f^i\ot m_i\,,$$
o\`u l'indice de sommation $i$ parcourt un sous-ensemble fini de
$\nn$. L'\'equation (\ref{pdt1}) donne
$$
[D,a]= [D_E,f^i]\ot m_i + f^i\Gamma_E \ot [D_I,m_i]\,.
$$
En multipliant \`a droite et \`a gauche par l'op\'erateur
unitaire $\Gamma_E\ot \ii_I$, on peut \'ecrire
$$
\norm{[D_E, f^i]\ot m_i + f^i\Gamma_E\ot[D_I,m_i]}=\norm{
-[D_E,f^i] \ot m_i + f^i\Gamma_E\ot[D_I,m_i]}
$$
o\`u on utilise le fait que $\Gamma_E=\Gamma_E^*$ commute avec
$f^i$ et anticommute avec $D_E$. Quels que soient deux
\'el\'ements $u,v$ d'un espace norm\'e,
$$2\norm{u}\leq\norm{u+v}+\norm{u-v},$$
d'o\`u
\begin{eqnarray}
\label{gen1}
& &\norm{ [D_E, f^i]\ot m_i} \leq \norm{[D,a]} \,\\
\nonumber &\text{et}&\,  \norm{f^i\Gamma_E\ot[D_I,m_i]} \leq
\norm{[D,a]}\,.
\end{eqnarray}
En factorisant le terme de gauche de cette derni\`ere\
in\'egalit\'e par $\Gamma_E\ot \ii_I$,
\begin{equation}
\label{gen2} \norm{f^i\ot[D_I,m_i]} \leq  \norm{[D,a]}\,.
\end{equation}

Pour chaque $\oe\in \pp(\aa_E)$ et $a\in\aa_+$, on d\'efinit
l'\'el\'ement $m_E$ de $\aa_I$ par
$$
m_E\doteq\te (f^i)m_i.
$$
$m_E$ est autoadjoint. En effet, puisque $a$ est positif il
existe $b\doteq f^p\ot m_p$ tel que
$$
a= b^*b = \frac{1}{2} ( f^{pq} \ot m_{pq} + {f^{pq}}^{*}\ot
{m_{pq}}^{*})
$$
o\`u $f^{pq}\doteq {f^p}^{*}f^q$ et $m_{pq}=m_p^{*}m_q$; ainsi
$$
m_E=\frac{1}{2}( \te (f^{pq})m_{pq} +  \te
({f^{pq}}^{*}){m_{pq}}^{*})= m_E^{*}.
$$
Ainsi l'\'el\'ement de $\bb(\hh_I)$
$$i[D_I,m_E]= i \lp \te \ot \ii_I\rp \lp f^i\ot  [D_I,  m_i]\rp$$
est normal. En notant  ${\cal S}_I\supset \ss(\aa_I)$ l'ensemble
des \'etats de $\bb(\hh_{I})$ et une notation $S_E$ similaire, il
vient d'apr\`es l'\'equation (\ref{normetat})
\begin{eqnarray*}\displaystyle
\norm{[D_I,m_E]} &  = & \displaystyle \suup{\tau_I\in\, {\cal S}_I} \abs{{\tau_I} ([D_I,m_E])},\\
                 &\leq& \displaystyle \suup{(\tau_E, \tau_I) \in \, {\cal S}_E\times{\cal S}_I} \abs { \lp \tau_E \ot \tau_I \rp \lp f^i\ot [D_I,m_i] \rp},\\
 &\leq& \displaystyle\norm{f^i\ot [D_I,m_i]},
\end{eqnarray*}
o\`u l'on a remarqu\'e que $if^i\ot [D_I,m_i]\in\bb(\hh)$ est
\'egalement normal. Avec (\ref{gen2}) on trouve
$$\norm{[D_I,m_E]}\leq \norm{[D,a]}.$$
Comme $(\tei)(a)-(\tei')(a)=\ti(a_E) - \ti'(m_E)$, on obtient
finalement
$$d(\tei,\tei')\leq d_I(\ti,\ti').$$
Cette borne sup\'erieure est atteinte par $\ii_E\ot a_I$ o\`u
$a_I\in\aa_I$ r\'ealise le supr\'emum pour $T_I$,
$$
d_I(\ti,\ti')\doteq\abs{(\ti-\ti')(a_I)}\, \text{ et }
\norm{[D_I, a_I]= 1 }.$$

La preuve pour $d(\tei,\tei')$ est similaire, en utilisant
(\ref{gen1}) au lieu de (\ref{gen2}). \hfill $\blacksquare$
\newline

\noindent Ce th\'eor\`eme ne donne pas une description compl\`ete
de la g\'eom\'etrie. Dans les mod\`eles du type discret $\times$
continu, il indique que la distance sur chaque copie de la
vari\'et\'e est la distance g\'eod\'esique alors que la distance
\`a l'int\'erieur d'une fibre ne d\'epend pas de la fibre choisie
et est compl\`etement d\'etermin\'ee par la partie interne du
triplet spectral. Ceci ne donne aucune information sur la
distance {\it crois\'ee}, c'est \`a dire la distance entre
\'etats correspondant \`a diff\'erents  points de la fibre et
diff\'erents points de la vari\'et\'e. A noter \'egalement que la
discussion sur la distance de Gromov entre vari\'et\'es munies de
m\'etriques diff\'erentes\cite{connes} n'est pas transposable ici
car de telles
 vari\'et\'es \' echappent \`a la description par un produit tensoriel de triplets spectraux.

\subsection{Distance crois\'ee}

Les points cl\'es du th\'eor\`eme \ref{sansfluct} sont les
\'equations (\ref{gen1}) et (\ref{gen2}). La premi\`ere oublie la
partie interne du commutateur et fait sens pour la distance entre
\'etats de m\^eme partie interne. Au contraire (\ref{gen2}) ne
prend pas en compte la partie externe du commutateur et suffit
\`a d\' eterminer la distance entre \'etats de m\^eme partie
externe. Le calcul de la distance crois\'ee $d\lp \tei,\, \te'\ot
\ti'\rp$ n\'ecessite de prendre en consid\'eration \`a la fois la
partie interne et la parti externe du commutateur, ce qui rend le
calcul beaucoup plus d\'elicat. Cependant lorsque $T_E$ d\'ecrit
une vari\'et\'e et $\aa_I$ est une $W^*$-\alg alors la distance
crois\'ee entre \'etats dont les parties internes sont des
\'etats purs normaux, en somme directe (cf. la d\'efinition
ci-dessous)  et dont la somme des supports commute avec
l'op\'erateur de Dirac interne, s'interp\`ete en terme de
mod\`ele de Kaluza-Klein discret. Bien que l'espace interne soit
disconnexe, il apparait que la distance non commutative coincide
avec la distance g\'eod\'esique d'une vari\'et\'e connexe
compacte de dimension $(4{+}1)$. Cette vari\'et\'e demeure
"virtuelle" en ceci que les points entre les couches de dimension
$4$ ne font pas partie de la g\'eom\'etrie. Le plongement de
l'espace non commutatif dans un espace continu de dimension
sup\'erieur est un artifice de calcul. Il ne doit pas masquer une
propri\'et\'e fondamentale, \`a savoir que deux parties d'un
espace non commutatif n'ont pas besoin d'\^etre connect\'ees pour
\^etre \`a distance finie.

Ce r\'esultat s'applique \`a une toute petite classe d'\'etats,
n\'eammoins il est significatif puisque les distances du mod\`ele
standard entre dans ce cadre.
  Tout le calcul repose sur l'observation suivante: si dans une g\'eom\'etrie $(\aa, \hh, D, J, \Gamma)$ deux \'etats purs normaux $\ou$, $\od$ sont en somme
directe et que la somme de leur support commute avec
l'op\'erateur de Dirac, alors $d(\ou, \od)$ coincide avec la
distance d'une g\'eom\'etrie o\`u $\aa = \cc^2$.

Pour traiter le mod\`ele standard il est important que les
r\'esultats de ce chapitre soient vrais pour des alg\`ebres
r\'eelles. Comme on utilise seulement la propri\'et\'e
(\ref{projecteur}), il n'est pas n\'ecessaire d'\'etudier la
th\'eorie des $W^*$-alg\`ebres r\'eelles. On appelle simplement,
par abus de langage, {\it \'etat normal d'une alg\`ebre r\'eelle}
tout \'etat r\'eel auquel est associ\'e un projecteur de l'\alg
satisfaisant (\ref{projecteur}). Dans ce chapitre, $\aa$
d\'esigne indiff\'eremment une $W^*$-\alg complexe ou une \alg
r\'eelle admettant des \'etats r\'eels normaux. On d\'esigne
g\'en\'eriquement par $\kk$ les corps $\cc$ et $\rr$.

{\defi Deux \'etats normaux $\tu$, $\td$ de $\aa$ sont dits {\it
en somme directe} si $s_1 a s_2 = 0$ pour tout $a\in\aa$.}
\newline

\noindent Cette d\'efinition se justifie en remarquant que si
$s_1\aa s_2 = 0$, alors les id\'eaux bilat\`eres principaux $\aa
s_1 \aa$ et $\aa s_2 \aa$ sont en somme directe. Rappelons que
pour tout \'el\'ement $s$ de $\aa$, l'id\'eal bilat\`ere
principal\cite{lang} $\aa s \aa$ est l'ensemble des sommes $a^i s
b_i$ o\`u $a^i, b_i \in \aa$. Si $s_1 \aa s_2 = 0$, alors $\aa
s_1 \aa$ et $\aa s_2 \aa$ sont en somme directe en ce sens que
leur intersection est vide. En effet si  $c \in \aa s_1 \aa \cap
\aa s_2 \aa$ alors il existe $a^i$, $b_i$, $p^k$ et $q_k$ dans
$\aa$ tels que $$c = a^i s_1 b_i = p^k s_2 q_k.$$ En multipliant
\`a gauche par ${a^j}^*s_1$, on trouve que ${a^j}^*s_1 a^i s_1
b_i =0$ pour tout $j$. Autrement dit $c^* c = {b_j}^* s_1
{a^j}^*  a^i s_1 b_j = 0$, d'o\`u $\norm{c} = 0=c$.

{\prop \label{reduction0} Soient $s_1, s_2$ les supports  de deux
\'etats purs normaux $\ou$, $\od$ d'une alg\`ebre $\aa$ sur
$\kk$. Soit $(\aa, \hh, D, \pi)$ un triplet spectral dans
laquelle $[D, \pi(s_1) + \pi(s_2)]= 0$. Si $\ou, \od$ sont en
somme directe alors
$$d (\ou, \od) = d_e(\omega_k, \omega_k') = \frac{1}{\norm{M}}$$
o\`u $ \omega_k$, $\omega_k'$ sont les deux \'etats purs de
$\aa_e\doteq\kk^2$ et $d_e$ est la distance associ\'ee au triplet
 $T_e \doteq (\aa_e, \hh_e, D_e, \pi_e)$ dans lequel
 $$\hh_e\doteq \hh_1 \oplus \hh_2\, ,\quad D_e\doteq \dm{cc}0 & M\\ M^* & 0 \fm, \quad
\pi_e\doteq \pi\big|_{\hh_e} $$ o\`u $\hh_1\doteq \pi(s_1)\hh$,
$\hh_2\doteq \pi(s_2)\hh$ et $M$ est une application lin\'eaire
born\'ee de $\hh_2$ dans $\hh_1$.}
\newline

\noindent{\it Preuve.} La preuve d\'ecoule de la proposition
\ref{projectionlem} avec $e\doteq s_1 + s_2$. Les \'etats sont en
somme directe donc $s_1\aa s_1 \simeq \kk$ par
(\ref{projecteur}), inclus dans $\aa s_1\aa$, est en somme
directe avec $s_2\aa s_2\simeq \kk$. Comme  $s_1 s_2 = 0$ il est
imm\'ediat que
$$\aa_e = \alpha_e(\aa) = (s_1 + s_2)\aa(s_1 + s_2) \simeq \kk^2.$$
On obtient un isomorphisme explicite en identifiant $s_1$, $s_2$
\`a la base canonique de $\kk^2$
\begin{equation}
\label{alphaedea} \alpha_e(a) = \ou(a) s_1 \oplus \od(a) s_2 =
(\ou(a), \od(a)).
\end{equation}
Les deux \'etats purs $\omega_k$, $\omega_k'$ de $\aa_e$
extraient respectivement les premi\`ere et deuxi\`eme composantes
du doublet de nombres complexes $\alpha_e(a)$, de sorte que
$$\omega_k\circ\alpha_e = \ou \text{ et }\, \omega_k'\circ\alpha_e = \od,$$
d'o\`u $d_e(\omega_k , \omega_k') = d( \ou, \od)$ par la
proposition \ref{projectionlem}.

Naturellement $\hh_1$ est en somme directe avec $\hh_2$ puisque,
si $\phi= \pi(s_1)\xi= \pi(s_2)\zeta$, alors $\pi(s_1)\phi = \phi
= \pi(s_1 s_2)\zeta = 0$. D'o\`u
$$\hh_e \doteq (\pi(s_1) + \pi(s_2))\hh = \hh_1 \oplus \hh_2.$$
Par d\'efinition $D_e$ est la projection sur $\hh_e$ de la
restriction de $D$ \`a $\hh_e$,
$$D_e = \dm{cc} U & M \\ M^* & W\fm$$
o\`u $M$ est une application de $\hh_2$ dans  $\hh_1$ et $V$, $W$
des endomorphismes de $\hh_1$, $\hh_2$ respectivement (les
parties antidiagonales sont adjointes l'une de l'autre car $D$
est autoadjoint).

On note $\pi_e(s_1)= \ii_1$ l'identit\'e de $\bb(\hh_1)$ et
$\pi_e(s_2)= \ii_2$ l'identit\'e de $\bb(\hh_2)$. Pour tout
$\alpha_e(a)\in \aa_e$,
\begin{equation}
\label{piea} \pi_e(\alpha_e(a)) = \ou(a)\pi_e(s_1) \oplus \od(a)
\pi_e(s_2) = \dm{cc} \ou(a)\ii_1 & 0 \\ 0 & \od(a)\ii_2 \fm.
\end{equation}
Ainsi $\pi_e$ commute avec la partie diagonale de $D_e$ et la
distance $d_e$  coincide avec celle calcul\'ee en prenant
$V=W=0$, tout autre chose \'egale. On a alors $\norm{[D_e,
\pi_e(a)]} = \abs{\ou(a) - \od(a)}\norm{M}$ d'o\`u
$$d_e(\omega_k, \omega_k') \leq \frac{1}{\norm{M}},$$
cette borne sup\'erieure \'etant atteinte par $a =
\norm{M}^{-1}s_1$.

Soulignons que, quoique $D$ puisse \^etre non born\'e, $M$ est
n\'ecessairement born\'e. Pour s'en convaincre rappelons que si
$B$ est un op\'erateur born\'e sur un espace de Hilbert $\hh$ et
$p$ un projecteur de rang $1$, alors $pB$ est born\'e (on
v\'erifie par l'in\'egalit\'e de Cauchy-Schwarz que pour tout
$\phi\in \hh$, $\norm{pB\phi} \leq \norm{B\phi}$). Par
d\'efinition d'un triplet spectral, $[D,\pi(s_2)]$ est born\'e
donc $\pi(s_1)[D,\pi(s_2)] = \pi(s_1)D\pi(s_2)$ est born\'ee, de
m\^eme que sa restriction $M$ \`a $\hh_e$.  \hfill $\blacksquare$
\newline

\noindent Naturellement ce r\'esultat n'a d'int\'er\^et que pour
des \'etats purs dont la somme des supports commutent avec
l'op\'erateur de Dirac sans qu'aucun des supports pris
individuellement ne commute avec $D$. On sait en effet, par le
corollaire \ref{finitude2}, qu'un tel \'etat est infiniment
distant des autres \'etats purs.

La proposition \ref{reduction0} s'\'etend imm\'ediatement \`a des
produits de g\'eom\'etries.

{\defi Un \'etat $\tau_E\ot\omega_I$ de $\aa = \aa_E\ot \aa_I$
est dit "semi-normal"  quand $\oi\in\pp(\aa_I)_*$.}
\newline

\noindent Quelle que soit la g\'eom\'etrie externe, si la somme
des parties internes de deux \'etats semi-normaux en somme
directe commute avec l'op\'erateur de Dirac interne, la distance
est identique \`a la distance calcul\'ee en projetant la
g\'eom\'etrie interne sur $\kk^2$.

{\cor \label{reduction} Soient $\oeu$, $\oed$ deux \'etats
semi-normaux de $\aa$, $T= T_E\ot T_I$ une g\'eom\'etrie dans
laquelle $[D_I, s_1 + s_2]=0$ et $p\doteq \ii_E\ot e \doteq
\ii_E\ot  (s_1 + s_2)$. Si, en tant qu'\'etats normaux de
$\aa_I$,  $\ou$ et $\od$ sont en somme directe alors
$$d(\oeu, \oed) = d_p ( \omega_E\ot \omega_k, \omega_E\ot \omega_k')$$
o\`u $\omega_k$, $\omega_k'$ sont les \'etats purs de $\kk^2$ et
$d_p$ est la distance associ\'e au triplet $T_E\ot T_e$, $T_e$
\'etant le triplet d\'efini par application de la proposition
pr\'ec\'edente \`a $T_I$.}
\newline

\noindent {\it Preuve.} Puisque $\aa_p = \alpha_p(\aa) = \cinf\ot
\kk^2$,
$$(\oeu) \circ \alpha_p = \oe \ot (\ou \circ\alpha_e) = \omega_E\ot \omega_k$$
et $(\oed)\circ\alpha_p = \omega_E\ot \omega_k'$. Comme $[D, p] =
\Gamma_E \ot [D_I, e] = 0$, le r\'esultat est imm\'ediat. \hfill
$\blacksquare$ \vspace{1truecm}

Supposons maintenant que $T_E$, donn\'e par
(\ref{triplettvariete}), d\'ecrive une vari\'et\'e riemannienne
compacte \`a spin $\mm$ de dimension quatre. On note
$$x_1 \doteq \omx \ot \ou \text{ et } y_2 \doteq \omy \ot \od$$
deux \'etats semi-normaux de $\aa$. Comme $\aa_E$ est
ab\'elienne, $x_1$ et $y_2$ sont purs selon la proposition
\ref{etatproduit}. Si $\aa_I$ est une alg\`ebre r\'eelle, on
prend $\aa_E=C^\infty_\rr(M)$ et  $\re(\omega_x)$ comme \'etat
pur externe. La preuve suivante est \'ecrite pour le cas
complexe, son adaptation au cas r\'eelle est imm\'ediate.

Si $\ou$ et $\od$ sont en somme directe et que la somme de leur
support commute avec l'op\'erateur de Dirac interne, alors
l'espace interne est orthogonal \`a la vari\'et\'e au sens du
th\'eor\`eme de Pythagore.

{\thm \label{pythagore} Soient $\ou, \od \in \pp(\aa_I)_*$ deux
\'etats purs normaux en somme directe, de supports $s_1$, $s_2$
tels que $[D_I, s_1 + s_2]= 0.$  Pour tous points $x,y$ de $\mm$
$$d(x_1, y_2)^2 = d(x_1, y_1)^2 + d(y_1, y_2)^2.$$}
\noindent {\it Preuve.} La preuve se divise en trois \'etapes.
Tout d'abord la g\'eom\'etrie $(\aa, \hh, D)$ est
isom\'etriquement projet\'ee sur un mod\`ele \`a deux couches.
Ensuite on montre que la distance co\"{\i}ncide avec la distance
g\'eod\'esique d'une vari\'et\'e riemannienne compacte de
dimension $4+1$ et, enfin, qu'elle v\'erifie le th\'eor\`eme de
Pythagore.

1) Avec les notations du corollaire \ref{reduction}, en posant
$x_k\doteq \omx \ot \omega_k$ et $x_k'\doteq \omy \ot \omega_k'$,
\begin{equation}
\label{etapeun} d(x_1, y_2)= d_p(x_k, y_k').
\end{equation}
Comme $\aa_p = \cinf \ot \kk^2$, la distance $d_p$ est celle d'un
mod\`ele \`a deux couches. Dans la g\'eom\'etrie r\'eduite
$(\aa_p, \hh_p, D_p)$ un \'el\'ement g\'en\'erique de $\aa_p$
s'\'ecrit,  d'apr\`es (\ref{alphaedea}),
%
$$a = f^i \ot \ou(m_i) \, \oplus \, f^i \ot \od(m_i) = f \oplus g,$$
o\`u $ m_i\in \aa_I$ et $f^i, f\doteq f^i\ou(m_i), g\doteq
f^i\od(m_i) \in \aa_E$. Conform\'ement au lemme \ref{positif}, on
suppose que $f\oplus g$ est positif, c'est \`a dire que $f$ et
$g$ sont des fonctions r\'eelles (r\'eelles positives si $\aa$
est vue comme alg\`ebre r\'eelle).

Selon (\ref{piea}), $a$ est repr\'esent\'e par $f\ii_E\ot \ii_1
\, \oplus \, g\ii_E \ot \ii_2.$ L'op\'erateur de Dirac $D_p =
-i\ds \ot \ii_I - \gamma^5 \ot D_e$, o\`u $D_e$ est donn\'e par
la proposition \ref{reduction0}, est tel que
\begin{equation}
\label{co} [D_p, a] = - \dm{cc}
i\ds f \ot\ii_1 & (g-f)\gamma^5\ot M \\[1ex]
\overline{(f-g)}\gamma^5\ot M^{*} & i\ds \ot\ii_2\fm.
\end{equation}

Les \'etats purs $x_k$ et $y_k'$ agissent selon
$$x_k(a) =f(x), \qquad y_k'(a) = g(y).$$

2) Montrons que $d_p$ coincide avec la distance g\'eod\'esique
d'une vari\'et\'e compacte
$$
\mm'\doteq [0,1] \times \mm
$$
munies  des coordonn\'ees ${x'}^\tau=(t, x^\mu)$, de la m\'etrique
\begin{equation*}
\{g^{\tau\kappa}(x')\}\doteq\dm{cc} \norm{M}^2 &0\\ 0 &
g^{\mu\nu}(x)\fm
\end{equation*}
et d'une structure de spin par l'ajout aux matrices gamma
pr\'ec\'edentes de
$$
\gamma^t=\norm{M} \gamma^5.
$$
D'apr\`es la section 1 de ce chapitre,  il suffit de montrer que
$d_p$ coincide avec la distance non commutative $L'$ du triplet
$$
\aa'=C^{\infty}(\mm'),\qquad \hh'=L_2(\mm',S), \qquad
D'=-i\gamma^\tau\partial_\tau=-i\gamma^t\partial_t- i\ds$$ (pour
\'eviter toute confusion, pr\'ecisons que la notation $\aa'$ n'a
aucun lien avec le commutant).

Pour se faire, on note $\aa''$ le sous-ensemble de ${\aa'}_+$
compos\'e de toutes les fonctions du type
$$
\phi (t,x)\doteq (1-t) f(x) + tg(x)
$$
o\`u $f$ et $g$ sont des fonctions r\'eelles sur $\mm$. Alors
\begin{eqnarray*}
\nonumber \norm{[D',\phi]}^2 &=& \suup{(t,x)\in
M'}\norm{(\grad{f})(t,x)} = \sup_{(t,x)\in\mm'} \, \left\{
g^{\tau\kappa}(t,x)\,\partial_\tau\phi(t,x)\,
\partial_\kappa\phi(t,x) \right\}
\\
           &\leq&\label{ameq3} \sup_{x\in\mm} \left\{ \,
\abs{(f-g)(x)}^2\norm{M}^2  + \suup{t\in[0,1]}\,  P(t,x)\,
\right\},
\end{eqnarray*}
o\`u
$$
P(t,x)\doteq t^2 \norm{d{(f-g)(x)}}^2 +
2tg^{\mu\nu}(x)\,\partial_\mu (f-g)(x)\,\partial_\nu
g(x)+\norm{d{g(x)}}^2
$$
est une parabole en $t$ positive et de coefficient directeur
positif, c'est \`a dire que $P(t)$  atteint son maximum sur ses
bords, en $t=0$ ou $1$. Remarquons que
$$P(0,x)= \norm{(\grad{g})(x)}^2,\qquad  P(1,x)=\norm{(\grad{f})(x)}^2$$
et, par (\ref{co})
\begin{eqnarray*}
 &\displaystyle\norm{ \dm{cc} \ii_E \ot \ii_{\alpha_k} &~0 \\ 0&~0 \fm  [D_p,a]\dm{cc}
\ii_E\ot \ii_{\alpha_k}&0
\\ 0 & \gamma^5\ot\ii_{\alpha_{k'}}\fm }^2=\norm{
\dm{cc} i\ds f\ot \ii_{\alpha_k} ~& (g-f)\ii_E\ot M\\ 0 ~&
0\fm}^2 &
\\
&\displaystyle=\suup{x\in\mm}\;\left\{\,\norm{(\grad{f})(x)}^2
+\, \abs{f(x)-g(x)}^2 \norm{M}^2\, \right\} \leq
\norm{[D_e,a]}^2.&
\end{eqnarray*}
De m\^eme
$$\;\suup{x\in\mm}\;\left\{\,\norm{(\grad{g})(x)}^2 +\, \abs{f(x)-g(x)}^2
\norm{M}^2\, \right\}\leq \norm{[D_e,a]}^2$$ d'o\`u
$$
\norm{[D',\phi]}\leq \norm{[D_e,a]}.
$$
En cons\'equence, puisque $x_k(a) - y_k'(a) =
\phi(0,x)-\phi(1,y)$,
\begin{equation}
\label{etapedeux} d_p(x_k, y_k')\leq \sup_{\phi\in{\aa''}} \left\{
\abs{\phi(0,x)-\phi(1,y)}\,/\,\norm{[D',\phi]\leq1} \right\} \leq
L'\lp(0,x),(1,y)\rp\,.
\end{equation}

La d\'emonstration de l'in\'egalit\'e oppos\'ee requiert une
connaissance plus appronfondie de la g\'eometrie de $\mm'$. Comme
$\{g^{\tau\kappa}(x')\}$ est diagonale par blocs et ne d\'epend
pas de $t$,
 les coefficients de la connexion de Levi-Civita sont
$$
\Gamma^t_{t\mu}=\Gamma^t_{\mu t}= \frac{1}{2}g^{tt}\partial_{\mu}
g_{tt}\,,\quad \Gamma^\mu_{tt}=-\frac{1}{2} g^{\mu\nu}
\partial_\nu g_{tt}\,, \quad
\Gamma^{\mu}_{t\nu}=\Gamma^{\mu}_{\nu t} =
\Gamma^{t}_{tt}=\Gamma^t_{\mu\nu}=0
$$
o\`u l'on pose $g_{tt}=(g^{tt})^{-1}=\norm{M}^{-2}.$ On note
$d\tau$ l'\'el\'ement de longueur dans $M'$. Les \'equations des
g\'eod\'esiques s'\'ecrivent
\begin{eqnarray}
&\displaystyle \frac{d^2 t}{d\tau^2} + g^{tt}(\partial_\mu g_{tt})
\frac{dt}{d\tau}\frac{dx^{\mu}}{d\tau}= 0\,,& \label{geo1}
\\
\label{geo2} &\displaystyle \frac{d^2 x^\mu}{d\tau^2} -
\frac{1}{2} g^{\mu\nu}(\partial_\nu g_{tt})
\frac{dt}{d\tau}\frac{dt}{d\tau}
+\Gamma^{\mu}_{\lambda\rho}\frac{d x^\lambda}{d\tau}\frac{d
x^\rho}{d\tau} = 0&
\end{eqnarray}
et, parce que $g_{tt}$ ne d\'epend pas non plus de $x^\mu$, elles
se r\'eduisent \`a
\begin{eqnarray}
\label{constante}
&\displaystyle \frac{dt}{d\tau}= \text{constant}\doteq g^{tt}K &\\
\label{geo2} &\displaystyle\frac{d^2 x^\mu}{d\tau^2}
+\Gamma^{\mu}_{\lambda\rho}\frac{dx^\lambda}{d\tau}\frac{d
x^\rho}{d\tau} = 0
\end{eqnarray}
o\`u $K$ est une constante r\'eelle. En d'autres termes, la
projection sur $\mm$ d'une g\'eod\'esique $\cal{G}'$ de $\mm'$
est une g\'eod\'esique $\cal{G}$ de $\mm$, et la projection de
$\cal{G'}$ sur  l'hyperplan de codimension $1$ contenant ${\cal
G}$ et orthogonal \`a $\mm$ est une ligne droite (c'est \`a dire
une  g\'eod\'esique de l'hyperplan).

Soit  $\{x^a (\tau)\}$ une g\'eod\'esique de $\mm'$
param\'etris\'ee par son \'el\'ement de longueur
 $d\tau$. En utilisant  (\ref{constante}),
\begin{equation}
\label{dessin} 1= \frac{d\tau^2}{d\tau^2} =
g_{\mu\nu}\frac{dx^\mu}{d\tau}\frac{dx^\nu}{d\tau}+ g^{tt}K^2.
\end{equation}
 Soit $ds$ l'\'el\'ement de longueur de $\mm$. En supposant que $g^{tt}K^2\neq1$ (ce point est discut\'e plus bas),
\begin{eqnarray}
\label{dtau1}
&\displaystyle d\tau^2 =  \frac{ds^2}{1 - g^{tt}K^2},&\\
\label{dtau2} &\displaystyle dt = \frac{dt}{d\tau}d\tau =
\frac{g^{tt}Kds}{\sqrt{1-g^{tt}K^2}}.
\end{eqnarray}
Pour tout point $q$ de $\mm$, on note ${\cal G}_q'$ une
g\'eod\'esique minimale de $\mm'$ entre les points $(0,q)$ et
$(1,y)$, et ${\cal G}_q$ sa projection sur $\mm$. On d\'efinit la
fonction  $f_0\in C(M)$,
$$
\label{gun} f_0(q) = \sqrt{1-g^{tt}K^2} L(q) =
\sqrt{1-g^{tt}K^2}\int_{{\cal G}_q}ds,$$ o\`u $L$ est d\'efinie
en (\ref{defl}). En prenant $a_0=(f_0,g_0)\in C(M)\ot\kk^2$ avec
$g_0 \doteq f_0 - K$,
\begin{equation}
\label{fg} x_k(a_0)-y_k'(a_0) = f_0(x) - g_0(y)= f_0(x) + K.
\end{equation}
L'\'equation (\ref{dtau2}) indique que
$$1=\int_{{\cal G}_x'} dt =  \frac{g^{tt}K}{\sqrt{1-g^{tt}K^2}}\int_{{\cal
G}_x}  ds.
$$
Inser\'e dans (\ref{fg}) sous la forme  $K1$, on obtient
$$
x_1(a_0)-y_2(a_0)=\sqrt{1-g^{tt}K^2}\int_{{\cal G}_x}
ds+\frac{g^{tt}K^2}{\sqrt{1-g^{tt}K^2}}\int_{{\cal G}_x} ds =
\frac{1}{\sqrt{1-g^{tt}K^2}}\int_{{\cal G}_x}ds.$$ En utilisant
(\ref{dtau1}),
\begin{equation}
x_k(a_0)-y_k'(a_0) = \int_{{\cal G}_x'} d\tau=
L'\lp(0,x),(1,y)\rp. \label{presque}
\end{equation}
Par ailleurs $\ds f_0=\ds g_0$ et $\partial_\mu f_0 = \sqrt{1 -
g^{tt}K^2}\partial_\mu L$,  de sorte que (\ref{co}) donne
\begin{eqnarray*}
\displaystyle \norm{[D_p, a_0]}^2 &=&\suup{q\in\mm}
\left\{g^{\mu\nu}(q)\partial_\mu f_0(q)
\partial_\nu f_0(q) +
g^{tt}K^2\right\}\\
& =&\displaystyle  \suup{q\in\mm} \left\{ (1 -
g^{tt}K^2)\norm{(\grad{L})(q)}^2 + g^{tt}K^2\right\}.
\end{eqnarray*}
Par (\ref{normedel}) on trouve alors $\norm{[D_p, a_0]}\leq 1$.
Comme dans le cas d'une vari\'et\'e, on suppose qu'il existe une
suite $f_n$ de fonctions lisses sur $M$ convergeant vers $f_0$
telle que, avec une notation \'evidente, la suite $a_n$
v\'erifie $\norm{[D_p, a_n]}\leq 1$ pour tout $n$.  Avec
(\ref{presque}), on obtient alors
$$d_p(x_k, y_k')\geq L'((0,x),(1,y)).$$
Associ\'ee \`a (\ref{etapedeux}) et (\ref{etapeun}),
\begin{equation}
\label{resultat1} d(x_1, y_2)= L'((0,x),(1,y))\,.
\end{equation}
Ce r\'esultat est vrai aussi longtemps que $g^{tt}K^2\neq 1$. Si
ce n'est pas le cas,  alors
$$TM\ni U\doteq \frac{dx^\mu}{d\tau}\partial_\mu = 0$$
car (\ref{dessin}) indique que $g(U,U)=0$ et par d\'efinition $g$
n'est pas d\'eg\'en\'er\'ee. En clair, $x^{\mu}(\tau)$ est une
constante. Une telle \'equation ne peut pas \^etre  l'\'equation
d'une g\'eod\'esique ${\cal G'}_x$ \`a moins que $x=y$. Par
cons\'equent (\ref{resultat1}) est vrai tant que $x\neq y$.

Lorsque $x=y$, (\ref{etapeun}) donne $d(y_1,y_2)= d_p(y_k,y_k')$.
En notant $d_e$ la distance associ\'ee au triplet $T_e$ seul, la
proposition \ref{sansfluct} garantit que
$d_p(y_k,y_k')=d_e(\omega_k, \omega_k').$ Cette distance est
calcul\'ee dans le corollaire \ref{reduction0} et vaut
\begin{equation}
\label{resultat3} d(y_1, y_2)=\frac{1}{\norm{M}}.
\end{equation}
La projection ${\cal G}_y$ de la g\'eod\'esique ${\cal
G'}_x={\cal G}'_y$ est, par (\ref{geo2}), une g\'eod\'esique
entre $y$ et $y$, c'est \`a dire un point. ${\cal G}'_y$ est une
ligne droite dans l'hyperplan. D\`es lors, $d\tau^2=g_{tt}dt^2$ et
$$L'\lp (0,y),(1,y)\rp= \sqrt{g_{tt}}\int_{{\cal G}_y'} dt
=\sqrt{g_{tt}}=\frac{1}{\norm{M}}.$$ Par cons\'equent $ d(y_1,
y_2)=L'\lp (0,y),(1,y)\rp $ et (\ref{resultat1}) est vrai m\^eme
si  $x=y$.
\bigskip

3)  La derni\`ere \'etape consiste \`a montrer que
(\ref{resultat1}) satisfait le th\'eor\`eme de Pythagore.
$g^{tt}$ \'etant constant, l'\'equation (\ref{dtau1}) signifie
que $d\tau$ et $ds$ sont \'egaux \`a une constante pr\`es. De
cette mani\`ere on peut param\'etriser une g\'eod\'esique de
$\mm'$ par $ds$ plut\^ot que par $d\tau$ et obtenir, gr\^ace aux
\'equations des g\'eod\'esiques,
$$dt= g^{tt}K'ds$$
o\`u $K'$ est une constante r\'eelle. Alors
$$d\tau^2 = g_{tt}dt^2 + ds^2 =ds^2(1 + g^{tt}{K'}^2),$$
d'o\`u
\begin{eqnarray}
\nonumber L'\lp(0,x),(1,y)\rp &=& \sqrt{1 +
g^{tt}{K'}^2}\int_{{\cal G}'_x} ds= \sqrt{1
+ g^{tt}{K'}^2}L(x,y)\\
\label{pythun} &=&\sqrt{L(x,y)^2 + g^{tt}{K'}^2L(x,y)^2}.
\end{eqnarray}
D'une part le th\'eor\`eme \ref{sansfluct} donne
$L(x,y)=d(\xox,\yox)$, d'autre part
$$g^{tt}{K'}^2L(x,y)^2= g_{tt}\lp\int_{{\cal G}_x'} g^{tt}{K'} ds\rp ^2=
g_{tt}\lp\int_{{\cal G}_x'} dt \rp
^2=g_{tt}=\frac{1}{\norm{M}^2}=d^2(y_1, y_2)$$ par
(\ref{resultat3}). Avec (\ref{resultat1}) et (\ref{pythun}), ceci
prouve que
\begin{equation*}
d(x_1, y_2)^2=d(x_1, y_1)^2 + d^2(y_1, y_2)\,.
\tag*{\mbox{$\blacksquare$}}
\end{equation*}
\newline

Avec ce th\'eor\`eme, toutes les distances du mod\`ele \`a deux
couches sont connues. Quand l'\alg interne est de dimension
finie, tous les \'etats appartenant \`a des composantes
diff\'erentes de la d\'ecomposition (\ref{algfini}) de $\aa_I$
sont en somme directe, donc susceptibles de relever de ce
th\'eor\`eme.

\section{Exemples}

Appliquons ces r\'esultats \`a des mod\`eles d'espace temps o\`u
l'alg\`ebre interne est l'une de celles d\'ecrites au chapitre
pr\'ec\'edent.

\subsection{Espace fini commutatif}

Soit $\aa_I = \cc^n$, $n\in\nn$, repr\'esent\'ee diagonalement
sur $\hh_I=\cc^n$. Le support du $i^{\text{\`eme}}$ \'etat de
$\aa$ est la matrice $e_{ii}$ de la base canonique de $\mn$. Tous
les \'etats sont en somme directe. Pour que $[D, s_i + s_j]= 0$,
il faut et il suffit que
\begin{equation}
\label{dnul} D_{il} = D_{li} = 0 \text{ et } D_{jl} = D_{lj} = 0
\text{ pour tout } l \text{ diff\'erent de } i \text{ et } j.
\end{equation}
Dans la repr\'esentation graphique de la section
II.\ref{espacefinicommut}, en se souvenant que la distance ne
d\'epend que des chemins reliant $i$ \`a $j$ (proposition
\ref{chemin}), la condition (\ref{dnul}) signifie que le seul
chemin entre les points $i$ et $j$ est pr\'ecis\'ement le lien
$i-j$.

Dans le cas plus simple, $n=2$,  le th\'eor\`eme \ref{pythagore}
muni le mod\`ele \`a deux couches d'une m\'etrique cylindrique.
Pour $n=3$, (\ref{dnul}) impose qu'un lien au moins soit coup\'e,
mettons $D_{13}$ pour fixer les notations. L'espace non
commutatif correspondant est un mod\`ele \`a trois couches.  On
d\'esigne les points par $x_i$. Les couches $1$ et $2$ forment un
mod\`ele \`a deux couches avec une m\'etrique cylindrique de
coefficient suppl\'ementaire $g^{tt} = D_{12}^2$. De m\^eme pour
les couches $2$ et $3$ avec un coefficient $D_{23}^2$. En
revanche le th\'eor\`eme \ref{pythagore} ne dit rien de la
distance crois\'ee entre $x_1$ et $y_3$. On sait seulement, en
vertu du th\'eor\`eme \ref{sansfluct} et des r\'esultats de la
section II.\ref{troispoints}, que
$$d(x_1, x_3) = \sqrt{\frac{D_{12}^2 + D_{32}^2}{D_{12}D_{32}}}.$$

Le sch\'ema est similaire pour ÒÒ$n=4$. L'espace non commutatif
est un mod\`ele \`a quatre couches. Les trois paires $i,i+1$ sont
munies d'une m\'etrique cylindrique et les autres distances
crois\'ees ne sont pas connues. L'espace r\'egulier ne satisfait
pas les conditions requises, aussi le prochain exemple sera non
commutatif.

\subsection{ Espace \`a deux points}

Les notations sont celles de la section II.\ref{deuxpoints}.
Rappelons simplement que $\aa = \mn \oplus \cc$ est
repr\'esent\'ee par une matrice diagonale par bloc sur $\hh =
\cc^{n+1}$. L'op\'erateur de Dirac est
$$D= \dm{cc} 0 & e_1\\ {e_1}^* & 0 \fm$$
o\`u $e_1$ d\'esigne le premier vecteur de la base canonique de
$\cc^n$. L'\'etat $\oc$ de $\cc$ est en somme directe avec tous
les \'etats purs $\ox$ de $\mn$. La somme des supports
$$s_c \oplus s_\xi = \dm{cc} s_\xi & ~0 \\ 0& ~1 \fm$$
commute avec $D$ si et seulement si $s_\xi e_1  = e_1,$ c'est \`a
dire $s_\xi = s_{e_1}$. En d'autres termes $\ox$ est le p\^ole
nord de la sph\`ere $S^2$ qui, rappelons le, est l'unique point
\`a distance finie de $\omega_c$.

Concernant la distance crois\'ee entre deux \'etats de $\mn$, on
serait tent\'e de relacher une des hypoth\`eses du th\'eor\`eme
\ref{pythagore} et s'int\'eresser \`a des \'etats $s_\xi$,
$s_\zeta$ qui ne sont pas en somme directe. Cette condition n'est
en effet pas n\'ecessaire pour effectuer la projection du triplet
spectral telle qu'elle est pr\'esent\'ee dans la proposition
\ref{reduction0}. Du moment que $s_\xi$ et $s_\zeta$ sont
orthogonaux, $e = s_\xi + s_\zeta$ est un projecteur et on trouve
que la distance est la m\^eme que celle calcul\'ee avec le
triplet $T_e$. Gr\^ace au lemme \ref{projecteurcroise} il
apparait que $\aa_e = e\aa e$ est isomorphe \`a $\mn$.

Pour $n=2$, deux \'etats purs orthogonaux sont \`a une distance
infinie l'un de l'autre puisque d'altitude $z_\xi$, $z_\zeta$
diff\'erentes (sauf \'eventuellement deux \'etats purs sur
l'\'equateur):
$$\bar{\xi_1}\zeta_1 = - \bar{\xi_2}\zeta_2 \longrightarrow z_\xi = \abs{\xi_1}^2 - \abs{\xi_2}^2 = \frac{\abs{\xi_2}^2}{\abs{\zeta_1}^2}\lp
\abs{\zeta_2}^2 - \abs{\zeta_1}^2\rp = -
\frac{\abs{\xi_2}^2}{\abs{\zeta_1}^2}z_\zeta.$$ Par
l'in\'egalit\'e triangulaire il est imm\'ediat que toute distance
$d(\omx\ot \ox, \omy\ot\oz)$ est \'egalement infinie. Les cas
$n\geq 3$ n'ont pas \'et\'e envisag\'es dans le chapitre
pr\'ec\'edent car la d\'etermination du supr\'emum n'est pas
ais\'ee.

\chapter{Fluctuation de la m\'etrique}

\section{Connexion et perturbation de la m\'etrique}

Les th\'eories de jauge, du type Yang-Mills, sont construites sur
un fibr\'e vectoriel o\`u les fibres sont le support d'une
repr\'esentation du {\it groupe de jauge} de l'interaction. De la
m\^eme mani\`ere qu'\`a un espace compact $X$
 est associ\'ee l'alg\`ebre $C(X)$
de ses fonctions continues, \`a tout fibr\'e vectoriel
$E\rightarrow X$ est associ\'e le module de ses sections
continues $\Gamma(E)$ d\'efini en (\ref{modulesection}). C'est un
module sur $C(X)$ qui est fini et projectif [\citelow{jgb}, {\it
Prop. 2.9}]. La d\'efinition d'un module projectif fini est
donn\'ee dans la section I.II.4  (\'enonc\'e de la condition de
finitude); de toutes ses propri\'et\'es nous retiendrons
celle-ci: tout module projectif fini sur $C(X)$  est le module
des sections continues d'un fibr\'e vectoriel sur $X$. Ce
th\'eor\`eme, du \`a  Serre et Swan, est le pendant pour les
fibr\'es vectoriels du th\'eor\`eme de Gelfand. Comme pour le
couple espace compact/$C^*$-alg\`ebre commutative, on montre que
la cat\'egorie des fibr\'es vectoriels sur un espace compact $X$
est \'equivalente \`a la cat\'egorie des modules projectifs sur
$C(X)$. Ainsi un module projectif fini sur l'alg\`ebre $\aa$ d'un
triplet spectral r\'eel $(\aa, \hh, D, \Gamma, J)$ est un bon
candidat pour jouer le r\^ole de fibr\'e vectoriel pour la
g\'eom\'etrie en question, et servir de support \`a la
formulation non commutative d'une th\'eorie de jauge.

\subsection{Transformation de jauge}
Dans une th\'eorie de jauge, le {\it potentiel de jauge} \-- le
quadrivecteur potentiel pour l'\'electroma\-gn\'etisme par
exemple \-- est la forme locale d'une connexion, une
transformation de jauge correspondant \`a un changement de
connexion. En g\'eom\'etrie non commutative, la
connexion\cite{gravity} est d\'efinie par analogie avec la
formule (\ref{connexion}). Au lieu d'une vari\'et\'e $M$, on se
donne un triplet spectral $(\aa, \hh, D)$. $\ginf(E)$ est
remplac\'e par un $\aa$-module projectif fini $\ee$. La
proposition \ref{commutprop} sugg\`ere que les $1$-formes de la
g\'eom\'etrie $(\aa, \hh, D)$ soient g\'en\'er\'ees par des
\'el\'ements du type $[D,a]$. L'ensemble $\Omega^1(M)$ des
sections de $T^*M$ est un $C(M)$-module. On demande donc que
l'ensemble $\Omega^1_D$ des $1$-formes de la g\'eom\'etrie $(\aa,
\hh, D)$ soit un $\aa$-module. Autrement dit
\begin{equation}
\label{omega1d} \Omega^1_D \doteq \left\{ a^i [D, b_i]\,, \; a^i,
b_i \in \aa\right\}.
\end{equation}
{\defi Soit $(\aa, \hh, D)$ un triplet spectral. Une {\it
connexion} sur un $\aa$-module projectif fini $\ee$ est une
application $\aa$-lin\'eaire $\del:\, \ee \mapsto \ee\ot_{\aa}
\Omega^1_D$ satisfaisant la r\`egle de Leibniz
$$\del (sa) = (\del s) a + s\ot [D,a]$$
pour tout $a\in\aa, s\in \ee$.}
\newline

Lorsque qu'un fibr\'e vectoriel $E\rightarrow X$ est muni d'un
produit scalaire fibre \`a fibre, le module $\Gamma(E)$ h\'erite
d'une structure hermitienne \`a valeur dans $C(X)$:
$$(\sigma_1 \lvert \sigma_2) (x) = \scl{\sigma_1(x)}{\sigma_2(x)}.$$ Adapt\'ee \`a un module (par convention \`a droite)
sur une $C^*$-alg\`ebre $\aa$ quelconque, la structure
hermitienne d\'efini un {\it $C^*$-module.}

{\defi Un $C^*$-module sur une $C^*$-alg\`ebre $\aa$ est un
espace vectoriel $\ee$ qui est aussi un $\aa$-module (pas
forc\'ement projectif fini) muni d'un couplage $\ee\times \ee
\rightarrow \aa$ tel que
\begin{eqnarray*}
(r \vert s+t) &=& (r \vert s) + (r \vert t),\\
(r \vert s a) &=& (r\vert s) a,\\
(r\vert s)    &=& (s \vert r)^*,\\
(s\vert s)    &>& 0 \text{ pour } s\neq 0
\end{eqnarray*}
o\`u $r,s,t \in \ee$ et $a\in\aa$, tel que $\ee$ soit complet
pour la norme
$$\norm{s}\doteq\sqrt{\norm{(s\vert s)}}.$$}

\noindent Les modules plein de la d\'efinition \ref{morita} de
l'\'equivalence de Morita sont des $C^*$-modules.

Quand un $\aa$-module projectif fini $\ee$ est aussi un
$C^*$-module \-- $\aa$ est une $C^*$-alg\`ebre \-- se pose la
question de la compatibilit\' e de la connexion avec la structure
hermitienne. L'\'equivalent non commutatif de la connexion de
Levi-Civita est une {\it connection hermitienne}, ie. une
connexion satisfaisant la version non commutative de (\ref{lvg}),
\`a savoir
\begin{equation}
\label{connexhermit} ( s \lvert \del r) - (\del s \lvert r) = [D,
(s\lvert r)].
\end{equation}
Pr\'ecisons que si $\del s = s^i \ot \varpi_i$, $s^i\in \ee$,
$\varpi^i\in\Omega^1_D$, alors
$$(\del s \lvert r) \doteq {\varpi_i}^* (s^i\lvert r )\; \text{ et }\; ( r \lvert \del s) \doteq (r \lvert s^i)\varpi_i.$$
La diff\'erence d'un signe $-$ entre (\ref{lvg}) et
(\ref{connexhermit}) provient de la d\'efinition $d a \doteq
[D,a]$, puisqu'alors $d(a^*) = - (da)^*$. Un th\'eor\`eme
fondamental de la g\'eom\'etrie riemannienne indique que pour
toute vari\'et\'e (pseudo)-riemannienne, il existe une unique
connexion compatible avec la m\'etrique et de torsion nulle. Pour
les $C^*$-modules projectifs fini, on un th\'eor\`eme du m\^eme
ordre, qui repose sur le fait que tout module projectif fini sur
$\aa$ est de la forme
\begin{equation}
\label{empf} \ee = e\aa^N
\end{equation}
o\`u $\aa^N$ d\'esigne le $\aa$-module des vecteurs colonnes de
dimension $N$ \`a entr\'ee dans $\aa$, et $e = e^2\in M_N(\aa)$.
Tout \'el\'ement $s$ d'un $\aa$-module projectif fini est un
$\aa$-vecteur colonne et, puisque $\Omega^1_D$ est un
$\aa$-module, $\del s\in \ee\ot_\aa\Omega^1_D$ est un vecteur \`a
entr\'ee dans $\Omega^1_D$. On note $\xi\in\aa^N$ le vecteur de
composante $\xi_j\in\aa$ tel que $s=e\xi$, et $d\xi$ le vecteur
de composante $[D, \xi_i]\in \Omega^1_D$. On montre alors que
l'ensemble des connexions hermitiennes est un espace affine.

{\prop\label{connexhermiti} Soit $\ee\simeq e\aa^N$ un
$C^*$-module projectif fini. La structure hermitienne de $\ee$
est induite par la structure hermitienne canonique
 de $\aa^N$. Sur ce module, toutes les connexions hermitiennes sont donn\'ees par
$$\del (e\xi) = d(e\xi) + eAe\xi$$
o\`u $A\in M_N(\Omega^1_D)$ est une matrice hermitienne.}
\newline

Toute endomorphisme inversible $\alpha$ de $\ee$ d\'efinit un
endomorphisme de l'espace des connexions
\begin{equation}
\label{connexend} \del \mapsto (\alpha\ot \ii) \del \alpha^{-1}.
\end{equation}
On peut choisir de faire agir un endomorphisme de $\ee$ sur
l'espace des connexions autrement, mais l'action (\ref{connexend})
permet de caract\'eriser facilement un certain type
d'endomorphisme qui pr\'eserve l'hermicit\'e. Un endomorphisme
$\aa$-lin\'eaire $\alpha$ de $\ee$ poss\`ede un adjoint s'il
existe un endomorphisme $\alpha^*$ tel que
$$(r \lvert \alpha s ) = (\alpha^* r \lvert s)$$
pour tout $r,s \in\ee$. On note $\text{End}_A
(E)$\index{endea@$\text{End}_A (E)$} l'alg\`ebre des
endomorphismes
 avec adjoint (c'est une $C^*$-alg\`ebre pour la norme d'op\'erateur [\citelow{jgb},{\it Th. 3.1}]. Un tel endomorphisme est {\it
unitaire} s'il pr\'eserve la structure hermitienne
$$(\alpha r \lvert \alpha s) = (\ r \lvert s),$$
c'est \`a dire si $\alpha^*\alpha = \alpha\alpha^* = \ii_{\ee}$
(l'endomorphisme identit\'e). Le groupe des endomorphismes
unitaire est not\'e ${\cal U}(\ee)$.\index{ue@${\cal U}(\ee)$} On
montre alors\cite{connes} que si $\del$ est une connexion
hermitienne sur $\ee$ et $u\in{\cal U}(\ee)$, alors
$(u\ot\ii)\del u^*$ est une connexion hermitienne. D'o\`u la
d\'efinition d'une {\it transformation de jauge}.

{\defi L'action de ${\cal U}(\ee)$ sur les connexions
hermitiennes est appel\'ee transformation de jauge.}
\newline

\noindent La matrice $A$ de la proposition \ref{connexhermiti}
est l'\'equivalent non commutatif du potentiel de jauge.

\subsection{Op\'erateur de Dirac covariant}

Etant donn\'es une g\'eom\'etrie $(\aa, \hh, D, J, \Gamma)$ et un
$\aa$-module projectif fini $\ee$, on peut construire des
connexions sur $\ee$. L'interpr\'etation g\'eom\'etrique de ces
connexions, c'est \`a dire leur influence sur la g\'eom\'etrie
$(\aa, \hh, D)$, passe par la construction d'un nouveau triplet
spectral.

Tout \'el\'ement $s$ d'un $\aa$-module projectif fini est un
$\aa$-vecteur colonne. On note $\bar{s}$ le $\aa$-vecteur ligne
correspondant. L'ensemble des $\bar{s}$ pour $s\in \ee$ est un
$\aa$-module projectif \`a gauche, not\'e $\bar{\ee}$, o\`u
l'action de $\aa$ est
$$a\bar{s} \doteq \overline{sa^*}.$$

{\prop Soit $(\aa, \hh, D, \Gamma)$ un triplet spectral r\'eel de
dimension $n$ et $\del$ une connexion hermitienne sur un
$\aa$-module projectif finie $\ee$. Soit
\begin{eqnarray*}
\tilde{\aa}&\doteq& \text{ End}_A(\ee),\\
\tilde{\hh}&\doteq& \ee \ot_\aa \hh \ot_\aa \bar{\ee}
\end{eqnarray*}
et l'op\'erateur $\tilde{D}$ agissant sur $\tilde{\hh}$ par
$$\tilde{D} (s\ot \psi\ot \bar{r}) \doteq (\del s)\psi\ot\bar{r} + s\ot D\psi\ot\bar{r} + s\ot \psi\overline{\del r}.$$
Alors $(\tilde{\aa}, \tilde{\hh}, \tilde{D},  \tilde{J},
\tilde{\Gamma})$ avec
\begin{eqnarray*}
\tilde{J}(s\ot\psi\ot\bar{r}) &\doteq& r\ot J\psi \ot \bar{s},\\
\tilde{\Gamma}(s\ot\psi\ot\bar{r}) &\doteq& s\ot \Gamma\psi \ot
\bar{r}
\end{eqnarray*}
est un triplet spectral r\'eel de dimension $n$.}
\newline

\noindent L'action de $\del s = s^i\ot\varpi_i$ sur $\hh$ est
d\'efini en voyant $\varpi_i$ comme un op\'erateur sur $\hh$ via
la d\'efinition (\ref{omega1d}) de $\Omega^1_D$
$$(\del s)\psi = s^i \ot \varpi_i\psi.$$
De m\^eme on d\'efinit $\psi\overline{\del s} = \psi
\overline{s^i\ot\varpi_i} \doteq  J\varpi_i J^{-1}
\psi\ot\bar{s^i}.$

Quand $\tilde{\aa}\neq\aa$ les deux g\'eom\'etries sont
difficilement comparables puisqu'elles ne reposent pas sur le
m\^eme espace des \'etats. En revanche, si on choisit le
$\aa$-module trivial $\ee=\bar{\aa}=\aa$,  on obtient
$\tilde{\aa}= \aa$, $\tilde{\hh}=\hh$ et $\tilde{D}= D + A +
JAJ^{-1}.$
{\defi L'op\'erateur $D_A\doteq D + A + JAJ^{-1}$ est appel\'e
op\'erateur de Dirac covariant.}
\newline

\noindent L'emploi du terme covariant se justifie en remarquant
que l'action d'un unitaire $u\in{\cal U}(\aa)$, par la
modification de la connexion, induit une transformation de $D_A$
en
$$
D_{A'} = D + A' + JA'J^{-1},
$$
o\`u $A'\doteq u Au^* + u[D,  u^*] $. Autrement dit sous une
transformation de jauge, $A$ se transforme selon
$$A\mapsto   u Au^*+ u[D, u^*].$$
qui est bien la  loi de transformation du potentiel vecteur en
\'electromagn\'etisme
$$A\mapsto uAu^{-1} + udu^{-1}.$$

Comme a priori $[D_A, a]\neq [D,a]$ pour un $a$ quelconque de
$\aa$, le remplacement de $D$ par $D_A$, c'est \`a dire le
passage d'une th\'eorie \`a connexion nulle \`a une th\'eorie
covariante, induit une perturbation de la m\'etrique appel\'ee
{\it fluctuation interne de la m\'etrique}. Par analogie avec la
connexion de Levi-Civita qui est nulle si et seulement si
l'espace est plat, les fluctuations internes de la m\'etrique
rendent compte d'une courbure de l'espace non commutatif qui n'a
pas d'\'equivalent commutatif puisqu'alors $A$ est nul.

Le reste de ce chapitre est consacr\'e \`a l'adaptation des
r\'esultats des chapitres pr\'ec\'edents en pr\'esence d'une
connexion non nulle.

\section{Fluctuations de la m\'etrique dans les produits de g\'eom\'etrie}

Soit $(\aa, \hh, D)$ un triplet spectral r\'eel. Pour all\'eger
les notations, on note $\Omega^1$ au lieu de $\Omega^1_D$
l'espace des $1$-formes.

{\lem \label{trace1forme} $[a, J\omega J^{-1}]=0,\; \forall
\omega\in\Omega^1,a\in\aa$.}
\newline

\noindent {\it Preuve.} $[J^{-1}aJ,[D,b_i]]=0$ (axiome du premier
ordre)  et $[a,J a^iJ^{-1}]=0$ (r\'ealit\'e) garantissent que
\begin{align*}
[a, J\omega J^{-1}]&= [a,J a^i[D,b_i]J^{-1}] \\
                        &= aJ a^iJ^{-1}J[D,b_i]J^{-1}-Ja^i[D,b_i]J^{-1}a \\
                        &= J a^i[D,b_i]J^{-1}a - Ja^i[D,b_i]J^{-1}a=0 \,.
\tag*{\mbox{$\blacksquare$}}
\end{align*}
Comme cons\'equence imm\'ediate,
\begin{equation}
\label{ca} [D_A,a]=[D+ A,a]\,.
\end{equation}

Soit $T_E\ot T_I$ un produit de g\'eom\'etries tel que d\'efini
au chapitre II. Les 1-formes sont donn\'ees  par\cite{kt,schucker}
$$
\Omega^1=\Omega^1_E\ot \Omega^0_I +
\chi_E\Omega^0_E\ot\Omega^1_I\,,
$$
o\`u $\Omega^0_E=\aa_E$ est l'ensemble des 0-formes de $\aa_E$,
les autres termes \'etant d\'efinis de mani\`ere analogue.
 Quand $T_E$ est le triplet spectral d'une vari\'et\'e,
$$
\Omega_E^1\ni f^j[-i\ds,g_j\ii_E]= -if^j(\gamma^m \partial_\mu
g_j)=-i\gamma^m f_\mu\,,
$$
o\'u $f^j,g_j,f_\mu\doteq f^j\partial_\mu g_j \in\cinf$. Une
1-forme du triplet total est
$$
\Omega^1\ni -i\gamma^m f_\mu^i \ot a_i - \gamma^5 h^j\ot m_j$$
o\`u $a_i\in\aa_I$, $h^j\in\cinf$, $m_j\in\Omega_I^1$. Un
potentiel vecteur est donn\'e par
\begin{equation}
\label{h} A = -i\gamma^m \ot A_\mu - \gamma^5\ot H
\end{equation}
avec $A_\mu\doteq {f^i}_\mu a_i$ un champ de vecteur (sur $\mm$)
\`a valeur dans les \'el\'ements anti-adjoints de $\aa_I$ et
$H\doteq h^j m_j$ un champs scalaire \`a valeur dans $\Omega^1_I$.
Pour une alg\`ebre de matrices (ou une somme directe d'alg\`ebres
de matrices), les \'el\'ements anti-adjoints forment l'alg\`ebre
de Lie du groupe des unitaires. Ce groupe de Lie repr\'esente le
groupe de jauge de la th\'eorie, donc  $A_\mu$ est un potentiel
de jauge. Dans [\citelow{gravity}] une formule est donn\'ee  pour
les fluctuations de la m\'etrique dues \`a  $A_\mu$. Ici nous
nous int\'eressons aux fluctuations provenant uniquement du champ
scalaire $H$, et {\bf on suppose que
 ${\mathbf A_\mu=0}$.} Alors (\ref{ca}) devient
\begin{equation}
\label{daa} [D_A,a]= [D - \gamma^5\ot H,a].
\end{equation}

Dor\'enavant on \'ecrit  $D_A\doteq D - \gamma^5\ot H$. Pour ne
pas alourdir les notations, on d\'esigne toujours par $d$ la
distance associat\'ee au triplet $(\aa, \hh, D_A)$. Selon
(\ref{pdt1}), une fluctuation scalaire substitue
$$D_H\doteq D_I+ H$$
\`a $D_I$. La diff\'erence essentielle est que maintenant
l'op\'erateur de Dirac $D_H$ d\'epend de $x$, de sorte que tout
point de $\mm$ d\'efinit un triplet spectral interne
$$
T_I^x\doteq (\aa_I, \hh_I, D_H(x))\,.
$$
Cette interpr\'etation de la fluctuation scalaire permet une
adaption facile du th\'eor\`eme \ref{sansfluct}.

\subsection{Distance dans le continu et dans le discret}

\noindent {\it {\bf Th\'eor\`eme \ref{sansfluct}'.} \label{fluct}
Soit $L$ la distance g\'eod\'esique dans $\mm$ et $d_x$ la
distance associ\'ee au triplet spectral $T_I^x$. Pour tout $x, y
\in \mm$ ($\omega_x$, $\omega_y$ d\'esignent les \'etats purs
associ\'es) et $\tau, \tau' \in \ss(\aa_I)$,
\begin{eqnarray*}
d(\omega_x\ot\tau, \omega_x\ot\tau' ) &=& d_x(\tau,\tau'),\\
d(\omega_x\ot\tau, \omega_y\ot\tau) &=& L(x,y).
\end{eqnarray*}}

\noindent {\it Preuve.} La preuve du th\'eor\`eme \ref{sansfluct}
s'adapte facilement. Les notations sont identiques except\'ees
que $\tau_E$ est un \'etat pur et s'\'ecrit $\omega_x$. $a_E$ est
remplac\'e par $a_x$. Avec $H= h^j m_j$,
\begin{eqnarray}
\nonumber
[D_H(x), a_x] &=& [D_I + \omega_x(h^j) m_j, \omega_x(f^i)m_i ]\\
\nonumber
              &=& \omega_x(f^i)[D_I, m_i] + \omega_x(h^j)\omega_x(f^i) [
m_j, m_i ]\\
\label{remarque}
              &=& ( \omega_x \ot \ii_I ) \lp f^i\ot [D_I, m_i] + h^j f^i
\ot [  m_j, m_i ]\rp\\
\nonumber
              &=& ( \omega_x \ot \ii_I ) \lp f^i\ot [D_H, m_i ]\rp.
\end{eqnarray}
Comme $i[D_H(x), a_x]$ est normal,
\begin{eqnarray*}
\norm{[D_H(x), a_x]} &=& \suup{\tau_I\in {\cal S}_I}\abs{
\tau_I\lp[D_H(x),a_x]\rp}\\
                     &=& \suup{\tau_I\in {\cal S}_I}\abs{
(\omega_x\ot\tau_I)\lp
f^i\ot [D_H, m_i]\rp}\\
                  &\leq& \suup{\te\ot \tau_I\in \ss_E\ot
{\cal S}_I} \abs{ (\te\ot\ti)\lp
f^i\ot [D_H, m_i ]\rp}\\
                  &\leq& \norm{ f^i\ot [D_H, m_i] }.
\end{eqnarray*}
L'\'equation (\ref{gen2}) \'etant remplac\'ee par $\norm{f^i \ot
[D_H, m_i]}\leq \norm{[D_A,a]}$, on obtient
$$\norm{[D_H(x), a_x]}\leq\norm{[D_A,a]}.$$
La suite  de la preuve est identique \`a celle du th\'eor\`eme
\ref{sansfluct}. \hfill $\blacksquare$
\newline
\newline
A noter que dans (\ref{remarque}) on utilise que $\omega_x$ est
un caract\`ere (i.e. que c'est un \'etat pur et que $\aa_E$ est
commutative). Le th\'eor\`eme \ref{pythagore} est modifi\'e plus
s\'erieusement car la fluctuation introduit une d\'ependance en
$x$ du coefficient suppl\'ementaire
 dans la m\'etrique de Kaluza-Klein.
\newline

\subsection{Distance crois\'ee}

Les notations sont celles du th\'eor\`eme \ref{pythagore}.

\noindent {\it {\bf Theor\`eme \ref{pythagore}'.} Soient
$\ou,\od$ deux \'etats purs normaux de $\aa_I$ en somme directe
tels que la somme de leurs supports commute avec $D_H(x)$ en tout
$x$. Alors
$$d(x_1, y_2)= L'((0,x),(1,y))$$
o\`u $L'$ est la distance g\'eod\'esique de la vari\'et\'e \`a
spin $\mm'\doteq [0,1]\times \mm$ munie de la m\'etrique
$$\dm{cc} \norm{M(x)}^2 & 0 \\
0 & g^{\mu\nu}(x) \fm$$ dans laquelle $g^{\mu\nu}$ est la
m\'etrique de la vari\'et\'e initiale et $M$ est la restriction
\`a la repr\'esentation de $\aa_I s_1\aa_I$ de la projection de
$D_H$ sur la repr\'esentation de $\aa_Is_2\aa_I$.}
\newline

\noindent {\it Preuve.} Sauf mention contraire, les notations
sont celles du th\'eor\`eme \ref{pythagore}. La premi\`ere partie
de la preuve est \`a peine modifi\'ee. Soit
$\psi^r\ot\xi_r\in\hh$. Gr\^ace \`a (\ref{daa}) et \`a la
d\'efinition (\ref{h}) de $H$,
$$
[D_A, a] \psi^r\ot\xi_r = \gamma^5 \psi_r \ot [D_I, p] \xi_r +
\gamma^5 h^j\psi_r \ot [m_j,p]\xi_r\in \hh.$$ Evalu\'ee en
$x\in\mm$, l'expression ci-dessus donne
$$[D_A, a] \psi^r (x) \ot\xi_r = \gamma^5 \psi_r(x) \ot [D_I +  H(x),
p]\xi_r = 0
$$
par hypoth\`eses, indiquant que $[D_A , a]$ est l'endomorphisme
nul de $\hh$ si bien que par le lemme \ref{reduction},
$$d(\xox,\yoz)=d_e(x_1,y_2).$$
La diff\'erence avec le th\'eor\`eme \ref{pythagore} est que
$D_r$ d\'epend de $x$. Plus pr\'ecis\'ement, $M$ est une matrice
dont les entr\'ees sont des champs scalaires sur $\mm$.

D\'esormais $g^{tt}(x)\doteq\norm{M(x)}^2$ d\'epend de $x$ mais
est constant par rapport \`a $t$. Les \'equations des
g\'eod\'esiques (\ref{geo1}, \ref{geo2}) ne se r\'eduisent pas \`a
(\ref{constante}) mais \`a
\begin{eqnarray*}
\frac{d}{d\tau}
(g_{tt}\frac{dt}{d\tau})&=&(\frac{d}{d\tau}g_{tt})\frac{dt}{d\tau}+
g_{tt}\frac{d}{d\tau}(\frac{dt}{d\tau})\\
                                        &=&(\partial_\mu
g_{tt})\frac{dt}{d\tau}\frac{dx^\mu}{d\tau}+ g_{tt}\frac{d^2t}{d\tau^2}\\
                                        &=&g_{tt}\lp g^{tt}(\partial_\mu
g_{tt})\frac{dt}{d\tau}\frac{dx^\mu}{d\tau}+
\frac{d^2t}{d\tau^2}\rp=0
 \end{eqnarray*}
par (\ref{geo1}). Ainsi $g_{tt}\frac{dt}{d\tau}=K$ est une
constante. La seule diff\'erence avec la premi\`ere \'equation
(\ref{constante}) est que
\begin{equation}
\label{constanteh} \frac{dt}{d\tau}=Kg^{tt}(x)
\end{equation}
est maintenant fonction de $x$. On d\'efinit $a_0= (f_0, g_0)$ par
\begin{equation}
\label{gunh} f_0(q) \doteq
\int_{{\cal G}_q}\sqrt{1-K^2g^{tt}}ds\,,\qquad g_0 \doteq f_0
-K\,,
\end{equation}
o\`u ${\cal G}_q'$ est une g\'eod\'esique minimale entre $(0,q)$
et le point fixe $(1,y)$. ${\cal G}_q$ d\'esigne sa projection
sur $\mm$ (noter que ${\cal G}_q$ n'est pas une g\'eod\'esique de
$\mm$). En supposant que
\begin{equation}
\label{restriction3} K^2g^{tt}(p)\neq 1
\end{equation}
pour tout $p \in {\cal G}_q$, on \'ecrit $d\tau =
\frac{ds}{\sqrt{ 1 - K^2 g^{tt}} }$ et
\begin{equation}
\label{dtauh} 1=\int_{{\cal G}_q'} dt = \int_{{\cal G}_q'}
\frac{dt}{d\tau} d\tau = \int_{{\cal G}_q}
\frac{Kg^{tt}}{\sqrt{1-K^2g^{tt}}}ds\,.
\end{equation}
 Si (\ref{restriction3}) n'est pas v\'erifi\'ee, on note $G$ l'ensemble des points $p$
de ${\cal G}_q$ pour lesquels $1 - K^2g^{tt}(p)=0$. $G'$
d\'esigne l'ensemble  des points correspondants dans ${\cal
G}'_q$. Pour tout $p'\in {\cal G'}_q$, (\ref{constanteh}) donne
\begin{eqnarray*}
\frac{dt}{d\tau}d\tau=K^{-1}d\tau\,,&
\end{eqnarray*}
et au lieu de (\ref{dtauh}),
$$
1 = \int_{{\cal G}_q/G}
\frac{Kg^{tt}}{\sqrt{1-K^2g^{tt}}}ds+\int_{G'}K^{-1} d\tau\,.
$$
Ins\'er\'e dans $x_1(a_0) -y_2(a_0) = f_0(x) + K$ en tant que $K
\times 1$, cette expression garantit que
\begin{eqnarray*}
\nonumber x_1(a_0) -y_2(a_0) &=&\int_{{\cal G}_x}
\sqrt{1-K^2g^{tt}} ds + \int_{{\cal G}_x/G}
\frac{K^2g^{tt}}{\sqrt{1-K^2g^{tt}(x)}}ds
+ \int_{G'} d\tau \\
&=& \int_{G} \sqrt{1-K^2g^{tt}}ds  + \int_{{\cal G}_x/G}
\frac{ds}{\sqrt{1-K^2g^{tt}(x)}}  + \int_{G'} d\tau \\
&=&\int_{{\cal G}_x'/G'} d\tau + \int_{G'} d\tau =
L'\lp(0,x),(1,y)\rp.
\end{eqnarray*}
La fonction $f_0$ est, par d\'efinition (\ref{gunh}), constante
sur un hyperplan dans un voisinage de $q$. Dans un r\'ef\'erentiel
ad\'equat \--$\{x^1,x^2,x^3\}$ d\'esignant les coordonn\'ees de
l'hyperplan et $x^0$ la coordonn\'ee suppl\'ementaire \-- on
\'ecrit $ds(q) = \sqrt{g_{00}(q)} dx^0$ et $\partial_\mu
f_0(q)=\delta_\mu^0
\partial_0 f_0 (q)$. Ainsi
\begin{eqnarray*}
\partial_\mu f_0(q) &=& \delta_\mu^0 \sqrt{1-K^2g^{tt}(q)}\sqrt{g_{00}(q)}\,,
\\
g^{\mu\nu}(q)\partial_\mu f_0(q)\partial_\nu f_0(q) &=& g^{00} (1
- g^{tt}K^2) g_{00} = 1 - g^{tt}K^2 \,,
\end{eqnarray*}
d'o\`u $\norm{[D_e,a_0]}=1$ et le r\'esultat. \hfill
$\blacksquare$
\newline

Quelques pr\'ecisions sur ce th\'eor\`eme. Tout d'abord, puisque
tous les coefficients de la m\'etrique sont d\'ependants en $x$,
la distance g\'eod\'esique ne peut en aucun cas satisfaire le
th\'eor\`eme de Pythagore. Ensuite, rappelons que par
d\'efinition une m\'etrique n'est pas d\'eg\'en\'er\'ee et,
implicitement, nous avons supposer que $M(x)$ ne s'annule en
aucun point. C'\'etait n\'ecessaire dans le th\'eor\`eme
\ref{pythagore} afin que la distance reste finie. Ici la question
est plus subtile dans la mesure o\`u $M$ peut tr\`es bien
 n'\^etre nul que pour certains points $x$. Soit $\ker(M)\subset\mm$ l'ensemble de ces points. Pour tout
$q\in \ker(M)$, $d\lp (0,q),(1,q)\rp=+\infty$ par la proposition
2'. De plus
\begin{eqnarray*}
d \lp (0,q),(1,q)\rp &\leq&  d\lp (0,q),(0,x) \rp + d \lp
(0,x),(1,y)\rp + d\lp (1,y),(1,q) \rp\\
                       &\leq&   L(p,x) + d\lp (0,x),(1,y) \rp + L(y,q)\,,
\end{eqnarray*}
donc $d\lp (0,x),(1,y)\rp=+\infty$ pour tout $x,y\in\mm$, ce qui
contredit le th\'eor\`eme 4' d\`es que $x=y\notin \ker(M)$. Une
solution est de consid\'erer que $(t,q)$ avec $q\in \ker(M)$ est
un point isol\'e \`a distance infinie de tout autre point et de
d\'efinir $\mm'$ comme $[0,1]\times\mm/\ker(M)$. Si tout chemin
entre $x$ et $y$ traverse $\ker(M)$, cette op\'eration d\'ecoupe
$\mm'$ en morceaux disconnexes. Une meilleure solution consiste
\`a prendre en compte la partie non scalaire de la fluctuation.
Faute de temps, cet aspect n'est pas \'etudi\'e dans cette
th\`ese. On renvoie \`a [\citelow{gravity}] pour l'\'etude du
champ de jauge vu
 comme m\'etrique.

\section{Exemples}

Dans cette derni\`ere section, on \'etudie la m\'etrique d'espace
produit du discret par le continu dont la partie interne est
l'une de celles d\'ecrites dans le chapitre II. On donne
\'egalement un r\'esultat concernant la distance dans le mod\`ele
standard.

\subsection{Espaces commutatifs - le mod\`ele \`a deux couches}

Avec les notations de la section 3.III.1, on v\'erifie que
$[D_H,s_i + s_j]=0$ d\`es que  $[D_I, s_i + s_j]=0$. Cette
condition n'est pas n\'ecessaire et la fluctuation peut \^etre
telle que le  th\'eor\`eme \ref{pythagore}' s'applique alors que
le th\'eor\`eme \ref{pythagore} ne s'appliquait pas dans la
th\'eorie de jauge nulle. Bien entendu le cas le plus simple,
$k=2$, muni le mod\`ele \`a deux couches d'une m\'etrique
cylindrique ou le coefficient suppl\'ementaire de la m\'etrique
est une fonction de la vari\'et\'e.

 \subsection{ Le mod\`ele standard.}

Le triplet spectral du mod\`ele standard (cf.
[\citelow{connes,gravity,spectral}] et [\citelow{bridge}] pour le
calcul d\'etaill\'e de la masse du boson de Higgs) est le produit
du triplet spectral r\'eel (\ref{td}), not\'e ici $T_E$,  par une
g\'eom\'etrie interne o\`u l'alg\`ebre r\'eelle
$$\aa_I=\hhh \oplus \cc \oplus M_3(\cc)$$
est represent\'ee sur
$$\hh_I=\cc^{90}=\hh^P \oplus \hh^A= \hh_L^P \oplus \hh_R^P \oplus \hh_L^A
\oplus \hh_R^A\,.$$ La base de $\hh_L^P=\cc^{24}$ est donn\'ee
par les fermions gauches
$$
\dm{c} u\\d \fm_L,\;\dm{c} c\\s \fm_L,\;\dm{c} t\\b \fm_L,\;\dm{c}
\nu_e\\e \fm_L,\;\dm{c} \nu_\mu\\\mu \fm_L,\;\dm{c} \mu_\tau\\
\tau \fm_L,
$$
 et la base de $\hh_R^P=\cc^{21}$ est form\'ee des fermions droits
$u_R,\, d_r,\, c_R,\, s_R,\, t_R,\, b_R \text{ et }\, e_R,\,
\mu_R,\, \tau_R$ (le mod\`ele a \'et\'e construit du temps o\`u
les neutrinos n'avaient pas de masse). L'indice de couleur des
quarks est omis.
 $\hh_R^A$ et $\hh_L^A$ correspondent  aux antiparticules. $(a\in\hhh,\; b\in\cc,
c\in M_3(\cc))$ est repr\'esent\'e par
\begin{equation*}
\label{repms} \pi_I(a,b,c)\doteq \pi^P (a,b) \oplus \pi^A (b,c)
\doteq \pi_L^P(a) \oplus \pi_R^P(b) \oplus \pi_L^A(b,c) \oplus
\pi_R^A (b,c)
\end{equation*}
o\`u, en \'ecrivant $B\doteq\dm{cc} b~&~0\\0~&~\bar{b}\fm\in\hhh$
et $N$ le nombre de g\'en\'erations de fermions,
\begin{eqnarray*}
\pi_L^P(a)\doteq a\ot\ii_N\ot\ii_3\,  \oplus\,
a\ot\ii_N\,,\qquad& &
\pi_R^P(b) \doteq B \ot \ii_N \ot \ii_3\, \oplus \, \bar{b}\ot \ii_N\,,\\
\pi_L^A(b,c)\doteq \ii_2\ot\ii_N\ot c \,\oplus\,
\bar{b}\ii_2\ot\ii_N\,,\qquad & &
\pi_R^A(b,c)\doteq\ii_2\ot\ii_N\ot c \, \oplus \, \bar{b}\ii_n\,.
\end{eqnarray*}
On d\'efinit une structure r\'eelle
$$
J_I= \dm{cc} 0~ & ~\ii_{15N}\\ \ii_{15N}~ & ~0\fm \circ\, C
$$
et un op\'erateur de Dirac interne
$$
D_I\doteq\dm{cc} D_P & 0\\ 0& \;\bar{D_P} \fm= \dm{cc} D_P~ & 0
\\ 0&0\fm + J_I \dm{cc} D_P~ & 0 \\ 0&0\fm J^{-1}_I$$ dont les
entr\'ees sont les matrices $15N\times 15N$
$$D_P \doteq \dm{cc} 0&M\\ M^{*}& 0\fm,
$$
o\`u $M$ est la matrice $8N \times 7N$
\def\masseproj{\lp e_{11} \ot M_u + e_{22} \ot M_d \rp}
\def\masse{ \dm{cc}\masseproj \ot \ii_3& 0 \\0&e_2\ot M_e\fm}
\begin{equation}
\label{m} M\doteq\masse.
\end{equation}
Ici, $\{e_{ij}\}$ et $\{e_i\}$ d\'esignent les bases canoniques
de $\m2$ et $\cc^2$ respectivement. $M_u$, $M_d$, $M_e$ sont les
matrices de masse
$$M_u=\dm{ccc} m_u & 0&0 \\ 0&m_c&0\\ 0&0&m_t\fm,\quad
M_d=C_{KM}\dm{ccc} m_d & 0&0 \\ 0&m_s&0\\ 0&0&m_b\fm ,\quad
M_e=\dm{ccc} m_e & 0&0 \\ 0&m_\mu&0\\ 0&0&m_\tau\fm
$$
 dont les coefficients sont les masses des fermions \'el\'ementaires,
 \'eventuellement pond\'er\'ees par la matrice unitaire de Cabibbo-Kobayashi-Maskawa. La chiralit\'e,
 dernier \'el\'ement du triplet spectral r\'eel, est
$$
\Gamma_I= (-\ii_{8N}) \oplus \ii_{7N} \oplus (-\ii_{8N}) \oplus
\ii_{7N}\,.
$$

La pr\'esence de la repr\'esentation conjugu\'ee $\bar{b}$ dans
$\pi_I$ oblige \`a voir $\cc$ comme une alg\`ebre r\'eelle. Par
cons\'equent, l'\'etat pur $\omega_c$ de $\cc$ n'est plus
l'identit\'e mais la partie r\'eelle. L'\'etat pur $\omega_h$ de
$\hhh$ est d\'ecrit dans le chapitre I. Concernant
$\ss(M_3(\cc))$, on remarquera simplement qu'\'etant lin\'eaires,
deux \'etats purs ayant m\^eme noyau sont proportionnels; comme
ils co\"{\i}ncide sur l'identit\'e, ils sont alors \'egaux.
L'alg\`ebre interne \'etant r\'eelle, par produit tensoriel
l'alg\`ebre externe doit \^etre vue comme l'alg\`ebre r\'eelle
$C^\infty_\rr(M)$ et un \'etat pur externe est $\re(\omega_x)$.

La g\'eom\'etrie non commutative donne une interpr\'etation du
champ de Higgs comme $1$-forme de la g\'eom\'etrie interne. Par
fluctuation scalaire,  les $1$-formes sont \'etroitement li\'ees
\`a la m\'etrique et le champ de Higgs s'interpr\`ete en effet
comme coefficient d'une m\'etrique.

Le calcul suivant est men\'e en jauge nulle $A_\mu = 0$. {\prop
\label{distancems} La partie finie de la g\'eom\'etrie du
mod\`ele standard avec fluctuation interne scalaire de la
m\'etrique en jauge nulle est un  mod\`ele \`a deux couches
index\'ees par les \'etats de $\cc$ et $\hhh$. Chacune des couches
est une copie de la vari\'et\'e  riemannienne \`a spin initiale,
munie de sa m\'etrique.  La composante suppl\'ementaire de la
m\'etrique, correspondant \`a la dimension discr\`ete,  est
$$
g^{tt}(x)= \lp\abs{1+h_1(x)}^2+\abs{h_2(x)}^2\rp m_t^2
$$
o\`u $\dm{c} h_1\\h_2\fm$ est le doublet de Higgs et $m_t$ la
masse du quark top.}
\newline

\noindent {\it Preuve.} $\pi_I$ signifie $\pi_I(a,b,c)$ et
$\Delta\doteq\dm{cc} D_P~&0\\0&0\fm$ afin que $D_I=\Delta+
J\Delta J^{-1}$.

Puisque $\Delta$ est une $1$-forme\cite{krajew}, le lemme
\ref{trace1forme} donne  $[J_I\Delta J^{-1}_I, \pi_I] =0$ et on
peut prendre $D_H= \Delta + H$. Par un calcul
explicite\cite{iochum},
\begin{equation*}
\label{hms}
H=\dm{cccc}0&\pi^P_L(h)M~&~0~&~0~\\
M^{*}\pi^P_L(h^{*})&0&0&0\\0&0&0&0\\0&0&0&0\fm
\end{equation*}
o\`u $h$ est un champ scalaire \`a valeur quaternionique.  Alors
\begin{equation}
\label{massems}
D_H = \dm{cccc}0&\Phi M~&~0~&~0~\\
M^{*}\Phi^{*}&0&0&0\\0&0&0&0\\0&0&0&0\fm,
\end{equation}
o\`u
\begin{equation*}
\label{Phi}
\Phi\doteq (h + \ii_\hhh )\ot\ii_{4 N}=\dm{cc} 1 + h_1~ & ~h_2\\
-\bar{h}_2~& ~1+ \bar{h}_1\fm\, \ot \ii_{ 4N},
\end{equation*}
avec $h_1$, $h_2$ deux champs scalaires complexes.

Par (\ref{distance}), les distances dans le mod\`ele standard
sont identiques \`a celles du triplet $ (\aa_s,\, \hh, \,D)$, o\`u
$\aa_s = {\cinf}_s \ot {\aa_I}_s$ est la sous-alg\`ebre des
\'el\'ements auto-adjoints  de $\aa$, avec
$${\aa_I}_s = \cc_s \oplus \hhh_s \oplus M_3(\cc)_s =  \rr\,
\oplus\,\rr\oplus M_3(\cc)_s.$$ La repr\'esentation $\pi_s$ de ce
triplet est la  restriction de $\pi$ \`a $\aa_s$. Pour les
quaternions, $\pi_s$ substitue
$$\dm{cc} \theta & ~0\\ 0 & ~\theta \fm \; \text{ \`a } \;
\dm{cc} \theta &~\bar{\rho} \\ -\bar{\rho} & ~\bar{\theta}\fm.$$
En d'autres termes, \`a chaque repr\'esentation de $\hhh$
correspond la somme directe de la repr\'esentation fondamentale
de $\rr = \hhh_s$ avec elle-m\^eme. Dans ce cas, $\omega_h$ vu
comme \'etat pur de $\hhh_s$ est bien l'identit\'e. La projection
associ\'ee  $s_h\in\hhh_s$ est simplement  le nombre $1$ qui
satisfait naturellement (\ref{projecteur}). Il en va de m\^eme
pour $\omega_c$ vu comme \'etat pur de $\rr=\cc_s$ (on note $s_c
= 1$ le projecteur associ\'e). On obtient alors que
$$\pi_s(s_c \oplus s_h) = \dm{cc} \ii_{15 N} & 0 \\ 0&
\dm{cccc} 0_{6N} &       &        & \\
                 & \ii_{2 N} &        & \\
                 &       &0_{6N}& \\
                 &       &        &\ii_{N}
\fm \fm
$$
commute avec $D_H$ d\'efini en (\ref{massems}).  Le th\'eor\`eme
\ref{pythagore}' s'applique pour les \'etats de $\aa$ dont la
partie interne est $\omega_c$ ou $\omega_h$. Puisque
$$
\pi_s(s_h\hh_I) = \hh^P_L\, \text{ et }\, \pi_s(s_c\hh_I) =
\hh^P_R \oplus \hh^A_{lep}\, ,
$$
o\`u $\hh^A_{lep}= \cc^{3 N}$ est le sous-ensemble de $\hh^A$
g\'en\'er\'e par les anti-leptons, le coefficient de m\'etrique
suppl\'ementaire est
$$
g^{tt}(x) = \norm{\Phi(x) M}^2.
$$
Comme attendu, $\Phi M$ est une matrice $2 \alpha^\hhh \times \lp
\alpha^\cc + \alpha^{\overline{\cc}}\, \rp$,  o\`u $\alpha^\hhh=
4N$ est la d\'eg\'en\'erescence de la repr\'esentation de
$\hhh_s$ dans $\pi^P_L$, et $\alpha^\cc=3N$,
$\alpha^{\overline{\cc}}= 4 N$ sont d\'efinis de mani\`ere
similaire. En utilisant la forme explicite (\ref{m}),
\begin{eqnarray*}
\norm{\Phi(x) M}^2 &=& \max \left\{ \,\norm{ (\Phi(x)\ot\ii_3) (
e_{11}\ot M_u + e_{22}\ot M_d)}^2,
\norm{(\Phi(x)\ot\ii_3)(e_2\ot M_e)}^2\, \right\}\\
&=& \lp\abs{1+h_1(x)}^2+\abs{h_2(x)}^2 \rp \max \left\{ \,
{m_t}^2 , {m_\tau}^2
\right\}\\
&=& \lp \abs{1+h_1(x)}^2+\abs{h_2(x)}^2 \rp {m_t}^2.
\end{eqnarray*}

Les autres distances mettent en oeuvre les \'etats purs de
$M_3(\cc)$ et sont infinies. En effet,
$$\norm{[D_H, \pi_I(a, b, c)]}= \norm{
\Big[\dm{cc} 0 & \Phi M \\ M^* \Phi^* & 0 \fm, \pi^P(a,b)\Big]}$$
ne met aucune contrainte sur $c$, si bien que pour
$\omega'\in\ss\lp M_3(\cc)\rp$ et $\omega\in\ss(\aa_I)$,
$$
d_I(\omega',\omega) \geq \suup{c \in M_3(\cc)} \abs{\omega'(c) -
\omega(c)}.
$$
Pour $\omega=\omega_c$, $c=\lambda\ii_3$ avec
$\lambda\rightarrow\infty$ garantit que la distance $d_I(\omega',
\omega_c)$ est infinie. Alors
$$
d_I(\omega_2, \omega_0) = d(x_2, x_0) \leq d (x_2, y_0) + d(y_0,
x_0) \leq d (x_2, y_0) + L(x,y)
$$
par le th\'eor\`eme \ref{sansfluct}', et $d (x', y_c) = +\infty
$. La preuve est rigoureusement la m\^eme pour $\omega=\omega_h$.
Pour $\omega\in\ss(M_3(\cc))$, il suffit de se rappeler qu'il
existe $c'\in ker({\omega'})$ tel que  $c'\notin ker(\omega)$, ce
qui  rend $d_I(\omega_2, \omega)$ infinie.  \hfill $\blacksquare$

{\renewcommand{\thechapter}{}\renewcommand{\chaptername}{}
\addtocounter{chapter}{-1}
\chapter{Conclusion}\markboth{\sl Conclusion}{\sl Conclusion}}

Calculer explicitement les distances par la formule
(\ref{distance}) permet d'avoir une image "intuitive" des espaces
non commu\-tatifs. Dans cette th\`ese, ces calculs ont pu \^etre
men\'es \`a terme parce que la g\'eom\'etrie pr\'esentait des
propri\'et\'es particuli\`eres: deux des coefficients de
l'op\'erateur de Dirac nuls pour l'espace \`a quatre points,
\'etats en somme directe dont la somme des supports commute avec
$D$ pour les produits de g\'eom\'etrie. Un premier axe de
recherche est de s'affranchir de ces contraintes qui sont
purement techniques et n'ont pas de justification physique.

Dans le cas commutatif fini, une piste est d'ajouter une
condition sur le triplet spectral. En exigeant que les distances
entre $n$ points satisfassent, outre l'in\'egalit\'e
triangulaire,  les m\^emes propri\'et\'es que les distances entre
$n$ points d'un espace euclidien de dimension donn\'ee, on peut
esp\'erer caract\'eriser des op\'erateurs de Dirac pour lesquels
les calculs soient  possibles.   .

Concernant les produits de g\'eom\'etrie, il est probable que le
mod\`ele \`a deux couches muni d'une m\'etrique cylindrique n'est
plus significatif d\`es lors que les \'etats ne sont plus en
somme directe et/ou  que la somme de leurs supports ne commutent
pas avec $D$. On pouvait esp\'erer que le fibr\'e en sph\`ere
construit par projection de l'alg\`ebre interne sur $M_2(\cc)$
(quand les \'etats ne sont pas en somme directe mais seulement
orthogonaux) \'etait une bonne piste. Malheureusement, comme
discut\'e \`a la fin du chapitre III, de tels \'etats sont \`a
distance infinie (sauf peut-\^etre ceux sur l'\'equateur, i.e.
d'altitude $z=0$ dans la fibration de Hopf).  Pour deux \'etats
non orthogonaux, la somme des supports n'est pas un projecteur et
on ne peut plus simplifier le calcul en projetant l'alg\`ebre
interne sur une alg\`ebre de dimension inf\'erieure. L'exemple du
cas \`a quatre point laisse augurer qu'un calcul direct sera vite
impraticable. Plut\^ot que de r\'esoudre des cas acad\'emiques,
il semble donc pr\'ef\'erable d'entreprendre des calculs au cas
par cas, pour les mod\`eles dict\'es par la physique.
\newline

Une autre question ouverte est la d\'efinition non commutative
d'une g\'eod\'esique. En particulier dans l'espace fini
commutatif, doit-on consid\'erer que tout \'etat appartenant \`a
un chemin $(i,j)$  est \'el\'ement d'une "g\'eod\'esique" entre
$i$ et $j$ (car si on coupe tous les liens attach\'es \`a cet
\'etat, la distance entre $i$ et $j$ augmente) ou faut-il
raffiner cette d\'efinition en trouvant un \'equivalent de la
propri\'et\'e \'el\'ementaire v\'erifi\'ee pour la droite
r\'eelle: la g\'eod\'esique est l'ensemble des points o\`u la
d\'eriv\'ee de la fonction r\'ealisant le supr\'emum dans la
formule de la distance est partout \'egale \`a $1$ ?
\newline

Plus urgent, car en relation avec la physique, est de prendre en
compte la partie non scalaire de la fluctuation. Une formule
existe\cite{gravity} pour une connexion de jauge $A_\mu$ non
nulle mais en l'absence de fluctuation scalaire. Le r\'esultat,
via l'holonomie de la connexion, permet de caract\'eriser la
finitude de la distance. Il est donc d'une grande importance
physique puisqu'il peut rendre finie la m\'etrique du secteur
interaction forte du mod\`ele standard. Ecrire la preuve de ce
r\'esultat est une priorit\'e, en l'associant ensuite aux
r\'esultats pr\'esent\'es dans cette th\`ese afin d'obtenir une
formule pour une fluctuation compl\`ete comprenant \`a la fois une
partie scalaire et une partie de jauge. Ce r\'esulat trouverait
une application directe dans le tore non commutatif dont les
fluctuations internes de la m\'etrique, contrairement au cas
commutatif o\`u $A + JAJ^{-1}$ est nul,  sont hautement non
triviales.

Alors seulement on pourra esp\'erer interpr\'eter physiquement
les distances non commutatives du mod\`ele standard et proposer
des tests exp\'erimentaux. D\'eceler une structure discr\`ete \`a
l'\'echelle de Planck n'est peut-\^etre pas hors de port\'e de
l'exp\'erience, comme le sugg\`erent les travaux sur les jets de
rayons gamma\cite{giac} selon lesquels une structure discr\`ete
de la g\'eom\'etrie peut se r\'ev\`eler par addition d'effets
minuscules sur une longue trajectoire.
\newline

Deux points enfin restent en suspend. Tout d'abord
l'impossibilit\'e pour l'instant de traiter des m\'etriques
pseudo-riemanniennes (cf [\citelow{DS4}] pour des propositions
sur cette question) qui rend d\'elicate l'interpr\'etation
physique mentionn\'ee ci-dessus, et les z\'eros du doublet de
Higgs avec le lien qu'ils \'etablissent entre la
d\'eg\'en\'er\'escence de la m\'etrique au sens non commutatif et
des probl\`emes subtils de th\'eorie des champs tels que le
probl\`eme de Gribov\cite{itzub}.

\chapter*{Annexe.}

\subsection*{Coefficients de $V_{eff}(x,y)$ pour le cas g\'en\'eral \`a quatre points}

\begin{eqnarray*}
V_4(x)&=& {\frac{4{{(d_3 d_4-d_2d_5)}^2}
          ({{d_4}^2}d_6^2+ d_2^2 ({{d_4}^2}+ d_6^2))}{{ d_2^4}d_3^2d_4^4d_5^2d_6^2}},\\
V_3(x) &=&{\frac{8x(d_2d_5-d_3d_4)( d_3d_4d_5 d_6( d_2^2+ d_4^2)
          + d_1(d_2d_3d_4(d_4^2+d_6^2) -d_4^2d_5d_6^2 -d_2^2d_5(d_4^2+2d_6^2)))}{ d_1{ d_2^3} d_3^2
          { d_4^4} d_5^2 d_6^2}},\\
V_2(x)&=& {\frac{4{x^2}}{d_1^2d_2^2d_3^2d_4^4 d_5^2 d_6^2}}
[{d_4^2  (d_3^2d_4^2d_5^2+d_1^2{{(d_3d_4-d_2 d_5)}^2}
+d_2^2( d_3^2d_4^2 + (d_3^2+d_4^2)d_6^2))}\\
      & &- 2d_1d_4d_6( d_2 d_4d_5(d_4^2-2d_3^2)+d_3d_4^2d_5^2
+d_2^2 d_3 (d_4^2 + 3d_5^2))\\
      & &+ d_6^2(d_4^2 {{(d_3 d_4-d_2d_5)}^2}+d_1^2(d_4^2(d_3^2+d_4^2+d_6^2)-6d_2d_3d_4d_5 +d_2^2(d_4^2+6d_6^2)))]\\
      &-& {\frac{4 (d_2^2(d_3^2 (d_4^2 +d_5^2 +d_6^2)+d_5^2 (d_4^2+2d_6^2))-2d_2d_3d_4d_5d_6^2  + d_4^2 (  d_5^2  d_6^2 +  d_3^2 (  d_5^2
+ 2  d_6^2 )  )  ) }{ d_2^2  d_3^2
          d_4^2 d_6^2 d_6^2}},\\
V_1(x)
&=&{\frac{8{x^3}(d_1d_6-d_3d_4)(d_1d_2d_3d_4(d_4^2+d_6^2)-(d_1^2d_2d_4^2-d_3d_4(d_1^2+d_4^2)d_5+d_2(2d_1^2+d_4^2)d_6^2)d_6)}{{d_1^3}d_2d_3^2
{ d_4^4}
d_6^2d_6^2}}\\
       &+&{\frac{8x(d_3d_4d_5d_6(d_3 d_4-d_2d_5) +d_1 ( d_2  d_3^2 (  d_4^2 +  d_6^2 )  +
       d_6^2(d_2(d_3^2+2d_5^2)-d_3d_4d_5))) }{d_1 d_2  d_3^2  d_4^2  d_6^2  d_6^2}},\\
V_0(x)&=& 4 ({d_3^{-2}}+{d_6^{-2}}+{{d_6}^{-2}}) +{\frac{4{x^4}(
d_4^2  d_5^2 +  d_1^2(d_4^2+d_5^2))
      {{( d_3 d_4 - d_1  d_6 ) }^2}}{{ d_1^4}  d_3^2 { d_4^4}  d_6^2  d_6^2}} \\
      &-&{\frac{4{x^2}(d_4^2(2d_3^2d_5^2 +  d_6^2(d_3^2+d_5^2 )) -2d_1d_3d_4 d_5^2d_6  +
         d_1^2( d_6^2(d_4^2 + 2d_5^2)+d_3^2 (d_4^2+d_5^2+d_6^2)))}{d_1^2d_3^2d_4^2d_6^2d_6^2}}.
\end{eqnarray*}
\subsection*{Calcul de $d(1,2)$ lorsque $\frac{1}{d_2}=\frac{1}{d_5}=\infty$}
\begin{eqnarray*}
y_1 =\text{sign}(d_1d_6 - d_3d_4) d_6 \sqrt{\frac{d_3^2 +
d_6^2}{d_4^2 + d_6^2}} &,&
z_1 =d_3\frac{(d_1d_3 + d_4d_6)}{\sqrt{(d_3d_4-d_1d_6)^2}}\sqrt{ \frac{d_4^2 + d_6^2}{d_3^2 + d_6^2}},\\
y_{2} = {\frac{d_1|{d_3}-{d_4}|\pm |d_4({d_1}+{d_6})|}{{\sqrt{{{(
{d_3} - {d_4}) }^ 2} + {{( {d_1} + {d_6}) }^2}}}}}&,&
z_{2} = \pm {\frac{{d_3}\,\left( {d_1} + {d_6} \right) }{{\sqrt{{{\left( {d_3} - {d_4} \right) }^2} +{{\left( {d_1} + {d_6} \right) }^2}}}}},\\
y_{3} = {\frac{d_1|{d_3}+{d_4}|\pm|d_4({d_1}-{d_6})|}{{\sqrt{{{(
{d_3}+{d_4}) }^ 2} + {{( {d_1}-{d_6}) }^2}}}}}&,& z_{3} = \pm
{\frac{{d_3}\,\left( {d_1}-{d_6} \right) }{{\sqrt{{{\left(
{d_3}+{d_4} \right) }^2} +{{\left( {d_1}-{d_6} \right) }^2}}}}}.
\end{eqnarray*}
Le choix des signes est dict\'e par le signe de l'expression dans
la valeur absolue.

\subsection*{Calcul de $d(1,3)$ quand $\frac{1}{d_2}=\frac{1}{d_5}=\infty$}
\begin{eqnarray*}
x_0 &=&  d_1d_6\sqrt{\frac{d_3^2 + d_6^2}{d_1d_6-d_3d_4}}, \; z_0 =\frac{d_3^2}{\sqrt{d_3^2 +d_6^2}}, \\
x_1 &=& \frac{d_1^2}{\sqrt{d_1^2 +d_4^2}}                 , \; z_1 ={\frac{d_3d_4\sqrt{d_1^2 +d_4^2}}{d_3d_4 - d_1d_6}},\\
x_2 &=& \text{sign} (d_1d_3+d_4 d_6){\frac{d_1\,( d_3 + d_4 ) }{{\sqrt{{{( d_3 + d_4 ) }^2} + {{( d_1 - d_6 ) }^2}}}}},\\
z_{2}&=&{\frac{d_3(d_3{\sqrt{{{(d_1d_3+d_4d_6)}^2}}}\pm
d_6(d_3(d_3+d_4)-d_1d_6+{{d_6}^2}))}
{{\sqrt{{{(d_3+d_4)}^2}+{{(d_1-d_6)}^2}}}({{d_3}^2}+{{d_6}^2})}},\\
x_3 &=& \text{sign}(d_1d_3+d_4d_6){\frac{{d_1}({d_3}-{d_4})}
{{\sqrt{{{({d_3}-{d_4})}^2}+{{({d_1}+{d_6})}^2}}}}},\\
z_{3} &=&{\frac{d_3(d_4{\sqrt{{{(d_1d_3+d_4d_6)}^2}}}\pm
d_6(d_4(d_4-d_3)+d_1(d_1+d_6)))}{(d_3 d_4-d_1d_6){\sqrt{{{(
d_3-d_4)}^2}+{{(d_1+d_6)}^2}}}}}.
\end{eqnarray*}
Le choix des signes est dict\'e par le signe de l'expression dans
la valeur absolue. \vfill \pagebreak

\printindex \thispagestyle{empty}

\noindent{\bf R\'esum\'e:} Cette th\`ese \'etudie l'aspect
m\'etrique de la g\'eom\'etrie non commutative \`a travers la
formulation de Connes de la distance entre \'etats d'une
alg\`ebre.

La d\'efinition d'un espace non commutatif est l'objet du premier
chapitre. Des propri\'et\'es g\'en\'erales de la formule de la
distance sont mises en \'evidence ainsi que d'importantes
simplifications quand l'alg\`ebre est de von Neumann.

Dans le deuxi\`eme chapitre, les distances sont calcul\'ees pour
des alg\`ebres de dimension finie. Les cas $\cc^n$ et $M_n(\cc)$
sont envisag\'es.

Dans la  troisi\`eme chapitre, on \'etudie la distance pour  des
g\'eom\'etries obtenues par produit de l'espace -temps riemannien
avec une g\'eom\'etrie discr\`ete. Des conditions sont \'etablies
garantissant que l'espace discret soit orthogonal, au sens du
th\'eor\`eme de Pythagore, \`a l'espace continu. On obtient ainsi
une description compl\`ete de la m\'etrique pour un exemple de
base de la g\'eom\'etrie non commutative, le mod\`ele \`a deux
couches. On montre \'egalement en toute g\'en\'eralit\'e que la
m\'etrique d'une g\'eom\'etrie n'est pas perturb\'ee quand on
r\'ealise son produit avec une autre g\'eom\'etrie.

Le dernier chapitre \'etudie l'\'evolution de la m\'etrique
lorsque la g\'eom\'etrie est perturb\'ee par des champs de
jauges. En se limitant \`a la partie scalaire de ces champs, on
calcule les distances dans la  g\'eom\'etrie du mod\`ele
standard. Il apparait que le champ de Higgs est le coefficient
d'une m\'etrique riemannienne dans un espace de dimension 4
(continues) + 1 (discr\`ete).
\newline

\noindent{\bf Summary:}
 The aim of this thesis is the metric aspect of noncommutative
geometry as defined by Connes.

The first chapter exposes the definition of a noncommutative
space as well as general properties of the distance formula
between states of an algebra. Simplifications occur when dealing
with von Neumann algebras.

Some distances for finite dimensional algebras are explicitly
computed in the second chapter, including $\cc^n$ and $M_n(\cc)$.

Third part studies the distance for product of geometries.
Conditions are found making the internal discrete space
orthogonal to the continuous riemannian spacetime. This gives a
complete description of the metric of a basic noncommutative
example: the two sheet model.

In the last chapter, one studies the fluctuations of the metric
when the geometry is gauged. Limiting ourselves to the scalar
part of the gauge field, one compute the distances in the
geometry of the standard model and find that the Higgs field is
the component of a metric in a fifth discrete dimension.
\newline

\vspace{0.5truecm}

\noindent{\bf Mots cl\'es:} g\'eom\'etrie non commutative,
\'etats, distance, th\'eorie de jauge, champ de Higgs.
\newline

 \vspace{0.5truecm}

\noindent{\bf Laboratoire:} Centre de physique th\'eorique, CNRS
Luminy case 907, 13288 Marseille cedex 9.

\end{document}